\documentclass[sigconf]{acmart}

\usepackage{makecell}
\usepackage{bm}
\usepackage{enumitem}
\setlist{nolistsep}
\usepackage{multirow}
\usepackage{float}
\usepackage{tabularx}
\usepackage{pifont}
\usepackage{booktabs}
\usepackage{caption}
\usepackage{colortbl}
\usepackage[subrefformat=parens,skip=0pt]{subcaption}
\usepackage{overpic}
\usepackage{marginnote}
\usepackage[export]{adjustbox}
\usepackage[detect-all]{siunitx}
\usepackage{xspace}

\newcommand\ie{i.\,e.\xspace}
\newcommand\eg{e.\,g.\xspace}

\newcommand\US{U.S.\xspace}

\newcommand{\var}[1]{\mathit{#1}}

\def\sym#1{\ifmmode^{#1}\else\(^{#1}\)\fi}

\AtBeginDocument{%
  }

\copyrightyear{2025}
\acmYear{2025}
\setcopyright{cc}
\setcctype{by}
\acmConference[CHI '25]{CHI Conference on Human Factors in Computing Systems}{April 26-May 1, 2025}{Yokohama, Japan}
\acmBooktitle{CHI Conference on Human Factors in Computing Systems (CHI '25), April 26-May 1, 2025, Yokohama, Japan}
\acmDOI{10.1145/3706598.3713909}
\acmISBN{979-8-4007-1394-1/25/04}

\begin{document}

\title{Community Fact-Checks Trigger Moral Outrage in Replies to Misleading Posts on Social Media}

\author{Yuwei Chuai}
\affiliation{
  \institution{University of Luxembourg}
  \country{Luxembourg}}
\email{yuwei.chuai@uni.lu}

\author{Anastasia Sergeeva}
\affiliation{
  \institution{University of Luxembourg}
  \country{Luxembourg}}
\email{anastasia.sergeeva@uni.lu}

\author{Gabriele Lenzini}
\affiliation{
  \institution{University of Luxembourg}
  \country{Luxembourg}}
\email{gabriele.lenzini@uni.lu}

\author{Nicolas Pr{\"o}llochs}
\affiliation{
 \institution{JLU Giessen}
 \country{Germany}}
\email{nicolas.proellochs@wi.jlug.de}


\begin{abstract}
Displaying community fact-checks is a promising approach to reduce engagement with misinformation on social media. However, how users respond to misleading content \emph{emotionally} after community fact-checks are displayed on posts is unclear. Here, we employ quasi-experimental methods to \emph{causally} analyze changes in sentiments and (moral) emotions in replies to misleading posts following the display of community fact-checks. Our evaluation is based on a large-scale panel dataset comprising $\var{N}=\num{2225260}$ replies across \num{1841} source posts from X's Community Notes platform. We find that informing users about falsehoods through community fact-checks significantly increases negativity (by 7.3\%), anger (by 13.2\%), disgust (by 4.7\%), and moral outrage (by 16.0\%) in the corresponding replies. These results indicate that users perceive spreading misinformation as a violation of social norms and that those who spread misinformation should expect negative reactions once their content is debunked. We derive important implications for the design of community-based fact-checking systems.
\end{abstract}





\maketitle

\section{Introduction}

% Problem statement
Social media platforms like X (formerly Twitter), Facebook, and TikTok have become increasingly influential sources for news consumption and social engagement, especially among younger generations \cite{van2024social,deloitte2021are}. However, the fast-paced and interactive nature of social media has also compromised the quality of the information shared, as content can be posted rapidly without undergoing the thorough fact-checking essential to journalistic ethics \cite{spj2014code}. As a result, misinformation spreads easily on social media platforms and has become a concerning issue in the digital age \cite{lazer2018science,vosoughi2018spread,wef2024global}. The negative effects of user engagement with online misinformation have been repeatedly observed across various domains \cite{ecker2022psychological,allcott2017social,enders2022relationship,green2022online,pierri2022online,ecker2024misinformation}. For instance, political misinformation is especially troubling, as engagement with false information during elections can undermine democracy and destabilize societies \cite{ecker2024misinformation,aral2019protecting,vosoughi2018spread,green2022online}. Given the significant risks and threats posed by online misinformation, social media providers are thus urged to establish content moderation policies and implement effective interventions to identify misinformation and inform users on their platforms \cite{donovan2020social,lazer2018science,drolsbach2024content,bar2023new}.

% Community notes are an ideal source to examine how users respond to misleading posts after being flagged
As a countermeasure, the social media platform X/Twitter has recently adopted a crowdsourced (community-based) fact-checking approach, ``Community Notes,'' to fact-check misleading posts and communicate veracity to users on its platform \cite{prollochs2022community}. ``Community Notes'' represents the first attempt by a major social media platform to apply a crowdsourced approach to fact-checking on a large scale. It allows users to add context or correct information to posts via \emph{community notes}, providing additional perspectives or clarifications. Once community notes are rated helpful by other community contributors, they will be directly displayed on the corresponding misleading posts. While the accuracy of crowdsourced fact-checking approaches has been thoroughly validated by previous research \cite{bhuiyan2020investigating,epstein2020will,pennycook2019fighting,saeed2022crowdsourced,drolsbach2023diffusion}, a causal understanding of how users react to social media content that has been subject to community-based fact-checking is still in its infancy.

%% Research Gap
Previous research studying the effect of community fact-checks on social media users has primarily focused on analyzing how it affects \emph{engagement} with misinformation (\eg, shares, comments, replies, or likes) \cite{chuai2024roll,drolsbach2024community,drolsbach2023diffusion,chuai2024community.new,renault2024collaboratively}. For instance, studies have shown that community notes can reduce users' belief in false content and their intentions to share misleading posts \cite{chuai2024community,chuai2024community.new,renault2024collaboratively,drolsbach2023believability}. However, little is known about how social media users \emph{emotionally} react to social media posts that have been subject to fact-checking. Emotions are highly influential for human decision-making and judgment \cite{lerner2015emotion}. It is thus conceivable that fact-checking social media posts triggers emotional backlash towards the authors of such content.\footnote{Note that an alternative rationale posits that some users (\eg, those with strong partisan alignments) may respond emotionally to a post from their ``side'' being labeled as misleading, rather than reacting to the misinformation itself. However, previous research suggests that replies are typically aimed at the content or the post authors \cite{zade2024reply}; and users are more likely to reply to authors from ideologically opposing groups \cite{gaisbauer2021ideological}. In line with this notion, our later analysis and manual validation imply that community notes primarily elicit replies focused on the content of misleading posts or their authors (see details in Section \ref{sec:moral_outrage_analysis}).} For example, users may feel a sense of urgency to correct the record or call out the spreaders of misinformation. In this scenario, the debunking of misinformation via community notes might provoke strong emotional reactions (\eg, negative sentiment, anger, disgust) from social media users. Note that such emotional backlash from the user base may be seen as a double-edged sword. On the one hand, it can be a desirable reaction in the fight against misinformation as it may signal to the content creator and other users that spreading misinformation is unacceptable and subject to public scrutiny. Also, collective emotional backlash may help pressure authors to correct or delete misleading content and discourage others from sharing or engaging with it. On the other hand, however, social media platforms also need to maintain a healthy and constructive dialogue. In this regard, the emotional reactions from the user base pose risks of fueling political polarization, escalating conflicts, and entrenching divisions. Yet, empirical evidence on how fact-checking emotionally resonates with the audience on social media is missing.

% Spreading online misinformation => social norm and moral violations
Furthermore, fact-checking social media posts may trigger pronounced moral reactions from the user base. Content moderation of online speech is a moral minefield, especially when two key values come into conflict: upholding freedom of expression and preventing harm caused by the violation of social norms and spreading misinformation \cite{kozyreva2023resolving}. From a theoretical perspective, spreading online misinformation may be perceived as immoral behavior, motivating users' moral reasoning and judgment and triggering their moral emotions \cite{neumann2024morality}. Previous work has identified the phenomenon of \emph{moral outrage}, which occurs when individuals perceive the actions of another person or an entire situation as a serious violation of their moral principles or ethical standards. Moral outrage is commonly associated with a higher likelihood of inter-group hostility and can contribute to the further escalation of conflicts on the platforms \cite{crockett2017moral,neumann2024morality,van2024social}. Emotionally, this often results in a combination of anger and disgust, prompting individuals to express these emotions in an attempt to restore what they perceive as the proper moral norm \cite{salerno2013interactive}. This response typically involves blaming the violator and demanding corrective actions \cite{van2024social,neumann2024morality}. Based on this rationale, we hypothesize that social media users might react with expressions of moral outrage to misleading posts and their authors once they are informed about the falsehoods via community notes.

\textbf{Research questions:} In this study, we \emph{causally} analyze whether the display of community notes changes expressions of sentiments and (moral) emotions in social media users' replies to misleading posts. Specifically, we address two main research questions. Our first research question is to examine the effects of community notes on sentiments and basic emotions in replies (RQ 1.1). Previous research further suggests that emotions may play a particularly pronounced role in the context of political (mis-)information \cite{brady2017emotion,rathje2021out}. We thus additionally analyze how the emotional reactions differ across misleading posts that cover a political vs. a non-political topic (RQ1.2):
\begin{itemize}[leftmargin=*]
    \item \textbf{RQ1.1:} Which sentiments (positivity, negativity) and basic emotions (\eg, anger, disgust) in replies to misleading posts are triggered by the display of community notes?
    \item \textbf{RQ1.2:} Do the effects of community notes on sentiments and basic emotions in replies differ across political and non-political misleading posts?
\end{itemize}
Furthermore, when anger and disgust are mixed, they generate ``moral outrage,'' a more intense emotional reaction to perceived moral transgressions than either anger or disgust alone \cite{salerno2013interactive,crockett2017moral,brady2021social}. Our second research question is to examine whether community notes trigger moral outrage in replies (RQ2.1), and analyze the sensitivity across political and non-political misleading posts (RQ2.2):
\begin{itemize}[leftmargin=*]
    \item \textbf{RQ2.1:} Does the display of community notes trigger moral outrage in replies to misleading posts?
    \item \textbf{RQ2.2:} Do the effects of community notes on moral outrage in replies differ across political and non-political misleading posts?
\end{itemize}

\textbf{Data \& methods:} To address our research questions, we collect a dataset of $\var{N}=$ \num{2225260} replies to \num{1841} misleading posts that have been fact-checked via community notes over a 4-month period since the roll-out of ``Community Notes'' feature on X/Twitter. The dataset includes all the replies that are directly toward misleading posts before and after the display of community notes. Subsequently, we employ methods from Natural Language Processing (NLP) in combination with Regression Discontinuity Design (RDD) to estimate the causal effects of community notes on the sentiments and (moral) emotions in the replies to the fact-checked posts. 

\textbf{Contributions:} To the best of our knowledge, this study is the first to causally analyze the emotional reactions of social media users in response to (community) fact-checks. We find that informing users about falsehoods through the display of community fact-checks significantly increases the expression of negativity in the corresponding replies by 7.3\%. Furthermore, we provide a fine-granular analysis of emotions. Here, the largest effect sizes are observed for (negative) emotions from the moral emotions family. Specifically, displaying community notes increases anger in the corresponding replies by 13.2\%, disgust by 4.7\%, and moral outrage by 16.0\%. The observed effects are significantly more pronounced for political vs. non-political posts. Overall, these results imply that social media users perceive misinformation as moral transgressions and that authors of misinformation should expect negative reactions once their content is debunked. Based on our findings, we derive important implications for behavioral theory, media literacy, and the design of community-based fact-checking systems on social media.

\section{Background and Related Work}
\subsection{Misinformation on Social Media}
% News consumption: social media as a main source
Social media platforms, as a source for news consumption and social engagement, have become increasingly popular. As of July 2024, there is a large population base of $\sim$5.17 billion users on social media, and the number continues to grow \cite{van2024social,statista2024number}. Especially among younger users aged 18 to 24, up to 65\% prefer social media as one of their preferred news sources \cite{deloitte2021are}. Together, these data indicate a trend toward partially substituting traditional media with social media channels that provide fast access to information \cite{brannon2024speed} and offer a more interactive way to engage with content \cite{sang2020signalling}.

% Misinformation problem: easy to spread and the viewpoints of the prevalence
% With the increasing use of social media for news consumption and social engagement, 
Due to the lack of editorial oversight, social media platforms have also become fertile soil for spreading misinformation. Recently, there have been some critics arguing that the prevalence of misinformation on social media is overestimated, and the exposure is concentrated among narrow fringe groups  \cite{altay2023misinformation,budak2024misunderstanding}. However, given the large population base on social media platforms, its negative effects can be profound. For example, the spread of online health misinformation has posed severe risks to public health in both the Global North and the Global South \cite{pierri2022online,enders2022relationship,varanasi2022accost,karusala2022towards}. Additionally, the rise of AI makes generating misinformation faster, which poses new challenges and further undermines the credibility of social media \cite{zhou2023synthetic,gamage2022deepfakes,zhang2023what,feuerriegel2023research}.

\textbf{The spread of political misinformation on social media:}
The spread of political misinformation on social media has become a tool to pursue political agenda, and its negative effects are continually observed in threatening the stability of societies and eroding trust in democracy  \cite{allcott2017social,ecker2022psychological,ecker2024misinformation}. For example, conspiracy theories claiming that the 2020 \US presidential election was ``stolen'' were widely shared on social media. It is found that users who were exposed to at least one misinformation website are 17.3\% more likely to believe that the election was fraudulent \cite{dahlke2024effect}. As surveyed, almost 40\% of the \US public and, especially, almost 70\% of Republicans negated the legitimacy of the 2020 presidential election outcome \cite{ssrs2023cnn}. The reduced faith in electoral institutions have had real-world consequences in affecting voters' behaviors in subsequent elections and hate campaigns \cite{green2022online,ecker2024misinformation}. Similar patterns have been observed in countries of the Global South, such as in Brazil \cite{recuero2020hyperpartisanship,haque2020combating}. Moreover, political misinformation is often interconnected with the politicization of other social issues, such as health crises and climate change, further contributing to polarized attitudes toward these topics \cite{hart2020politicization,chuai2025political,chuai2024news}.

Altogether, with the increasing use of social media for news consumption and the continuously observed negative effects of misinformation across various domains, the spread of misinformation on social media has become a significant societal issue in the digital age \cite{vosoughi2018spread,van2024social}. Particularly, political misinformation has been found to spread significantly faster than non-political misinformation on social media \cite{vosoughi2018spread}. Previous studies also show that social media users share information to single their political affiliation and moral stance, even knowing that it is inaccurate \cite{osmundsen2021partisan,mcloughlin2024misinformation}. To protect users from being affected by misinformation and prevent harm, it is both warranted and required to implement effective countermeasures to fight against its spread \cite{aral2019protecting,ecker2024misinformation}.

\subsection{Content Moderation for Misinformation}
Social media providers are increasingly urged to establish content moderation policies to regulate online behavior and combat the spread of misinformation on their platforms \cite{donovan2020social,lazer2018science}. The content moderation process includes two main aspects: fact-checking misinformation and handling identified misinformation. 

% Approaches for identifying misinformation
To identify misinformation, social media platforms are widely collaborating with professional third-party fact-checkers (\eg, the International Fact-Checking Network) \cite{meta2024metas}. While fact-checking experts can accurately access veracity \cite{micallef2022true,lee2023fact}, the implementation of this approach faces two main challenges. First, professional fact-checking is a lengthy and resource-consuming process, leading to a situation where only a small proportion of posts are actually checked \cite{micallef2022true,wilner2023attending,mcclure2022bridging,pilarski2024community}. Second, professional fact-checkers face suspicions of (political) bias, \eg, by selectively fact-checking content \cite{chuai2025political,tang2024knows}. Therefore, the trust of the public in professional fact-checking is discounted \cite{chuai2024community,drolsbach2024community,flamini2019most}. To overcome these challenges, researchers have proposed to fact-check and identify misleading posts on social media platforms at scale based on the non-expert fact-checkers from the crowd \cite{allen2021scaling,micallef2020role,pennycook2019fighting}. While the individual assessments can have bias and noise \cite{allen2022birds}, it has been shown that the accuracy of aggregated judgments, even from relatively small crowds, is reliable and comparable to the accuracy of expert fact-checkers \cite{bhuiyan2020investigating,epstein2020will,pennycook2019fighting,saeed2022crowdsourced}. Additionally, the crowdsourced assessments are perceived as more trustworthy by the users than fact-checks from experts \cite{allen2021scaling,zhang2024profiling,drolsbach2024community}.

% Content moderation options and moral dilemma
Subsequently, social media platforms generally have three options for handling identified misleading posts: removing them, reducing their visibility, or informing users about the falsehoods \cite{meta2024metas,juneja2023assessing,wojcik2022birdwatch}. The first two options -- removal and visibility reduction -- are forms of ``hard moderation,'' where the platforms limit or entirely cut off the spread of misinformation. In contrast, informing other users that a post is misleading represents ``soft moderation,'' where the platforms refrain from imposing direct controls and instead allow users to make their own decisions on how they process the informed misleading post. The hard moderation approaches face a dilemma, where the two key values of protecting freedom of expression and preventing harm from the spread of misinformation come into conflict \cite{kozyreva2023resolving,schaffner2024community}. Given this, soft moderation approaches can be regarded as a trade-off between freedom of expression and misinformation intervention. Previous lab/survey experiments consistently show that communicating the falsehood to users (especially with explanations) is effective in reducing users' beliefs and sharing intentions \cite{clayton2020real,yaqub2020effects,porter2021global,zhang2024profiling,heuer2022comparative,jahanbakhsh2024browser,allen2021scaling,hassoun2023practicing}.

\subsection{Sentiments, Basic Emotions, and Moral Outrage on Social Media}

% Sentiments (valence): definition and effects on information sharing
As an essential element of human communication, sentiments and emotions play a pivotal role in shaping interactions on social media platforms \cite{lerner2015emotion,stieglitz2013emotions,berger2012makes,jakubik2023online}. They can be understood as two affective dimensions: valence (sentiment) and arousal (emotion). Sentiment, \ie, the valence of content (positive vs. negative), is an attitude, thought, or judgment promoted by a feeling \cite{yadollahi2017current}. Research has shown that negativity is a particularly significant characteristic of online engagement \cite{zollo2015emotional,del2016echo,robertson2023negativity}. Negative sentiment tends to evoke stronger reactions, promoting users to comment and share online posts, especially among polarized communities \cite{zollo2015emotional,del2016echo}.

% Basic emotions (affect ladenness)
The relationship between sentiments and social interaction is more nuanced than simply the valence (positive vs. negative) of the content alone. From the arousal dimension, specific emotions (\ie, affect ladenness) are embedded within the content, adding intensity to the positive or negative feelings expressed \cite{pang2008opinion,yadollahi2017current}. With various models classifying basic emotions, Ekman's six basic emotion model is widely used and commonly shared across other frameworks \cite{yadollahi2017current}. Specifically, Ekman's model identifies six universal emotions: anger, disgust, fear, joy, sadness, and surprise \cite{ekman1992argument,sauter2010cross,ekman2013emotion}. These emotions are considered fundamental across different cultures and can combine to form more complex emotional experiences \cite{sauter2010cross}. Emotions also vary on the level of physiological arousal or activation they evoke \cite{berger2012makes}. For example, anger and sadness are often expressed as negative emotions. However, anger is a state of heightened arousal and makes online content more viral, while sadness is low arousal and makes online content less viral \cite{berger2012makes,chuai2022really}. Particularly, users' reliance on emotions can promote their belief in misinformation \cite{martel2020reliance}. Several studies have found that emotions can explain the spread of true vs. false information on social media, with anger being the most significant emotion in driving the spread of misinformation \cite{prollochs2021emotions,chuai2022anger}.

% Moral emotions
Additionally, \emph{moral emotions} can arise in response to perceived moral violations or standards and motivate moral behavior \cite{haidt2003moral}. One large family of moral emotions is other-condemning emotions that often lead to behavioral outcomes, that is, actions to correct the other's behavior and increase justice-relevant judgment \cite{haidt2003moral,van2015testing}. Specifically, anger and disgust are the main other-condemning emotions used in coordinating and constraining the behavior of the group member who made the morally unacceptable actions \cite{haidt2003moral}. During the COVID-19 pandemic, other-condemning emotions are significantly associated with the higher virality of misinformation on social media \cite{solovev2022moral}. 
%% Out-group animosity -- politics and polarization
Within political contexts, social media posts containing moral-emotional language have been shown to be 20\% more likely to be shared, particularly within polarized partisan networks (liberal or conservative), with stronger effects within these groups than between them \cite{brady2017emotion}. This highlights the role of moral emotions in amplifying online engagement, particularly through expressions of out-group animosity. Such expressions, especially within partisan networks, have been shown to drive higher engagement than in-group favoritism, further entrenching political polarization \cite{rathje2021out}.

% Moral outrage -- mixture of anger and disgust
Previous studies highlight \emph{moral outrage} -- a mixture of anger and disgust triggered by a perceived moral norm violation -- as a powerful emotion that motivates people to shame and punish wrongdoers, especially in the digital age \cite{salerno2013interactive,crockett2017moral,brady2023overperception,solovev2023moralized}. Social media platforms, such as X/Twitter, inherently amplify expressions of moral outrage over time, as users who learn such language get rewarded with an increased number of likes and reposts \cite{brady2021social}. Moral outrage is also contagious. Knowing that a person or entity is being publicly condemned can provoke moral outrage in others \cite{shah2020perceived}. This contagious has the potential to ignite online firestorms -- large waves of insults, criticism, or swearing directed at persons and organizations on social media \cite{rost2016digital,strathern2020against}. In addition, users frequently overestimate the level of outrage and hostility in social media posts, perceiving emotional expressions as more extreme than the authors actually report. This overperception can lead to escalated responses and increase the spread of moral outrage \cite{brady2023overperception}. The consequences of heightened moral outrage on social media are double-edged. On the one hand, it can fuel inter-group conflict and polarization, exacerbating online tensions and divisions. On the other hand, it can also amplify constructive facets of morality, such as social support and prosociality \cite{van2024social,crockett2017moral}. This dual role underscores the importance of understanding how moral outrage operates in specific circumstances to mitigate its negative impacts while promoting its potential positive effects.

% Summary -- sentiments and emotions are influential in shaping online engagement with misinformation. However, how community fact-checks causally resonate with users emotionally toward online misinformation remains unclear. Which states the aim of this study.
Overall, the analysis of sentiments and (moral) emotions is a crucial and growing area of research for understanding online engagement and its consequences \cite{feuerriegel2025using}. A large research body has documented that sentiments, (moral) emotions, and moral outrage play a pronounced role in shaping engagement with (mis-)information on social media. However, an understanding of how users emotionally react to social media posts after they are informed of the falsehoods is missing, which presents the aim of our study.

\section{Data and Methods}

\subsection{Datasets}
\begin{figure*}
\centering
\begin{subfigure}[t]{0.425\textwidth}
    \caption{}
    \fbox{\includegraphics[width=\textwidth]{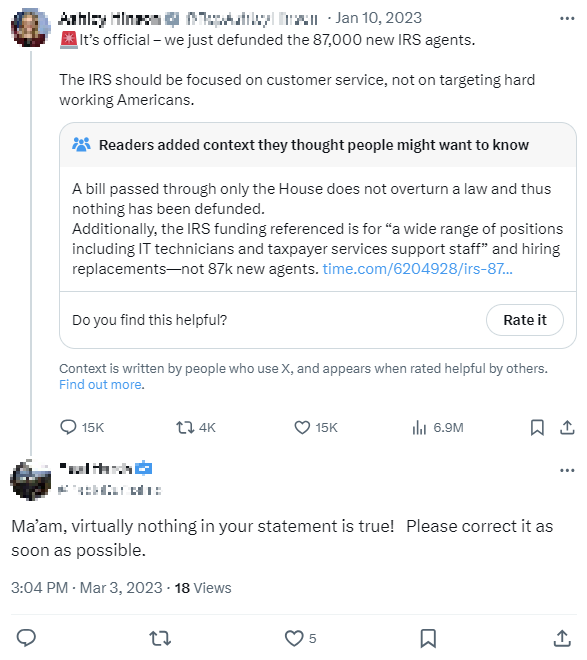}}
    \label{fig:cn_example}
\end{subfigure}
\hspace{2em}
\begin{subfigure}[t]{0.5\textwidth}
    \caption{}
    \includegraphics[width=\textwidth]{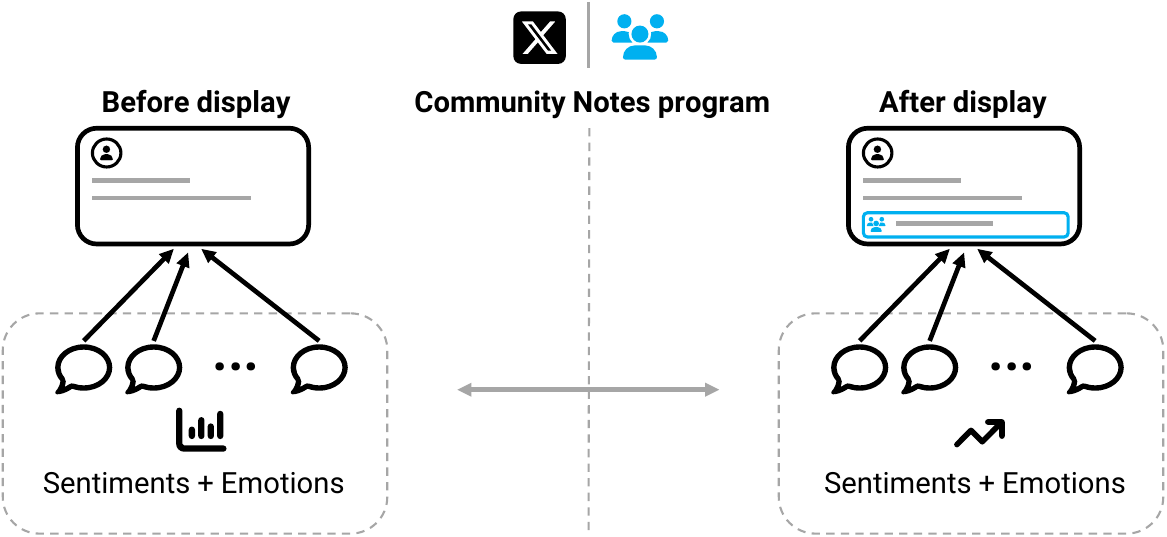}
    \label{fig:before_after_display}
\end{subfigure}
\caption{Research overview. \subref{fig:cn_example} An example of a misleading post with a displayed community note and one direct reply after note display. \subref{fig:before_after_display} Illustration of our research setup. Before note display, users who reply to the source posts can only see the original post content. After note display, users who reply to the source posts can see both original post content and displayed community notes flagging the post content as misleading. Using the sentiments and emotions in replies before note display as a baseline, we can examine the changes in sentiments and emotions in replies after the display of community notes.}
\label{fig:cn_illustration}
\end{figure*}

In this study, we analyze the emotional reactions of social media users in response to (community) fact-checks. For this purpose, we collect a large dataset of posts that have been fact-checked on X's (formerly Twitter) ``Community Notes'' platform (formerly Birdwatch). ``Community Notes''  was officially rolled out to users on December 11, 2022, and is a crowdsourced fact-checking program that allows users to identify posts that they believe are misleading and write short textual notes that provide context to the posts. After a community note is created, other users can rate its helpfulness and, if it reaches a certain level of helpfulness, it is displayed prominently beneath the original post. An example of a misleading post with a displayed community note and one direct reply is shown in Fig. \ref{fig:cn_example}. Importantly, before note display, users who reply to the source posts can only see the original post content (as shown in Fig. \ref{fig:before_after_display}). After note display, users who reply to the source posts can see both original post content and displayed community notes flagging the post content as being misleading. 

As detailed in the following, we collect data from three different sources: (i) note dataset, (ii) post dataset, and (iii) reply dataset. Subsequently, we combine these three data sources into a longitudinal dataset that allows us to analyze the changes in reply emotions before and after the display of community notes.

\subsubsection{Note dataset.} 
X publishes all community notes and their associated status histories on a dedicated website and updates them daily.\footnote{All community notes are publicly available at \url{https://x.com/i/communitynotes/download-data}.} We downloaded the community note datasets on April 8, 2023, which include all note contributions up until April 6, 2023. Our note dataset contains the details for each note, including its unique ID, creation time, classification (misleading or not misleading), referenced source post ID, and the note status history (\ie, when the notes were rated helpful and displayed on the source posts). In our data, there are \num{68235} community notes for \num{45664} source posts. However, only \num{4686} source posts (10.3\%) received displayed community notes until the date of data collection.

\subsubsection{Post dataset.} 
We use X's Academic Research API to search for the \num{4686} source posts with displayed community notes and successfully collect \num{3872} (82.6\%) posts. Notably, given that the ``Community Notes'' program was rolled out on December 11, 2022, and mainly developed in the \US, we only consider the source posts that were created after the roll-out and are written in English. There are \num{1953} source posts remaining after this filtering step. 

\subsubsection{Reply dataset.} 
We retrieve the entire conversation thread for each source post via X's Academic Research API.\footnote{See the documentation for X's API at \url{https://developer.x.com/en/docs/twitter-api/conversation-id}.} Each conversation thread contains all the replies to the original posts, and they share one unique conversation ID. In this study, we analyze the changes of sentiments and emotions in replies to the original misleading posts after the display of community notes. Therefore, we only consider direct replies to the source posts and omit indirect replies (\ie, replies to replies). We successfully pull $N=$ \num{2225260} direct replies to \num{1841} misleading source posts (94.3\%) that were fact-checked via community notes during a period of 4 months\footnote{The collection of our dataset was completed over two months, from April to June 2023. Subsequently, X deprecated the Academic Research API, restricting our ability to access additional data for analysis. Notwithstanding, previous research implies that our dataset covering the initial four-month period provides a representative data basis for the Community Notes program \cite{chuai2024community,chuai2024community.new,renault2024collaboratively}.} since its roll-out (Table \ref{tab:data_summary}). The remaining 5.7\% (\num{112}) misleading source posts have no (English) replies and are thus omitted.

\begin{table*}
\centering
\caption{Dataset overview. Reported are mean values or count numbers for the variables (standard deviations in parentheses).}
\begin{tabular}{l*{3}{c}}
\toprule
&{(1)} & {(2)}& {(3)}\\
&{All}&{Political}&{Non-political}\\
\midrule
\#Replies&{2,225,260}&{1,068,732}&{1,156,528}\\
\#Source posts&{1841}&{477}&{1364}\\
Date range of replies&{2022-12-11 -- 2023-05-28}&{2022-12-11 -- 2023-05-27}&{2022-12-11 -- 2023-05-28}\\
Date range of source posts&{2022-12-11 -- 2023-04-05}&{2022-12-11 -- 2023-04-05}&{2022-12-11 -- 2023-04-05}\\
\underline{Sentiments in replies}\\
\quad Positive&{0.112 (0.222)}&{0.098 (0.209)}&{0.124 (0.233)}\\
\quad Negative&{0.565 (0.336)}&{0.608 (0.324)}&{0.525 (0.342)}\\
\underline{Sentiments in source posts}\\
\quad Source positive&{0.114 (0.222)}&{0.084 (0.184)}&{0.125 (0.233)}\\
\quad Source negative&{0.443 (0.334)}&{0.513 (0.315)}&{0.419 (0.337)}\\
\underline{Emotions in replies}\\
\quad Anger&{0.181 (0.260)}&{0.205 (0.272)}&{0.160 (0.246)}\\
\quad Disgust&{0.097 (0.183)}&{0.098 (0.181)}&{0.096 (0.185)}\\
\quad Fear&{0.050 (0.146)}&{0.047 (0.138)}&{0.054 (0.153)}\\
\quad Joy&{0.066 (0.170)}&{0.061 (0.163)}&{0.069 (0.176)}\\
\quad Sadness&{0.107 (0.192)}&{0.109 (0.193)}&{0.104 (0.192)}\\
\quad Surprise&{0.172 (0.247)}&{0.164 (0.241)}&{0.180 (0.252)}\\
\underline{Emotions in source posts}\\
\quad Source anger&{0.159 (0.243)}&{0.195 (0.260)}&{0.147 (0.236)}\\
\quad Source disgust&{0.041 (0.117)}&{0.041 (0.112)}&{0.040 (0.119)}\\
\quad Source fear&{0.173 (0.276)}&{0.151 (0.242)}&{0.181 (0.287)}\\
\quad Source joy&{0.077 (0.176)}&{0.068 (0.158)}&{0.080 (0.182)}\\
\quad Source sadness&{0.126 (0.211)}&{0.140 (0.216)}&{0.122 (0.209)}\\
\quad Source surprise&{0.175 (0.245)}&{0.174 (0.249)}&{0.175 (0.244)}\\
\bottomrule
\end{tabular}
\label{tab:data_summary}
\end{table*}

\subsection{Sentiments and Basic Emotions}
% Sentiments: positive and negative
% Emotions: anger, disgust, fear, joy, sadness, and surprise

We extract sentiment and emotion features in the source posts and their replies using two state-of-the-art machine learning models. Specifically, we compute the probabilities of positive and negative sentiments for each source post and reply using the Twitter-roBERTa-base model (2022 updated). It was trained on 124 million posts created from January 2018 to December 2021, and fine-tuned for sentiment analysis.\footnote{The sentiment model is available at \url{https://huggingface.co/cardiffnlp/twitter-roberta-base-sentiment-latest}.} This model achieves a high prediction performance compared to other sentiment models \cite{camacho2022tweetnlp,loureiro2022timelms}. Subsequently, we compute the probabilities of Ekman's six basic emotions (\ie, anger, disgust, fear, joy, sadness, and surprise) \cite{ekman1992argument,sauter2010cross} for each source post and reply using a distilled version of the RoBERTa-base model.\footnote{The emotion model is available at \url{https://huggingface.co/j-hartmann/emotion-english-distilroberta-base}.} The model was pre-trained on OpenWebTextCorpus, a reproduction of OpenAI's WebText dataset with more than 8 million documents, and fine-tuned for emotion analysis \cite{hartmann2022emotionenglish}. The performance of this emotion analysis model is comparable to human annotations and has been used in several previous studies \cite{butt2022goes,rozado2022longitudinal}. Summary statistics for the sentiments and emotions in the source posts and replies are reported in Table \ref{tab:data_summary}. 

\subsection{Political vs. Non-Political Posts}

Misleading posts in the political domain are particularly concerning due to their potential to undermine democratic processes and destabilize societies \cite{ecker2024misinformation}. As observed by previous studies \cite{vosoughi2018spread,osmundsen2021partisan,mcloughlin2024misinformation}, political misinformation spreads faster than non-political misinformation and is particularly resistant to corrections. Therefore, we additionally conduct a sensitivity analysis examining changes in sentiments and emotions in replies to misleading posts that cover a political vs. a non-political topic. To do so, we extract topic domains from the context annotations attached to each source post and classify them as political or non-political. X performs entity recognition and semantic analysis (\eg, keywords, hashtags, and handles) for all posts and assigns corresponding domain labels based on a topic domain list.\footnote{The topic domains are curated by domain experts and have undergone refinement over several years. X's internal team audits this approach quarterly through the manual review of \num{10000} posts across all the domains. The precision scores are consistently $\sim$80\%. The topic domain list is available at \url{https://developer.x.com/en/docs/twitter-api/annotations/overview}.} Within the domain list, we consider ``Politician,'' ``Political Body,'' and ``Political Race'' as political topic domains. Posts containing one of the political topic domains are classified as political; otherwise, they are classified as non-political. As a result, there are \num{477} political misleading posts in our data, accounting for 25.9\% of all misleading posts. However, nearly half of the replies are directed at political misleading posts (see Table \ref{tab:data_summary}).

\subsection{Empirical Models}
In this study, we longitudinally examine the direct replies to the source posts before and after the display of community notes. To address our research questions, we use Regression Discontinuity Design (RDD) to causally estimate the effects of the display of community notes in terms of sentiments and basic emotions. RDD is a quasi-experimental method to estimate the average treatment effect of an intervention by exploiting a discontinuity in the relationship between a running variable and an outcome variable at a specific threshold or cutoff point \cite{angrist2009mostly,chuai2024roll,imbens2008regression,cattaneo2022regression}. In the following, we introduce the dependent variables (Section \ref{sec:dependent_vars}), display indicator and running variable (Section \ref{sec:display_indicator}), and additional independent variables (Section \ref{sec:independent_vars}) in our analysis. Subsequently, we detail the specification of our empirical RDD models (Section \ref{sec:reg_models}). 

\subsubsection{Dependent variables.} 
\label{sec:dependent_vars}
We define the following dependent variables that measure expressions of sentiments, basic emotions, and moral outrage in replies to social media posts: 
\begin{itemize}[leftmargin=*]
    \item Dependent variables for positive and negative sentiments
    \begin{itemize}
        \item $\bm{\var{Positive}}$: A continuous variable from 0 to 1 that indicates the probability of positive sentiment in a reply.
        \item $\bm{\var{Negative}}$: A continuous variable from 0 to 1 that indicates the probability of negative sentiment in a reply.
    \end{itemize}
    \item Dependent variables for basic emotions
    \begin{itemize}
        \item $\bm{\var{Anger}}$: A continuous variable from 0 to 1 that indicates the probability of anger in a reply.
        \item $\bm{\var{Disgust}}$: A continuous variable from 0 to 1 that indicates the probability of disgust in a reply.
        \item $\bm{\var{Fear}}$: A continuous variable from 0 to 1 that indicates the probability of fear in a reply.
        \item $\bm{\var{Joy}}$: A continuous variable from 0 to 1 that indicates the probability of joy in a reply.
        \item $\bm{\var{Sadness}}$: A continuous variable from 0 to 1 that indicates the probability of sadness in a reply.
        \item $\bm{\var{Surprise}}$: A continuous variable from 0 to 1 that indicates the probability of surprise in a reply.
    \end{itemize}
    \item Dependent variable for moral outrage
    \begin{itemize}
        \item $\bm{\var{MoralOutrage}}$: A continuous variable from 0 to 1 that indicates the probability of moral outrage in a reply. Moral outrage represents a unique combination of anger and disgust in which the two emotions interact so that each exacerbates the other’s influence on judgments \cite{salerno2013interactive}. This interaction extends beyond the isolated effects of anger and disgust, thereby offering important theoretical insights into moral outrage \cite{salerno2013interactive,crockett2017moral,brady2021social}. Accordingly, we use the product of anger and disgust (\ie, $\var{Anger} \times \var{Disgust}$) as a measure of moral outrage to explore their interactive effect.\footnote{As a robustness check, we repeated our analysis with a dedicated classifier for digital outrage developed by previous research \cite{brady2021social}. Here, we observe quantitatively identical results (see details in Suppl. \ref{sec:sm_moral_outrage}).}
    \end{itemize}
\end{itemize}

\subsubsection{Display indicator and running variable.}
\label{sec:display_indicator}
Based on RDD, we aim to examine whether there are significant discontinued changes in sentiments and emotions after the display of community notes and estimate the effects. The key rationale behind this is that the sentiments and emotions in replies should change smoothly (or keep stable) over time if without the display of community notes. Given this, we define the display indicator of community notes (binary treatment indicator) as $\bm{\var{Displayed}}$. It is equal to 1 if a reply is created after note display; otherwise, it is equal to 0. Additionally, we examine the changes in sentiments and emotions over time before and after the display of community notes. Therefore, we define the running variable as $\bm{\var{HoursFromDisplay}}$, which is a continuous variable indicating the hour(s) from note display to reply creation.

\subsubsection{Additional independent variables.}
\label{sec:independent_vars} 
Given that connected social media users may share similar emotions \cite{goldenberg2020digital,kramer2014experimental,rosenbusch2019multilevel}, the sentiments and emotions in the source posts are likely to affect the sentiments and emotions in the replies from users who are exposed to these source posts (see correlation analysis in Suppl. \ref{sec:sm_corr_reply_post}). Thus, we incorporate source sentiments and emotions into the regression models as independent variables:
\begin{itemize}[leftmargin=*]
    \item Independent variables for source sentiments in misleading posts
    \begin{itemize}
        \item $\bm{\var{SourcePositive}}$: A continuous variable from 0 to 1 that indicates the probability of positive sentiment in a source post.
        \item $\bm{\var{SourceNegative}}$: A continuous variable from 0 to 1 that indicates the probability of negative sentiment in a source post.
    \end{itemize}
    \item Independent variables for source emotions in misleading posts
    \begin{itemize}
        \item $\bm{\var{SourceAnger}}$: A continuous variable from 0 to 1 that indicates the probability of anger in a source post.
        \item $\bm{\var{SourceDisgust}}$: A continuous variable from 0 to 1 that indicates the probability of disgust in a source post.
        \item $\bm{\var{SourceFear}}$: A continuous variable from 0 to 1 that indicates the probability of fear in a source post.
        \item $\bm{\var{SourceJoy}}$: A continuous variable from 0 to 1 that indicates the probability of joy in a source post.
        \item $\bm{\var{SourceSadness}}$: A continuous variable from 0 to 1 that indicates the probability of sadness in a source post.
        \item $\bm{\var{SourceSurprise}}$: A continuous variable from 0 to 1 that indicates the probability of surprise in a source post.
    \end{itemize} 
\end{itemize}
In addition, the times when community notes are displayed on the source post are not fixed and depend on note helpfulness scores that are recalculated every hour. Therefore, changes in sentiments and emotions after note display might be confounded by the post age from post creation to reply creation. Given this, we use $\bm{\var{PostAge}}$ to control for the possible confounding effects of post age. Descriptive statistics of sentiments and emotions in replies and source posts are summarized in Table \ref{tab:data_summary}.

\subsubsection{Regression models.}
\label{sec:reg_models}
We specify two RDD models to estimate the effects of the display of community notes on sentiments and emotions in replies, respectively. For positive and negative sentiments, the RDD model is specified as:
\begin{equation}
\begin{aligned}
    \var{Sentiment_{i}}=\, &\beta_{0} + \beta_{1}\var{Displayed_{i}} + \beta_{2}\var{HoursFromDisplay_{i}} \\
    &+ \beta_{3}\var{SourcePositive_{i}} + \beta_{4}\var{SourceNegative_{i}} \\
    &+ \beta_{5}\var{PostAge_{i}} + \epsilon_{i},
\end{aligned}
\end{equation}
where $\var{Sentiment}$ denotes the dependent variable of $\var{Positive}$ or $\var{Negative}$, $\var{\beta}_{0}$ is the intercept, and $\epsilon$ is the residual. $\var{\beta}_{1}$ to $\var{\beta}_{5}$ are coefficient estimates for the independent variables. 

For the six basic emotions, the RDD model is specified as:
\begin{equation}
\begin{aligned}
    \var{Emotion_{i}} =\, & \beta_{0} + \beta_{1}\var{Displayed_{i}} + \beta_{2}\var{HoursFromDisplay_{i}} \\
    &+ \beta_{3}\var{SourceAnger_{i}} + \beta_{4}\var{SourceDisgust_{i}} \\
    &+ \beta_{5}\var{SourceFear_{i}} + \beta_{6}\var{SourceJoy_{i}} \\
    &+ \beta_{7}\var{SourceSadness_{i}} + \beta_{8}\var{SourceSurprise_{i}} \\
    &+ \beta_{9}\var{PostAge_{i}} + \epsilon_{i},
\end{aligned}
\end{equation}
where $\var{Emotion}$ denotes the dependent variable of one of the six basic emotions: $\var{Anger}$, $\var{Disgust}$, $\var{Fear}$, $\var{Joy}$, $\var{Sadness}$, or $\var{Surprise}$. Additionally, we further take $\var{MoralOutrage}$ (\ie, $\var{Anger} \times \var{Disgust}$) as the dependent variable to explore the changes in moral outrage after the display of community notes. The independent variables for source sentiments are replaced with the variables for source emotions. 

To balance the scales of variables, we further $z$-standardize all continuous variables when fitting the RDD models. The standardization is based on:
\begin{equation}
    \var{z_{i}} = \frac{\var{x_{i}}-\mu}{\sigma},
\end{equation}
where $\var{x_{i}}$ denotes original value for a continuous variable $\var{X}$, and $\var{z_{i}}$ is the standardized value. $\mu$ and $\sigma$ are mean and standard deviation of the variable $\var{X}$, respectively. Therefore, the coefficient estimates for $\var{Displayed}$, \ie, $\var{\beta_{1}}$, in the models cannot be directly interpreted as the effects of note display. Given this, we define that $\var{Y_{before}}$ is the outcome of a dependent variable before the display of community notes and $\var{Y_{after}}$ is the outcome of a dependent variable after the display of community notes. Assuming that all the other independent variables are equal for the outcomes of $\var{Y_{before}}$ and $\var{Y_{after}}$, we have:
\begin{equation}
\begin{aligned}
&\var{Y_{after}} - \var{Y_{before}} = \var{\beta_{1}}\\
\Rightarrow \, &\frac{\var{y_{after}}-\mu}{\sigma} - \frac{\var{y_{before}}-\mu}{\sigma} = \var{\beta_{1}}\\
\Rightarrow \, &\var{y_{after}} - \var{y_{before}} = \sigma\var{\beta_{1}}.\\
\end{aligned}
\end{equation}
The value of $\sigma\var{\beta_{1}}$ can represent the estimated (predicted) effects of the display of community notes on the sentiments and emotions, with all else being equal.

\textbf{Considerations for the adoption of RDD:} 
Previous studies evaluating the effect of community notes on \emph{engagement} (\eg, the number of reposts) have typically employed Difference-in-Differences (DiD) methods by constructing a counterfactual control group that evolves in parallel with the treatment group prior to the display of community notes \cite{chuai2024community,chuai2024roll,chuai2024community.new,renault2024collaboratively}. However, constructing such control groups often omits a significant portion of posts. While our dataset includes a large volume of replies, the limited number of source posts makes it infeasible to conduct such matching at the post level. Additionally, previous studies focusing on engagement typically analyze repost counts at an hourly level to examine engagement reductions attributed to the display of community notes. Therefore, they do not require the collection of every individual repost with a precise timestamp, and can count the number of reposts as zero for intervals without repost activity. In contrast, this paper focuses on analyzing sentiments and emotions in individual replies to misleading posts before and after the display of community notes. This analysis necessitates collecting every reply along with its associated timestamp and content. As sentiments and emotions cannot be assumed to be zero in intervals without replies, gaps in time units are likely to arise. These considerations lead us to adopt RDD with a continuous running variable in time \cite{imbens2008regression,cattaneo2022regression}.

Further, research has discussed the use of time as a running variable in RDD, \ie, Regression Discontinuity in Time (RDiT), and suggested considering the time-series nature of the underlying data-generating process \cite{hausman2018regression,keele2006dynamic}. On social media platforms, the autoregressive time-series structure originates from the creation of original posts, extends to direct replies, and further encompasses indirect replies (\ie, replies to replies), with each direct reply initiating a separate reply thread. To minimize complications from autoregressive processes within reply threads, our study focuses exclusively on direct replies to the original posts and omits indirect replies. Nevertheless, direct replies to the same source post across different observational units are still likely to exhibit correlations in terms of sentiments and emotions. However, according to X's algorithmic rules \cite{x2024replies}, replies in the conversations associated with the same source post are typically not displayed in chronological order. Thus, the correlations among direct replies are likely to primarily arise from shared factors, \eg, source sentiments and emotions, in the original posts (as illustrated in Fig. \ref{fig:before_after_display}).\footnote{In our later robustness checks, we validate that the correlations of sentiments and emotions among direct replies are stable and consistent across different temporal orders. The results confirm that the correlations arise from the influence of original posts rather than from temporal dependencies between direct replies. Furthermore, we repeated our analysis with lagged-dependent covariates as an additional check. Again, the results remain robust and support our findings (see details in Suppl. \ref{sec:sm_autocorr}).} Following prior research \cite{vosoughi2018spread,hausman2018regression}, we cluster standard errors in all regression models by unique source posts to address the potential violation of independence assumption. By clustering the standard errors, we account for the potential within-cluster correlations and provide more precise and reliable standard error estimates. In addition to the cluster-robust method, the incorporation of source sentiments and emotions in the regression models explicitly controls for the correlations among replies linked to the same source posts.

\section{Results}
\subsection{Data Overview}
\label{sec:data_overview}

\begin{figure*}
\centering
\begin{subfigure}{0.32\textwidth}
\caption{}
\includegraphics[width=\textwidth]{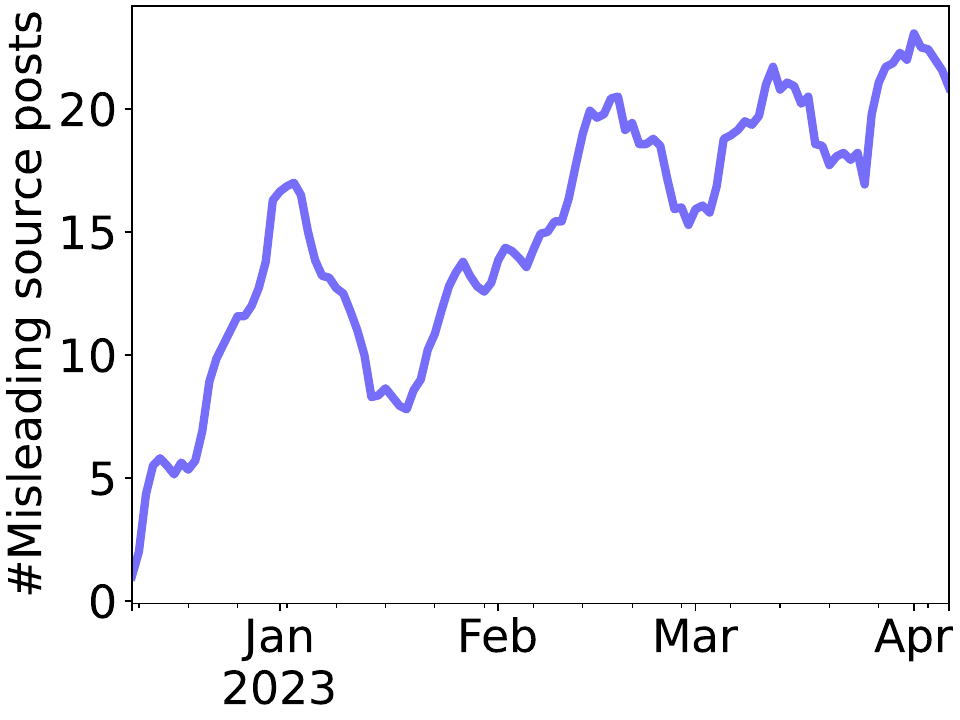}
\label{fig:post_count}
\end{subfigure}
\hfill
\begin{subfigure}{0.32\textwidth}
\caption{}
\includegraphics[width=\textwidth]{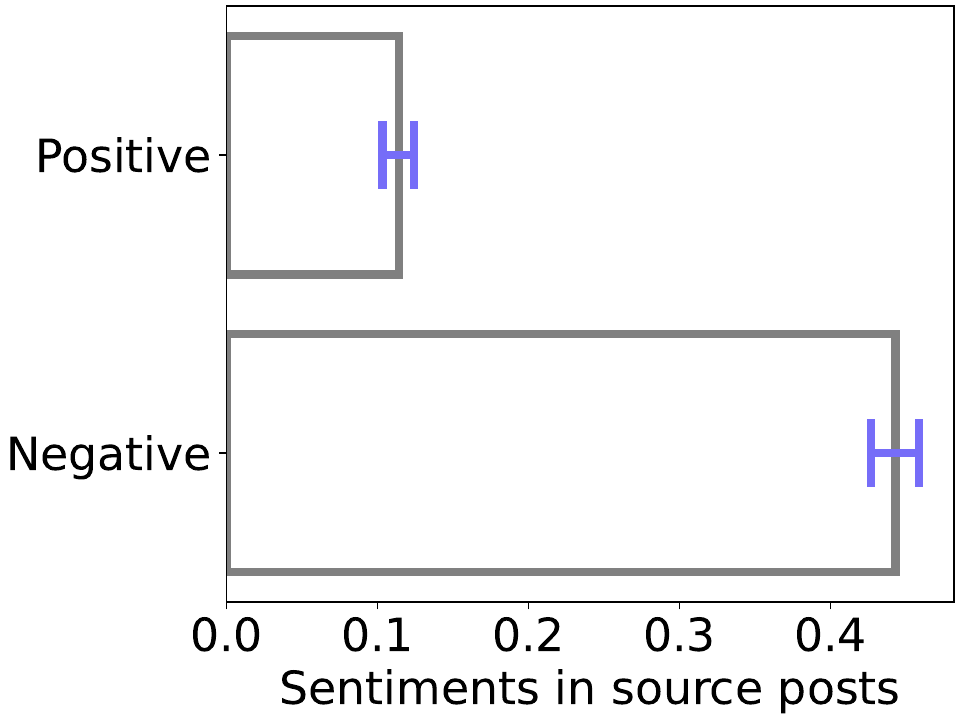}
\label{fig:source_sentiments}
\end{subfigure}
\hfill
\begin{subfigure}{0.32\textwidth}
\caption{}
\includegraphics[width=\textwidth]{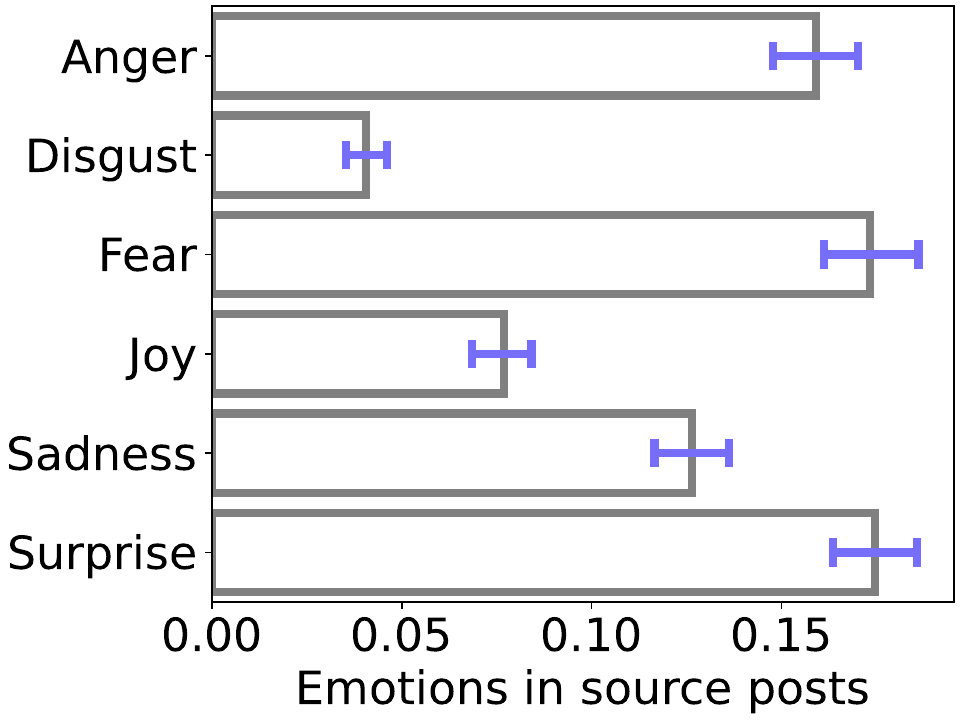}
\label{fig:source_emotions}
\end{subfigure}

\begin{subfigure}{0.32\textwidth}
\caption{}
\includegraphics[width=\textwidth]{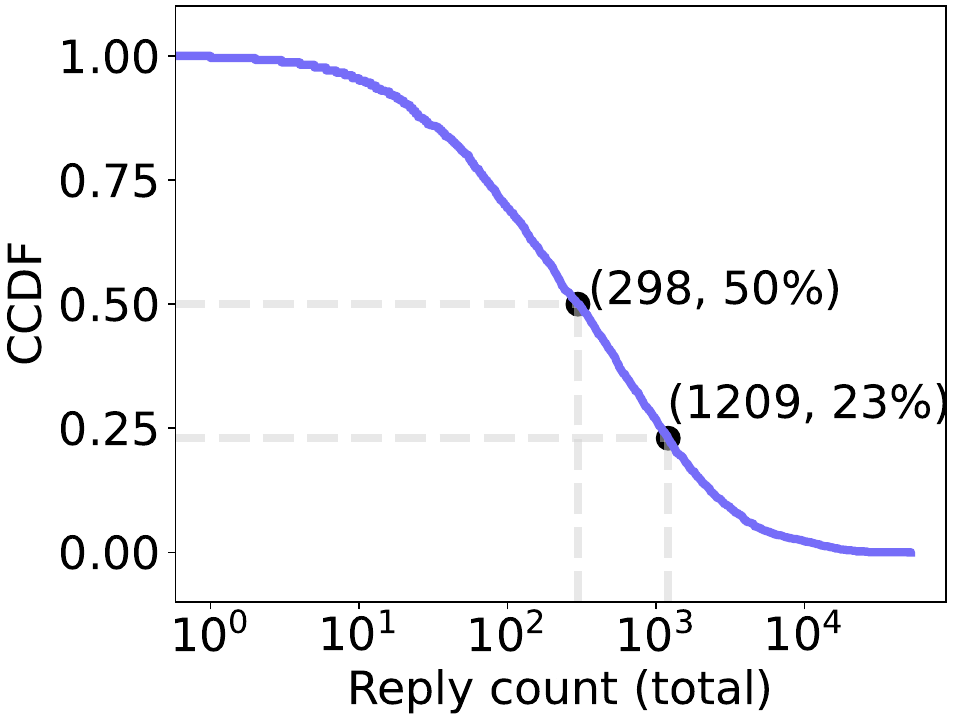}
\label{fig:reply_count}
\end{subfigure}
\hfill
\begin{subfigure}{0.32\textwidth}
\caption{}
\includegraphics[width=\textwidth]{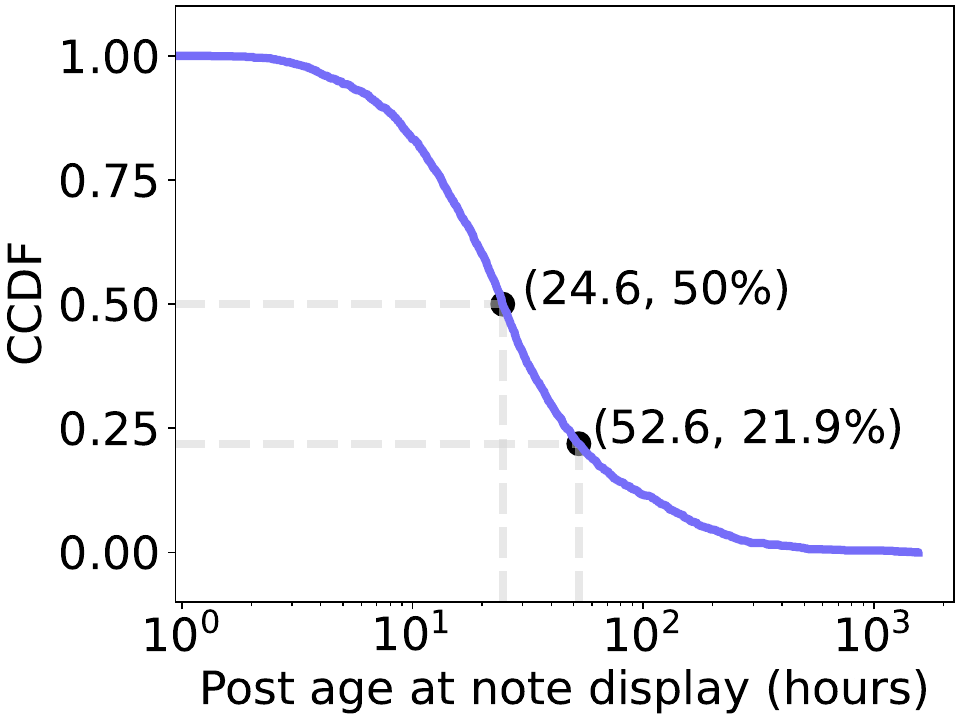}
\label{fig:hours_to_display}
\end{subfigure}
\hfill
\begin{subfigure}{0.32\textwidth}
\caption{}
\includegraphics[width=\textwidth]{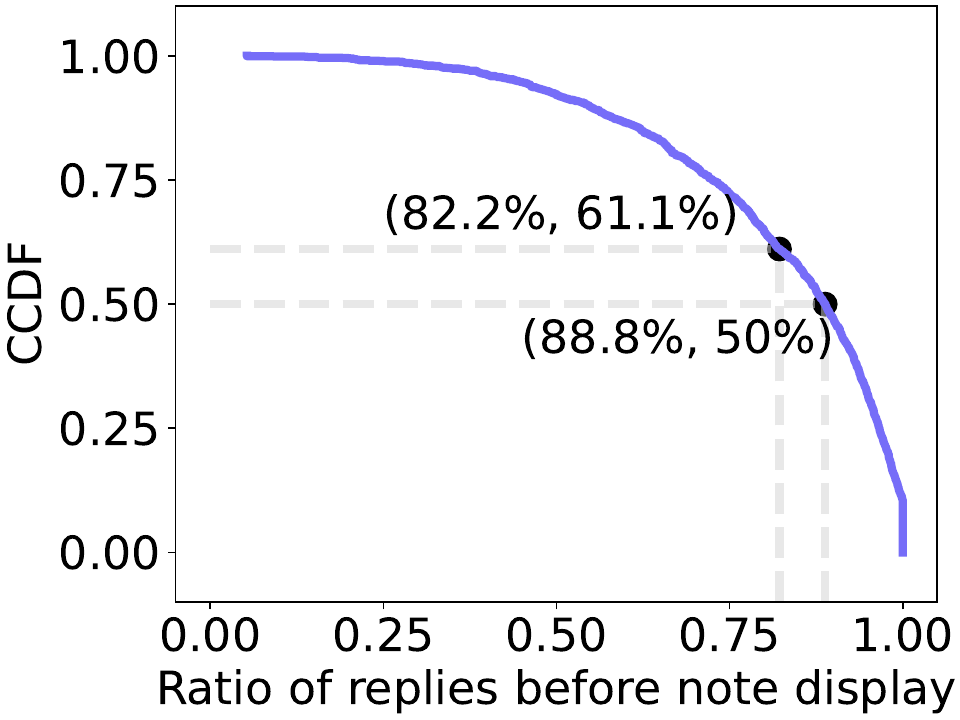}
\label{fig:reply_ratio}
\end{subfigure}

\begin{subfigure}{0.32\textwidth}
\caption{}
\includegraphics[width=\textwidth]{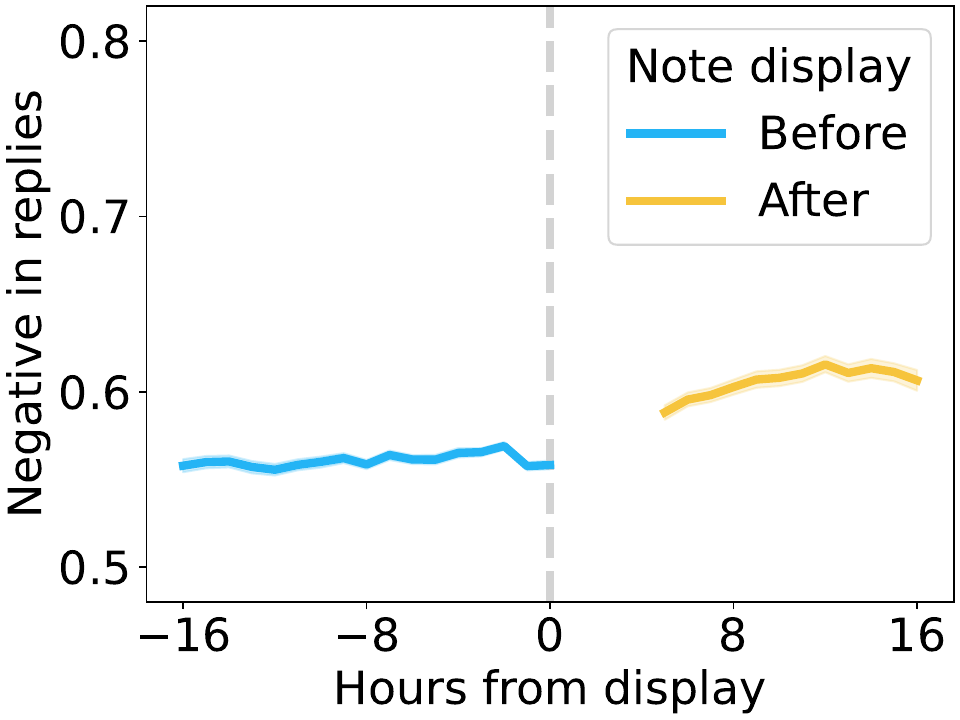}
\label{fig:negative_over_time}
\end{subfigure}
\hfill
\begin{subfigure}{0.32\textwidth}
\caption{}
\includegraphics[width=\textwidth]{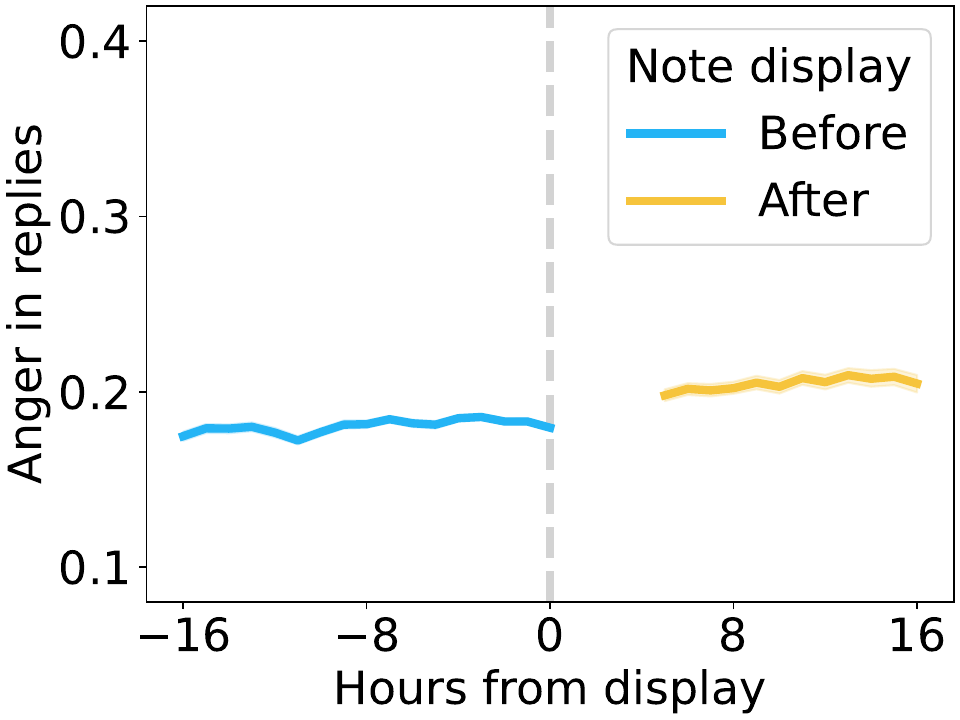}
\label{fig:anger_over_time}
\end{subfigure}
\hfill
\begin{subfigure}{0.32\textwidth}
\caption{}
\includegraphics[width=\textwidth]{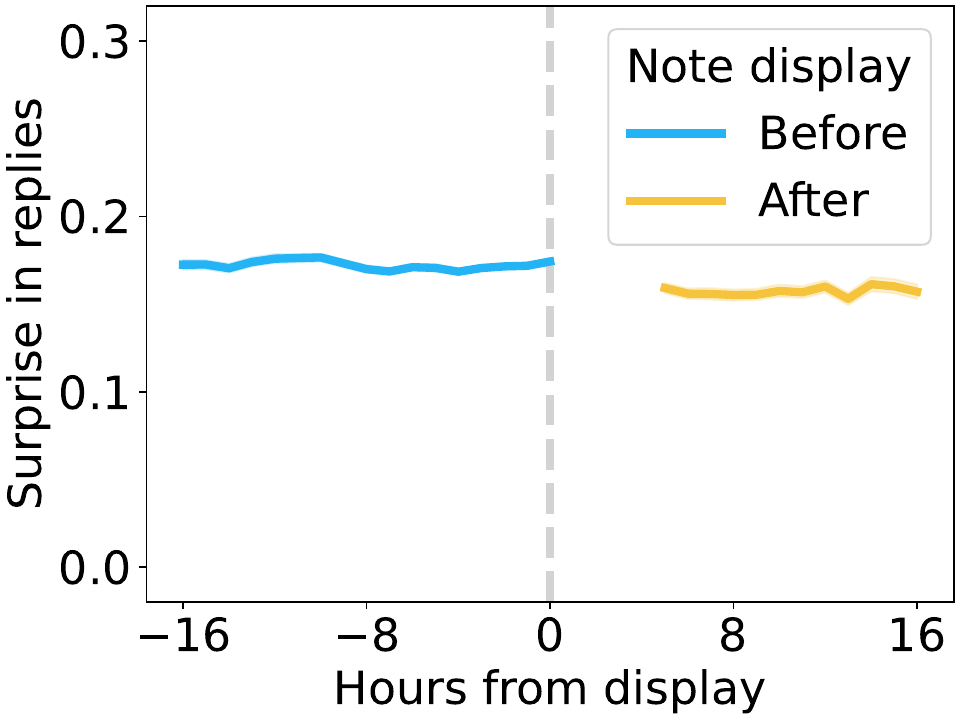}
\label{fig:surprise_over_time}
\end{subfigure}
\caption{Summary statistics for misleading source posts, direct replies, and display of community notes. \subref{fig:post_count} The 2-week rolling average daily number of misleading source posts that are attached with displayed community notes. \subref{fig:source_sentiments} The means of positive and negative sentiments in the misleading source posts. \subref{fig:source_emotions} The means of anger, disgust, fear, joy, sadness, and surprise in the misleading source posts. \subref{fig:reply_count} The Complementary Cumulative Distribution Function (CCDF) for the number of replies that are directed at each misleading source post. \subref{fig:hours_to_display} The CCDF for hour(s) from post creation to note display. \subref{fig:reply_ratio} The CCDF for the ratio of replies before note display to the total replies. \subref{fig:negative_over_time} The hourly averages of negative sentiment in replies across hours from note display. \subref{fig:anger_over_time} The hourly averages of anger in replies across hours from note display. \subref{fig:surprise_over_time} The hourly averages of surprise in replies across hours from note display. The hourly averages of positive sentiment and other emotions are shown in Fig. \ref{fig:emotions_over_time} in Suppl. \ref{sec:rdd_observation}. The error bars (bands) represent 95\% Confidence Intervals (CIs). Notably, previous studies consistently suggest a potential cold-start period of 4 hours for community notes to reach their full effect \cite{chuai2024community,renault2024collaboratively}. Therefore, we visually omit reply points within the initial four hours after the display of community notes for better readability.}
\label{fig:summary_statistics}
\end{figure*}

Our dataset covers four months since the roll-out of the ``Community Notes'' program on X in December of 2022. Fig. \ref{fig:post_count} shows that the daily number of misleading posts on which community notes are displayed has an increasing trend, with an average of 15.7. Fig. \ref{fig:source_sentiments} shows that misleading source posts tend to have higher negative sentiment (mean of 0.443), compared to positive sentiment (mean of 0.114; $t=-28.810$, $p<0.001$). In terms of emotions, anger (mean of 0.159), fear (mean of 0.173), and surprise (mean of 0.175) are most common in the fact-checked misleading source posts (Fig. \ref{fig:source_emotions}).

To examine the effects of community notes on the sentiments and emotions in replies, we collect a large-scale longitudinal dataset of replies to misleading posts. Fig. \ref{fig:reply_count} shows that each misleading source post receives an average of \num{1209} (median of \num{298}) replies in total (\ie, before and after the display of community notes). On average, community notes are displayed 52.5 hours (median of 24.6 hours) after the creation of the source misleading posts (Fig. \ref{fig:hours_to_display}). Notably, most of the replies (mean of 82.2\%, median of 88.8\%) are created before the display of community notes (Fig. \ref{fig:reply_ratio}). Therefore, some misleading posts may not have sufficient reply samples after note display to perform comparisons with the replies before note display. Given this, we restrict our main analysis to misleading posts that have at least ($\geq$) 5 replies both before and after the display of community notes. This criterion\footnote{This criterion is primarily applied to improve statistical inference. Our results remain robust and consistent when this restriction is removed (see details in Suppl. \ref{sec:sm_no5limit}).} results in \num{1339} misleading posts being included in the comparison of sentiments and emotions in replies before and after note display.

The key assumption of RDD in this paper is that a discontinuity exists around the cut-off point, \ie, the time when community notes are displayed \cite{smith2017strategies,imbens2008regression,cattaneo2022regression}. In the absence of this intervention, the sentiments and emotions in replies are expected to remain continuous or stable. To validate this assumption, we first examine changes in sentiments and emotions over time before the display of community notes. Our analysis shows that, in the absence of community notes, the sentiments and emotions in replies remain relatively stable and exhibit no statistically significant relationship with the age of misleading posts (see details in Suppl. \ref{sec:rdd_observation}). Subsequently, we use an interrupted time series design to analyze potential discontinuities in sentiments and emotions in replies following the display of community notes. The results reveal statistically significant discontinuities for negative sentiment, anger, and surprise in replies after the display of community notes (each $p<0.05$). Additionally, aside from the identified discontinuity at the cut-off point, the sentiments and emotions in replies after note display show trends similar to those in replies before note display and remain stable over time (see details in Suppl. \ref{sec:rdd_observation}). These findings confirm that RDD is appropriate for examining the effects of community notes on sentiments and emotions in replies. 

Another important consideration for RDD is to include as many samples as possible within a relatively short bandwidth, aiming to mitigate potential confounding factors associated with larger windows around the display of community notes \cite{imbens2008regression,cattaneo2022regression,hausman2018regression}. In our dataset, 97.8\% of total replies are captured within a one-week window before and after the display of community notes. We consider 10\% of this window (\ie, 16 hours) as the bandwidth, in which 61.3\% of total replies are included, to examine the effect sizes of community notes on sentiments and emotions in our main analysis.\footnote{Given the stable trends of sentiments and emotions in replies over time, we verify the robustness of RDD estimates across different bandwidths and find qualitatively identical results throughout the entire reply lifespan (see Fig. \ref{fig:rdd_coefs_bandwidths}).} To further illustrate the discontinuity around the display of community notes, we plot the hourly averages of sentiments and emotions in replies within the 16-hour bandwidth before and after note display. Figs. \ref{fig:negative_over_time}--\ref{fig:surprise_over_time} demonstrate the significant discontinuities in negative sentiment, anger, and surprise at the cut-off point with stable levels at both sides (see Fig. \ref{fig:emotions_over_time} in Suppl. \ref{sec:rdd_observation} for the visualizations of positive sentiment and other emotions).

To shed light on the relationships between sentiments and basic emotions in our reply dataset, we examine their cross-correlations based on Pearson's $r$ correlation coefficients. We find that anger ($r=0.406$, $p<0.001$), disgust ($r=0.323$, $p<0.001$), fear ($r=0.089$, $p<0.001$), and sadness ($r=0.192$, $p<0.001$) have significantly positive correlations with negative sentiment, while joy ($r=0.572$, $p<0.001$) and surprise ($r=0.044$, $p<0.001$) are positively correlated with positive sentiment (see Table \ref{tab:corr_sentiment_emotion} in Suppl. \ref{sec:sm_descriptive} for all cross-correlations). This suggests that, in replies to misleading posts, negative feelings are often expressed through anger, disgust, fear, and sadness. Notably, anger and disgust have the strongest correlations with negative sentiment. We further examine the cross-correlations among the six basic emotions in each reply. We find that only anger and disgust have a significantly positive correlation with each other ($r=0.111$, $p<0.001$). Other emotions all show negative correlations (reported in Table \ref{tab:corr_emotions} in Suppl. \ref{sec:sm_descriptive}). This suggests that replies to misleading posts often express anger and disgust together, which is a strong indicator of moral outrage \cite{salerno2013interactive}.

\subsection{Analysis of Sentiments and Basic Emotions (RQ1.1)}
\subsubsection{Descriptive analysis.}
We start by using $t$-tests to descriptively report changes in sentiments and emotions in replies after the display of community notes, compared to before note display.\footnote{We use the $t$-test rather than non-parametric alternatives such as the Mann–Whitney $U$-test or the Kolmogorov–Smirnov test due to its advantage in reporting confidence intervals, which enhances the robustness of statistical testing for large-scale data. Although the distributions of emotions and sentiments are not strictly normal, their sample means are approximately normally distributed in our dataset. Therefore, the $t$-test is valid in our study based on the central limit theorem \cite{lumley2002importance,fagerland2012t}.} For sentiments, positive sentiment in replies after the display of community notes (mean of 0.115) is significantly lower than before the note display (mean of 0.127; $t=10.249$, $p<0.001$; Fig. \ref{fig:positive_roberta_in_replies}). In contrast, negative sentiment in replies after the display of community notes (mean of 0.566) is significantly higher than before the note display (mean of 0.524; $t=-23.748$, $p<0.001$; Fig. \ref{fig:negative_roberta_in_replies}). In terms of basic emotions, anger (mean of 0.179) and disgust (mean of 0.093) in replies after the display of community notes are significantly higher than anger (mean of 0.155; $t=-18.531$, $p<0.001$; Fig. \ref{fig:anger_roberta_in_replies}) and disgust (mean of 0.090; $t=-2.717$, $p<0.01$; Fig. \ref{fig:disgust_roberta_in_replies}) in replies before the note display, respectively. Fear, joy, and sadness have no statistically significant changes after the display of community notes (each $p>0.05$; see Figs. \ref{fig:fear_roberta_in_replies}--\ref{fig:sadness_roberta_in_replies}). Surprise in replies after the display of community notes (mean of 0.179) is significantly lower than that in replies before the note display (mean of 0.185; $t=5.325$, $p<0.001$; Fig. \ref{fig:surprise_roberta_in_replies}). Taken together, our descriptive analysis suggests that replies after note display tend to be more negative and embed higher anger and disgust, compared to replies before the display of community notes (see details in Suppl. \ref{sec:descriptive_analysis}).

\begin{figure*}
\centering
\begin{subfigure}{0.32\textwidth}
\caption{}
\includegraphics[width=\textwidth]{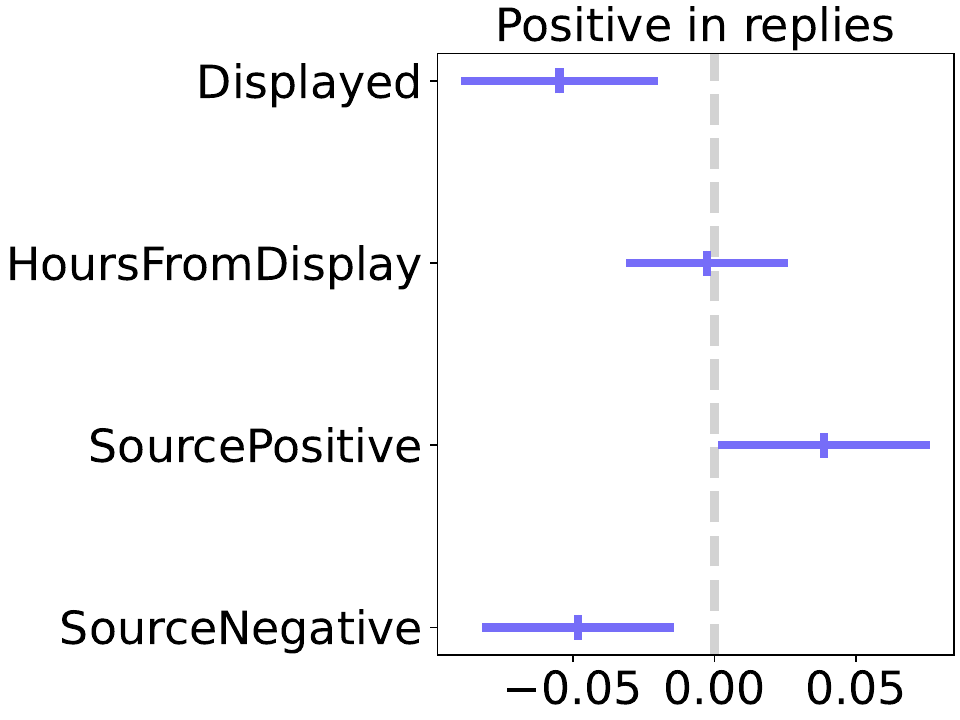}
\label{fig:Positive_coefs}
\end{subfigure}
\hspace{1cm}
\begin{subfigure}{0.32\textwidth}
\caption{}
\includegraphics[width=\textwidth]{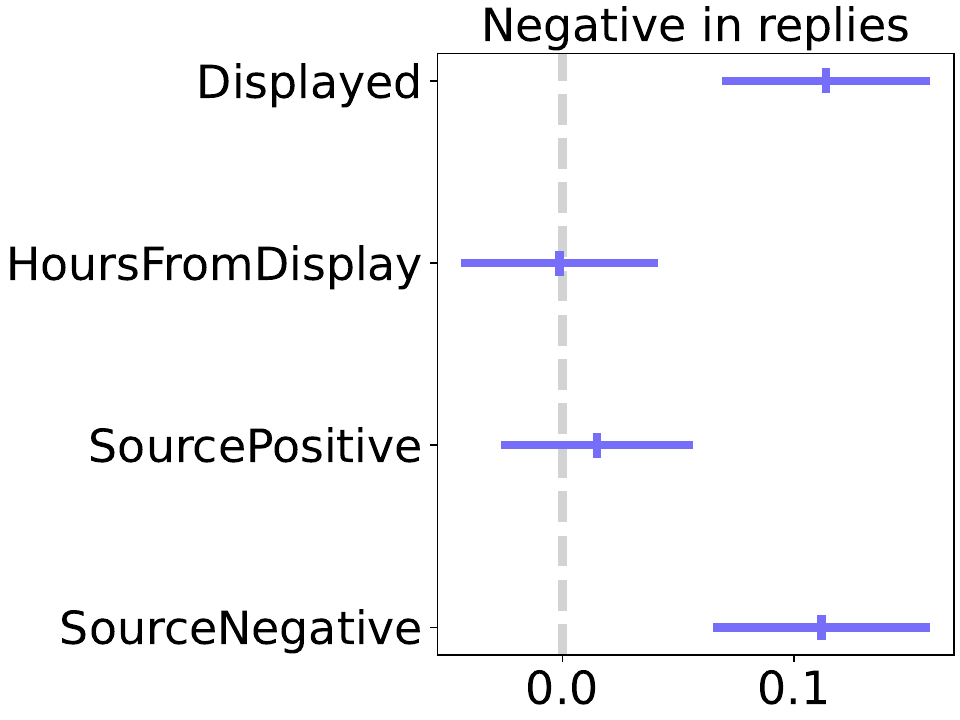}
\label{fig:Negative_coefs}
\end{subfigure}

\begin{subfigure}{0.32\textwidth}
\caption{}
\includegraphics[width=\textwidth]{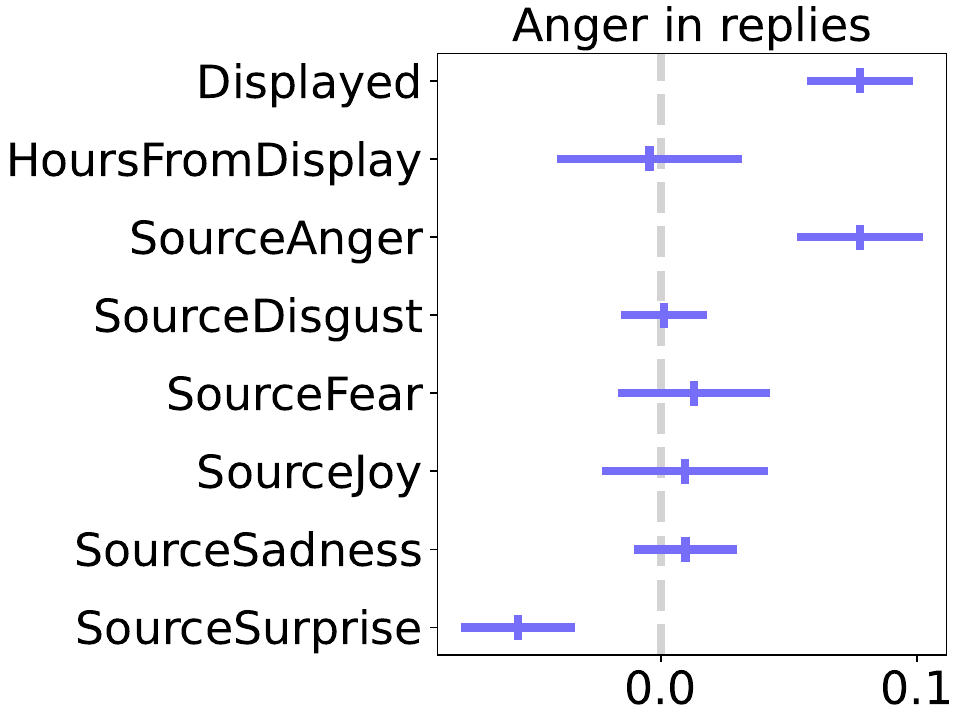}
\label{fig:Anger_coefs}
\end{subfigure}
\hfill
\begin{subfigure}{0.32\textwidth}
\caption{}
\includegraphics[width=\textwidth]{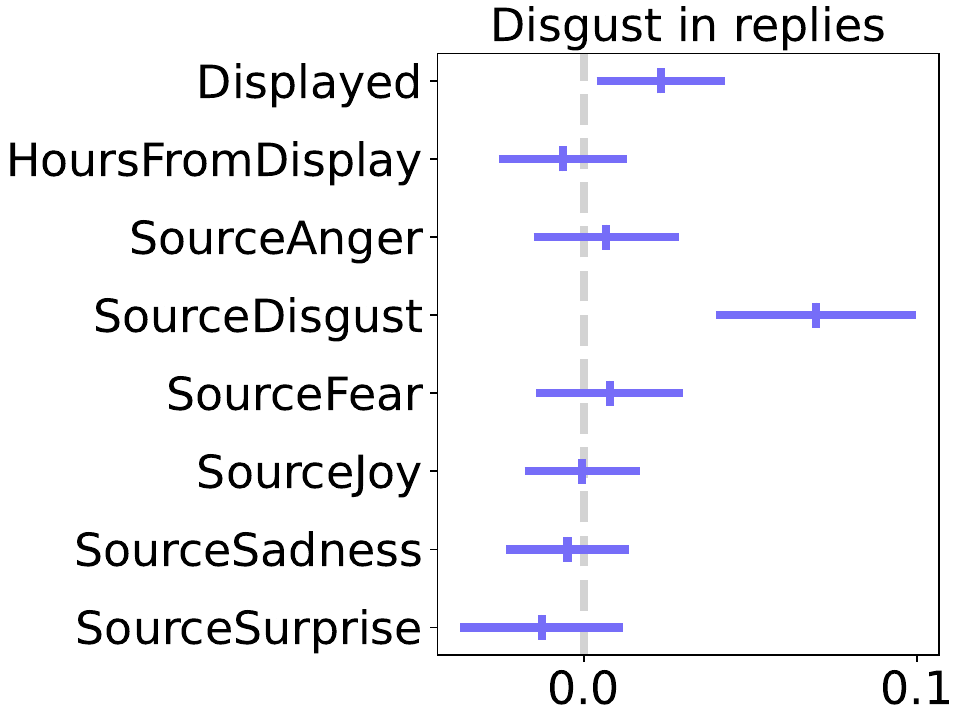}
\label{fig:Disgust_coefs}
\end{subfigure}
\hfill
\begin{subfigure}{0.32\textwidth}
\caption{}
\includegraphics[width=\textwidth]{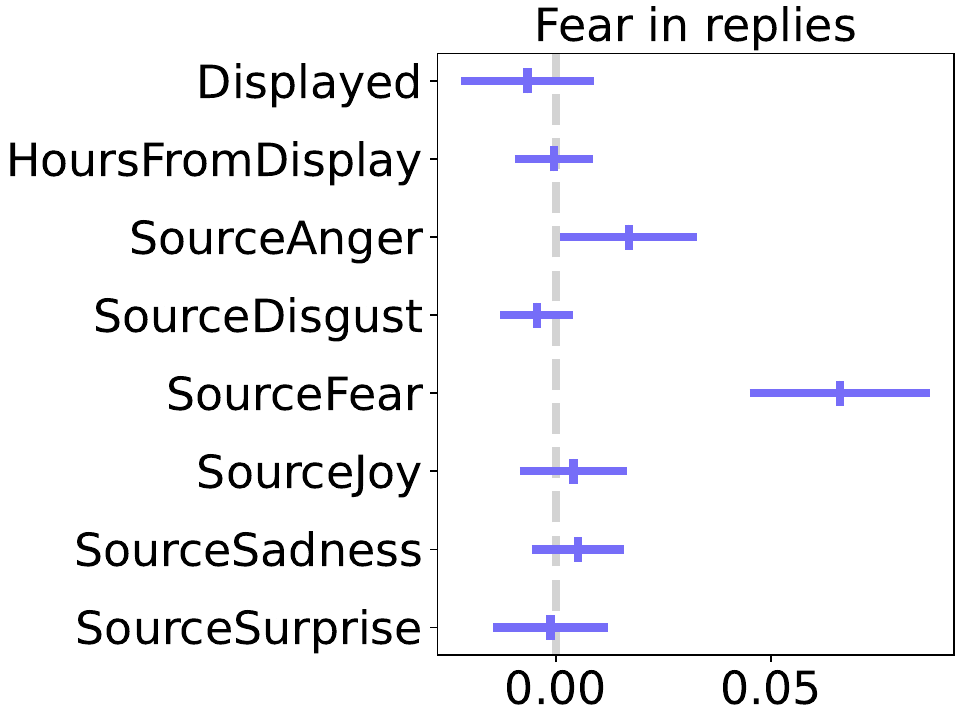}
\label{fig:Fear_coefs}
\end{subfigure}

\begin{subfigure}{0.32\textwidth}
\caption{}
\includegraphics[width=\textwidth]{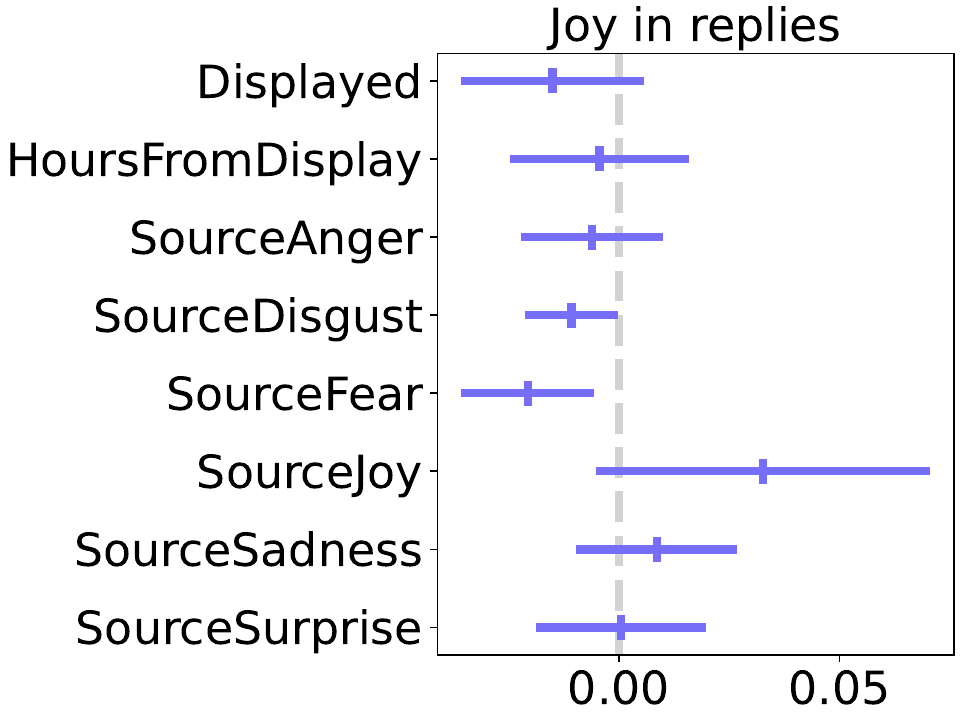}
\label{fig:Joy_coefs}
\end{subfigure}
\hfill
\begin{subfigure}{0.32\textwidth}
\caption{}
\includegraphics[width=\textwidth]{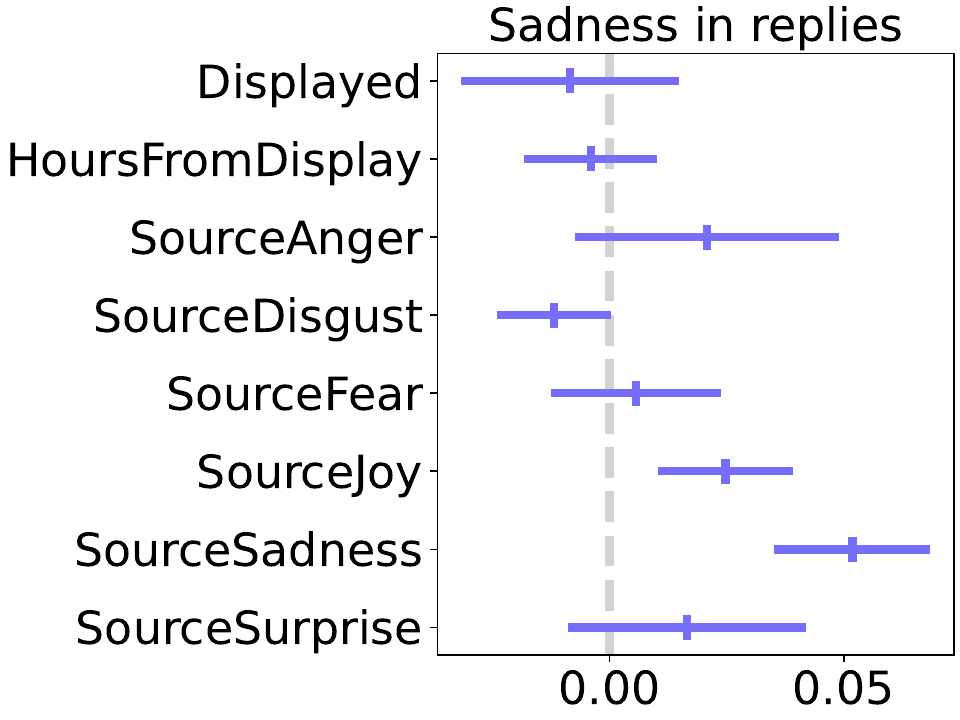}
\label{fig:Sadness_coefs}
\end{subfigure}
\hfill
\begin{subfigure}{0.32\textwidth}
\caption{}
\includegraphics[width=\textwidth]{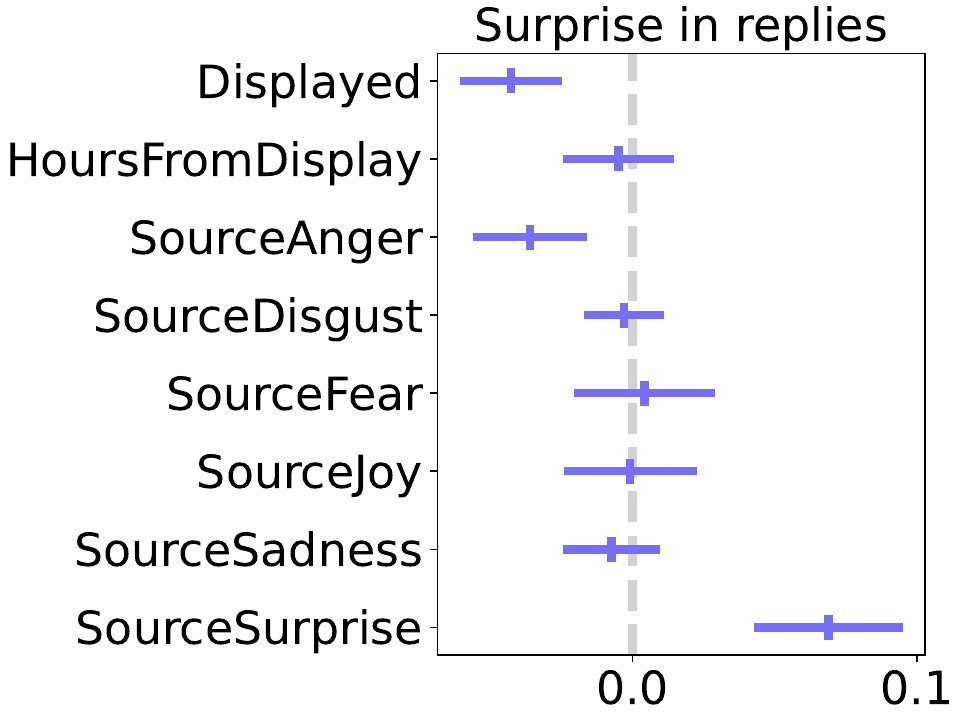}
\label{fig:Surprise_coefs}
\end{subfigure}
\caption{The estimated coefficients for the independent variables -- $\bm{\var{Displayed}}$, $\bm{\var{HoursFromDisplay}}$, and source sentiments (or emotions). The independent variable $\bm{\var{PostAge}}$ is included during estimation but omitted in the visualization for better readability. Shown are mean values with error bars representing 95\% CIs. Standard errors are clustered by source posts. The dependent variables are \subref{fig:Positive_coefs} positive sentiment in replies, \subref{fig:Negative_coefs} negative sentiment in replies, \subref{fig:Anger_coefs} anger in replies, \subref{fig:Disgust_coefs} disgust in replies, \subref{fig:Fear_coefs} fear in replies, \subref{fig:Joy_coefs} joy in replies, \subref{fig:Sadness_coefs} sadness in replies, and \subref{fig:Surprise_coefs} surprise in replies, respectively. The full estimation results are reported in Suppl. \ref{sec:sm_main_analysis}.}
\label{fig:emotion_coefs}
\end{figure*}

% Coefficient estimates of treatment effects
\subsubsection{Causal analysis.} 
\begin{figure*}
\centering
\begin{subfigure}{0.32\textwidth}
\caption{}
\includegraphics[width=\textwidth]{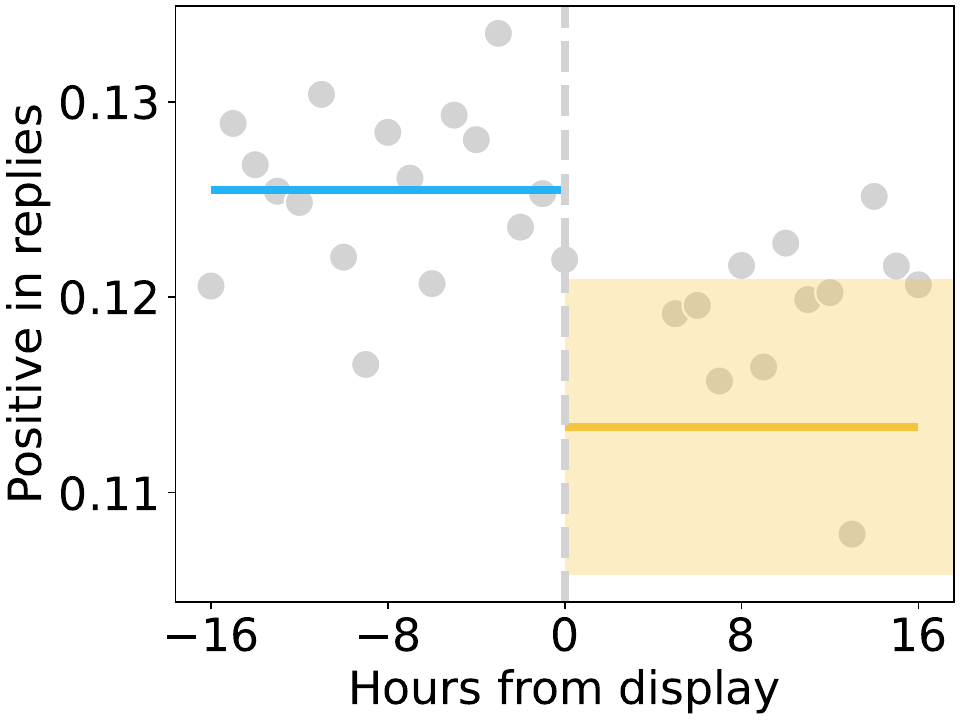}
\label{fig:positive_hours_from_display}
\end{subfigure}
\hspace{1cm}
\begin{subfigure}{0.32\textwidth}
\caption{}
\includegraphics[width=\textwidth]{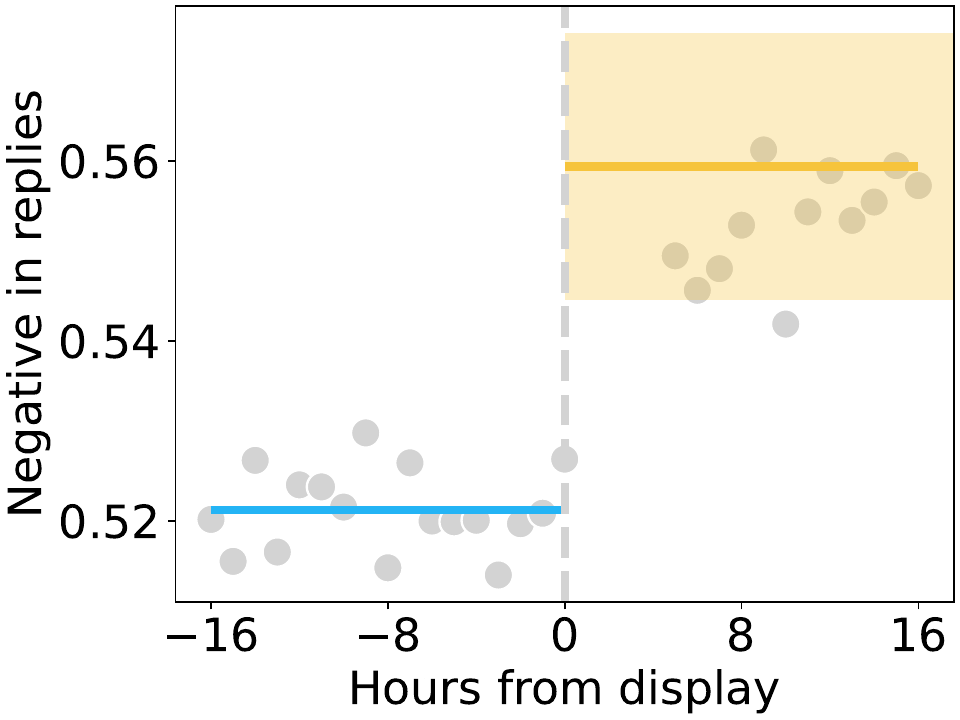}
\label{fig:negative_hours_from_display}
\end{subfigure}

\begin{subfigure}{0.32\textwidth}
\caption{}
\includegraphics[width=\textwidth]{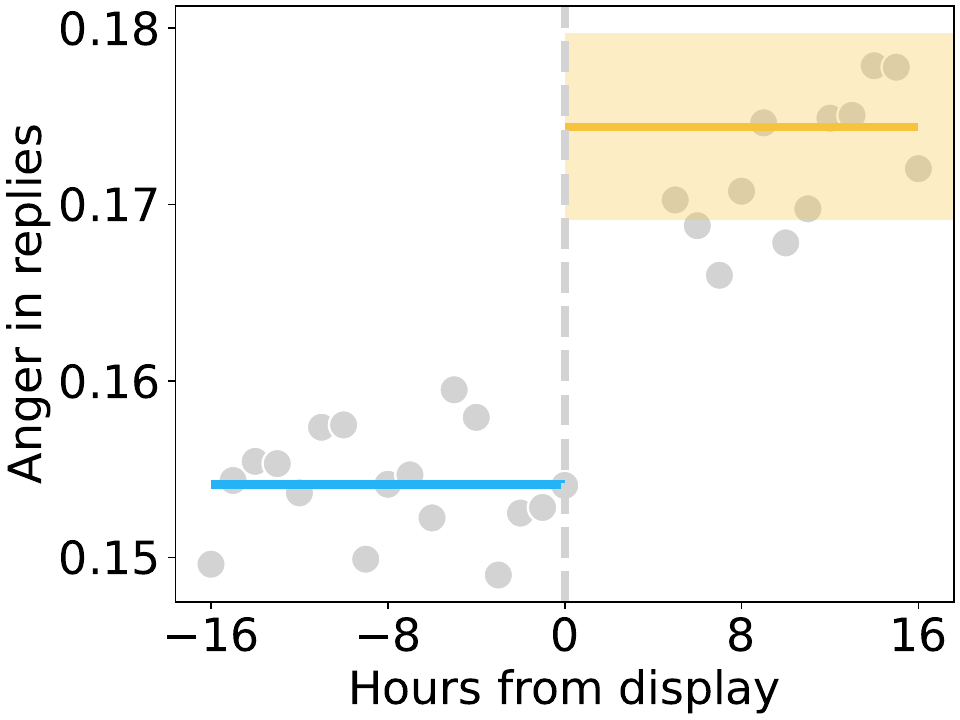}
\label{fig:anger_hours_from_display}
\end{subfigure}
\hfill
\begin{subfigure}{0.32\textwidth}
\caption{}
\includegraphics[width=\textwidth]{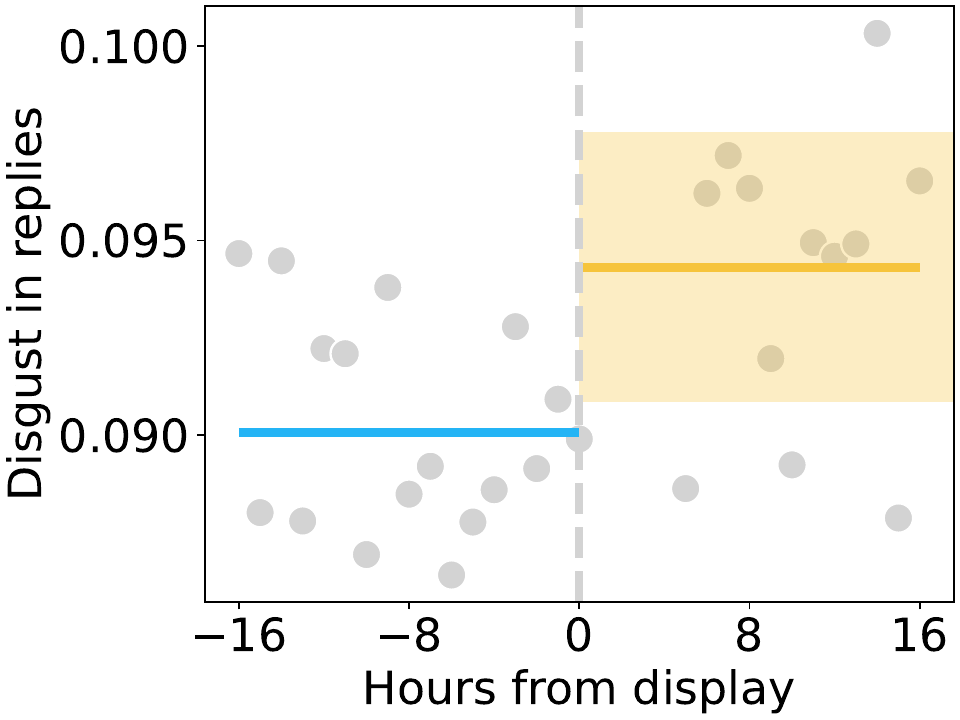}
\label{fig:disgust_hours_from_display}
\end{subfigure}
\hfill
\begin{subfigure}{0.32\textwidth}
\caption{}
\includegraphics[width=\textwidth]{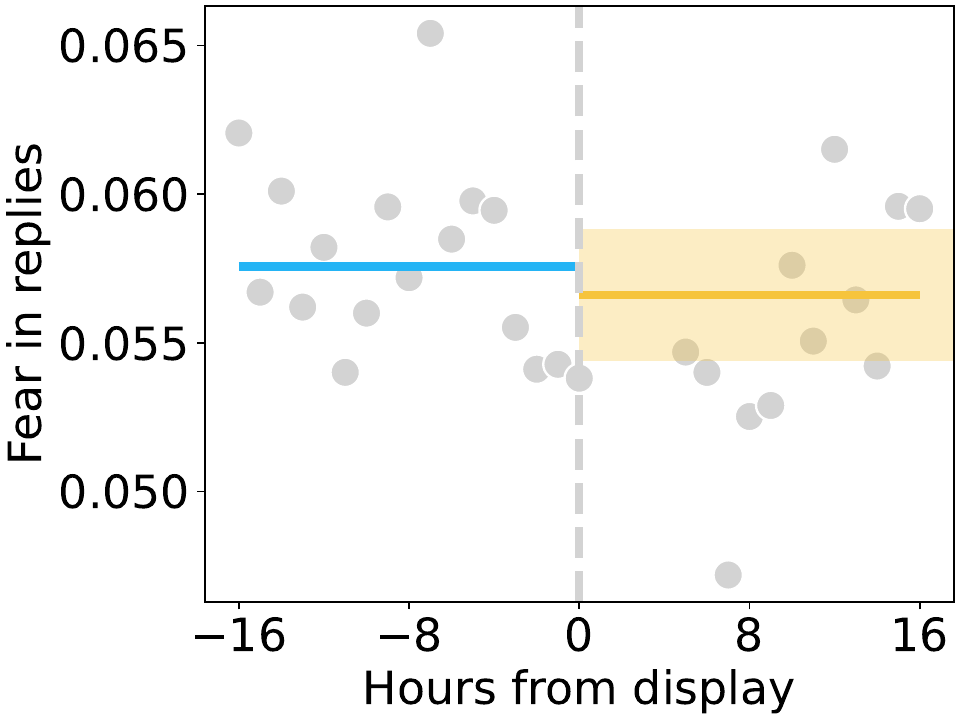}
\label{fig:fear_hours_from_display}
\end{subfigure}

\begin{subfigure}{0.32\textwidth}
\caption{}
\includegraphics[width=\textwidth]{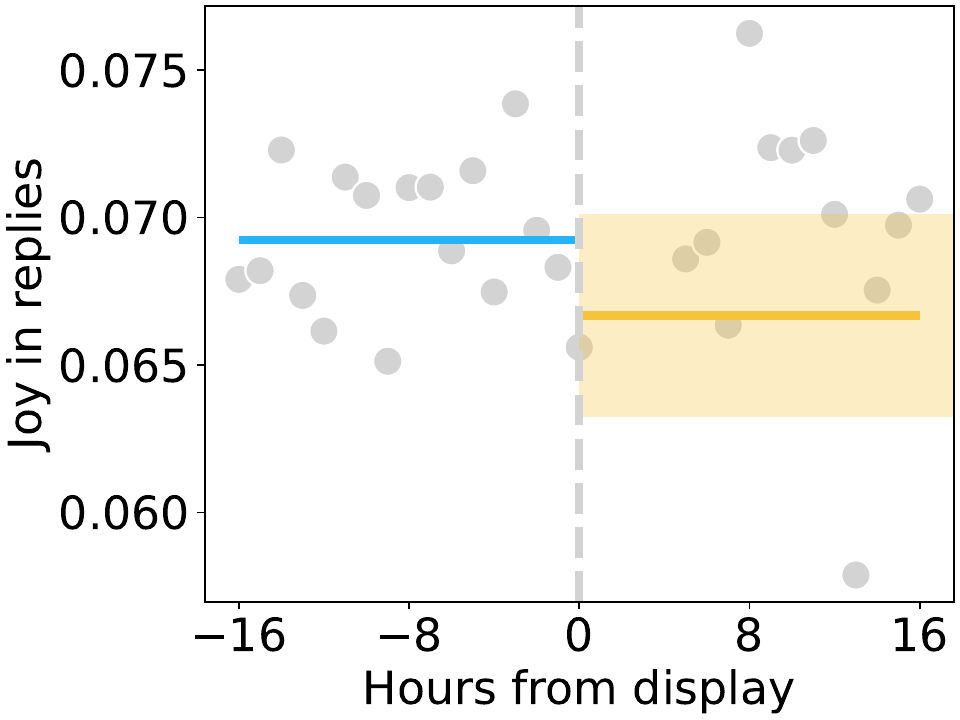}
\label{fig:joy_hours_from_display}
\end{subfigure}
\hfill
\begin{subfigure}{0.32\textwidth}
\caption{}
\includegraphics[width=\textwidth]{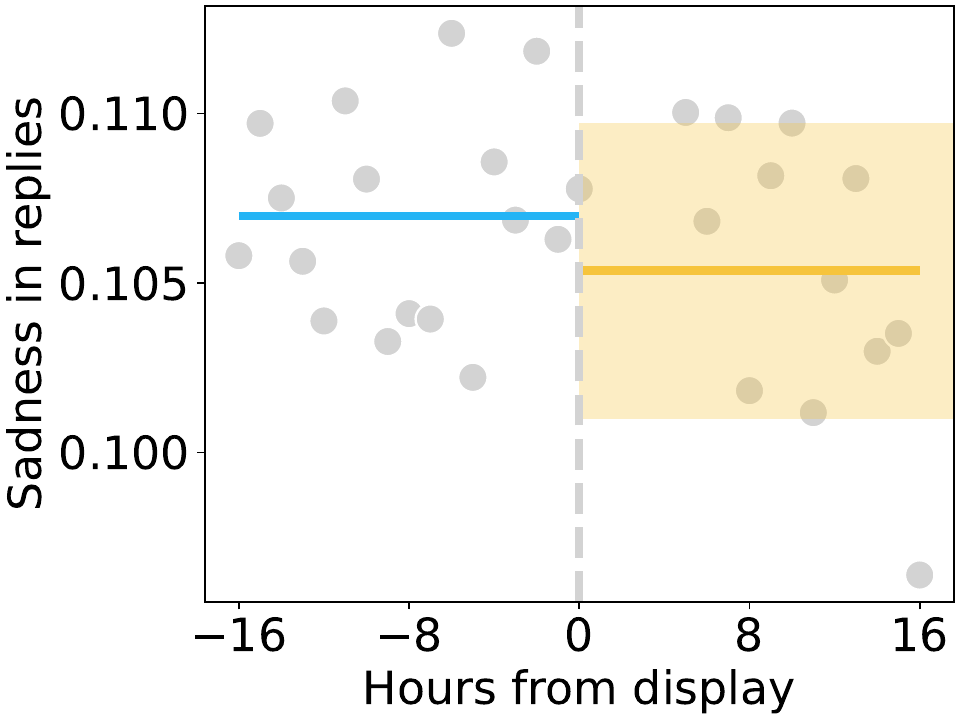}
\label{fig:sadness_hours_from_display}
\end{subfigure}
\hfill
\begin{subfigure}{0.32\textwidth}
\caption{}
\includegraphics[width=\textwidth]{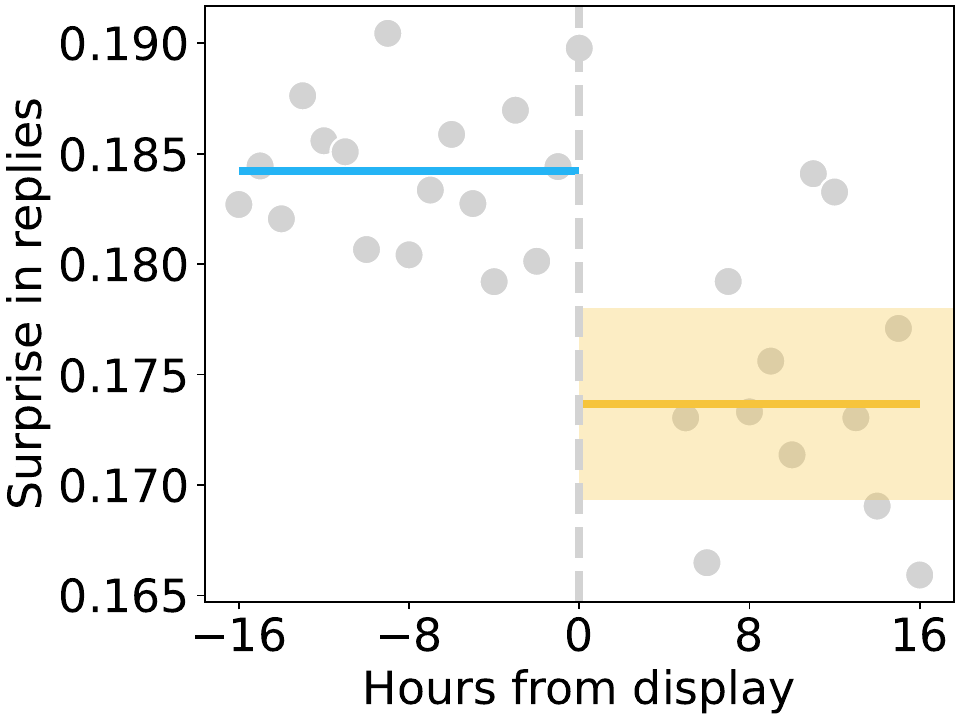}
\label{fig:surprise_hours_from_display}
\end{subfigure}
\caption{The predicted effects of the display of community notes on sentiments and emotions in replies. \subref{fig:positive_hours_from_display} The predicted effect on positive sentiment in replies. \subref{fig:negative_hours_from_display} The predicted effect on negative sentiment in replies. \subref{fig:anger_hours_from_display} The predicted effect on anger in replies. \subref{fig:disgust_hours_from_display} The predicted effect on disgust in replies. \subref{fig:fear_hours_from_display} The predicted effect on fear in replies. \subref{fig:joy_hours_from_display} The predicted effect on joy in replies. \subref{fig:sadness_hours_from_display} The predicted effect on sadness in replies. \subref{fig:surprise_hours_from_display} The predicted effect on surprise in replies. The grey points indicate the hourly averages of sentiments or emotions between 16 hours before and 16 hours after the display of community notes. Similar to Fig. \ref{fig:summary_statistics}, we visually omit reply points within the initial four hours after the display of community notes for better readability. The blues lines indicate the averages of sentiments or emotions in replies over the 16 hours before note display and represent the baselines during before-display period. The yellow lines indicate the averages of sentiments or emotions over the 16 hours after the display of community notes. The average at the yellow line is the sum of the predicted effect and the corresponding baseline at the blue line in each figure. The yellow bands represent 95\% CIs. The predicted effects are transformed based on the coefficient estimates of $\bm{\var{Displayed}}$ (see Fig. \ref{fig:emotion_coefs}). The full estimation results are reported in Suppl. \ref{sec:sm_main_analysis}.}
\label{fig:predicted_effects}
\end{figure*}

We now use regression models based on RDD to estimate the causal changes in sentiments and emotions in replies to source misleading posts that are attributed to the display of community notes. The coefficient estimates for the regressions across sentiments and basic emotions are visualized in Fig. \ref{fig:emotion_coefs}. Additionally, to interpret the effect sizes of the display of community notes on the specific sentiments and emotions in replies to misleading posts, we examine the extent to which the sentiments and basic emotions in replies change compared to their corresponding baselines before the display of community notes. The changes in sentiments and emotions in replies to misleading posts within the 16-hour window, along with the predicted effects of displaying community notes, are visualized in Fig. \ref{fig:predicted_effects}.

\textbf{Effect of community notes display:}
For positive sentiment in replies, the coefficient estimate of $\var{Displayed}$ is significantly negative (Fig. \ref{fig:Positive_coefs}: $\var{coef.} = -0.055$, $p<0.01$; 95\% CI: $[-0.089, -0.021]$). This means that the display of community notes decreases the positive sentiment in replies by 0.012 (95\% CI: $[-0.020, -0.005]$). As shown in Fig. \ref{fig:positive_hours_from_display}, the baseline of positive sentiment in replies before note display is 0.125, and the positive sentiment in replies after note display is predicted to be 0.113 (95\% CI: $[0.106, 0.121]$). This suggests that positive sentiment in replies decreases by 9.7\% after the display of community notes compared to that before note display. In contrast, for negative sentiment in replies, the coefficient estimate of $\var{Displayed}$ is significantly positive (Fig. \ref{fig:Negative_coefs}: $\var{coef.} = 0.114$, $p<0.001$; 95\% CI: $[0.070, 0.158]$). This means that the display of community notes increases the negative sentiment in replies by 0.038 (95\% CI: $[0.023, 0.053]$). As shown in Fig. \ref{fig:negative_hours_from_display}, the baseline of negative sentiment in replies before note display is 0.521, and the negative sentiment in replies after note display is predicted to be 0.559, (95\% CI: $[0.545, 0.574]$). This suggests that negative sentiment in replies increases by 7.3\% after the display of community notes compared to that before note display. 

We further examine the effects of community notes display on changes in basic emotions in replies to misleading posts before and after the display of community notes. The results are as follows:
\begin{itemize}[leftmargin=*]
    \item For anger in replies, the coefficient estimate of $\var{Displayed}$ is significantly positive (Fig. \ref{fig:Anger_coefs}: $\var{coef.} = 0.078$, $p<0.001$; 95\% CI: $[0.058, 0.098]$). This means that the display of community notes increases anger in replies by 0.020 (95\% CI: $[0.015, 0.026]$). As shown in Fig. \ref{fig:anger_hours_from_display}, the baseline of anger in replies before note display is 0.154, and the predicted anger in replies after note display is 0.174 (95\% CI: $[0.169, 0.180]$). This suggests that anger in replies increases by 13.2\% after the display of community notes compared to that before note display.
    \item For disgust in replies, the coefficient estimate of $\var{Displayed}$ is significantly positive (Fig. \ref{fig:Disgust_coefs}: $\var{coef.} = 0.023$, $p<0.05$; 95\% CI: $[0.004, 0.042]$). This means that the display of community notes increases disgust in replies by 0.004 (95\% CI: $[0.001, 0.008]$). As shown in Fig. \ref{fig:disgust_hours_from_display}, the baseline of disgust in replies before note display is 0.090, and the predicted disgust in replies after note display is 0.094 (95\% CI: $[0.091, 0.098]$). This suggests that disgust in replies increases by 4.7\% after the display of community notes compared to that before note display.
    \item For fear, joy, and sadness in replies, the coefficient estimates of $\var{Displayed}$ are all not statistically significant (each $p>0.05$; see Figs. \ref{fig:Fear_coefs}--\ref{fig:Sadness_coefs}). This means that the display of community notes has no significant effect on these three emotions in replies to misleading posts.
    \item For surprise in replies, the coefficient estimate of $\var{Displayed}$ is significantly negative (Fig. \ref{fig:Surprise_coefs}: $\var{coef.} = -0.043$, $p<0.001$; 95\% CI: $[-0.060, -0.025]$). This means that the display of community notes decreases surprise in replies by 0.011 (95\% CI: $[-0.015, -0.006]$). As shown in Fig. \ref{fig:surprise_hours_from_display}, the baseline of surprise in replies before note display is 0.184, and the predicted surprise in replies after note display is 0.174 (95\% CI: $[0.169, 0.178]$). This suggests that surprise in replies decreases by 5.7\% after the display of community notes compared to that before note display.
\end{itemize}
Taken together, the display of community notes increases 7.3\% more negative sentiment and triggers 13.2\% more anger and 4.7\% more disgust in replies to misleading posts, respectively. The effects of community notes display on sentiments and emotions in replies are concordant with the observed changes of sentiments and emotions in replies after the display of community notes (see Fig. \ref{fig:emotions_in_replies} in Suppl. \ref{sec:descriptive_analysis}).

% Emotion transfers
\textbf{Effect of sentiments and emotions in source posts:}
We further examine the link between the original sentiments and emotions in the source misleading posts and the subsequent sentiments and emotions in the direct replies. For positive sentiment in replies, the coefficient estimate of $\var{SourcePositive}$ in Fig. \ref{fig:Positive_coefs} is significantly positive ($\var{coef.} = 0.039$, $p<0.05$; 95\% CI: $[0.002, 0.075]$); for negative sentiment in replies, the coefficient estimate of $\var{SourceNegative}$ in Fig. \ref{fig:Negative_coefs} is also significantly positive ($\var{coef.} = 0.112$, $p<0.001$; 95\% CI: $[0.066, 0.158]$). This means that the positive and negative sentiments in replies are positively linked to the original positive and negative sentiments in the source misleading posts, respectively. In terms of emotions, the coefficient estimate of $\var{SourceAnger}$ in Fig. \ref{fig:Anger_coefs} ($\var{coef.} = 0.078$, $p<0.001$; 95\% CI: $[0.054, 0.102]$), the coefficient estimate of $\var{SourceDisgust}$ in Fig. \ref{fig:Disgust_coefs} ($\var{coef.} = 0.070$, $p<0.001$; 95\% CI: $[0.040, 0.099]$), the coefficient estimate of $\var{SourceFear}$ in Fig. \ref{fig:Fear_coefs} ($\var{coef.} = 0.066$, $p<0.001$; 95\% CI: $[0.045, 0.087]$), the coefficient estimate of $\var{SourceSadness}$ in Fig. \ref{fig:Sadness_coefs} ($\var{coef.} = 0.052$, $p<0.001$; 95\% CI: $[0.036, 0.068]$), and the coefficient estimate of $\var{SourceSurpirse}$ in Fig. \ref{fig:Surprise_coefs} ($\var{coef.} = 0.069$, $p<0.001$; 95\% CI: $[0.043, 0.094]$) are all significantly positive. This suggests that the expressions of anger, disgust, fear, sadness, and surprise in source misleading posts have significantly positive effects on anger, disgust, fear, sadness, and surprise in the subsequent replies, respectively. However, the coefficient estimate of $\var{SourceJoy}$ in Fig. \ref{fig:Joy_coefs} is not statistically significant ($\var{coef.} = 0.033$, $p=0.084$; 95\% CI: $[-0.004, 0.070]$). Thus, we find no evidence that the expression of joy in replies is significantly linked to the expression of joy in the source misleading posts.

% Emotion stability over time
\textbf{Effect of reply timing:}
We examine whether the sentiments and emotions in replies change over time without the display of community notes. We find that the coefficient estimates of $\var{HoursFromDisplay}$ are not statistically significant across all dependent variables (each $p>0.05$; see Figs. \ref{fig:Positive_coefs}--\ref{fig:Surprise_coefs}). The coefficient estimates of $\var{PostAge}$ are also not statistically significant (see details in Suppl. \ref{sec:sm_main_analysis}). This means that the sentiments and emotions in replies are relatively stable over time if community notes are not displayed on the corresponding misleading posts.

\begin{figure*}
    \centering
    \includegraphics[width=\textwidth]{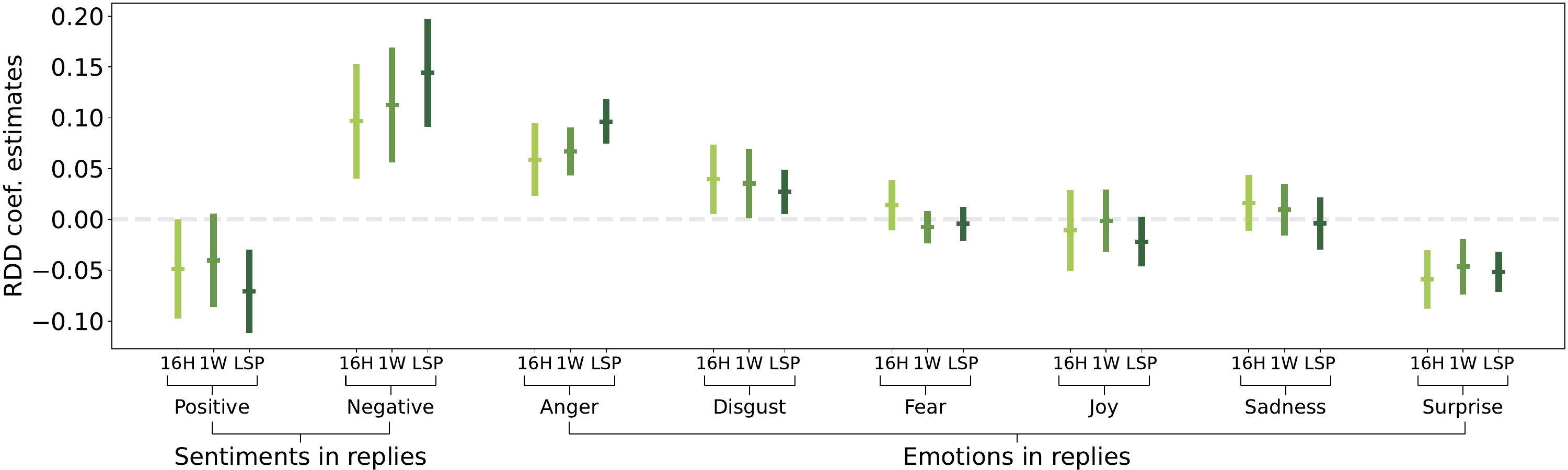}
    \caption{The RDD coefficient estimates (\ie, $\bm{\var{Displayed}}$) for sentiments and emotions in replies across three different bandwidths. The three bandwidths are 16 hours (16H), 1 week (1W), and reply lifespan (LSP), respectively. Shown are mean values with error bars representing 95\% CIs. Standard errors are clustered by source posts. The full estimation results are reported in Suppl. \ref{sec:rdd_coefs_bandwidths}.}
    \label{fig:rdd_coefs_bandwidths}
\end{figure*}

\textbf{RDD estimates across different bandwidths:} Based on the previous analysis of stable trends in sentiments and emotions within the reply lifespan (see Section \ref{sec:data_overview}), the RDD estimates across different bandwidths should remain robust and have no statistically significant differences. To validate this, we examine RDD estimates using three different bandwidths: 16 hours (16H), one week (1W), and the entire reply lifespan (LSP). As shown in Fig. \ref{fig:rdd_coefs_bandwidths}, the RDD coefficient estimates across sentiments and emotions in replies have no statistically significant differences within the three bandwidths (see full estimation results in Suppl. \ref{sec:rdd_coefs_bandwidths}). Additionally, we conduct $\var{SUEST}$ tests to compare RDD models across the three bandwidths, further confirming the equivalence of the common coefficient estimates within these models (each $p>0.05$, see Table \ref{tab:suest_test_bandwidths}).

\begin{table*}
\centering
\caption{$\bm{\var{SUEST}}$ tests for the equality of common coefficients across RDD models with different bandwidths. The tests are conducted separately for each sentiment/emotion in replies.}
\begin{tabular}{l*{7}{S}}
\toprule
&\multicolumn{3}{c}{(1)}&\multicolumn{3}{c}{(2)}\\
&\multicolumn{3}{c}{16H vs. LSP}&\multicolumn{3}{c}{1W vs. LSP}\\
\cmidrule(lr){2-4}\cmidrule(lr){5-7}
&$\chi^{2}$&$p$&{\#Variables}&$\chi^{2}$&$p$&{\#Variables}\\
\midrule
\underline{Sentiments in replies}\\
\quad Positive&8.47&0.132&5&10.74&0.057&5\\
\quad Negative&3.97&0.554&5&4.30&0.507&5\\
\underline{Emotions in replies}\\
\quad Anger&12.87&0.168&9&15.89&0.069&9\\
\quad Disgust&9.03&0.434&9&8.25&0.509&9\\
\quad Fear&6.23&0.717&9&8.16&0.518&9\\
\quad Joy&8.10&0.524&9&8.45&0.490&9\\
\quad Sadness&13.68&0.134&9&6.70&0.668&9\\
\quad Surprise&6.49&0.690&9&7.54&0.581&9\\
\bottomrule
\end{tabular}
\label{tab:suest_test_bandwidths}
\end{table*}

% Summary for main analysis
\textbf{Summary of findings:} In sum, we find that replies are more negative and express more anger and disgust after the display of community notes compared to those before the display. Specifically, the display of community notes triggers 13.2\% more anger and 4.7\% more disgust in replies to source misleading posts. Additionally, the sentiments and almost all emotions (except for joy) in replies are positively linked to those expressed in the source misleading posts. Furthermore, we find that the expressions of sentiments and emotions in replies are relatively stable over time if without the display of community notes.

\subsection{Analysis Across Politic and Non-Political Misleading Posts (RQ1.2)}
\begin{figure*}
\centering
\begin{subfigure}{0.32\textwidth}
\caption{}
\includegraphics[width=\textwidth]{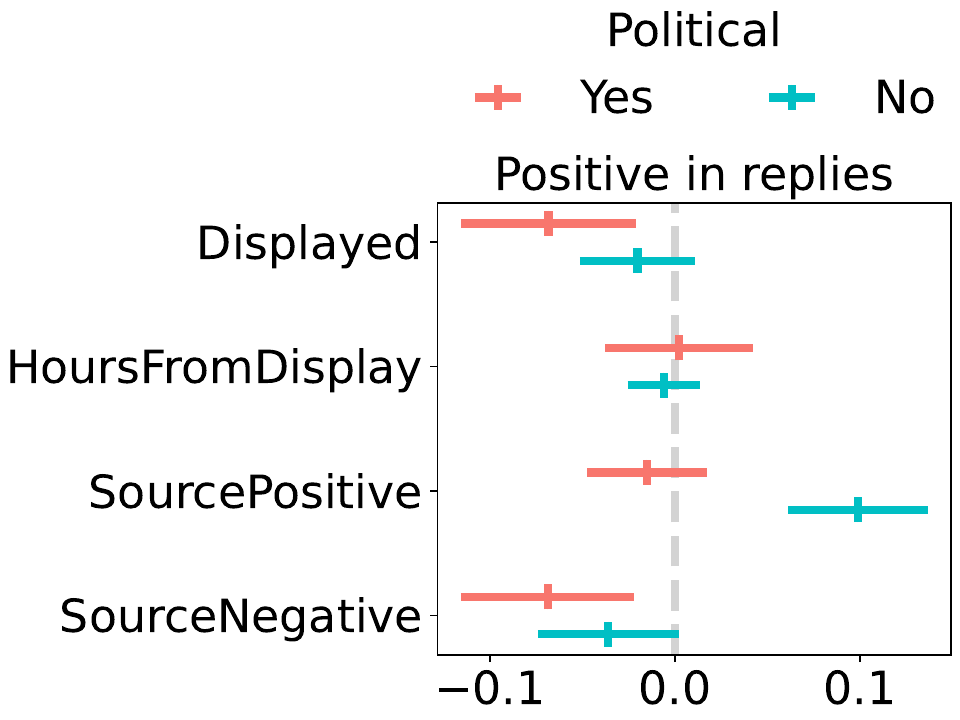}
\label{fig:Positive_coefs_politics}
\end{subfigure}
\hspace{1cm}
\begin{subfigure}{0.32\textwidth}
\caption{}
\includegraphics[width=\textwidth]{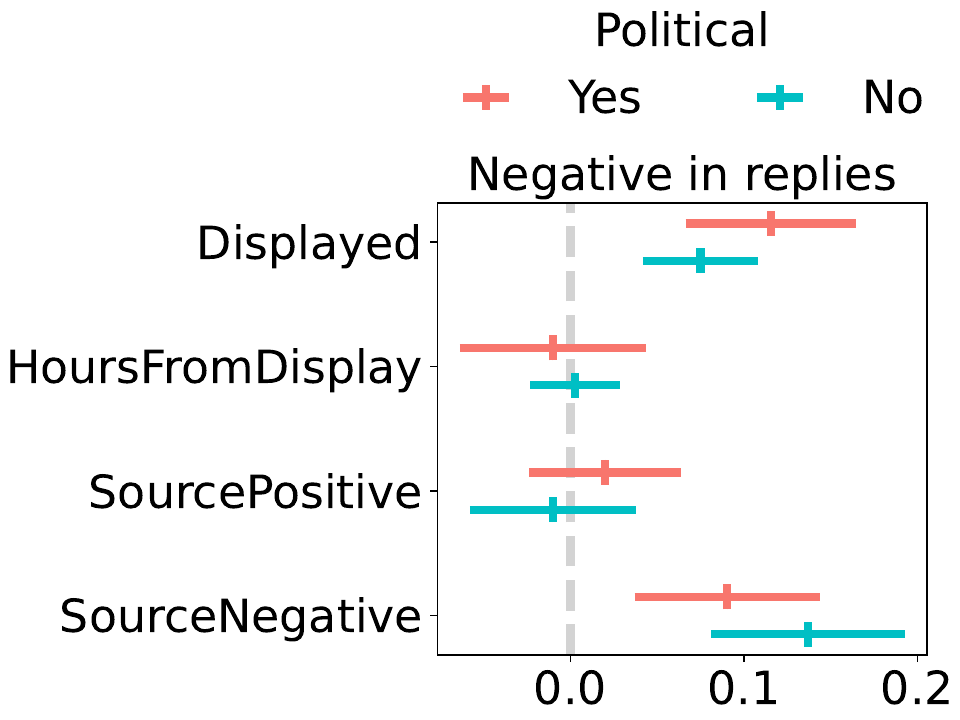}
\label{fig:Negative_coefs_politics}
\end{subfigure}

\begin{subfigure}{0.32\textwidth}
\caption{}
\includegraphics[width=\textwidth]{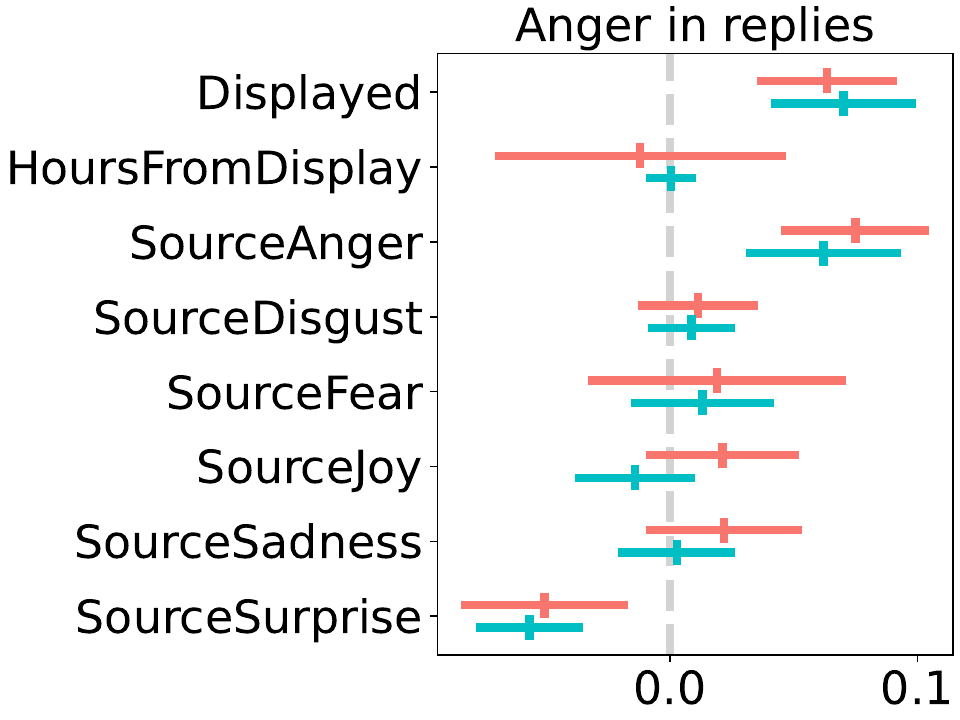}
\label{fig:Anger_coefs_politics}
\end{subfigure}
\hfill
\begin{subfigure}{0.32\textwidth}
\caption{}
\includegraphics[width=\textwidth]{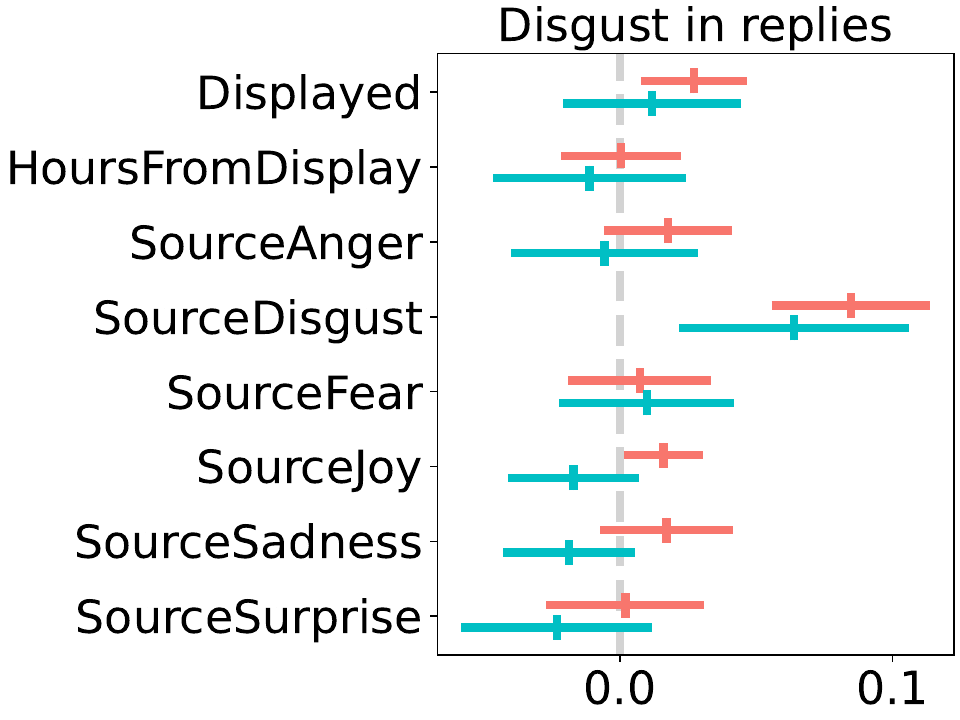}
\label{fig:Disgust_coefs_politics}
\end{subfigure}
\hfill
\begin{subfigure}{0.32\textwidth}
\caption{}
\includegraphics[width=\textwidth]{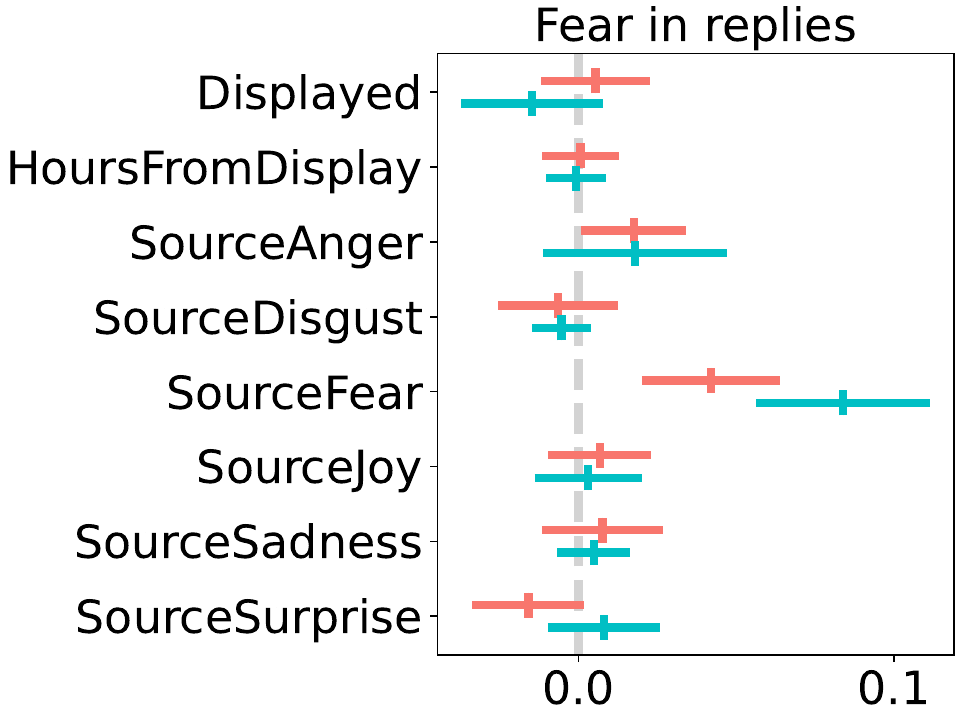}
\label{fig:Fear_coefs_politics}
\end{subfigure}

\begin{subfigure}{0.32\textwidth}
\caption{}
\includegraphics[width=\textwidth]{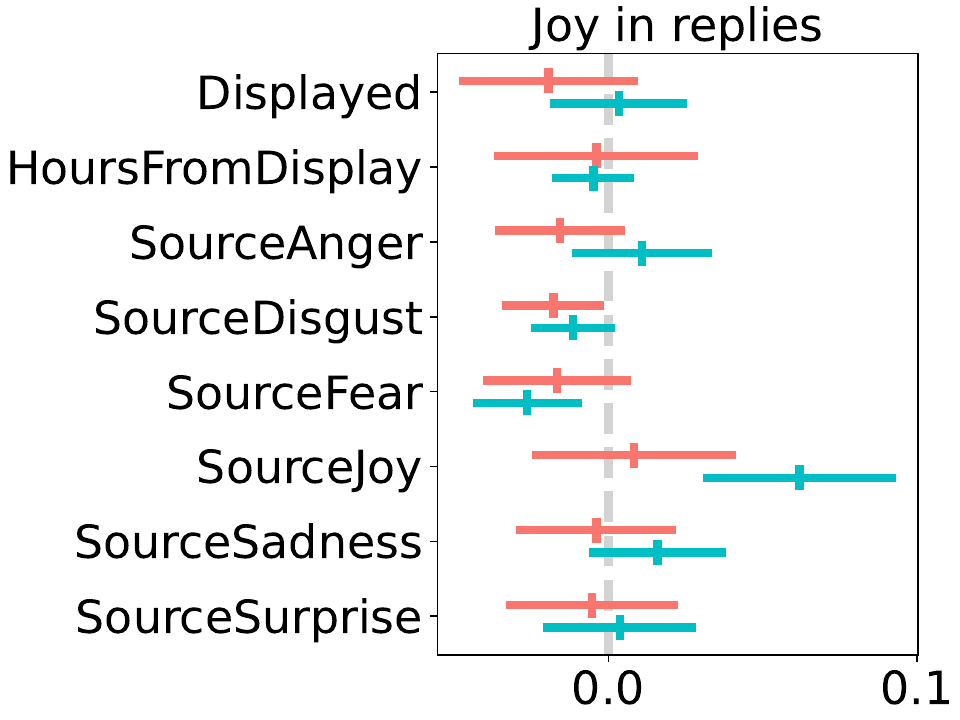}
\label{fig:Joy_coefs_politics}
\end{subfigure}
\hfill
\begin{subfigure}{0.32\textwidth}
\caption{}
\includegraphics[width=\textwidth]{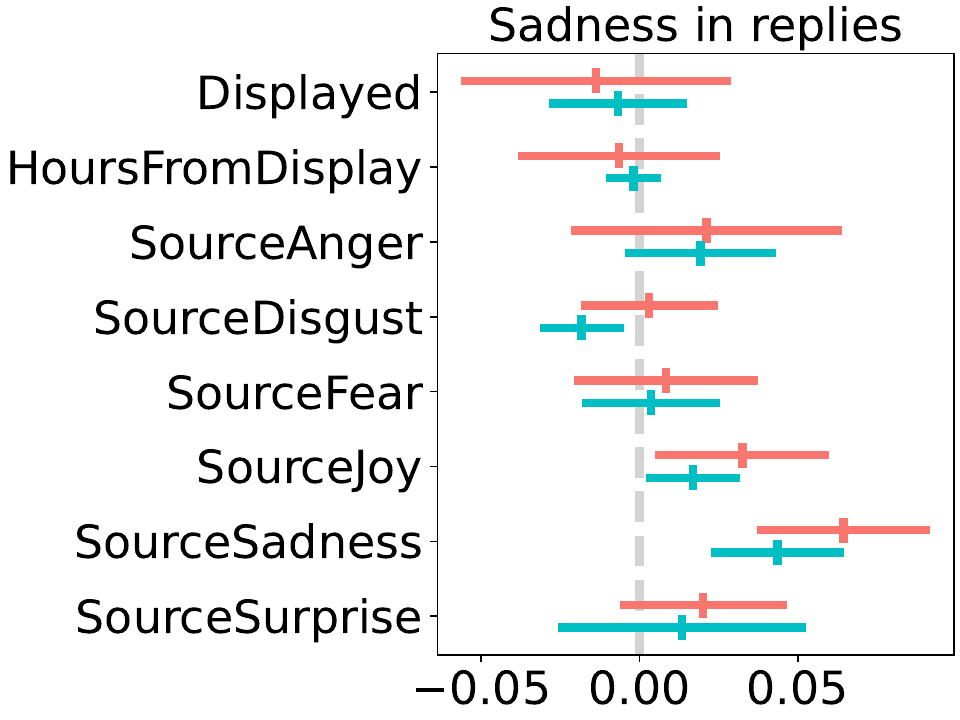}
\label{fig:Sadness_coefs_politics}
\end{subfigure}
\hfill
\begin{subfigure}{0.32\textwidth}
\caption{}
\includegraphics[width=\textwidth]{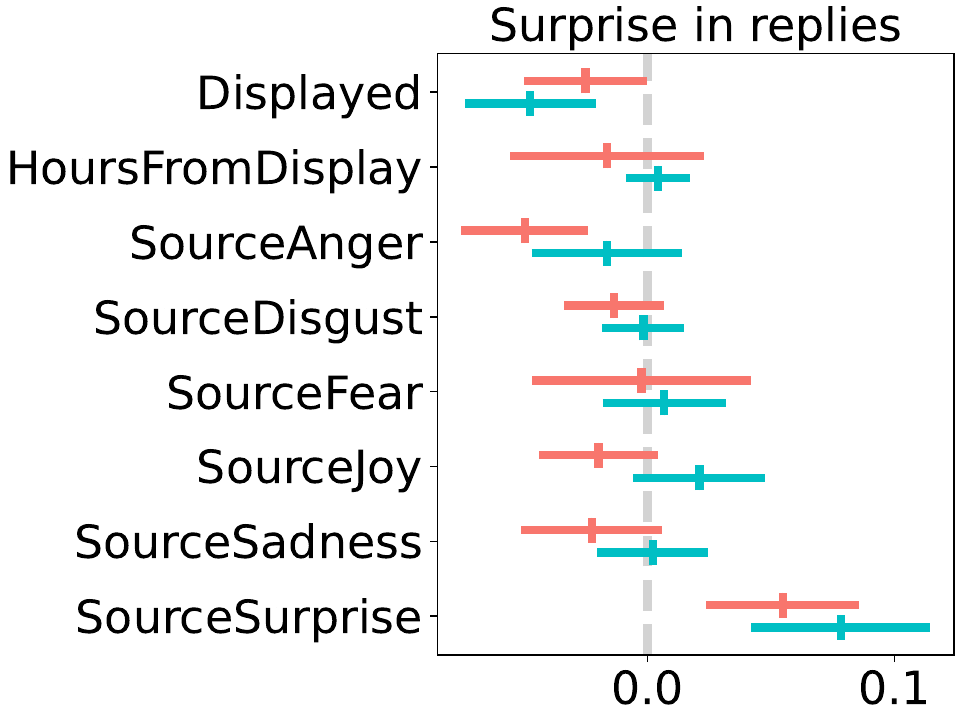}
\label{fig:Surprise_coefs_politics}
\end{subfigure}
\caption{The estimated coefficients for the independent variables -- $\bm{\var{Displayed}}$, $\bm{\var{HoursFromDisplay}}$, and source sentiments (or emotions) across political and non-political misleading posts. The independent variable $\bm{\var{PostAge}}$ is included during estimation but omitted in the visualization for better readability. Shown are mean values with error bars representing 95\% CIs. Standard errors are clustered by source posts. The dependent variables are \subref{fig:Positive_coefs_politics} positive sentiment in replies, \subref{fig:Negative_coefs_politics} negative sentiment in replies, \subref{fig:Anger_coefs_politics} anger in replies, \subref{fig:Disgust_coefs_politics} disgust in replies, \subref{fig:Fear_coefs_politics} fear in replies, \subref{fig:Joy_coefs_politics} joy in replies, \subref{fig:Sadness_coefs_politics} sadness in replies, and \subref{fig:Surprise_coefs_politics} surprise in replies, respectively. The full estimation results are reported in Suppl. \ref{sec:sm_sensitivity_analysis}.}
\label{fig:emotion_coefs_politics}
\end{figure*}

% Treatment effects across political and non-political misleading posts
We conduct sensitivity analysis to examine whether the effects of community notes on sentiments and emotions in replies vary across political and non-political misleading posts. The coefficient estimates for the sensitivity analysis across political and non-political misleading posts are reported in Fig. \ref{fig:emotion_coefs_politics}.

We first examine the sensitivity of the effects of community notes on positive and negative sentiments. In Fig. \ref{fig:Positive_coefs_politics}, the coefficient estimate of $\var{Displayed}$ for positive sentiment in replies to political misleading posts is significant and negative ($\var{coef.} = -0.068$, $p<0.01$; 95\% CI: $[-0.115, -0.022]$). However, the coefficient estimate of $\var{Displayed}$ for positive sentiment in replies to non-political misleading posts is not statistically significant ($\var{coef.} = -0.020$, $p=0.197$; 95\% CI: $[-0.051, 0.010]$). This means that the negative effect of community notes on positive sentiment in replies is limited to political misleading posts. Regarding the negative sentiment in replies (Fig. \ref{fig:Negative_coefs_politics}), the coefficient estimates of $\var{Displayed}$ are significantly positive in both political ($\var{coef.} = 0.116$, $p<0.001$; 95\% CI: $[0.067, 0.164]$) and non-political misleading posts ($\var{coef.} = 0.075$, $p<0.001$; 95\% CI: $[0.042, 0.107]$). This means that the effects of community notes generally apply to both political and non-political misleading posts. 

Next, we examine the sensitivity of the effects of community notes on basic emotions: 
\begin{itemize}[leftmargin=*]
    \item In terms of anger in replies (Fig. \ref{fig:Anger_coefs_politics}), the coefficient estimates of $\var{Displayed}$ are significantly positive in both political ($\var{coef.} = 0.063$, $p<0.001$; 95\% CI: $[0.036, 0.091]$) and non-political misleading posts ($\var{coef.} = 0.070$, $p<0.001$; 95\% CI: $[0.042, 0.099]$). This suggests that the display of community notes triggers more anger in replies to both political and non-political misleading posts.
    \item In Fig. \ref{fig:Disgust_coefs_politics}, the coefficient estimate of $\var{Displayed}$ for disgust in replies to political misleading posts is significantly positive ($\var{coef.} = 0.027$, $p<0.01$; 95\% CI: $[0.008, 0.046]$). However, the coefficient estimate of $\var{Displayed}$ for disgust in replies to non-political misleading posts is statistically not significant ($\var{coef.} = 0.012$, $p=0.473$; 95\% CI: $[-0.020, 0.044]$). This indicates that the effect of community notes on increasing disgust in replies is only applicable to political misleading posts.
    \item As shown in Figs. \ref{fig:Fear_coefs_politics}--\ref{fig:Sadness_coefs_politics}, the coefficient estimates of $\var{Displayed}$ for fear, joy, and sadness in replies to both political and non-political misleading posts are consistently not significant (each $p>0.05$). This means that the display of community notes has no significant effects on fear, joy, and sadness in replies to either political or non-political misleading posts.
    \item In Fig. \ref{fig:Surprise_coefs_politics}, the coefficient estimates of $\var{Displayed}$ for surprise in replies to both political ($\var{coef.} = -0.025$, $p<0.05$; 95\% CI: $[-0.050, -0.001]$) and non-political misleading posts ($\var{coef.} = -0.048$, $p<0.001$; 95\% CI: $[-0.074, -0.022]$) are significantly negative. This means that the display of community notes generally leads to a decrease of surprise in replies to both political and non-political misleading posts.
\end{itemize}

% Emotion transfers across politcal and non-political misleading posts
Additionally, we examine the associations between the sentiments and emotions in replies and the source sentiments and emotions across political and non-political misleading posts. Fig. \ref{fig:Positive_coefs_politics}) shows that the coefficient estimate of $\var{SourcePositive}$ is not statistically significant for positive sentiment in replies to political misleading posts ($\var{coef.} = -0.015$, $p=0.350$; 95\% CI: $[-0.047, 0.017]$), while it is significantly positive for positive sentiment in replies to non-political misleading posts ($\var{coef.} = 0.099$, $p<0.001$; 95\% CI: $[0.062, 0.136]$). Similarly, in Fig. \ref{fig:Joy_coefs_politics}, the coefficient estimate of $\var{SourceJoy}$ is statistically not significant for joy in replies to political misleading posts ($\var{coef.} = 0.008$, $p=0.614$; 95\% CI: $[-0.024, 0.041]$), while it is significantly positive for joy in replies to non-political misleading posts ($\var{coef.} = 0.062$, $p<0.001$; 95\% CI: $[0.031, 0.092]$). This suggests that the positive association between source positive sentiment (joy) in original posts and positive sentiment (joy) in replies disappears for political misleading posts. In contrast, for negative sentiment, anger, disgust, fear, sadness, and surprise in replies, there are consistently positive associations with their corresponding emotions in the source misleading posts, regardless of whether the posts are political or non-political (each $p<0.05$, see Figs. \ref{fig:Negative_coefs_politics}--\ref{fig:Fear_coefs_politics} and Figs. \ref{fig:Sadness_coefs_politics}--\ref{fig:Surprise_coefs_politics}).

% Stability of sentiments and emotions over time across political and non-political misleading posts
Finally, we find that the coefficient estimates of $\var{HoursFromDisplay}$ for all sentiment/emotions in replies to either political or non-political misleading posts are consistently not significant (each $p>0.05$; see Fig. \ref{fig:emotion_coefs_politics}). This suggests that the stability of sentiments and emotions over time remains robust, irrespective of whether the misleading posts are political or non-political.

% Summary for sensitivity analysis
\textbf{Summary of findings:} Taken together, we observe two aspects of our findings that are specific to political misleading posts: (i) The effects of community notes on the decrease of positive sentiment and the increase of disgust in replies only apply to political misleading posts; (ii) The transfer of positive sentiment, particularly joy, from source posts to replies is disrupted in discussions on political topics.

\subsection{Analysis of Moral Outrage (RQ2.1 \& RQ2.2)}
\label{sec:moral_outrage_analysis}

\begin{figure*}
\centering
\begin{subfigure}{0.32\textwidth}
\caption{}
\includegraphics[width=\textwidth]{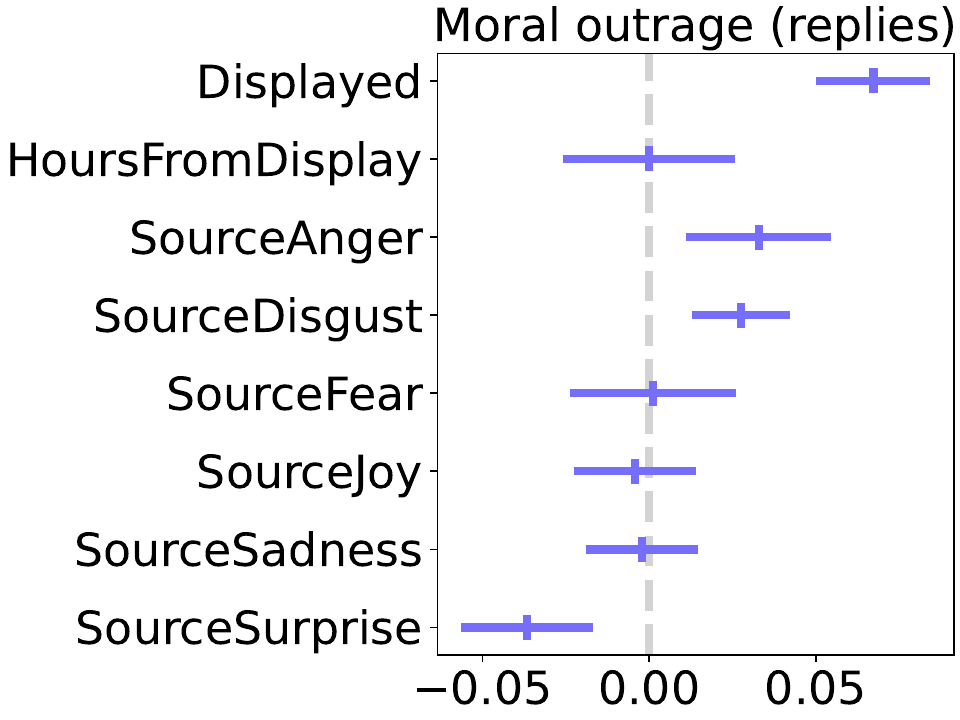}
\label{fig:outrage_coefs}
\end{subfigure}
\hspace{1cm}
\begin{subfigure}{0.32\textwidth}
\caption{}
\includegraphics[width=\textwidth]{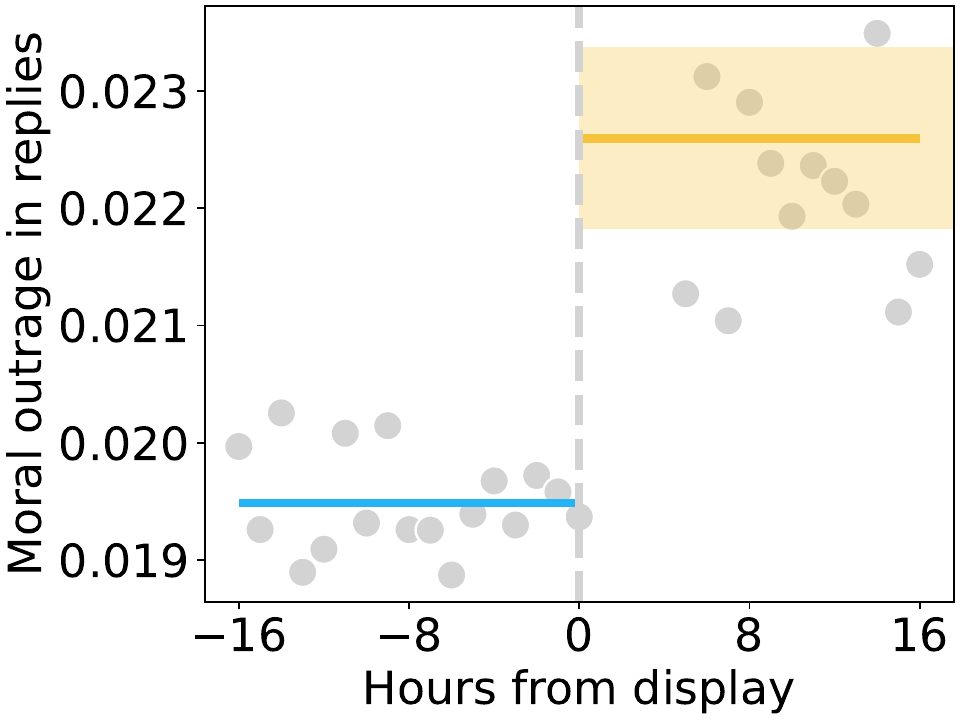}
\label{fig:outrage_hours_from_display}
\end{subfigure}

\begin{subfigure}{0.32\textwidth}
\caption{}
\includegraphics[width=\textwidth]{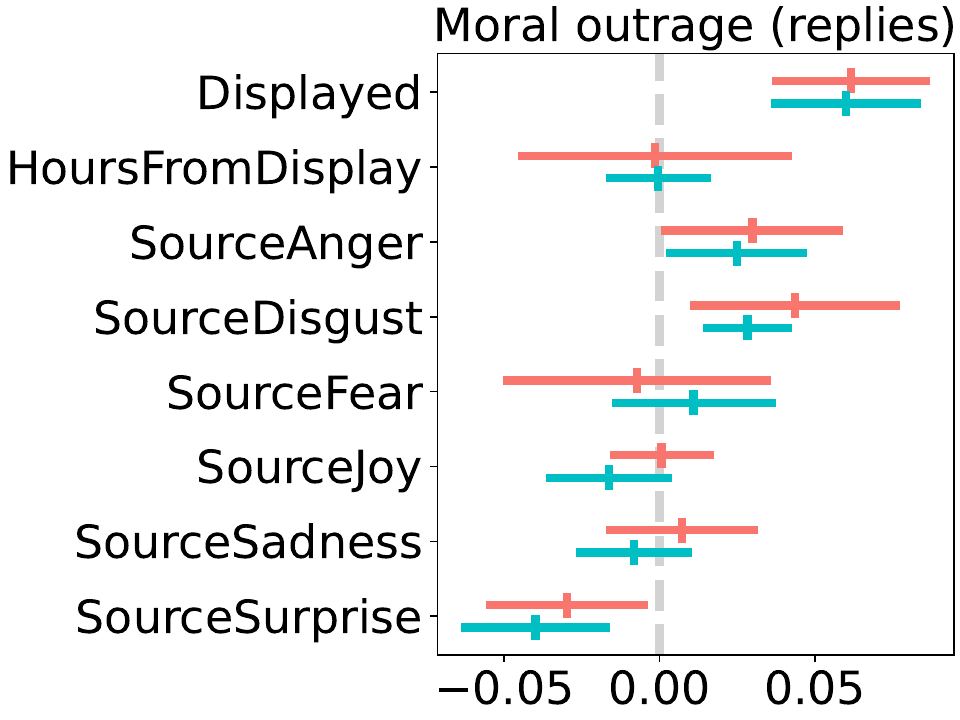}
\label{fig:outrage_coefs_politics}
\end{subfigure}
\hspace{1cm}
\begin{subfigure}{0.32\textwidth}
\caption{}
\includegraphics[width=\textwidth]{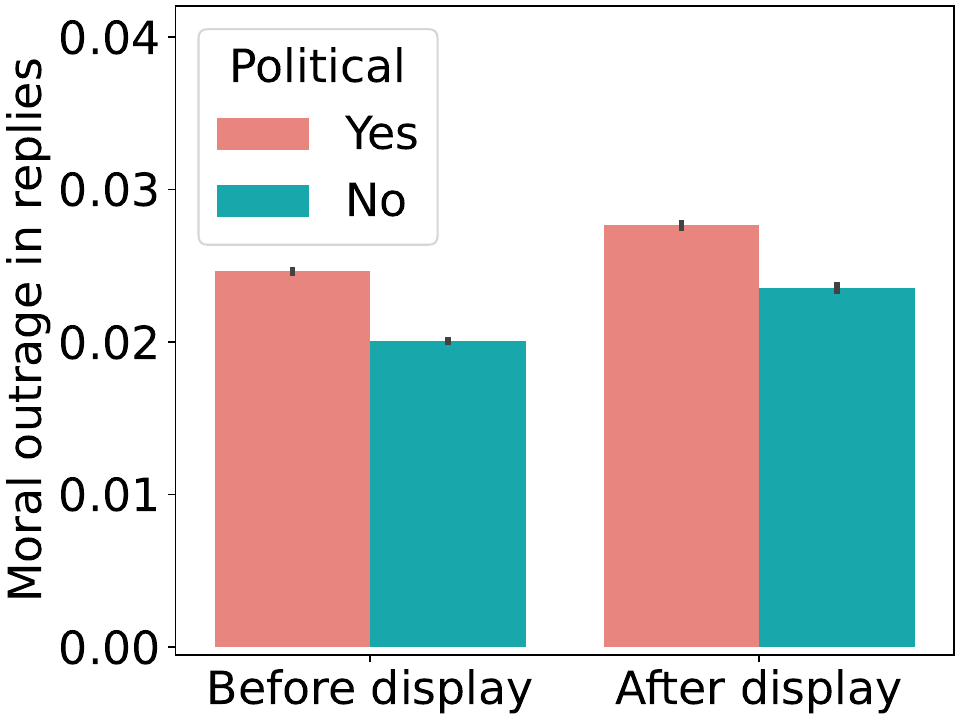}
\label{fig:moral_outrage_bar}
\end{subfigure}

\begin{subfigure}{\textwidth}
\caption{}
\includegraphics[width=\textwidth]{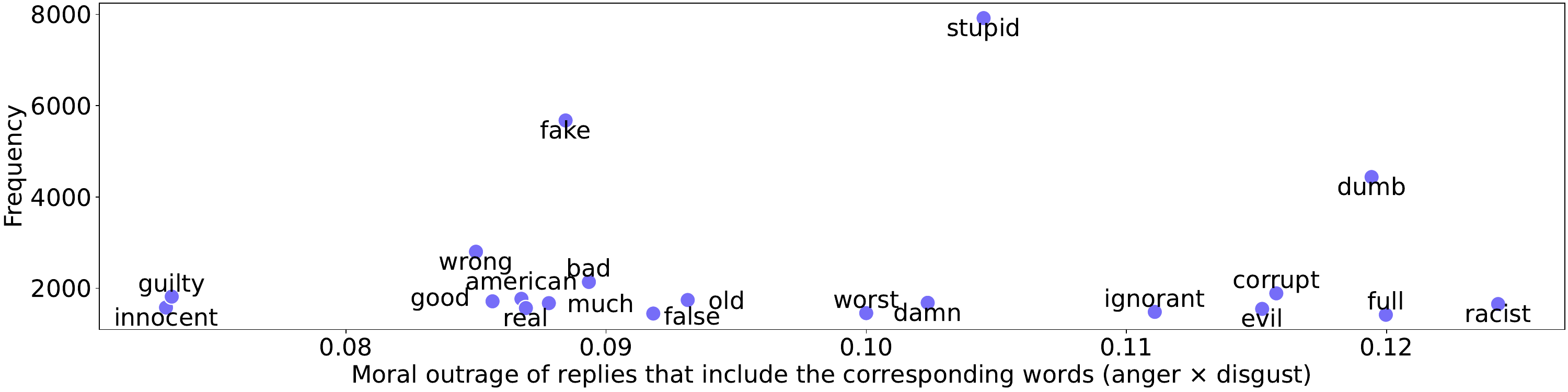}
\label{fig:outrage_word_freq}
\end{subfigure}
\caption{Moral outrage in replies to misleading posts increases after the display of community notes. \subref{fig:outrage_coefs} The estimated coefficients for the independent variables -- $\bm{\var{Displayed}}$, $\bm{\var{HoursFromDisplay}}$, and source emotions. \subref{fig:outrage_hours_from_display} The predicted effect of the display of community notes on moral outrage in replies to misleading posts. The error band represents 95\% CIs. \subref{fig:outrage_coefs_politics} The estimated coefficients for the independent variables -- $\bm{\var{Displayed}}$, $\bm{\var{HoursFromDisplay}}$ across political and non-political misleading posts. \subref{fig:moral_outrage_bar} The averages of moral outrage in replies to political and non-political misleading posts before and after the display of community notes. The error bars represent 95\% CIs. \subref{fig:outrage_word_freq} The twenty most frequent adjectives within replies with high moral outrage. For \subref{fig:outrage_coefs} and \subref{fig:outrage_coefs_politics}, the independent variable $\bm{\var{PostAge}}$ is included during estimation but omitted in the visualization for better readability. Shown are mean values with error bars representing 95\% CIs. Standard errors are clustered by source posts. The full estimation results are reported in Suppl. \ref{sec:sm_moral_outrage}.
}
\label{fig:moral_outrage}
\end{figure*}

Given that moral outrage is typically expressed through anger and disgust, we examine whether the display of community notes triggers the moral outrage of users in their replies to misleading posts. Analogous to previous research \cite{crockett2017moral}, we define the dependent variable moral outrage as $\var{MoralOutrage} = \var{Anger} \times \var{Disgust}$. We first examine the changes of moral outrage in replies to all misleading posts. As shown in Fig. \ref{fig:outrage_coefs}, the coefficient estimate of $\var{Displayed}$ is significantly positive ($coef.=0.067$, $p<0.001$; 95\% CI: $[0.050, 0.084]$). This means that the display of community notes increases moral outrage in replies to misleading posts by 0.003 (95\% CI: $[0.002, 0.004]$). Additionally, using the baseline of moral outrage during a period of 16 hours before note display (mean of 0.019, Fig. \ref{fig:outrage_hours_from_display}), we find that the display of community notes triggers 16\% more moral outrage in replies and increases it to a value of 0.023 (95\% CI: $[0.022, 0.023]$).

\textbf{Robustness with alternative classifiers:} 
Previous research developed a dedicated classifier to predict moral outrage, namely, the Digital Outrage Classifier (DOC) \cite{brady2021social}. As a robustness check, we repeat our analysis with this classifier and find that the coefficient estimate of $\var{Displayed}$ remains significantly positive ($coef.=0.067$, $p<0.001$; 95\% CI: $[0.041, 0.092]$). A $\var{SUEST}$ test indicates no statistically significant difference in the effect of $\var{Displayed}$ between the operationalization of moral outrage in our main analysis and the operationalization of moral outrage via the DOC classifier ($\var{\chi^2}=0.00, p=0.957$). Additionally, anger and disgust are central components of other-condemning moral emotions \cite{solovev2022moral}. Using a dedicated moral emotion classifier developed by prior research \cite{kim2024moral}, we repeat our analysis and find that the effect of $\var{Displayed}$ on other-condemning moral emotions is also significantly positive ($coef.=0.132$, $p<0.001$; 95\% CI: $[0.103, 0.161]$). Altogether, these results consistently support that community notes trigger moral outrage beyond the isolated effects on anger and disgust (see details in Suppl. \ref{sec:sm_moral_outrage}).

\textbf{Analysis across political and non-political misleading posts:}
We further examine the effects of community notes display on moral outrage in replies to political and non-political misleading posts separately. As shown in Fig. \ref{fig:outrage_coefs_politics}, the coefficient estimates of $\var{Displayed}$ are significantly positive within both political ($coef.=0.062$, $p<0.001$; 95\% CI: $[0.037, 0.087]$) and non-political misleading posts ($coef.=0.060$, $p<0.001$; 95\% CI: $[0.036, 0.084]$). This indicates that the effects of community notes in triggering moral outrage are robust across political and non-political misleading posts. Additionally, we analyze the average levels of moral outrage in replies to political and non-political misleading posts before and after the display of community notes (Fig. \ref{fig:moral_outrage_bar}). Before note display, the averages of moral outrage in replies to political and non-political misleading posts are 0.025 and 0.020, respectively ($t=63.709$, $p<0.001$). After note display, these averages increase to 0.028 for political posts and 0.024 for non-political posts ($t=31.042$, $p<0.001$). This suggests that the moral outrage in replies to political misleading posts is significantly higher than that in replies to non-political misleading posts, regardless of the display of community notes. However, the averages of moral outrage in the source posts between political (mean of 0.010) and non-political ones (mean of 0.009; $t=0.575$, $p=0.566$) have no significant difference. This indicates that the heightened moral outrage in replies to political posts, compared to non-political posts, is not simply a transfer from the source posts’ moral outrage.

\textbf{Validation of targets of moral outrage:}
We conduct multiple tests to validate that the measured moral outrage in replies is primarily directed at the misleading posts or their authors (rather than reflecting emotional reactions from users whose ``side'' has been labeled as misleading).

First, we consider two scenarios where replies may not be directed toward original posts: mentions of other users (not authors) in replies and mentions of community notes in replies. If the level of moral outrage in these two scenarios is lower than in replies directed solely toward original posts, this would suggest that the moral outrage in replies is primarily focused on misleading content. We find that the ratio of mentioning other users in replies after note display (10.4\%) is lower than that in replies before note display (12.3\%). Meanwhile, after note display, moral outrage in replies that do not mention others (mean of 0.028) is 180.0\% more than that in replies that mention others (mean of 0.010; $t=-82.018$, $p<0.001$). Additionally, after the display of community notes, replies may discuss community notes rather than the original misleading posts. Given this, we check how many replies mention community notes after their display through the keywords: ``community note'' and ``CN.'' We find that, out of \num{533495} replies created after note display, only \num{2986} (0.6\%) replies mention community notes. Moreover, moral outrage in replies that do not mention community notes (mean of 0.026) is 73.3\% higher than the moral outrage in replies that mention community notes (mean of 0.015; $t=-11.791$, $p<0.001$). Taken together, these validations suggest that the majority of replies, especially those with high moral outrage, are directed toward source misleading posts.

Second, we randomly select 100 misleading posts with displayed community notes and manually evaluate the reply with the highest outrage score after the display of community note for each selected post. Specifically, two authors of this study independently reviewed these 100 replies, arriving at identical results: all 100 replies are consistently directed toward the content of the misleading posts or their authors.

Third, we analyze the most frequently used words in replies that express moral outrage. We extract all words from replies posted after note display, focusing on those with a higher level of moral outrage (\ie, above the mean value). The most frequent word in these replies is ``you,'' often indicating a direct reference to the post author. Subsequently, we remove stop words and use Spacy, a Python library for advanced natural language processing, to extract the twenty most frequent adjectives. Then, we calculate the average level of moral outrage for each word based on the replies that include it. Fig. \ref{fig:outrage_word_freq} shows that ``stupid,'' ``fake,'' and ``dumb'' are the most frequent adjectives in replies expressing moral outrage, with mean values of 0.105, 0.088, and 0.119, respectively. Additionally, words associated with higher levels of moral outrage (greater than 0.100) tend to be more critical (\eg, racist, dumb, and corrupt), while words associated with lower levels of moral outrage (below 0.100) are more related to the veracity of the content (\eg, wrong, fake, and false). This suggests that higher levels of moral outrage are associated with the use of more critical or aggressive words.

\textbf{Summary of findings:} Altogether, our analysis suggests that the display of community notes triggers moral outrage in replies to the corresponding misleading posts. Additionally, political misleading posts tend to receive more moral outrage compared to non-political misleading posts. However, the effects of community notes on moral outrage in replies are robust across political and non-political misleading posts.

\subsection{Robustness Checks}
We conduct a wide range of additional robustness checks to ensure the reliability of our findings:  
\begin{itemize}[leftmargin=*]
    \item \textbf{Multicollinearity check:} We check for possible multicollinearity issues using Variance Inflation Factors (VIFs). The VIFs are all close to one; and, thus well below the critical threshold of four (see details in Suppl. \ref{sec:sm_vifs}).
    \item \textbf{Analysis with lagged-dependent variables:} To address the possibility that users replying to original posts may be influenced by sentiments and emotions in previous direct replies, we conduct an auto-correlation test and repeat our analysis with lagged-dependent variables. The results remain robust, consistently supporting our main findings (see details in Suppl. \ref{sec:sm_autocorr}).
    \item \textbf{Shorter bandwidth and placebo test:} To further ensure that the increase in moral outrage in replies to misleading posts is specific to the display of community notes, we repeat our analysis within a shorter bandwidth, \ie, 2 hours, around the display of community notes and conduct a placebo test at a different cut-off point before note display for comparison. We find that all results are robust and consistent with our main analysis (see details in Suppl. \ref{sec:short_band_placebo}). 
    \item \textbf{Analysis without reply restriction:} We repeat the analysis without the restriction of 5 replies before and after note display. The results remain robust and consistent with our main findings (see details in Suppl. \ref{sec:sm_no5limit}).
    \item \textbf{Month-year fixed effects:} We incorporate month-year fixed effects into our regression models and repeat our analysis. The results remain robust and consistent with our main findings (see details in Suppl. \ref{sec:sm_month_year}).
    \item \textbf{Validation with dictionary-based approaches:} We repeat our analysis using sentiment and emotion lexicons. Specifically, we use the VADER lexicon for sentiments and the NRC lexicon for emotions \cite{mohammad2013crowdsourcing,hutto2014vader}. The results remain robust across machine learning models and lexicons (see details in Suppl. \ref{sec:sm_alternatives}).
    \item \textbf{Role of sentiments and emotions in community notes:} To explore whether sentiments and emotions in community notes can potentially moderate the effects of community notes display, we incorporate sentiments and emotions in community notes and their interactions with $\var{Displayed}$ into our regression models. The coefficient estimates of $\var{Displayed}$ across the dependent variables remain robust. Additionally, the majority of the coefficient estimates for the interactions between sentiments (emotions) in community notes and $\var{Displayed}$ are not statistically significant. This suggests that users' emotional reactions are not significantly affected by the sentiments or emotions in community notes (see details in Suppl. \ref{sec:sm_note_emotions}).
    \item \textbf{Robustness across helpfulness scores:} Misleading posts that garner significant public concern may be more likely to be quoted by influential accounts on X, thereby reaching a broader audience. This increased visibility may introduce confounding factors related to the helpfulness scores of community notes. For example, counter-partisan accounts can encourage other users to rate a community note as helpful, thereby increasing the helpfulness score of the note \cite{allen2022birds}. To account for such potential confounding factors, we incorporate $\var{NoteScore}$ and its interaction with $\var{Displayed}$ into our regression models. We find that the effect of community notes is robust across different helpfulness scores (see details in Suppl. \ref{sec:sm_note_score}).
    \item \textbf{Effect of community notes over the program's development:} Given that the Community Notes program continuously evolves, we incorporate $\var{MFRO}$ (\ie, months from the roll-out of Community Notes program) and its interaction with $\var{Displayed}$ into regression models to investigate how the effect of community notes changes over time from the roll-out of the program. We find that the effect of community notes on our dependent variables remains stable (see details in Suppl. \ref{sec:sm_mfro}).
\end{itemize}

% Summary of main findings
\subsection{Summary of Main Findings}
In summary, we conduct a comprehensive \emph{causal} analysis on the role of community notes in triggering sentiments, emotions, and moral outrage in replies to fact-checked social media posts. Furthermore, we extend our analysis by conducting sensitivity analysis across political and non-political misleading posts. Our main findings are as follows: 
\begin{itemize}[leftmargin=*]
    \item The display of community notes triggers 7.3\% more negativity, 13.2\% more anger, and 4.7\% more disgust in replies to misleading posts (RQ1.1).
    \item The effects of community notes on negativity and anger are robust across political and non-political misleading posts, while the effect on disgust only applies to political misleading posts (RQ1.2).
    \item The display of community notes triggers 16\% more moral outrage (RQ2.1), and this is robust across political and non-political misleading posts (RQ2.2).
\end{itemize}
In addition to our main findings, we find the evidence of emotion transfers and time-independence for sentiments and emotions in online social networks:
\begin{itemize}[leftmargin=*]
    \item The sentiments and emotions in replies to misleading posts are positively correlated with the sentiments and emotions of the corresponding source posts (with very few exceptions).
    \item The sentiments and emotions in replies are stable over time if the posts are not subject to fact-checking (\ie, do not receive a displayed community note). 
\end{itemize}

\section{Discussion}

\subsection{Relevance}
The spread of online misinformation has become a significant challenge that social media platforms have to tackle. However, traditional (expert-based) fact-checking approaches have limitations regarding their scalability and face trust issues among the user base. As a remedy, crowdsourced fact-checking represents a promising approach to identify misinformation at scale and increase users' trust in fact-checks. As the first large-scale attempt of crowdsourced fact-checking approach on a major social media platform, the ``Community Notes'' program has been shown to be effective in producing trustworthy community fact-checks and reducing engagement with misleading posts on X \cite{chuai2024community,drolsbach2024community}. It provides a reference for other social media platforms to implement similar crowdsourced fact-checking approaches (\eg, YouTube \cite{youtube2024testing}). However, to improve and extend the ``Community Notes'' program safely, it is important to understand how community notes affect the discussion environment after they are displayed on misleading posts, \ie, how users' emotional expressions change after they are informed about the falsehood. To this end, our study causally examines the changes in emotional expressions in replies to misleading posts before and after the display of community notes. As detailed in the next sections, our findings provide important insights into the effects of (community-based) fact-checking on the discussion environment on social media and may help to improve the design of crowdsourced fact-checking platforms. 

\subsection{Interpretations}
% Community notes increase anger and disgust and trigger moral outrage in replies to misleading posts.
\subsubsection{Community notes trigger negativity, anger, and disgust in replies to misleading posts.}
We find that the display of community notes leads users to post more negative replies, which is primarily due to heightened expressions of anger and disgust. These observations align with previous survey and lab studies in the field of psychology that people often report anger and disgust in response to moral violations of community, autonomy, and divinity ethics \cite{molho2017disgust,van2023expressions}. In the context of our study, the display of community notes signals to users that the posts they are exposed to spread misinformation, which may thereby trigger their perception of a moral violation. Anger and disgust, though both negative, differ in their aggressive tendencies when responding to moral violations. Anger is typically associated with high-cost, direct aggression (\eg, physical violence or confrontational responses), whereas disgust is typically associated with less costly indirect aggression (\eg, gossip and social exclusion). Additionally, previous research suggests that people tend to express more anger than disgust when moral violations directly affect themselves, while they express more disgust than anger when moral violations target others \cite{molho2017disgust}. Quantitatively, our findings reveal that the display of community notes increases anger in replies to misleading posts by 13.2\%, compared to a 4.7\% increase in disgust. Thus, users are more likely to express anger than disgust, which may indicate that users perceive the misleading content to primarily impact their own interests (rather than others'). Notably, our sensitivity analysis further reveals that the effect of community notes in increasing disgust is statistically significant only in replies to political misleading posts. This could partially explain why the display of community notes is less effective in reducing the spread of political misleading posts, compared to non-political ones, which has been observed in a previous study \cite{chuai2024community,chuai2024community.new}.
% needs to be further validated by more rigorous work in the future. 

\subsubsection{Community notes trigger moral outrage in replies to misleading posts.} 
When anger and disgust are mixed, they generate ``moral outrage,'' a more intense emotional reaction to perceived moral transgressions than either anger or disgust alone \cite{salerno2013interactive,crockett2017moral,brady2021social}. Our findings show that the display of community notes increases moral outrage by 16\% in replies to misleading posts, a greater increase than those observed for anger (13.2\%) and disgust (4.7\%) individually. This empirical evidence underscores the significant role of moral outrage in users' responses to the moral transgression of spreading misleading posts. Specifically, our findings suggest that displaying community notes on misleading posts is a strong stimulus that calls attention to the moral norm violation, thereby motivating expressions of moral outrage. Notably, research suggests that if users experience moral violations personally without such external stimuli, their intentions to express moral outrage might decrease \cite{crockett2017moral}. This phenomenon might be explained by social conformity that users are inclined to conform to the majority’s opinions on the quality of online posts \cite{wijenayake2020effect,colliander2019fake}. The display of community notes signals consensus among diverse users, thereby potentially encouraging exposed users' conformity and subsequent correcting actions.

The expression of moral outrage to spreading misinformation is a double-edged sword that can both mitigate the harm of online misinformation and exacerbate the polarization of online communities. On the one hand, moral outrage can strengthen debunking effects on exposed users and pressure authors to correct or delete their misleading posts \cite{peng2023rage,chuai2024community}. Previous research has shown that the display of community notes significantly increases the likelihood of deleting misleading posts by their authors \cite{chuai2024community}. Moreover, those who attempt to spread misinformation can expect strong negative reactions from the user base once their content is debunked, which may help to discipline users and encourage more responsible behavior on social media. On the other hand, social media platforms also need to maintain a healthy, inclusive, and balanced online environment. However, online moral outrage can deepen social divides and fuel political polarization \cite{crockett2017moral,brady2021social,ecker2024misinformation}. Additionally, the corrections and moral judgments from peer users may backfire and make authors of misleading posts feel defensive, promoting their subsequent shares of low-quality, partisan, and toxic content \cite{mosleh2021perverse}. In this light, while moral outrage can be a powerful tool for combating misinformation, it also risks escalating conflicts and entrenching divisions within online communities. This risk may partially explain why community notes lead to smaller reductions in replies compared to reposts \cite{renault2024collaboratively}. The potential negative effects of moral outrage triggered by community notes require a further rigorous examination to find a trade-off between mitigating the spread of misinformation and maintaining a positive discussion atmosphere.

\subsubsection{Emotion transfers and stable emotional responses over time.} 
Our study provides strong empirical evidence supporting the idea that connected users on social media platforms express similar sentiments/emotions. Specifically, we find that all sentiments and emotions in misleading posts, especially non-political ones, are effectively transferred to subsequent sentiments and emotions in replies. This emotion transfer phenomenon can be explained primarily through two mechanisms: emotion contagion and emotion alignment (\ie, homophily) among connected users. Here, emotion contagion refers to the spreading effect of emotions from one user to another \cite{rosenbusch2019multilevel,goldenberg2020digital}, and emotion homophily indicates that like-minded users tend to flock together, potentially forming echo chambers that dominate online interactions and reinforce biases and attitude polarization \cite{cinelli2021echo}. Our findings further reveal that the intensity of sentiments and emotions in replies remains stable over time, showing no signs of fading. Moreover, misleading posts fact-checked via community notes are predominantly characterized by negativity. The persistence of emotion transfers may further exacerbate these dynamics \cite{zollo2015emotional,del2016echo,chuai2022anger}.

\subsection{Implications for Further Design of Community-Based Fact-Checking}

Community-based fact-checking is a promising approach and can effectively reduce engagement with misinformation on social media \cite{allen2021scaling,jahanbakhsh2024browser,zhang2024profiling,chuai2024community,drolsbach2024community}. However, our study finds that communicating falsehoods to users via community fact-checks can also trigger moral outrage in response to the perceived moral violation of spread misinformation. This heightened moral outrage can have dual impacts. On the one hand, it may strengthen the effectiveness of community-based fact-checking by pressuring (disciplining) users to reconsider their actions. On the other hand, however, it may also backfire by increasing tension and polarization on social media \cite{mosleh2021perverse}. 

The dual role of moral outrage highlights the need to minimize its negative consequences without undermining its positive effects in the design and implementation of community-based fact-checking to maintain health and constructive dialogue on social media. Previous research suggests that temporarily deactivating social media accounts can reduce polarization on policy issues \cite{rathje2021out}. Additionally, a ``cooldown'' period for highly active and polarized threads has been proposed to ensure a deliberatively interactive environment \cite{efstratiou2022adherence}. Given this, social media platforms might consider implementing a \emph{temporary freezing period} for misleading posts once they receive helpful community fact-checks. During this period -- \eg, one hour -- the misleading posts would be invisible to others, while the authors are notified, giving them time to delete or correct the misleading content. If they do not respond within the given timeframe, the community fact-checks with associated misleading posts would then be publicly displayed on the platform. Similarly, platforms may consider \emph{temporarily limiting the ability to comment} on posts flagged by community fact-checks. This could take the form of a complete ban on posting responses during this period or a personal delay, where written comments are visible to others only after a certain period. In general, platforms may need to emphasize \emph{community guidelines} in heated/morally charged discussions (\eg, labels/brief prompts that appear when a user is about to post a highly emotional response; or additional monitoring by moderation teams) \cite{ribeiro2024post}. 

Importantly, however, any such design changes must be carefully evaluated through further research to assess their potential benefits and drawbacks. A core motivation behind community-based fact-checking is to reduce censorship on social media platforms and empower users with greater autonomy to improve trust in the fact-checking process \cite{drolsbach2024community}. Therefore, platforms must strike a delicate balance between fostering constructive discourse and preserving user autonomy to sustain trust and engagement in community-based fact-checking.

\subsection{Limitations and Future Research}
Our study has several limitations that could open up potential avenues for future research. First, our study is limited to a 4-month period since the roll-out of the ``Community Notes'' program. As the program is continually evolving, it would be valuable to examine the effects of community notes on sentiments, emotions, and moral outrage over a longer observation timeframe.\footnote{Notably, we collected our dataset via X's Academic Research API before it was deprecated. Currently, it is costly to collect such a high volume of data from X.} Furthermore, while we conduct extensive robustness checks to ensure the reliability of our findings, it is challenging for a quasi-experimental study to fully isolate the effect of community notes from other potential concurrent misinformation interventions or external factors. Therefore, future research is encouraged to validate our findings within controlled experimental designs.

Second, on the date of data collection, the ``Community Notes'' program was primarily well-established in the \US, and the fact-checked posts were mainly in English. Consequently, we only consider English posts in our study. In the future, a multilingual and cross-cultural evaluation would be beneficial to get a more comprehensive understanding of how community notes impact users' emotional reactions across languages and cultures. Additionally, our sensitivity analysis is limited to general political and non-political misleading posts. Future research could expand this scope by examining more fine-grained topics (\eg, health, business \& economics, celebrities, etc.).

Third, our measurement of moral outrage cannot definitively ascertain whether it is directed toward misleading posts or their authors. To validate our approach, we analyze two special cases: mentions of other users and mentions of community notes in replies. We find that the moral outrage in replies that mention other users or community notes is significantly lower than the moral outrage in replies that solely react to the misleading posts and do not mention other users or community notes. Additionally, replies with high moral outrage frequently use aggressive words (\eg, ``stupid,'' ``dumb''). All of these analyses support the notion that community notes trigger moral outrage and critical remarks to misleading posts or their authors. Future research could further validate our findings through more advanced subject-aware emotion detection models.

Finally, it has been found that the display of community notes increases the probability of post deletion by authors \cite{chuai2024community.new}. A possible contributing factor may be that expressions of moral outrage pressure authors to delete their misleading posts. This notion could be validated through the examination of moral outrage in replies to deleted and non-deleted posts. However, deleted misleading posts and corresponding replies cannot be collected through X's API. As a remedy, future research could conduct lab experiments to causally examine whether expressions of moral outrage increase authors' intentions to delete their misleading posts. Additionally, it would be interesting to further examine how the authors of misleading posts adapt their future behaviors on social media.

\section{Conclusion}

Community-based fact-checking is a promising approach to reduce engagement with misinformation on social media. Yet, an understanding of how community fact-checks affect users' emotional perception of misleading posts and their authors was missing. In this study, we analyze a large-scale panel dataset of replies to misleading posts that have been fact-checked via community notes. Our findings provide strong causal evidence that community fact-checks trigger moral outrage in replies to misleading posts. This suggests that social media users perceive spreading misinformation as moral transgressions, and that spreaders of misinformation should expect negative responses once their content is debunked. Our study highlights the need for further examination of both the positive and negative consequences of such emotional backlash.

\section{Ethics Statement}
This research has received ethical approval from the Ethics Review Panel of the University of Luxembourg (ref. ERP 23-053 REMEDIS). 
All analyses are based on publicly available data. To respect privacy, we explicitly do not publish usernames in our paper and only report aggregate results. We declare no competing interests.

\begin{acks}
This research is supported by the Luxembourg National Research Fund (FNR) and Belgian National Fund for Scientific Research (FNRS), as part of the project REgulatory Solutions to MitigatE DISinformation (REMEDIS), ref. INTER\_FNRS\_21\_16554939\_REMEDIS. Furthermore, this research is supported by a research grant from the German Research Foundation (DFG grant 492310022). 
\end{acks}

\bibliographystyle{ACM-Reference-Format}
\bibliography{refs}


\begin{thebibliography}{133}

%%% the Web address.

\ifx \showCODEN    \undefined \def \showCODEN     #1{\unskip}     \fi
\ifx \showDOI      \undefined \def \showDOI       #1{#1}\fi
\ifx \showISBNx    \undefined \def \showISBNx     #1{\unskip}     \fi
\ifx \showISBNxiii \undefined \def \showISBNxiii  #1{\unskip}     \fi
\ifx \showISSN     \undefined \def \showISSN      #1{\unskip}     \fi
\ifx \showLCCN     \undefined \def \showLCCN      #1{\unskip}     \fi
\ifx \shownote     \undefined \def \shownote      #1{#1}          \fi
\ifx \showarticletitle \undefined \def \showarticletitle #1{#1}   \fi
\ifx \showURL      \undefined \def \showURL       {\relax}        \fi
% invisible to TeX
\providecommand\bibfield[2]{#2}
\providecommand\bibinfo[2]{#2}
\providecommand\natexlab[1]{#1}
\providecommand\showeprint[2][]{arXiv:#2}

\bibitem[Allcott and Gentzkow(2017)]%
        {allcott2017social}
\bibfield{author}{\bibinfo{person}{Hunt Allcott} {and} \bibinfo{person}{Matthew Gentzkow}.} \bibinfo{year}{2017}\natexlab{}.
\newblock \showarticletitle{Social media and fake news in the 2016 election}.
\newblock \bibinfo{journal}{\emph{Journal of Economic Perspectives}} \bibinfo{volume}{31}, \bibinfo{number}{2} (\bibinfo{year}{2017}), \bibinfo{pages}{211--236}.
\newblock


\bibitem[Allen et~al\mbox{.}(2021)]%
        {allen2021scaling}
\bibfield{author}{\bibinfo{person}{Jennifer Allen}, \bibinfo{person}{Antonio~A Arechar}, \bibinfo{person}{Gordon Pennycook}, {and} \bibinfo{person}{David~G Rand}.} \bibinfo{year}{2021}\natexlab{}.
\newblock \showarticletitle{Scaling up fact-checking using the wisdom of crowds}.
\newblock \bibinfo{journal}{\emph{Science Advances}} \bibinfo{volume}{7}, \bibinfo{number}{36} (\bibinfo{year}{2021}), \bibinfo{pages}{eabf4393}.
\newblock


\bibitem[Allen et~al\mbox{.}(2022)]%
        {allen2022birds}
\bibfield{author}{\bibinfo{person}{Jennifer Allen}, \bibinfo{person}{Cameron Martel}, {and} \bibinfo{person}{David~G Rand}.} \bibinfo{year}{2022}\natexlab{}.
\newblock \showarticletitle{Birds of a feather don’t fact-check each other: Partisanship and the evaluation of news in Twitter’s Birdwatch crowdsourced fact-checking program}. In \bibinfo{booktitle}{\emph{Proceedings of the 2022 CHI Conference on Human Factors in Computing Systems}}. \bibinfo{pages}{1--19}.
\newblock


\bibitem[Altay et~al\mbox{.}(2023)]%
        {altay2023misinformation}
\bibfield{author}{\bibinfo{person}{Sacha Altay}, \bibinfo{person}{Manon Berriche}, {and} \bibinfo{person}{Alberto Acerbi}.} \bibinfo{year}{2023}\natexlab{}.
\newblock \showarticletitle{Misinformation on misinformation: {Conceptual} and methodological challenges}.
\newblock \bibinfo{journal}{\emph{Social Media + Society}} \bibinfo{volume}{9}, \bibinfo{number}{1} (\bibinfo{year}{2023}), \bibinfo{pages}{20563051221150412}.
\newblock


\bibitem[Angrist and Pischke(2009)]%
        {angrist2009mostly}
\bibfield{author}{\bibinfo{person}{Joshua~D Angrist} {and} \bibinfo{person}{J{\"o}rn-Steffen Pischke}.} \bibinfo{year}{2009}\natexlab{}.
\newblock \bibinfo{booktitle}{\emph{Mostly harmless econometrics: An empiricist's companion}}.
\newblock \bibinfo{publisher}{Princeton University Press}.
\newblock


\bibitem[Aral and Eckles(2019)]%
        {aral2019protecting}
\bibfield{author}{\bibinfo{person}{Sinan Aral} {and} \bibinfo{person}{Dean Eckles}.} \bibinfo{year}{2019}\natexlab{}.
\newblock \showarticletitle{Protecting elections from social media manipulation}.
\newblock \bibinfo{journal}{\emph{Science}} \bibinfo{volume}{365}, \bibinfo{number}{6456} (\bibinfo{year}{2019}), \bibinfo{pages}{858--861}.
\newblock


\bibitem[Berger and Milkman(2012)]%
        {berger2012makes}
\bibfield{author}{\bibinfo{person}{Jonah Berger} {and} \bibinfo{person}{Katherine~L Milkman}.} \bibinfo{year}{2012}\natexlab{}.
\newblock \showarticletitle{What makes online content viral?}
\newblock \bibinfo{journal}{\emph{Journal of Marketing Research}} \bibinfo{volume}{49}, \bibinfo{number}{2} (\bibinfo{year}{2012}), \bibinfo{pages}{192--205}.
\newblock


\bibitem[Bhuiyan et~al\mbox{.}(2020)]%
        {bhuiyan2020investigating}
\bibfield{author}{\bibinfo{person}{Md~Momen Bhuiyan}, \bibinfo{person}{Amy~X Zhang}, \bibinfo{person}{Connie~Moon Sehat}, {and} \bibinfo{person}{Tanushree Mitra}.} \bibinfo{year}{2020}\natexlab{}.
\newblock \showarticletitle{Investigating differences in crowdsourced news credibility assessment: {Raters}, tasks, and expert criteria}.
\newblock \bibinfo{journal}{\emph{Proceedings of the ACM on Human-Computer Interaction}} \bibinfo{volume}{4}, \bibinfo{number}{CSCW2} (\bibinfo{year}{2020}), \bibinfo{pages}{1--26}.
\newblock


\bibitem[Brady et~al\mbox{.}(2021)]%
        {brady2021social}
\bibfield{author}{\bibinfo{person}{William~J Brady}, \bibinfo{person}{Killian McLoughlin}, \bibinfo{person}{Tuan~N Doan}, {and} \bibinfo{person}{Molly~J Crockett}.} \bibinfo{year}{2021}\natexlab{}.
\newblock \showarticletitle{How social learning amplifies moral outrage expression in online social networks}.
\newblock \bibinfo{journal}{\emph{Science Advances}} \bibinfo{volume}{7}, \bibinfo{number}{33} (\bibinfo{year}{2021}), \bibinfo{pages}{eabe5641}.
\newblock


\bibitem[Brady et~al\mbox{.}(2023)]%
        {brady2023overperception}
\bibfield{author}{\bibinfo{person}{William~J Brady}, \bibinfo{person}{Killian~L McLoughlin}, \bibinfo{person}{Mark~P Torres}, \bibinfo{person}{Kara~F Luo}, \bibinfo{person}{Maria Gendron}, {and} \bibinfo{person}{MJ Crockett}.} \bibinfo{year}{2023}\natexlab{}.
\newblock \showarticletitle{Overperception of moral outrage in online social networks inflates beliefs about intergroup hostility}.
\newblock \bibinfo{journal}{\emph{Nature Human Behaviour}} \bibinfo{volume}{7}, \bibinfo{number}{6} (\bibinfo{year}{2023}), \bibinfo{pages}{917--927}.
\newblock


\bibitem[Brady et~al\mbox{.}(2017)]%
        {brady2017emotion}
\bibfield{author}{\bibinfo{person}{William~J Brady}, \bibinfo{person}{Julian~A Wills}, \bibinfo{person}{John~T Jost}, \bibinfo{person}{Joshua~A Tucker}, {and} \bibinfo{person}{Jay~J Van~Bavel}.} \bibinfo{year}{2017}\natexlab{}.
\newblock \showarticletitle{Emotion shapes the diffusion of moralized content in social networks}.
\newblock \bibinfo{journal}{\emph{Proceedings of the National Academy of Sciences}} \bibinfo{volume}{114}, \bibinfo{number}{28} (\bibinfo{year}{2017}), \bibinfo{pages}{7313--7318}.
\newblock


\bibitem[Brannon and Roy(2024)]%
        {brannon2024speed}
\bibfield{author}{\bibinfo{person}{William Brannon} {and} \bibinfo{person}{Deb Roy}.} \bibinfo{year}{2024}\natexlab{}.
\newblock \showarticletitle{The speed of news in Twitter (X) versus radio}.
\newblock \bibinfo{journal}{\emph{Scientific Reports}} \bibinfo{volume}{14}, \bibinfo{number}{1} (\bibinfo{year}{2024}), \bibinfo{pages}{11939}.
\newblock


\bibitem[Budak et~al\mbox{.}(2024)]%
        {budak2024misunderstanding}
\bibfield{author}{\bibinfo{person}{Ceren Budak}, \bibinfo{person}{Brendan Nyhan}, \bibinfo{person}{David~M Rothschild}, \bibinfo{person}{Emily Thorson}, {and} \bibinfo{person}{Duncan~J Watts}.} \bibinfo{year}{2024}\natexlab{}.
\newblock \showarticletitle{Misunderstanding the harms of online misinformation}.
\newblock \bibinfo{journal}{\emph{Nature}} \bibinfo{volume}{630}, \bibinfo{number}{8015} (\bibinfo{year}{2024}), \bibinfo{pages}{45--53}.
\newblock


\bibitem[Butt et~al\mbox{.}(2022)]%
        {butt2022goes}
\bibfield{author}{\bibinfo{person}{Sabur Butt}, \bibinfo{person}{Shakshi Sharma}, \bibinfo{person}{Rajesh Sharma}, \bibinfo{person}{Grigori Sidorov}, {and} \bibinfo{person}{Alexander Gelbukh}.} \bibinfo{year}{2022}\natexlab{}.
\newblock \showarticletitle{What goes on inside rumour and non-rumour tweets and their reactions: A psycholinguistic analyses}.
\newblock \bibinfo{journal}{\emph{Computers in Human Behavior}}  \bibinfo{volume}{135} (\bibinfo{year}{2022}), \bibinfo{pages}{107345}.
\newblock


\bibitem[Bär et~al\mbox{.}(2023)]%
        {bar2023new}
\bibfield{author}{\bibinfo{person}{Dominik Bär}, \bibinfo{person}{Nicolas Pröllochs}, {and} \bibinfo{person}{Stefan Feuerriegel}.} \bibinfo{year}{2023}\natexlab{}.
\newblock \showarticletitle{New threats to society from free-speech social media platforms}.
\newblock \bibinfo{journal}{\emph{Commun. ACM}} \bibinfo{volume}{66}, \bibinfo{number}{10} (\bibinfo{year}{2023}), \bibinfo{pages}{37--40}.
\newblock


\bibitem[Camacho-collados et~al\mbox{.}(2022)]%
        {camacho2022tweetnlp}
\bibfield{author}{\bibinfo{person}{Jose Camacho-collados}, \bibinfo{person}{Kiamehr Rezaee}, \bibinfo{person}{Talayeh Riahi}, \bibinfo{person}{Asahi Ushio}, \bibinfo{person}{Daniel Loureiro}, \bibinfo{person}{Dimosthenis Antypas}, \bibinfo{person}{Joanne Boisson}, \bibinfo{person}{Luis Espinosa~Anke}, \bibinfo{person}{Fangyu Liu}, {and} \bibinfo{person}{Eugenio Mart{\'\i}nez~C{\'a}mara}.} \bibinfo{year}{2022}\natexlab{}.
\newblock \showarticletitle{{T}weet{NLP}: Cutting-edge natural language processing for social media}. In \bibinfo{booktitle}{\emph{Proceedings of the 2022 Conference on Empirical Methods in Natural Language Processing: System Demonstrations}}. \bibinfo{pages}{38--49}.
\newblock


\bibitem[Cattaneo and Titiunik(2022)]%
        {cattaneo2022regression}
\bibfield{author}{\bibinfo{person}{Matias~D Cattaneo} {and} \bibinfo{person}{Rocio Titiunik}.} \bibinfo{year}{2022}\natexlab{}.
\newblock \showarticletitle{Regression discontinuity designs}.
\newblock \bibinfo{journal}{\emph{Annual Review of Economics}} \bibinfo{volume}{14}, \bibinfo{number}{1} (\bibinfo{year}{2022}), \bibinfo{pages}{821--851}.
\newblock


\bibitem[Chuai et~al\mbox{.}(2022)]%
        {chuai2022really}
\bibfield{author}{\bibinfo{person}{Yuwei Chuai}, \bibinfo{person}{Yutian Chang}, {and} \bibinfo{person}{Jichang Zhao}.} \bibinfo{year}{2022}\natexlab{}.
\newblock \showarticletitle{What really drives the spread of COVID-19 Tweets: A revisit from perspective of content}. In \bibinfo{booktitle}{\emph{2022 IEEE 9th International Conference on Data Science and Advanced Analytics}}. \bibinfo{pages}{1--10}.
\newblock


\bibitem[Chuai et~al\mbox{.}(2024a)]%
        {chuai2024community}
\bibfield{author}{\bibinfo{person}{Yuwei Chuai}, \bibinfo{person}{Moritz Pilarski}, \bibinfo{person}{Gabriele Lenzini}, {and} \bibinfo{person}{Nicolas Pr{\"o}llochs}.} \bibinfo{year}{2024}\natexlab{a}.
\newblock \showarticletitle{Community notes reduce the spread of misleading posts on {X}}.
\newblock \bibinfo{journal}{\emph{OSF}} (\bibinfo{year}{2024}).
\newblock


\bibitem[Chuai et~al\mbox{.}(2024b)]%
        {chuai2024community.new}
\bibfield{author}{\bibinfo{person}{Yuwei Chuai}, \bibinfo{person}{Moritz Pilarski}, \bibinfo{person}{Thomas Renault}, \bibinfo{person}{David Restrepo-Amariles}, \bibinfo{person}{Aurore Troussel-Cl{\'e}ment}, \bibinfo{person}{Gabriele Lenzini}, {and} \bibinfo{person}{Nicolas Pr{\"o}llochs}.} \bibinfo{year}{2024}\natexlab{b}.
\newblock \showarticletitle{Community-based fact-checking reduces the spread of misleading posts on social media}.
\newblock \bibinfo{journal}{\emph{ArXiv}} (\bibinfo{year}{2024}).
\newblock


\bibitem[Chuai et~al\mbox{.}(2024c)]%
        {chuai2024roll}
\bibfield{author}{\bibinfo{person}{Yuwei Chuai}, \bibinfo{person}{Haoye Tian}, \bibinfo{person}{Nicolas Pröllochs}, {and} \bibinfo{person}{Gabriele Lenzini}.} \bibinfo{year}{2024}\natexlab{c}.
\newblock \showarticletitle{Did the roll-out of {Community} {Notes} reduce engagement with misinformation on {X/Twitter}?}
\newblock \bibinfo{journal}{\emph{Proceedings of the ACM on Human-Computer Interaction}} \bibinfo{volume}{8}, \bibinfo{number}{CSCW2} (\bibinfo{year}{2024}), \bibinfo{pages}{1--52}.
\newblock


\bibitem[Chuai and Zhao(2022)]%
        {chuai2022anger}
\bibfield{author}{\bibinfo{person}{Yuwei Chuai} {and} \bibinfo{person}{Jichang Zhao}.} \bibinfo{year}{2022}\natexlab{}.
\newblock \showarticletitle{Anger can make fake news viral online}.
\newblock \bibinfo{journal}{\emph{Frontiers in Physics}}  \bibinfo{volume}{10} (\bibinfo{year}{2022}), \bibinfo{pages}{970174}.
\newblock


\bibitem[Chuai et~al\mbox{.}(2024d)]%
        {chuai2024news}
\bibfield{author}{\bibinfo{person}{Yuwei Chuai}, \bibinfo{person}{Jichang Zhao}, {and} \bibinfo{person}{Gabriele Lenzini}.} \bibinfo{year}{2024}\natexlab{d}.
\newblock \showarticletitle{From news sharers to post viewers: How topic diversity and conspiracy theories shape engagement with misinformation during a health crisis?}
\newblock \bibinfo{journal}{\emph{ArXiv}} (\bibinfo{year}{2024}).
\newblock


\bibitem[Chuai et~al\mbox{.}(2025)]%
        {chuai2025political}
\bibfield{author}{\bibinfo{person}{Yuwei Chuai}, \bibinfo{person}{Jichang Zhao}, \bibinfo{person}{Nicolas Pr{\"o}llochs}, {and} \bibinfo{person}{Gabriele Lenzini}.} \bibinfo{year}{2025}\natexlab{}.
\newblock \showarticletitle{Is fact-checking politically neutral? {Asymmetries} in how {U.S.} fact-checking organizations pick up false statements mentioning political elites}. In \bibinfo{booktitle}{\emph{Proceedings of the International AAAI Conference on Web and Social Media}}, Vol.~\bibinfo{volume}{forthcoming}.
\newblock


\bibitem[Cinelli et~al\mbox{.}(2021)]%
        {cinelli2021echo}
\bibfield{author}{\bibinfo{person}{Matteo Cinelli}, \bibinfo{person}{Gianmarco De~Francisci~Morales}, \bibinfo{person}{Alessandro Galeazzi}, \bibinfo{person}{Walter Quattrociocchi}, {and} \bibinfo{person}{Michele Starnini}.} \bibinfo{year}{2021}\natexlab{}.
\newblock \showarticletitle{The echo chamber effect on social media}.
\newblock \bibinfo{journal}{\emph{Proceedings of the National Academy of Sciences}} \bibinfo{volume}{118}, \bibinfo{number}{9} (\bibinfo{year}{2021}), \bibinfo{pages}{e2023301118}.
\newblock


\bibitem[Clayton et~al\mbox{.}(2020)]%
        {clayton2020real}
\bibfield{author}{\bibinfo{person}{Katherine Clayton}, \bibinfo{person}{Spencer Blair}, \bibinfo{person}{Jonathan~A Busam}, \bibinfo{person}{Samuel Forstner}, \bibinfo{person}{John Glance}, \bibinfo{person}{Guy Green}, \bibinfo{person}{Anna Kawata}, \bibinfo{person}{Akhila Kovvuri}, \bibinfo{person}{Jonathan Martin}, \bibinfo{person}{Evan Morgan}, {et~al\mbox{.}}} \bibinfo{year}{2020}\natexlab{}.
\newblock \showarticletitle{Real solutions for fake news? {Measuring} the effectiveness of general warnings and fact-check tags in reducing belief in false stories on social media}.
\newblock \bibinfo{journal}{\emph{Political Behavior}}  \bibinfo{volume}{42} (\bibinfo{year}{2020}), \bibinfo{pages}{1073--1095}.
\newblock


\bibitem[Colliander(2019)]%
        {colliander2019fake}
\bibfield{author}{\bibinfo{person}{Jonas Colliander}.} \bibinfo{year}{2019}\natexlab{}.
\newblock \showarticletitle{“This is fake news”: {Investigating} the role of conformity to other users’ views when commenting on and spreading disinformation in social media}.
\newblock \bibinfo{journal}{\emph{Computers in Human Behavior}}  \bibinfo{volume}{97} (\bibinfo{year}{2019}), \bibinfo{pages}{202--215}.
\newblock


\bibitem[Crockett(2017)]%
        {crockett2017moral}
\bibfield{author}{\bibinfo{person}{Molly~J Crockett}.} \bibinfo{year}{2017}\natexlab{}.
\newblock \showarticletitle{Moral outrage in the digital age}.
\newblock \bibinfo{journal}{\emph{Nature Human Behaviour}} \bibinfo{volume}{1}, \bibinfo{number}{11} (\bibinfo{year}{2017}), \bibinfo{pages}{769--771}.
\newblock


\bibitem[Dahlke and Hancock(2024)]%
        {dahlke2024effect}
\bibfield{author}{\bibinfo{person}{Ross Dahlke} {and} \bibinfo{person}{Jeffrey Hancock}.} \bibinfo{year}{2024}\natexlab{}.
\newblock \showarticletitle{Effects of misinformation on election beliefs: Disentangling motivated reasoning from selective exposure}.
\newblock \bibinfo{journal}{\emph{OSF}} (\bibinfo{year}{2024}).
\newblock


\bibitem[Del~Vicario et~al\mbox{.}(2016)]%
        {del2016echo}
\bibfield{author}{\bibinfo{person}{Michela Del~Vicario}, \bibinfo{person}{Gianna Vivaldo}, \bibinfo{person}{Alessandro Bessi}, \bibinfo{person}{Fabiana Zollo}, \bibinfo{person}{Antonio Scala}, \bibinfo{person}{Guido Caldarelli}, {and} \bibinfo{person}{Walter Quattrociocchi}.} \bibinfo{year}{2016}\natexlab{}.
\newblock \showarticletitle{Echo chambers: Emotional contagion and group polarization on facebook}.
\newblock \bibinfo{journal}{\emph{Scientific Reports}} \bibinfo{volume}{6}, \bibinfo{number}{1} (\bibinfo{year}{2016}), \bibinfo{pages}{37825}.
\newblock


\bibitem[Deloitte(2021)]%
        {deloitte2021are}
\bibfield{author}{\bibinfo{person}{Deloitte}.} \bibinfo{year}{2021}\natexlab{}.
\newblock \bibinfo{title}{Are younger generations moving away from traditional news sources?}
\newblock
\newblock
\newblock
\shownote{https://www.forbes.com/sites/kenrapoza/2017/02/26/can-fake-news-impact-the-stock-market/}.


\bibitem[Donovan(2020)]%
        {donovan2020social}
\bibfield{author}{\bibinfo{person}{Joan Donovan}.} \bibinfo{year}{2020}\natexlab{}.
\newblock \showarticletitle{Social-media companies must flatten the curve of misinformation.}
\newblock \bibinfo{journal}{\emph{Nature}} (\bibinfo{year}{2020}).
\newblock


\bibitem[Drolsbach and Pr{\"o}llochs(2023)]%
        {drolsbach2023diffusion}
\bibfield{author}{\bibinfo{person}{Chiara~Patricia Drolsbach} {and} \bibinfo{person}{Nicolas Pr{\"o}llochs}.} \bibinfo{year}{2023}\natexlab{}.
\newblock \showarticletitle{Diffusion of community fact-checked misinformation on {T}witter}.
\newblock \bibinfo{journal}{\emph{Proceedings of the ACM on Human-Computer Interaction}} \bibinfo{volume}{7}, \bibinfo{number}{CSCW2} (\bibinfo{year}{2023}), \bibinfo{pages}{1--22}.
\newblock


\bibitem[Drolsbach and Pr{\"o}llochs(2024)]%
        {drolsbach2024content}
\bibfield{author}{\bibinfo{person}{Chiara~Patricia Drolsbach} {and} \bibinfo{person}{Nicolas Pr{\"o}llochs}.} \bibinfo{year}{2024}\natexlab{}.
\newblock \showarticletitle{Content moderation on social media in the EU: Insights from the DSA Transparency Database}. In \bibinfo{booktitle}{\emph{Companion Proceedings of the ACM on Web Conference 2024}}. \bibinfo{pages}{939--942}.
\newblock


\bibitem[Drolsbach and Pröllochs(2023)]%
        {drolsbach2023believability}
\bibfield{author}{\bibinfo{person}{Chiara~Patricia Drolsbach} {and} \bibinfo{person}{Nicolas Pröllochs}.} \bibinfo{year}{2023}\natexlab{}.
\newblock \showarticletitle{Believability and harmfulness shape the virality of misleading social media posts}. In \bibinfo{booktitle}{\emph{Proceedings of the ACM Web Conference 2023}}. \bibinfo{pages}{4172--4177}.
\newblock


\bibitem[Drolsbach et~al\mbox{.}(2024)]%
        {drolsbach2024community}
\bibfield{author}{\bibinfo{person}{Chiara~Patricia Drolsbach}, \bibinfo{person}{Kirill Solovev}, {and} \bibinfo{person}{Nicolas Pr{\"o}llochs}.} \bibinfo{year}{2024}\natexlab{}.
\newblock \showarticletitle{Community notes increase trust in fact-checking on social media}.
\newblock \bibinfo{journal}{\emph{PNAS Nexus}} \bibinfo{volume}{3}, \bibinfo{number}{7} (\bibinfo{year}{2024}), \bibinfo{pages}{pgae217}.
\newblock


\bibitem[Ecker et~al\mbox{.}(2024)]%
        {ecker2024misinformation}
\bibfield{author}{\bibinfo{person}{Ullrich Ecker}, \bibinfo{person}{Jon Roozenbeek}, \bibinfo{person}{Sander van~der Linden}, \bibinfo{person}{Li~Qian Tay}, \bibinfo{person}{John Cook}, \bibinfo{person}{Naomi Oreskes}, {and} \bibinfo{person}{Stephan Lewandowsky}.} \bibinfo{year}{2024}\natexlab{}.
\newblock \showarticletitle{Misinformation poses a bigger threat to democracy than you might think}.
\newblock \bibinfo{journal}{\emph{Nature}} \bibinfo{volume}{630}, \bibinfo{number}{8015} (\bibinfo{year}{2024}), \bibinfo{pages}{29--32}.
\newblock


\bibitem[Ecker et~al\mbox{.}(2022)]%
        {ecker2022psychological}
\bibfield{author}{\bibinfo{person}{Ullrich~KH Ecker}, \bibinfo{person}{Stephan Lewandowsky}, \bibinfo{person}{John Cook}, \bibinfo{person}{Philipp Schmid}, \bibinfo{person}{Lisa~K Fazio}, \bibinfo{person}{Nadia Brashier}, \bibinfo{person}{Panayiota Kendeou}, \bibinfo{person}{Emily~K Vraga}, {and} \bibinfo{person}{Michelle~A Amazeen}.} \bibinfo{year}{2022}\natexlab{}.
\newblock \showarticletitle{The psychological drivers of misinformation belief and its resistance to correction}.
\newblock \bibinfo{journal}{\emph{Nature Reviews Psychology}} \bibinfo{volume}{1}, \bibinfo{number}{1} (\bibinfo{year}{2022}), \bibinfo{pages}{13--29}.
\newblock


\bibitem[Efstratiou and De~Cristofaro(2022)]%
        {efstratiou2022adherence}
\bibfield{author}{\bibinfo{person}{Alexandros Efstratiou} {and} \bibinfo{person}{Emiliano De~Cristofaro}.} \bibinfo{year}{2022}\natexlab{}.
\newblock \showarticletitle{Adherence to misinformation on social media through socio-cognitive and group-based processes}.
\newblock \bibinfo{journal}{\emph{Proceedings of the ACM on Human-Computer Interaction}} \bibinfo{volume}{6}, \bibinfo{number}{CSCW2} (\bibinfo{year}{2022}), \bibinfo{pages}{1--35}.
\newblock


\bibitem[Ekman(1992)]%
        {ekman1992argument}
\bibfield{author}{\bibinfo{person}{Paul Ekman}.} \bibinfo{year}{1992}\natexlab{}.
\newblock \showarticletitle{An argument for basic emotions}.
\newblock \bibinfo{journal}{\emph{Cognition \& Emotion}} \bibinfo{volume}{6}, \bibinfo{number}{3-4} (\bibinfo{year}{1992}), \bibinfo{pages}{169--200}.
\newblock


\bibitem[Ekman et~al\mbox{.}(2013)]%
        {ekman2013emotion}
\bibfield{author}{\bibinfo{person}{Paul Ekman}, \bibinfo{person}{Wallace~V Friesen}, {and} \bibinfo{person}{Phoebe Ellsworth}.} \bibinfo{year}{2013}\natexlab{}.
\newblock \bibinfo{booktitle}{\emph{Emotion in the human face: Guidelines for research and an integration of findings}}. Vol.~\bibinfo{volume}{11}.
\newblock \bibinfo{publisher}{Elsevier}.
\newblock


\bibitem[Enders et~al\mbox{.}(2022)]%
        {enders2022relationship}
\bibfield{author}{\bibinfo{person}{Adam~M. Enders}, \bibinfo{person}{Joseph Uscinski}, \bibinfo{person}{Casey Klofstad}, {and} \bibinfo{person}{Justin Stoler}.} \bibinfo{year}{2022}\natexlab{}.
\newblock \showarticletitle{On the relationship between conspiracy theory beliefs, misinformation, and vaccine hesitancy}.
\newblock \bibinfo{journal}{\emph{PLoS ONE}} \bibinfo{volume}{17}, \bibinfo{number}{10} (\bibinfo{year}{2022}), \bibinfo{pages}{e0276082}.
\newblock


\bibitem[Epstein et~al\mbox{.}(2020)]%
        {epstein2020will}
\bibfield{author}{\bibinfo{person}{Ziv Epstein}, \bibinfo{person}{Gordon Pennycook}, {and} \bibinfo{person}{David Rand}.} \bibinfo{year}{2020}\natexlab{}.
\newblock \showarticletitle{Will the crowd game the algorithm? Using layperson judgments to combat misinformation on social media by downranking distrusted sources}. In \bibinfo{booktitle}{\emph{Proceedings of the 2020 CHI Conference on Human Factors in Computing Systems}}. \bibinfo{pages}{1--11}.
\newblock


\bibitem[Fagerland(2012)]%
        {fagerland2012t}
\bibfield{author}{\bibinfo{person}{Morten~W Fagerland}.} \bibinfo{year}{2012}\natexlab{}.
\newblock \showarticletitle{T-tests, non-parametric tests, and large studies—a paradox of statistical practice?}
\newblock \bibinfo{journal}{\emph{BMC Medical Research Methodology}}  \bibinfo{volume}{12} (\bibinfo{year}{2012}), \bibinfo{pages}{1--7}.
\newblock


\bibitem[Feuerriegel et~al\mbox{.}(2023)]%
        {feuerriegel2023research}
\bibfield{author}{\bibinfo{person}{Stefan Feuerriegel}, \bibinfo{person}{Ren{\'e}e DiResta}, \bibinfo{person}{Josh~A Goldstein}, \bibinfo{person}{Srijan Kumar}, \bibinfo{person}{Philipp Lorenz-Spreen}, \bibinfo{person}{Michael Tomz}, {and} \bibinfo{person}{Nicolas Pr{\"o}llochs}.} \bibinfo{year}{2023}\natexlab{}.
\newblock \showarticletitle{Research can help to tackle {AI}-generated disinformation}.
\newblock \bibinfo{journal}{\emph{Nature Human Behaviour}} \bibinfo{volume}{7}, \bibinfo{number}{11} (\bibinfo{year}{2023}), \bibinfo{pages}{1818--1821}.
\newblock


\bibitem[Feuerriegel et~al\mbox{.}(2025)]%
        {feuerriegel2025using}
\bibfield{author}{\bibinfo{person}{Stefan Feuerriegel}, \bibinfo{person}{Abdurahman Maarouf}, \bibinfo{person}{Dominik B{\"a}r}, \bibinfo{person}{Dominique Geissler}, \bibinfo{person}{Jonas Schweisthal}, \bibinfo{person}{Nicolas Pr{\"o}llochs}, \bibinfo{person}{Claire~E Robertson}, \bibinfo{person}{Steve Rathje}, \bibinfo{person}{Jochen Hartmann}, \bibinfo{person}{Saif~M Mohammad}, \bibinfo{person}{Oded Netzer}, \bibinfo{person}{Alexandra Siegel}, \bibinfo{person}{Barbara Plank}, {and} \bibinfo{person}{Van~Bavel Jay}.} \bibinfo{year}{2025}\natexlab{}.
\newblock \showarticletitle{Using natural language processing to analyse text data in behavioural science}.
\newblock \bibinfo{journal}{\emph{Nature Reviews Psychology}}  \bibinfo{volume}{forthcoming} (\bibinfo{year}{2025}).
\newblock


\bibitem[Flamini(2019)]%
        {flamini2019most}
\bibfield{author}{\bibinfo{person}{Daniela Flamini}.} \bibinfo{year}{2019}\natexlab{}.
\newblock \bibinfo{title}{Most {Republicans} don’t trust fact-checkers, and most {Americans} don’t trust the media}.
\newblock
\newblock
\newblock
\shownote{https://www.poynter.org/ifcn/2019/most-republicans-dont-trust-fact-checkers-and-most-americans-dont-trust-the-media/}.


\bibitem[Forum(2024)]%
        {wef2024global}
\bibfield{author}{\bibinfo{person}{World~Economic Forum}.} \bibinfo{year}{2024}\natexlab{}.
\newblock \bibinfo{title}{Global risks report}.
\newblock
\newblock
\newblock
\shownote{https://www.weforum.org/publications/global-risks-report-2024/}.


\bibitem[Gaisbauer et~al\mbox{.}(2021)]%
        {gaisbauer2021ideological}
\bibfield{author}{\bibinfo{person}{Felix Gaisbauer}, \bibinfo{person}{Armin Pournaki}, \bibinfo{person}{Sven Banisch}, {and} \bibinfo{person}{Eckehard Olbrich}.} \bibinfo{year}{2021}\natexlab{}.
\newblock \showarticletitle{Ideological differences in engagement in public debate on {Twitter}}.
\newblock \bibinfo{journal}{\emph{PLoS ONE}} \bibinfo{volume}{16}, \bibinfo{number}{3} (\bibinfo{year}{2021}), \bibinfo{pages}{e0249241}.
\newblock


\bibitem[Gamage et~al\mbox{.}(2022)]%
        {gamage2022deepfakes}
\bibfield{author}{\bibinfo{person}{Dilrukshi Gamage}, \bibinfo{person}{Piyush Ghasiya}, \bibinfo{person}{Vamshi Bonagiri}, \bibinfo{person}{Mark~E Whiting}, {and} \bibinfo{person}{Kazutoshi Sasahara}.} \bibinfo{year}{2022}\natexlab{}.
\newblock \showarticletitle{Are deepfakes concerning? Analyzing conversations of deepfakes on reddit and exploring societal implications}. In \bibinfo{booktitle}{\emph{Proceedings of the 2022 CHI Conference on Human Factors in Computing Systems}}. \bibinfo{pages}{1--19}.
\newblock


\bibitem[Goldenberg and Gross(2020)]%
        {goldenberg2020digital}
\bibfield{author}{\bibinfo{person}{Amit Goldenberg} {and} \bibinfo{person}{James~J Gross}.} \bibinfo{year}{2020}\natexlab{}.
\newblock \showarticletitle{Digital emotion contagion}.
\newblock \bibinfo{journal}{\emph{Trends in Cognitive Sciences}} \bibinfo{volume}{24}, \bibinfo{number}{4} (\bibinfo{year}{2020}), \bibinfo{pages}{316--328}.
\newblock


\bibitem[Green et~al\mbox{.}(2022)]%
        {green2022online}
\bibfield{author}{\bibinfo{person}{Jon Green}, \bibinfo{person}{William Hobbs}, \bibinfo{person}{Stefan McCabe}, {and} \bibinfo{person}{David Lazer}.} \bibinfo{year}{2022}\natexlab{}.
\newblock \showarticletitle{Online engagement with 2020 election misinformation and turnout in the 2021 {Georgia} runoff election}.
\newblock \bibinfo{journal}{\emph{Proceedings of the National Academy of Sciences}} \bibinfo{volume}{119}, \bibinfo{number}{34} (\bibinfo{year}{2022}), \bibinfo{pages}{e2115900119}.
\newblock


\bibitem[Haidt et~al\mbox{.}(2003)]%
        {haidt2003moral}
\bibfield{author}{\bibinfo{person}{Jonathan Haidt} {et~al\mbox{.}}} \bibinfo{year}{2003}\natexlab{}.
\newblock \showarticletitle{The moral emotions}.
\newblock \bibinfo{journal}{\emph{Handbook of Affective Sciences}} \bibinfo{volume}{11}, \bibinfo{number}{2003} (\bibinfo{year}{2003}), \bibinfo{pages}{852--870}.
\newblock


\bibitem[Haque et~al\mbox{.}(2020)]%
        {haque2020combating}
\bibfield{author}{\bibinfo{person}{Md~Mahfuzul Haque}, \bibinfo{person}{Mohammad Yousuf}, \bibinfo{person}{Ahmed~Shatil Alam}, \bibinfo{person}{Pratyasha Saha}, \bibinfo{person}{Syed~Ishtiaque Ahmed}, {and} \bibinfo{person}{Naeemul Hassan}.} \bibinfo{year}{2020}\natexlab{}.
\newblock \showarticletitle{Combating misinformation in Bangladesh: Roles and responsibilities as perceived by journalists, fact-checkers, and users}.
\newblock \bibinfo{journal}{\emph{Proceedings of the ACM on Human-Computer Interaction}} \bibinfo{volume}{4}, \bibinfo{number}{CSCW2} (\bibinfo{year}{2020}), \bibinfo{pages}{1--32}.
\newblock


\bibitem[Hart et~al\mbox{.}(2020)]%
        {hart2020politicization}
\bibfield{author}{\bibinfo{person}{P~Sol Hart}, \bibinfo{person}{Sedona Chinn}, {and} \bibinfo{person}{Stuart Soroka}.} \bibinfo{year}{2020}\natexlab{}.
\newblock \showarticletitle{Politicization and polarization in COVID-19 news coverage}.
\newblock \bibinfo{journal}{\emph{Science Communication}} \bibinfo{volume}{42}, \bibinfo{number}{5} (\bibinfo{year}{2020}), \bibinfo{pages}{679--697}.
\newblock


\bibitem[Hartmann(2022)]%
        {hartmann2022emotionenglish}
\bibfield{author}{\bibinfo{person}{Jochen Hartmann}.} \bibinfo{year}{2022}\natexlab{}.
\newblock \bibinfo{title}{Emotion english DistilRoBERTa-base}.
\newblock \bibinfo{howpublished}{\url{https://huggingface.co/j-hartmann/emotion-english-distilroberta-base/}}.
\newblock


\bibitem[Hassoun et~al\mbox{.}(2023)]%
        {hassoun2023practicing}
\bibfield{author}{\bibinfo{person}{Amelia Hassoun}, \bibinfo{person}{Ian Beacock}, \bibinfo{person}{Sunny Consolvo}, \bibinfo{person}{Beth Goldberg}, \bibinfo{person}{Patrick~Gage Kelley}, {and} \bibinfo{person}{Daniel~M Russell}.} \bibinfo{year}{2023}\natexlab{}.
\newblock \showarticletitle{Practicing information sensibility: How Gen Z engages with online information}. In \bibinfo{booktitle}{\emph{Proceedings of the 2023 CHI Conference on Human Factors in Computing Systems}}. \bibinfo{pages}{1--17}.
\newblock


\bibitem[Hausman and Rapson(2018)]%
        {hausman2018regression}
\bibfield{author}{\bibinfo{person}{Catherine Hausman} {and} \bibinfo{person}{David~S Rapson}.} \bibinfo{year}{2018}\natexlab{}.
\newblock \showarticletitle{Regression discontinuity in time: {Considerations} for empirical applications}.
\newblock \bibinfo{journal}{\emph{Annual Review of Resource Economics}} \bibinfo{volume}{10}, \bibinfo{number}{1} (\bibinfo{year}{2018}), \bibinfo{pages}{533--552}.
\newblock


\bibitem[Heuer and Glassman(2022)]%
        {heuer2022comparative}
\bibfield{author}{\bibinfo{person}{Hendrik Heuer} {and} \bibinfo{person}{Elena~Leah Glassman}.} \bibinfo{year}{2022}\natexlab{}.
\newblock \showarticletitle{A comparative evaluation of interventions against misinformation: Augmenting the WHO checklist}. In \bibinfo{booktitle}{\emph{Proceedings of the 2022 CHI Conference on Human Factors in Computing Systems}}. \bibinfo{pages}{1--21}.
\newblock


\bibitem[Hutto and Gilbert(2014)]%
        {hutto2014vader}
\bibfield{author}{\bibinfo{person}{Clayton Hutto} {and} \bibinfo{person}{Eric Gilbert}.} \bibinfo{year}{2014}\natexlab{}.
\newblock \showarticletitle{{VADER}: {A} parsimonious rule-based model for sentiment analysis of social media text}. In \bibinfo{booktitle}{\emph{Proceedings of the International AAAI Conference on Web and Social Media}}, Vol.~\bibinfo{volume}{8}. \bibinfo{pages}{216--225}.
\newblock


\bibitem[Imbens and Lemieux(2008)]%
        {imbens2008regression}
\bibfield{author}{\bibinfo{person}{Guido~W Imbens} {and} \bibinfo{person}{Thomas Lemieux}.} \bibinfo{year}{2008}\natexlab{}.
\newblock \showarticletitle{Regression discontinuity designs: {A} guide to practice}.
\newblock \bibinfo{journal}{\emph{Journal of Econometrics}} \bibinfo{volume}{142}, \bibinfo{number}{2} (\bibinfo{year}{2008}), \bibinfo{pages}{615--635}.
\newblock


\bibitem[Jahanbakhsh and Karger(2024)]%
        {jahanbakhsh2024browser}
\bibfield{author}{\bibinfo{person}{Farnaz Jahanbakhsh} {and} \bibinfo{person}{David~R Karger}.} \bibinfo{year}{2024}\natexlab{}.
\newblock \showarticletitle{A browser extension for in-place signaling and assessment of misinformation}. In \bibinfo{booktitle}{\emph{Proceedings of the 2024 CHI Conference on Human Factors in Computing Systems}}. \bibinfo{pages}{1--21}.
\newblock


\bibitem[Jakubik et~al\mbox{.}(2023)]%
        {jakubik2023online}
\bibfield{author}{\bibinfo{person}{Johannes Jakubik}, \bibinfo{person}{Michael V{\"o}ssing}, \bibinfo{person}{Nicolas Pr{\"o}llochs}, \bibinfo{person}{Dominik B{\"a}r}, {and} \bibinfo{person}{Stefan Feuerriegel}.} \bibinfo{year}{2023}\natexlab{}.
\newblock \showarticletitle{Online emotions during the storming of the US Capitol: Evidence from the social media network Parler}. In \bibinfo{booktitle}{\emph{Proceedings of the International AAAI Conference on Web and Social Media}}, Vol.~\bibinfo{volume}{17}. \bibinfo{pages}{423--434}.
\newblock


\bibitem[Juneja et~al\mbox{.}(2023)]%
        {juneja2023assessing}
\bibfield{author}{\bibinfo{person}{Prerna Juneja}, \bibinfo{person}{Md~Momen Bhuiyan}, {and} \bibinfo{person}{Tanushree Mitra}.} \bibinfo{year}{2023}\natexlab{}.
\newblock \showarticletitle{Assessing enactment of content regulation policies: A post hoc crowd-sourced audit of election misinformation on YouTube}. In \bibinfo{booktitle}{\emph{Proceedings of the 2023 CHI Conference on Human Factors in Computing Systems}}. \bibinfo{pages}{1--22}.
\newblock


\bibitem[Karusala and Anderson(2022)]%
        {karusala2022towards}
\bibfield{author}{\bibinfo{person}{Naveena Karusala} {and} \bibinfo{person}{Richard Anderson}.} \bibinfo{year}{2022}\natexlab{}.
\newblock \showarticletitle{Towards conviviality in navigating health information on social media}. In \bibinfo{booktitle}{\emph{Proceedings of the 2022 CHI Conference on Human Factors in Computing Systems}}. \bibinfo{pages}{1--14}.
\newblock


\bibitem[Keele and Kelly(2006)]%
        {keele2006dynamic}
\bibfield{author}{\bibinfo{person}{Luke Keele} {and} \bibinfo{person}{Nathan~J Kelly}.} \bibinfo{year}{2006}\natexlab{}.
\newblock \showarticletitle{Dynamic models for dynamic theories: {The} ins and outs of lagged dependent variables}.
\newblock \bibinfo{journal}{\emph{Political Analysis}} \bibinfo{volume}{14}, \bibinfo{number}{2} (\bibinfo{year}{2006}), \bibinfo{pages}{186--205}.
\newblock


\bibitem[Kim et~al\mbox{.}(2024)]%
        {kim2024moral}
\bibfield{author}{\bibinfo{person}{Jaehong Kim}, \bibinfo{person}{Chaeyoon Jeong}, \bibinfo{person}{Seongchan Park}, \bibinfo{person}{Meeyoung Cha}, {and} \bibinfo{person}{Wonjae Lee}.} \bibinfo{year}{2024}\natexlab{}.
\newblock \showarticletitle{How do moral emotions shape political participation? {A} cross-cultural analysis of online petitions using language models}. In \bibinfo{booktitle}{\emph{Findings of the Association for Computational Linguistics ACL 2024}}. \bibinfo{pages}{16274--16289}.
\newblock


\bibitem[Kozyreva et~al\mbox{.}(2023)]%
        {kozyreva2023resolving}
\bibfield{author}{\bibinfo{person}{Anastasia Kozyreva}, \bibinfo{person}{Stefan~M Herzog}, \bibinfo{person}{Stephan Lewandowsky}, \bibinfo{person}{Ralph Hertwig}, \bibinfo{person}{Philipp Lorenz-Spreen}, \bibinfo{person}{Mark Leiser}, {and} \bibinfo{person}{Jason Reifler}.} \bibinfo{year}{2023}\natexlab{}.
\newblock \showarticletitle{Resolving content moderation dilemmas between free speech and harmful misinformation}.
\newblock \bibinfo{journal}{\emph{Proceedings of the National Academy of Sciences}} \bibinfo{volume}{120}, \bibinfo{number}{7} (\bibinfo{year}{2023}), \bibinfo{pages}{e2210666120}.
\newblock


\bibitem[Kramer et~al\mbox{.}(2014)]%
        {kramer2014experimental}
\bibfield{author}{\bibinfo{person}{Adam~DI Kramer}, \bibinfo{person}{Jamie~E Guillory}, {and} \bibinfo{person}{Jeffrey~T Hancock}.} \bibinfo{year}{2014}\natexlab{}.
\newblock \showarticletitle{Experimental evidence of massive-scale emotional contagion through social networks}.
\newblock \bibinfo{journal}{\emph{Proceedings of the National Academy of Sciences}} \bibinfo{volume}{111}, \bibinfo{number}{24} (\bibinfo{year}{2014}), \bibinfo{pages}{8788--8790}.
\newblock


\bibitem[Lazer et~al\mbox{.}(2018)]%
        {lazer2018science}
\bibfield{author}{\bibinfo{person}{David~MJ Lazer}, \bibinfo{person}{Matthew~A Baum}, \bibinfo{person}{Yochai Benkler}, \bibinfo{person}{Adam~J Berinsky}, \bibinfo{person}{Kelly~M Greenhill}, \bibinfo{person}{Filippo Menczer}, \bibinfo{person}{Miriam~J Metzger}, \bibinfo{person}{Brendan Nyhan}, \bibinfo{person}{Gordon Pennycook}, \bibinfo{person}{David Rothschild}, {and} \bibinfo{person}{{others}}.} \bibinfo{year}{2018}\natexlab{}.
\newblock \showarticletitle{The science of fake news}.
\newblock \bibinfo{journal}{\emph{Science}} \bibinfo{volume}{359}, \bibinfo{number}{6380} (\bibinfo{year}{2018}), \bibinfo{pages}{1094--1096}.
\newblock


\bibitem[Lee et~al\mbox{.}(2023)]%
        {lee2023fact}
\bibfield{author}{\bibinfo{person}{Sian Lee}, \bibinfo{person}{Aiping Xiong}, \bibinfo{person}{Haeseung Seo}, {and} \bibinfo{person}{Dongwon Lee}.} \bibinfo{year}{2023}\natexlab{}.
\newblock \showarticletitle{“Fact-checking” fact checkers: A data-driven approach}.
\newblock \bibinfo{journal}{\emph{Harvard Kennedy School Misinformation Review}} (\bibinfo{year}{2023}).
\newblock


\bibitem[Lerner et~al\mbox{.}(2015)]%
        {lerner2015emotion}
\bibfield{author}{\bibinfo{person}{Jennifer~S Lerner}, \bibinfo{person}{Ye Li}, \bibinfo{person}{Piercarlo Valdesolo}, {and} \bibinfo{person}{Karim~S Kassam}.} \bibinfo{year}{2015}\natexlab{}.
\newblock \showarticletitle{Emotion and decision making}.
\newblock \bibinfo{journal}{\emph{Annual Review of Psychology}} \bibinfo{volume}{66}, \bibinfo{number}{1} (\bibinfo{year}{2015}), \bibinfo{pages}{799--823}.
\newblock


\bibitem[Loureiro et~al\mbox{.}(2022)]%
        {loureiro2022timelms}
\bibfield{author}{\bibinfo{person}{Daniel Loureiro}, \bibinfo{person}{Francesco Barbieri}, \bibinfo{person}{Leonardo Neves}, \bibinfo{person}{Luis Espinosa~Anke}, {and} \bibinfo{person}{Jose Camacho-collados}.} \bibinfo{year}{2022}\natexlab{}.
\newblock \showarticletitle{{T}ime{LM}s: Diachronic language models from {T}witter}. In \bibinfo{booktitle}{\emph{Proceedings of the 60th Annual Meeting of the Association for Computational Linguistics: System Demonstrations}}. \bibinfo{pages}{251--260}.
\newblock


\bibitem[Lumley et~al\mbox{.}(2002)]%
        {lumley2002importance}
\bibfield{author}{\bibinfo{person}{Thomas Lumley}, \bibinfo{person}{Paula Diehr}, \bibinfo{person}{Scott Emerson}, {and} \bibinfo{person}{Lu Chen}.} \bibinfo{year}{2002}\natexlab{}.
\newblock \showarticletitle{The importance of the normality assumption in large public health data sets}.
\newblock \bibinfo{journal}{\emph{Annual Review of Public Health}} \bibinfo{volume}{23}, \bibinfo{number}{1} (\bibinfo{year}{2002}), \bibinfo{pages}{151--169}.
\newblock


\bibitem[Martel et~al\mbox{.}(2020)]%
        {martel2020reliance}
\bibfield{author}{\bibinfo{person}{Cameron Martel}, \bibinfo{person}{Gordon Pennycook}, {and} \bibinfo{person}{David~G Rand}.} \bibinfo{year}{2020}\natexlab{}.
\newblock \showarticletitle{Reliance on emotion promotes belief in fake news}.
\newblock \bibinfo{journal}{\emph{Cognitive Research: Principles and Implications}}  \bibinfo{volume}{5} (\bibinfo{year}{2020}), \bibinfo{pages}{1--20}.
\newblock


\bibitem[McClure~Haughey et~al\mbox{.}(2022)]%
        {mcclure2022bridging}
\bibfield{author}{\bibinfo{person}{Melinda McClure~Haughey}, \bibinfo{person}{Martina Povolo}, {and} \bibinfo{person}{Kate Starbird}.} \bibinfo{year}{2022}\natexlab{}.
\newblock \showarticletitle{Bridging contextual and methodological gaps on the “misinformation beat”: Insights from journalist-researcher collaborations at speed}. In \bibinfo{booktitle}{\emph{Proceedings of the 2022 CHI Conference on Human Factors in Computing Systems}}. \bibinfo{pages}{1--15}.
\newblock


\bibitem[McLoughlin et~al\mbox{.}(2024)]%
        {mcloughlin2024misinformation}
\bibfield{author}{\bibinfo{person}{Killian~L McLoughlin}, \bibinfo{person}{William~J Brady}, \bibinfo{person}{Aden Goolsbee}, \bibinfo{person}{Ben Kaiser}, \bibinfo{person}{Kate Klonick}, {and} \bibinfo{person}{MJ Crockett}.} \bibinfo{year}{2024}\natexlab{}.
\newblock \showarticletitle{Misinformation exploits outrage to spread online}.
\newblock \bibinfo{journal}{\emph{Science}} \bibinfo{volume}{386}, \bibinfo{number}{6725} (\bibinfo{year}{2024}), \bibinfo{pages}{991--996}.
\newblock


\bibitem[{Meta}(2024)]%
        {meta2024metas}
\bibfield{author}{\bibinfo{person}{{Meta}}.} \bibinfo{year}{2024}\natexlab{}.
\newblock \bibinfo{title}{Meta’s third-party fact-checking program}.
\newblock
\newblock
\urldef\tempurl%
\url{https://www.facebook.com/formedia/mjp/programs/third-party-fact-checking}
\showURL{%
\tempurl}


\bibitem[Micallef et~al\mbox{.}(2022)]%
        {micallef2022true}
\bibfield{author}{\bibinfo{person}{Nicholas Micallef}, \bibinfo{person}{Vivienne Armacost}, \bibinfo{person}{Nasir Memon}, {and} \bibinfo{person}{Sameer Patil}.} \bibinfo{year}{2022}\natexlab{}.
\newblock \showarticletitle{True or false: {Studying} the work practices of professional fact-checkers}.
\newblock \bibinfo{journal}{\emph{Proceedings of the ACM on Human-Computer Interaction}} \bibinfo{volume}{6}, \bibinfo{number}{CSCW1} (\bibinfo{year}{2022}), \bibinfo{pages}{1--44}.
\newblock


\bibitem[Micallef et~al\mbox{.}(2020)]%
        {micallef2020role}
\bibfield{author}{\bibinfo{person}{Nicholas Micallef}, \bibinfo{person}{Bing He}, \bibinfo{person}{Srijan Kumar}, \bibinfo{person}{Mustaque Ahamad}, {and} \bibinfo{person}{Nasir Memon}.} \bibinfo{year}{2020}\natexlab{}.
\newblock \showarticletitle{The role of the crowd in countering misinformation: {A} case study of the {COVID-19} infodemic}. In \bibinfo{booktitle}{\emph{2020 IEEE International Conference on Big Data}}. \bibinfo{pages}{748--757}.
\newblock


\bibitem[Mohammad and Turney(2013)]%
        {mohammad2013crowdsourcing}
\bibfield{author}{\bibinfo{person}{Saif~M. Mohammad} {and} \bibinfo{person}{Peter~D. Turney}.} \bibinfo{year}{2013}\natexlab{}.
\newblock \showarticletitle{Crowdsourcing a word–emotion association lexicon}.
\newblock \bibinfo{journal}{\emph{Computational Intelligence}} \bibinfo{volume}{29}, \bibinfo{number}{3} (\bibinfo{year}{2013}), \bibinfo{pages}{436--465}.
\newblock


\bibitem[Molho et~al\mbox{.}(2017)]%
        {molho2017disgust}
\bibfield{author}{\bibinfo{person}{Catherine Molho}, \bibinfo{person}{Joshua~M Tybur}, \bibinfo{person}{Ezgi G{\"u}ler}, \bibinfo{person}{Daniel Balliet}, {and} \bibinfo{person}{Wilhelm Hofmann}.} \bibinfo{year}{2017}\natexlab{}.
\newblock \showarticletitle{Disgust and anger relate to different aggressive responses to moral violations}.
\newblock \bibinfo{journal}{\emph{Psychological Science}} \bibinfo{volume}{28}, \bibinfo{number}{5} (\bibinfo{year}{2017}), \bibinfo{pages}{609--619}.
\newblock


\bibitem[Mosleh et~al\mbox{.}(2021)]%
        {mosleh2021perverse}
\bibfield{author}{\bibinfo{person}{Mohsen Mosleh}, \bibinfo{person}{Cameron Martel}, \bibinfo{person}{Dean Eckles}, {and} \bibinfo{person}{David Rand}.} \bibinfo{year}{2021}\natexlab{}.
\newblock \showarticletitle{Perverse downstream consequences of debunking: Being corrected by another user for posting false political news increases subsequent sharing of low quality, partisan, and toxic content in a Twitter field experiment}. In \bibinfo{booktitle}{\emph{Proceedings of the 2021 CHI Conference on Human Factors in Computing Systems}}. \bibinfo{pages}{1--13}.
\newblock


\bibitem[Neumann and Rhodes(2024)]%
        {neumann2024morality}
\bibfield{author}{\bibinfo{person}{Dominik Neumann} {and} \bibinfo{person}{Nancy Rhodes}.} \bibinfo{year}{2024}\natexlab{}.
\newblock \showarticletitle{Morality in social media: {A} scoping review}.
\newblock \bibinfo{journal}{\emph{New Media \& Society}} \bibinfo{volume}{26}, \bibinfo{number}{2} (\bibinfo{year}{2024}), \bibinfo{pages}{1096--1126}.
\newblock


\bibitem[of~Professional~Journalists(2014)]%
        {spj2014code}
\bibfield{author}{\bibinfo{person}{Society of Professional~Journalists}.} \bibinfo{year}{2014}\natexlab{}.
\newblock \bibinfo{title}{Code of ethics}.
\newblock
\newblock
\urldef\tempurl%
\url{https://www.spj.org/ethicscode.asp}
\showURL{%
\tempurl}


\bibitem[Osmundsen et~al\mbox{.}(2021)]%
        {osmundsen2021partisan}
\bibfield{author}{\bibinfo{person}{Mathias Osmundsen}, \bibinfo{person}{Alexander Bor}, \bibinfo{person}{Peter~Bjerregaard Vahlstrup}, \bibinfo{person}{Anja Bechmann}, {and} \bibinfo{person}{Michael~Bang Petersen}.} \bibinfo{year}{2021}\natexlab{}.
\newblock \showarticletitle{Partisan polarization is the primary psychological motivation behind political fake news sharing on {Twitter}}.
\newblock \bibinfo{journal}{\emph{American Political Science Review}} \bibinfo{volume}{115}, \bibinfo{number}{3} (\bibinfo{year}{2021}), \bibinfo{pages}{999--1015}.
\newblock


\bibitem[Pang et~al\mbox{.}(2008)]%
        {pang2008opinion}
\bibfield{author}{\bibinfo{person}{Bo Pang}, \bibinfo{person}{Lillian Lee}, {et~al\mbox{.}}} \bibinfo{year}{2008}\natexlab{}.
\newblock \showarticletitle{Opinion mining and sentiment analysis}.
\newblock \bibinfo{journal}{\emph{Foundations and Trends{\textregistered} in Information Retrieval}} \bibinfo{volume}{2}, \bibinfo{number}{1--2} (\bibinfo{year}{2008}), \bibinfo{pages}{1--135}.
\newblock


\bibitem[Peng et~al\mbox{.}(2023)]%
        {peng2023rage}
\bibfield{author}{\bibinfo{person}{Kun Peng}, \bibinfo{person}{Yu Zheng}, \bibinfo{person}{Yuewei Qiu}, {and} \bibinfo{person}{Qingrui Li}.} \bibinfo{year}{2023}\natexlab{}.
\newblock \showarticletitle{Rage of righteousness: Anger's role in promoting debunking effects during the COVID-19 pandemic}.
\newblock \bibinfo{journal}{\emph{Computers in Human Behavior}}  \bibinfo{volume}{148} (\bibinfo{year}{2023}), \bibinfo{pages}{107896}.
\newblock


\bibitem[Pennycook and Rand(2019)]%
        {pennycook2019fighting}
\bibfield{author}{\bibinfo{person}{Gordon Pennycook} {and} \bibinfo{person}{David~G Rand}.} \bibinfo{year}{2019}\natexlab{}.
\newblock \showarticletitle{Fighting misinformation on social media using crowdsourced judgments of news source quality}.
\newblock \bibinfo{journal}{\emph{Proceedings of the National Academy of Sciences}} \bibinfo{volume}{116}, \bibinfo{number}{7} (\bibinfo{year}{2019}), \bibinfo{pages}{2521--2526}.
\newblock


\bibitem[Pierri et~al\mbox{.}(2022)]%
        {pierri2022online}
\bibfield{author}{\bibinfo{person}{Francesco Pierri}, \bibinfo{person}{Brea~L. Perry}, \bibinfo{person}{Matthew~R. DeVerna}, \bibinfo{person}{Kai-Cheng Yang}, \bibinfo{person}{Alessandro Flammini}, \bibinfo{person}{Filippo Menczer}, {and} \bibinfo{person}{John Bryden}.} \bibinfo{year}{2022}\natexlab{}.
\newblock \showarticletitle{Online misinformation is linked to early {COVID}-19 vaccination hesitancy and refusal}.
\newblock \bibinfo{journal}{\emph{Scientific Reports}} \bibinfo{volume}{12}, \bibinfo{number}{1} (\bibinfo{year}{2022}), \bibinfo{pages}{5966}.
\newblock


\bibitem[Pilarski et~al\mbox{.}(2024)]%
        {pilarski2024community}
\bibfield{author}{\bibinfo{person}{Moritz Pilarski}, \bibinfo{person}{Kirill~Olegovich Solovev}, {and} \bibinfo{person}{Nicolas Pr{\"o}llochs}.} \bibinfo{year}{2024}\natexlab{}.
\newblock \showarticletitle{Community Notes vs. Snoping: How the crowd selects fact-checking targets on social media}. In \bibinfo{booktitle}{\emph{Proceedings of the International AAAI Conference on Web and Social Media}}, Vol.~\bibinfo{volume}{18}. \bibinfo{pages}{1262--1275}.
\newblock


\bibitem[Porter and Wood(2021)]%
        {porter2021global}
\bibfield{author}{\bibinfo{person}{Ethan Porter} {and} \bibinfo{person}{Thomas~J Wood}.} \bibinfo{year}{2021}\natexlab{}.
\newblock \showarticletitle{The global effectiveness of fact-checking: {Evidence} from simultaneous experiments in {Argentina}, {Nigeria}, {South Africa}, and the {United Kingdom}}.
\newblock \bibinfo{journal}{\emph{Proceedings of the National Academy of Sciences}} \bibinfo{volume}{118}, \bibinfo{number}{37} (\bibinfo{year}{2021}), \bibinfo{pages}{e2104235118}.
\newblock


\bibitem[Pr{\"o}llochs et~al\mbox{.}(2021)]%
        {prollochs2021emotions}
\bibfield{author}{\bibinfo{person}{Nicolas Pr{\"o}llochs}, \bibinfo{person}{Dominik B{\"a}r}, {and} \bibinfo{person}{Stefan Feuerriegel}.} \bibinfo{year}{2021}\natexlab{}.
\newblock \showarticletitle{Emotions explain differences in the diffusion of true vs. false social media rumors}.
\newblock \bibinfo{journal}{\emph{Scientific Reports}} \bibinfo{volume}{11}, \bibinfo{number}{1} (\bibinfo{year}{2021}), \bibinfo{pages}{22721}.
\newblock


\bibitem[Pröllochs(2022)]%
        {prollochs2022community}
\bibfield{author}{\bibinfo{person}{Nicolas Pröllochs}.} \bibinfo{year}{2022}\natexlab{}.
\newblock \showarticletitle{Community-based fact-checking on {Twitter}’s {Birdwatch} platform}. In \bibinfo{booktitle}{\emph{Proceedings of the International AAAI Conference on Web and Social Media}}, Vol.~\bibinfo{volume}{16}. \bibinfo{pages}{794--805}.
\newblock


\bibitem[Rathje et~al\mbox{.}(2021)]%
        {rathje2021out}
\bibfield{author}{\bibinfo{person}{Steve Rathje}, \bibinfo{person}{Jay~J. Van~Bavel}, {and} \bibinfo{person}{Sander van~der Linden}.} \bibinfo{year}{2021}\natexlab{}.
\newblock \showarticletitle{Out-group animosity drives engagement on social media}.
\newblock \bibinfo{journal}{\emph{Proceedings of the National Academy of Sciences}} \bibinfo{volume}{118}, \bibinfo{number}{26} (\bibinfo{year}{2021}), \bibinfo{pages}{e2024292118}.
\newblock


\bibitem[Recuero et~al\mbox{.}(2020)]%
        {recuero2020hyperpartisanship}
\bibfield{author}{\bibinfo{person}{Raquel Recuero}, \bibinfo{person}{Felipe~Bonow Soares}, {and} \bibinfo{person}{Anatoliy Gruzd}.} \bibinfo{year}{2020}\natexlab{}.
\newblock \showarticletitle{Hyperpartisanship, disinformation and political conversations on Twitter: The Brazilian presidential election of 2018}. In \bibinfo{booktitle}{\emph{Proceedings of the International AAAI Conference on Web and Social Media}}, Vol.~\bibinfo{volume}{14}. \bibinfo{pages}{569--578}.
\newblock


\bibitem[Renault et~al\mbox{.}(2024)]%
        {renault2024collaboratively}
\bibfield{author}{\bibinfo{person}{Thomas Renault}, \bibinfo{person}{David~Restrepo Amariles}, {and} \bibinfo{person}{Aurore Troussel}.} \bibinfo{year}{2024}\natexlab{}.
\newblock \showarticletitle{Collaboratively adding context to social media posts reduces the sharing of false news}.
\newblock \bibinfo{journal}{\emph{ArXiv}} (\bibinfo{year}{2024}).
\newblock


\bibitem[Ribeiro et~al\mbox{.}(2025)]%
        {ribeiro2024post}
\bibfield{author}{\bibinfo{person}{Manoel~Horta Ribeiro}, \bibinfo{person}{Robert West}, \bibinfo{person}{Ryan Lewis}, {and} \bibinfo{person}{Sanjay Kairam}.} \bibinfo{year}{2025}\natexlab{}.
\newblock \showarticletitle{Post guidance for online communities}.
\newblock \bibinfo{journal}{\emph{Proceedings of the ACM on Human-Computer Interaction}}  \bibinfo{volume}{forthcoming} (\bibinfo{year}{2025}).
\newblock


\bibitem[Robertson et~al\mbox{.}(2023)]%
        {robertson2023negativity}
\bibfield{author}{\bibinfo{person}{Claire~E Robertson}, \bibinfo{person}{Nicolas Pr{\"o}llochs}, \bibinfo{person}{Kaoru Schwarzenegger}, \bibinfo{person}{Philip P{\"a}rnamets}, \bibinfo{person}{Jay~J Van~Bavel}, {and} \bibinfo{person}{Stefan Feuerriegel}.} \bibinfo{year}{2023}\natexlab{}.
\newblock \showarticletitle{Negativity drives online news consumption}.
\newblock \bibinfo{journal}{\emph{Nature Human Behaviour}} \bibinfo{volume}{7}, \bibinfo{number}{5} (\bibinfo{year}{2023}), \bibinfo{pages}{812--822}.
\newblock


\bibitem[Rosenbusch et~al\mbox{.}(2019)]%
        {rosenbusch2019multilevel}
\bibfield{author}{\bibinfo{person}{Hannes Rosenbusch}, \bibinfo{person}{Anthony~M Evans}, {and} \bibinfo{person}{Marcel Zeelenberg}.} \bibinfo{year}{2019}\natexlab{}.
\newblock \showarticletitle{Multilevel emotion transfer on YouTube: Disentangling the effects of emotional contagion and homophily on video audiences}.
\newblock \bibinfo{journal}{\emph{Social Psychological and Personality Science}} \bibinfo{volume}{10}, \bibinfo{number}{8} (\bibinfo{year}{2019}), \bibinfo{pages}{1028--1035}.
\newblock


\bibitem[Rost et~al\mbox{.}(2016)]%
        {rost2016digital}
\bibfield{author}{\bibinfo{person}{Katja Rost}, \bibinfo{person}{Lea Stahel}, {and} \bibinfo{person}{Bruno~S Frey}.} \bibinfo{year}{2016}\natexlab{}.
\newblock \showarticletitle{Digital social norm enforcement: Online firestorms in social media}.
\newblock \bibinfo{journal}{\emph{PLoS ONE}} \bibinfo{volume}{11}, \bibinfo{number}{6} (\bibinfo{year}{2016}), \bibinfo{pages}{e0155923}.
\newblock


\bibitem[Rozado et~al\mbox{.}(2022)]%
        {rozado2022longitudinal}
\bibfield{author}{\bibinfo{person}{David Rozado}, \bibinfo{person}{Ruth Hughes}, {and} \bibinfo{person}{Jamin Halberstadt}.} \bibinfo{year}{2022}\natexlab{}.
\newblock \showarticletitle{Longitudinal analysis of sentiment and emotion in news media headlines using automated labelling with Transformer language models}.
\newblock \bibinfo{journal}{\emph{PLoS ONE}} \bibinfo{volume}{17}, \bibinfo{number}{10} (\bibinfo{year}{2022}), \bibinfo{pages}{e0276367}.
\newblock


\bibitem[Saeed et~al\mbox{.}(2022)]%
        {saeed2022crowdsourced}
\bibfield{author}{\bibinfo{person}{Mohammed Saeed}, \bibinfo{person}{Nicolas Traub}, \bibinfo{person}{Maelle Nicolas}, \bibinfo{person}{Gianluca Demartini}, {and} \bibinfo{person}{Paolo Papotti}.} \bibinfo{year}{2022}\natexlab{}.
\newblock \showarticletitle{Crowdsourced fact-checking at Twitter: How does the crowd compare with experts?}. In \bibinfo{booktitle}{\emph{Proceedings of the 31st ACM International Conference on Information \& Knowledge Management}}. \bibinfo{pages}{1736--1746}.
\newblock


\bibitem[Salerno and Peter-Hagene(2013)]%
        {salerno2013interactive}
\bibfield{author}{\bibinfo{person}{Jessica~M Salerno} {and} \bibinfo{person}{Liana~C Peter-Hagene}.} \bibinfo{year}{2013}\natexlab{}.
\newblock \showarticletitle{The interactive effect of anger and disgust on moral outrage and judgments}.
\newblock \bibinfo{journal}{\emph{Psychological Science}} \bibinfo{volume}{24}, \bibinfo{number}{10} (\bibinfo{year}{2013}), \bibinfo{pages}{2069--2078}.
\newblock


\bibitem[Sang et~al\mbox{.}(2020)]%
        {sang2020signalling}
\bibfield{author}{\bibinfo{person}{Yoonmo Sang}, \bibinfo{person}{Jee~Young Lee}, \bibinfo{person}{Sora Park}, \bibinfo{person}{Caroline Fisher}, {and} \bibinfo{person}{Glen Fuller}.} \bibinfo{year}{2020}\natexlab{}.
\newblock \showarticletitle{Signalling and expressive interaction: Online news users’ different modes of interaction on digital platforms}.
\newblock \bibinfo{journal}{\emph{Digital Journalism}} \bibinfo{volume}{8}, \bibinfo{number}{4} (\bibinfo{year}{2020}), \bibinfo{pages}{467--485}.
\newblock


\bibitem[Sauter et~al\mbox{.}(2010)]%
        {sauter2010cross}
\bibfield{author}{\bibinfo{person}{Disa~A Sauter}, \bibinfo{person}{Frank Eisner}, \bibinfo{person}{Paul Ekman}, {and} \bibinfo{person}{Sophie~K Scott}.} \bibinfo{year}{2010}\natexlab{}.
\newblock \showarticletitle{Cross-cultural recognition of basic emotions through nonverbal emotional vocalizations}.
\newblock \bibinfo{journal}{\emph{Proceedings of the National Academy of Sciences}} \bibinfo{volume}{107}, \bibinfo{number}{6} (\bibinfo{year}{2010}), \bibinfo{pages}{2408--2412}.
\newblock


\bibitem[Schaffner et~al\mbox{.}(2024)]%
        {schaffner2024community}
\bibfield{author}{\bibinfo{person}{Brennan Schaffner}, \bibinfo{person}{Arjun~Nitin Bhagoji}, \bibinfo{person}{Siyuan Cheng}, \bibinfo{person}{Jacqueline Mei}, \bibinfo{person}{Jay~L Shen}, \bibinfo{person}{Grace Wang}, \bibinfo{person}{Marshini Chetty}, \bibinfo{person}{Nick Feamster}, \bibinfo{person}{Genevieve Lakier}, {and} \bibinfo{person}{Chenhao Tan}.} \bibinfo{year}{2024}\natexlab{}.
\newblock \showarticletitle{"Community guidelines make this the best party on the internet": An in-depth study of online platforms' content moderation policies}. In \bibinfo{booktitle}{\emph{Proceedings of the 2024 CHI Conference on Human Factors in Computing Systems}}. \bibinfo{pages}{1--16}.
\newblock


\bibitem[Shah et~al\mbox{.}(2020)]%
        {shah2020perceived}
\bibfield{author}{\bibinfo{person}{Zakir Shah}, \bibinfo{person}{Jianxun Chu}, \bibinfo{person}{Sara Qaisar}, \bibinfo{person}{Zameer Hassan}, {and} \bibinfo{person}{Usman Ghani}.} \bibinfo{year}{2020}\natexlab{}.
\newblock \showarticletitle{Perceived public condemnation and avoidance intentions: The mediating role of moral outrage}.
\newblock \bibinfo{journal}{\emph{Journal of Public Affairs}} \bibinfo{volume}{20}, \bibinfo{number}{1} (\bibinfo{year}{2020}), \bibinfo{pages}{e2027}.
\newblock


\bibitem[Smith et~al\mbox{.}(2017)]%
        {smith2017strategies}
\bibfield{author}{\bibinfo{person}{Leah~M Smith}, \bibinfo{person}{Linda~E L{\'e}vesque}, \bibinfo{person}{Jay~S Kaufman}, {and} \bibinfo{person}{Erin~C Strumpf}.} \bibinfo{year}{2017}\natexlab{}.
\newblock \showarticletitle{Strategies for evaluating the assumptions of the regression discontinuity design: {A} case study using a human papillomavirus vaccination programme}.
\newblock \bibinfo{journal}{\emph{International Journal of Epidemiology}} \bibinfo{volume}{46}, \bibinfo{number}{3} (\bibinfo{year}{2017}), \bibinfo{pages}{939--949}.
\newblock


\bibitem[Solovev and Pr{\"o}llochs(2023)]%
        {solovev2023moralized}
\bibfield{author}{\bibinfo{person}{Kirill Solovev} {and} \bibinfo{person}{Nicolas Pr{\"o}llochs}.} \bibinfo{year}{2023}\natexlab{}.
\newblock \showarticletitle{Moralized language predicts hate speech on social media}.
\newblock \bibinfo{journal}{\emph{PNAS Nexus}} \bibinfo{volume}{2}, \bibinfo{number}{1} (\bibinfo{year}{2023}), \bibinfo{pages}{pgac281}.
\newblock


\bibitem[Solovev and Pröllochs(2022)]%
        {solovev2022moral}
\bibfield{author}{\bibinfo{person}{Kirill Solovev} {and} \bibinfo{person}{Nicolas Pröllochs}.} \bibinfo{year}{2022}\natexlab{}.
\newblock \showarticletitle{Moral emotions shape the virality of {COVID}-19 misinformation on social media}. In \bibinfo{booktitle}{\emph{Proceedings of the ACM Web Conference 2022}}. \bibinfo{pages}{3706--3717}.
\newblock


\bibitem[{SSRS}(2023)]%
        {ssrs2023cnn}
\bibfield{author}{\bibinfo{person}{{SSRS}}.} \bibinfo{year}{2023}\natexlab{}.
\newblock \bibinfo{title}{CNN Poll on Biden, economy and elections}.
\newblock
\newblock
\newblock
\shownote{https://www.documentcloud.org/documents/23895856-cnn-poll-on-biden-economy-and-elections}.


\bibitem[Statista(2024)]%
        {statista2024number}
\bibfield{author}{\bibinfo{person}{Statista}.} \bibinfo{year}{2024}\natexlab{}.
\newblock \bibinfo{title}{Number of internet and social media users worldwide as of July 2024}.
\newblock
\newblock
\newblock
\shownote{https://www.statista.com/statistics/617136/digital-population-worldwide/}.


\bibitem[Stieglitz and Dang-Xuan(2013)]%
        {stieglitz2013emotions}
\bibfield{author}{\bibinfo{person}{Stefan Stieglitz} {and} \bibinfo{person}{Linh Dang-Xuan}.} \bibinfo{year}{2013}\natexlab{}.
\newblock \showarticletitle{Emotions and information diffusion in social media—sentiment of microblogs and sharing behavior}.
\newblock \bibinfo{journal}{\emph{Journal of Management Information Systems}} \bibinfo{volume}{29}, \bibinfo{number}{4} (\bibinfo{year}{2013}), \bibinfo{pages}{217--248}.
\newblock


\bibitem[Strathern et~al\mbox{.}(2020)]%
        {strathern2020against}
\bibfield{author}{\bibinfo{person}{Wienke Strathern}, \bibinfo{person}{Mirco Schoenfeld}, \bibinfo{person}{Raji Ghawi}, {and} \bibinfo{person}{Juergen Pfeffer}.} \bibinfo{year}{2020}\natexlab{}.
\newblock \showarticletitle{Against the others! Detecting moral outrage in social media networks}. In \bibinfo{booktitle}{\emph{2020 IEEE/ACM International Conference on Advances in Social Networks Analysis and Mining}}. \bibinfo{pages}{322--326}.
\newblock


\bibitem[Tang et~al\mbox{.}(2024)]%
        {tang2024knows}
\bibfield{author}{\bibinfo{person}{Huiyun Tang}, \bibinfo{person}{Gabriele Lenzini}, \bibinfo{person}{Samuel Greiff}, \bibinfo{person}{Bj{\"o}rn Rohles}, {and} \bibinfo{person}{Anastasia Sergeeva}.} \bibinfo{year}{2024}\natexlab{}.
\newblock \showarticletitle{“{Who} knows? {Maybe} it really works”: {Analysing} users' perceptions of health misinformation on social media}. In \bibinfo{booktitle}{\emph{Proceedings of the 2024 ACM Designing Interactive Systems Conference}}. \bibinfo{pages}{1499--1517}.
\newblock


\bibitem[Van~Bavel et~al\mbox{.}(2024)]%
        {van2024social}
\bibfield{author}{\bibinfo{person}{Jay~J Van~Bavel}, \bibinfo{person}{Claire~E Robertson}, \bibinfo{person}{Kareena Del~Rosario}, \bibinfo{person}{Jesper Rasmussen}, {and} \bibinfo{person}{Steve Rathje}.} \bibinfo{year}{2024}\natexlab{}.
\newblock \showarticletitle{Social media and morality}.
\newblock \bibinfo{journal}{\emph{Annual Review of Psychology}} \bibinfo{volume}{75}, \bibinfo{number}{1} (\bibinfo{year}{2024}), \bibinfo{pages}{311--340}.
\newblock


\bibitem[Van~de Vyver and Abrams(2015)]%
        {van2015testing}
\bibfield{author}{\bibinfo{person}{Julie Van~de Vyver} {and} \bibinfo{person}{Dominic Abrams}.} \bibinfo{year}{2015}\natexlab{}.
\newblock \showarticletitle{Testing the prosocial effectiveness of the prototypical moral emotions: Elevation increases benevolent behaviors and outrage increases justice behaviors}.
\newblock \bibinfo{journal}{\emph{Journal of Experimental Social Psychology}}  \bibinfo{volume}{58} (\bibinfo{year}{2015}), \bibinfo{pages}{23--33}.
\newblock


\bibitem[van~der Eijk and Columbus(2023)]%
        {van2023expressions}
\bibfield{author}{\bibinfo{person}{Frances van~der Eijk} {and} \bibinfo{person}{Simon Columbus}.} \bibinfo{year}{2023}\natexlab{}.
\newblock \showarticletitle{Expressions of moral disgust reflect both disgust and anger}.
\newblock \bibinfo{journal}{\emph{Cognition and Emotion}} \bibinfo{volume}{37}, \bibinfo{number}{3} (\bibinfo{year}{2023}), \bibinfo{pages}{499--514}.
\newblock


\bibitem[Varanasi et~al\mbox{.}(2022)]%
        {varanasi2022accost}
\bibfield{author}{\bibinfo{person}{Rama~Adithya Varanasi}, \bibinfo{person}{Joyojeet Pal}, {and} \bibinfo{person}{Aditya Vashistha}.} \bibinfo{year}{2022}\natexlab{}.
\newblock \showarticletitle{Accost, accede, or amplify: Attitudes towards COVID-19 misinformation on WhatsApp in India}. In \bibinfo{booktitle}{\emph{Proceedings of the 2022 CHI Conference on Human Factors in Computing Systems}}. \bibinfo{pages}{1--17}.
\newblock


\bibitem[Vosoughi et~al\mbox{.}(2018)]%
        {vosoughi2018spread}
\bibfield{author}{\bibinfo{person}{Soroush Vosoughi}, \bibinfo{person}{Deb Roy}, {and} \bibinfo{person}{Sinan Aral}.} \bibinfo{year}{2018}\natexlab{}.
\newblock \showarticletitle{The spread of true and false news online}.
\newblock \bibinfo{journal}{\emph{Science}} \bibinfo{volume}{359}, \bibinfo{number}{6380} (\bibinfo{year}{2018}), \bibinfo{pages}{1146--1151}.
\newblock


\bibitem[Wijenayake et~al\mbox{.}(2020)]%
        {wijenayake2020effect}
\bibfield{author}{\bibinfo{person}{Senuri Wijenayake}, \bibinfo{person}{Danula Hettiachchi}, \bibinfo{person}{Simo Hosio}, \bibinfo{person}{Vassilis Kostakos}, {and} \bibinfo{person}{Jorge Goncalves}.} \bibinfo{year}{2020}\natexlab{}.
\newblock \showarticletitle{Effect of conformity on perceived trustworthiness of news in social media}.
\newblock \bibinfo{journal}{\emph{IEEE Internet Computing}} \bibinfo{volume}{25}, \bibinfo{number}{1} (\bibinfo{year}{2020}), \bibinfo{pages}{12--19}.
\newblock


\bibitem[Wilner et~al\mbox{.}(2023)]%
        {wilner2023attending}
\bibfield{author}{\bibinfo{person}{Tamar Wilner}, \bibinfo{person}{Kayo Mimizuka}, \bibinfo{person}{Ayesha Bhimdiwala}, \bibinfo{person}{Jason~C Young}, {and} \bibinfo{person}{Ahmer Arif}.} \bibinfo{year}{2023}\natexlab{}.
\newblock \showarticletitle{It’s about time: Attending to temporality in misinformation interventions}. In \bibinfo{booktitle}{\emph{Proceedings of the 2023 CHI Conference on Human Factors in Computing Systems}}. \bibinfo{pages}{1--19}.
\newblock


\bibitem[Wojcik et~al\mbox{.}(2022)]%
        {wojcik2022birdwatch}
\bibfield{author}{\bibinfo{person}{Stefan Wojcik}, \bibinfo{person}{Sophie Hilgard}, \bibinfo{person}{Nick Judd}, \bibinfo{person}{Delia Mocanu}, \bibinfo{person}{Stephen Ragain}, \bibinfo{person}{MB Hunzaker}, \bibinfo{person}{Keith Coleman}, {and} \bibinfo{person}{Jay Baxter}.} \bibinfo{year}{2022}\natexlab{}.
\newblock \showarticletitle{Birdwatch: Crowd wisdom and bridging algorithms can inform understanding and reduce the spread of misinformation}.
\newblock \bibinfo{journal}{\emph{ArXiv}} (\bibinfo{year}{2022}).
\newblock


\bibitem[X(2024)]%
        {x2024replies}
\bibfield{author}{\bibinfo{person}{X}.} \bibinfo{year}{2024}\natexlab{}.
\newblock \bibinfo{title}{About replies and mentions}.
\newblock
\newblock
\urldef\tempurl%
\url{https://help.x.com/en/using-x/mentions-and-replies}
\showURL{%
\tempurl}


\bibitem[Yadollahi et~al\mbox{.}(2017)]%
        {yadollahi2017current}
\bibfield{author}{\bibinfo{person}{Ali Yadollahi}, \bibinfo{person}{Ameneh~Gholipour Shahraki}, {and} \bibinfo{person}{Osmar~R Zaiane}.} \bibinfo{year}{2017}\natexlab{}.
\newblock \showarticletitle{Current state of text sentiment analysis from opinion to emotion mining}.
\newblock \bibinfo{journal}{\emph{ACM Computing Surveys (CSUR)}} \bibinfo{volume}{50}, \bibinfo{number}{2} (\bibinfo{year}{2017}), \bibinfo{pages}{1--33}.
\newblock


\bibitem[Yaqub et~al\mbox{.}(2020)]%
        {yaqub2020effects}
\bibfield{author}{\bibinfo{person}{Waheeb Yaqub}, \bibinfo{person}{Otari Kakhidze}, \bibinfo{person}{Morgan~L Brockman}, \bibinfo{person}{Nasir Memon}, {and} \bibinfo{person}{Sameer Patil}.} \bibinfo{year}{2020}\natexlab{}.
\newblock \showarticletitle{Effects of credibility indicators on social media news sharing intent}. In \bibinfo{booktitle}{\emph{Proceedings of the 2020 CHI Conference on Human Factors in Computing Systems}}. \bibinfo{pages}{1--14}.
\newblock


\bibitem[YouTube(2024)]%
        {youtube2024testing}
\bibfield{author}{\bibinfo{person}{YouTube}.} \bibinfo{year}{2024}\natexlab{}.
\newblock \bibinfo{title}{Testing new ways to offer viewers more context and information on videos}.
\newblock
\newblock
\urldef\tempurl%
\url{https://blog.youtube/news-and-events/new-ways-to-offer-viewers-more-context/}
\showURL{%
\tempurl}


\bibitem[Zade et~al\mbox{.}(2024)]%
        {zade2024reply}
\bibfield{author}{\bibinfo{person}{Himanshu Zade}, \bibinfo{person}{Spencer Williams}, \bibinfo{person}{Theresa~T Tran}, \bibinfo{person}{Christina Smith}, \bibinfo{person}{Sukrit Venkatagiri}, \bibinfo{person}{Gary Hsieh}, {and} \bibinfo{person}{Kate Starbird}.} \bibinfo{year}{2024}\natexlab{}.
\newblock \showarticletitle{To reply or to quote: {Comparing} conversational framing strategies on {Twitter}}.
\newblock \bibinfo{journal}{\emph{ACM Journal on Computing and Sustainable Societies}} \bibinfo{volume}{2}, \bibinfo{number}{1} (\bibinfo{year}{2024}), \bibinfo{pages}{1--27}.
\newblock


\bibitem[Zhang et~al\mbox{.}(2023)]%
        {zhang2023what}
\bibfield{author}{\bibinfo{person}{Yixuan Zhang}, \bibinfo{person}{Joseph~D Gaggiano}, \bibinfo{person}{Nutchanon Yongsatianchot}, \bibinfo{person}{Nurul~M Suhaimi}, \bibinfo{person}{Miso Kim}, \bibinfo{person}{Yifan Sun}, \bibinfo{person}{Jacqueline Griffin}, {and} \bibinfo{person}{Andrea~G Parker}.} \bibinfo{year}{2023}\natexlab{}.
\newblock \showarticletitle{What do we mean when we talk about trust in social media? A systematic review}. In \bibinfo{booktitle}{\emph{Proceedings of the 2023 CHI Conference on Human Factors in Computing Systems}}. \bibinfo{pages}{1--22}.
\newblock


\bibitem[Zhang et~al\mbox{.}(2024)]%
        {zhang2024profiling}
\bibfield{author}{\bibinfo{person}{Yixuan Zhang}, \bibinfo{person}{Yimeng Wang}, \bibinfo{person}{Nutchanon Yongsatianchot}, \bibinfo{person}{Joseph~D Gaggiano}, \bibinfo{person}{Nurul~M Suhaimi}, \bibinfo{person}{Anne Okrah}, \bibinfo{person}{Miso Kim}, \bibinfo{person}{Jacqueline Griffin}, {and} \bibinfo{person}{Andrea~G Parker}.} \bibinfo{year}{2024}\natexlab{}.
\newblock \showarticletitle{Profiling the dynamics of trust \& distrust in social media: A survey study}. In \bibinfo{booktitle}{\emph{Proceedings of the 2024 CHI Conference on Human Factors in Computing Systems}}. \bibinfo{pages}{1--24}.
\newblock


\bibitem[Zhou et~al\mbox{.}(2023)]%
        {zhou2023synthetic}
\bibfield{author}{\bibinfo{person}{Jiawei Zhou}, \bibinfo{person}{Yixuan Zhang}, \bibinfo{person}{Qianni Luo}, \bibinfo{person}{Andrea~G Parker}, {and} \bibinfo{person}{Munmun De~Choudhury}.} \bibinfo{year}{2023}\natexlab{}.
\newblock \showarticletitle{Synthetic lies: Understanding AI-generated misinformation and evaluating algorithmic and human solutions}. In \bibinfo{booktitle}{\emph{Proceedings of the 2023 CHI Conference on Human Factors in Computing Systems}}. \bibinfo{pages}{1--20}.
\newblock


\bibitem[Zollo et~al\mbox{.}(2015)]%
        {zollo2015emotional}
\bibfield{author}{\bibinfo{person}{Fabiana Zollo}, \bibinfo{person}{Petra~Kralj Novak}, \bibinfo{person}{Michela Del~Vicario}, \bibinfo{person}{Alessandro Bessi}, \bibinfo{person}{Igor Mozeti{\v{c}}}, \bibinfo{person}{Antonio Scala}, \bibinfo{person}{Guido Caldarelli}, {and} \bibinfo{person}{Walter Quattrociocchi}.} \bibinfo{year}{2015}\natexlab{}.
\newblock \showarticletitle{Emotional dynamics in the age of misinformation}.
\newblock \bibinfo{journal}{\emph{PLoS ONE}} \bibinfo{volume}{10}, \bibinfo{number}{9} (\bibinfo{year}{2015}), \bibinfo{pages}{e0138740}.
\newblock


\end{thebibliography}

\clearpage
\appendix

\onecolumn
\begin{center}
    \Large \textbf{Supplementary Materials for}
    \Large \textbf{``Community Fact-Checks Trigger Moral Outrage in Replies to Misleading Posts on Social Media''}
\end{center}
\renewcommand\thetable{S\arabic{table}}
\setcounter{table}{0}
\renewcommand\thefigure{S\arabic{figure}}
\setcounter{figure}{0}
\renewcommand\thesection{S\arabic{section}}
\setcounter{section}{0}

\section{Data Overview}
\label{sec:sm_descriptive}
\subsection{Correlations Between Sentiments/Emotions in Source Posts and Sentiments/Emotions in Replies}
\label{sec:sm_corr_reply_post}
\begin{figure}[H]
\centering
\begin{subfigure}{0.32\textwidth}
\includegraphics[width=\textwidth]{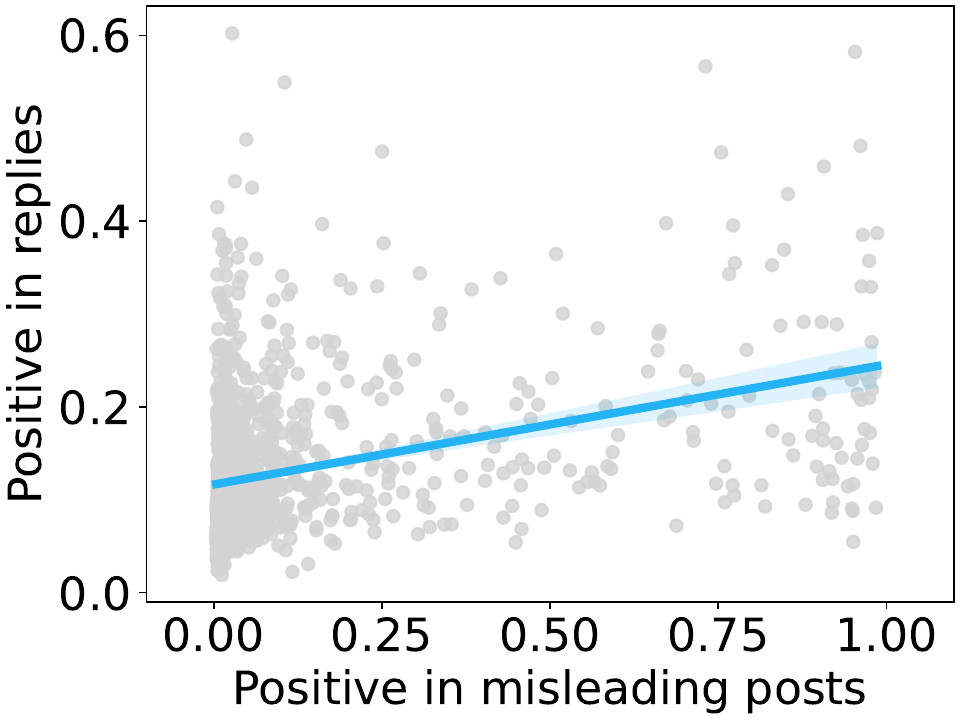}
\caption{}
\label{fig:positive_roberta}
\end{subfigure}
\hspace{1cm}
\begin{subfigure}{0.32\textwidth}
\includegraphics[width=\textwidth]{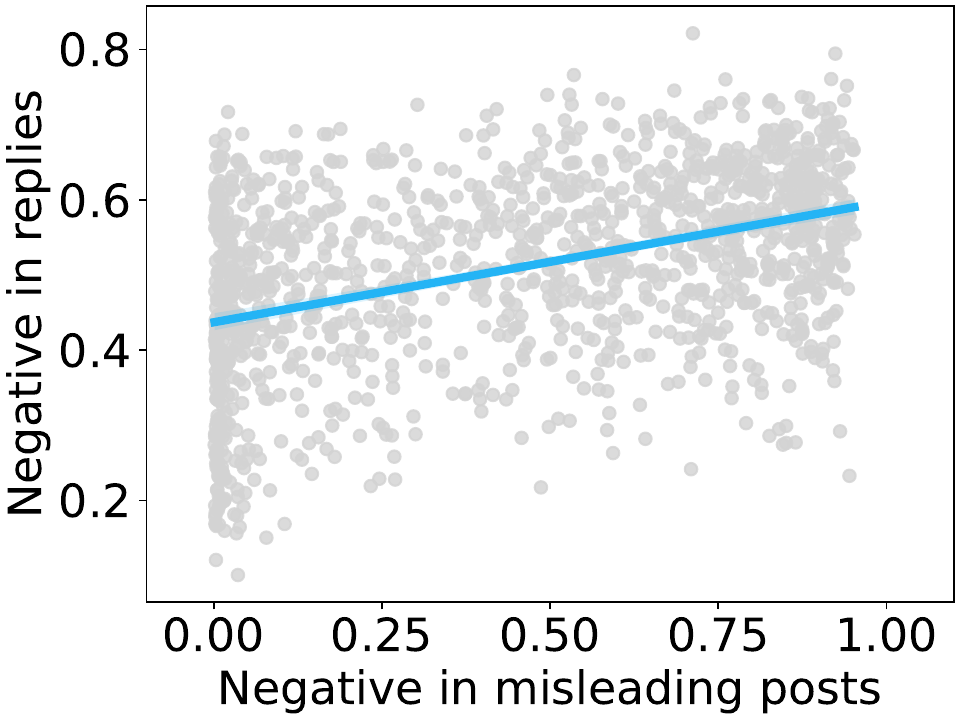}
\caption{}
\label{fig:negative_roberta}
\end{subfigure}

\begin{subfigure}{0.32\textwidth}
\includegraphics[width=\textwidth]{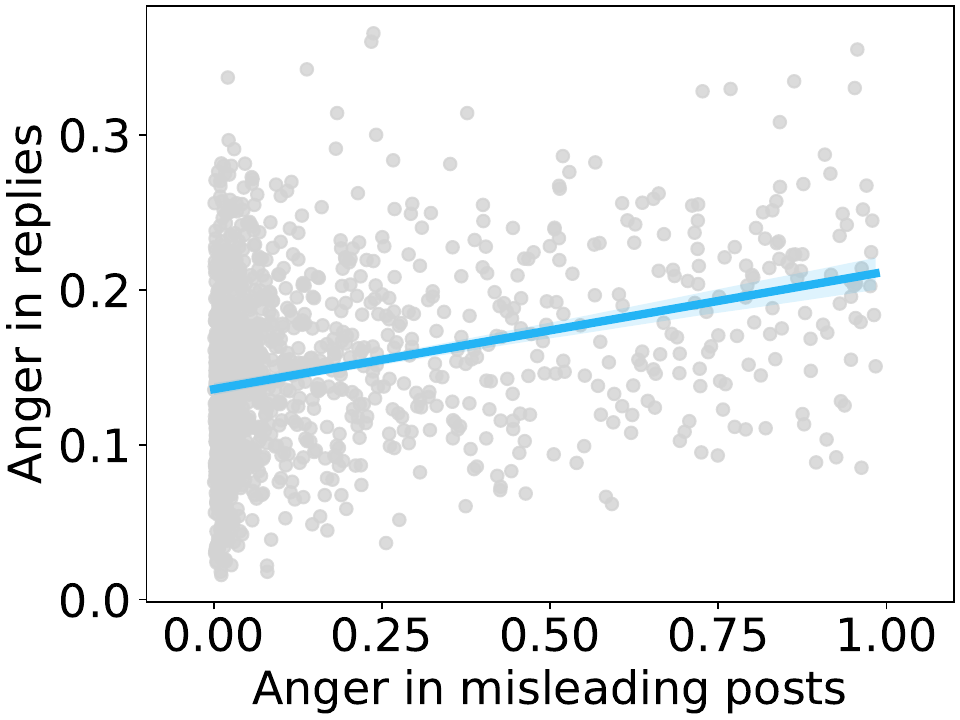}
\caption{}
\label{fig:anger_roberta}
\end{subfigure}
\hfill
\begin{subfigure}{0.32\textwidth}
\includegraphics[width=\textwidth]{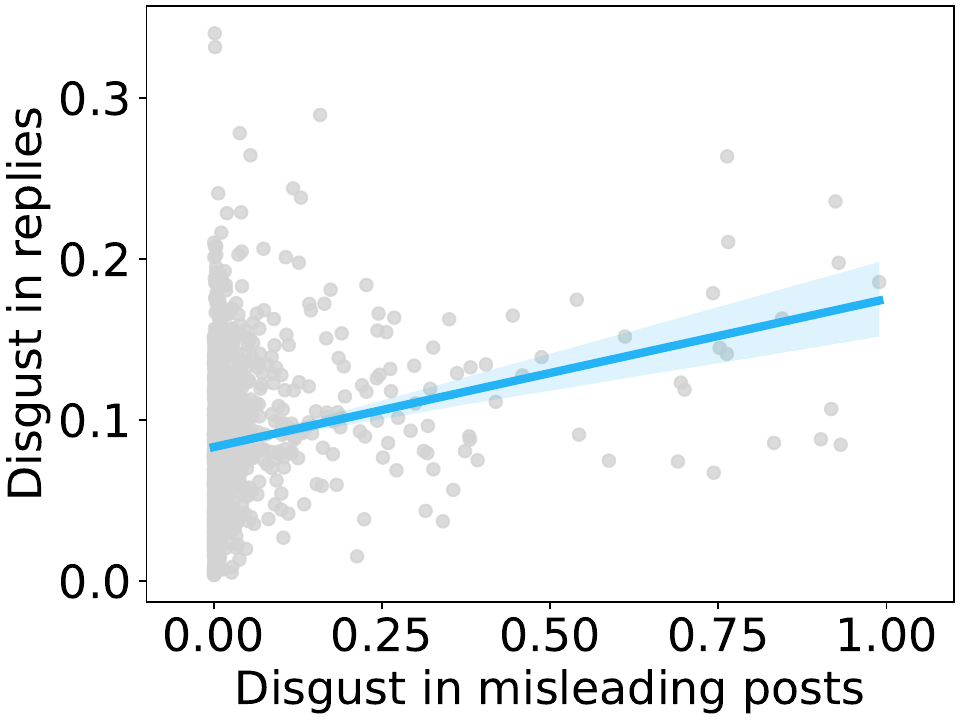}
\caption{}
\label{fig:disgust_roberta}
\end{subfigure}
\hfill
\begin{subfigure}{0.32\textwidth}
\includegraphics[width=\textwidth]{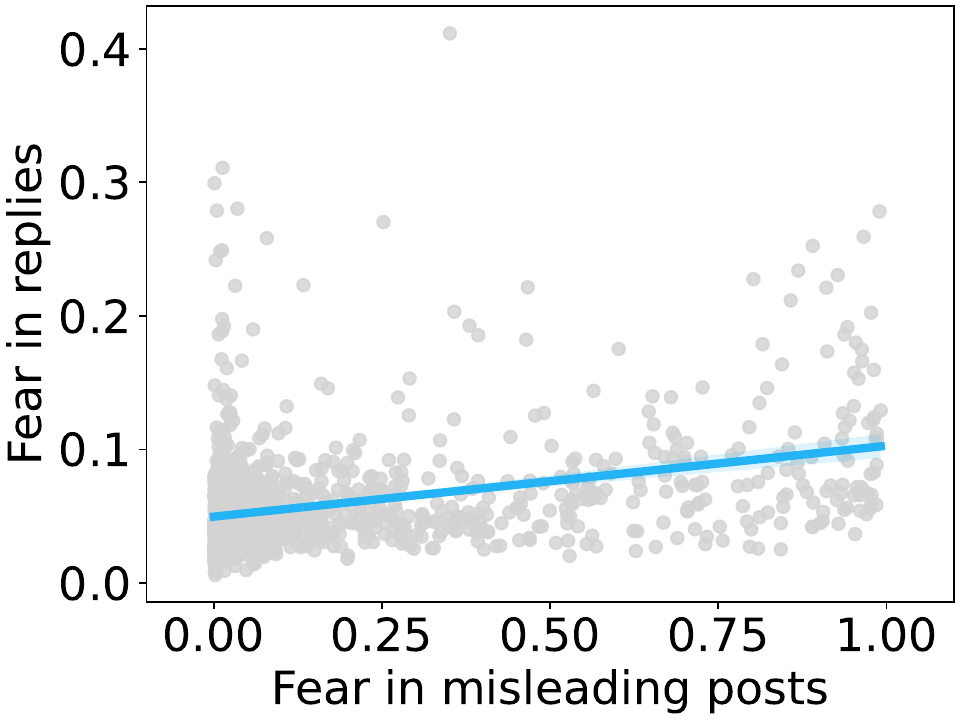}
\caption{}
\label{fig:fear_roberta}
\end{subfigure}

\begin{subfigure}{0.32\textwidth}
\includegraphics[width=\textwidth]{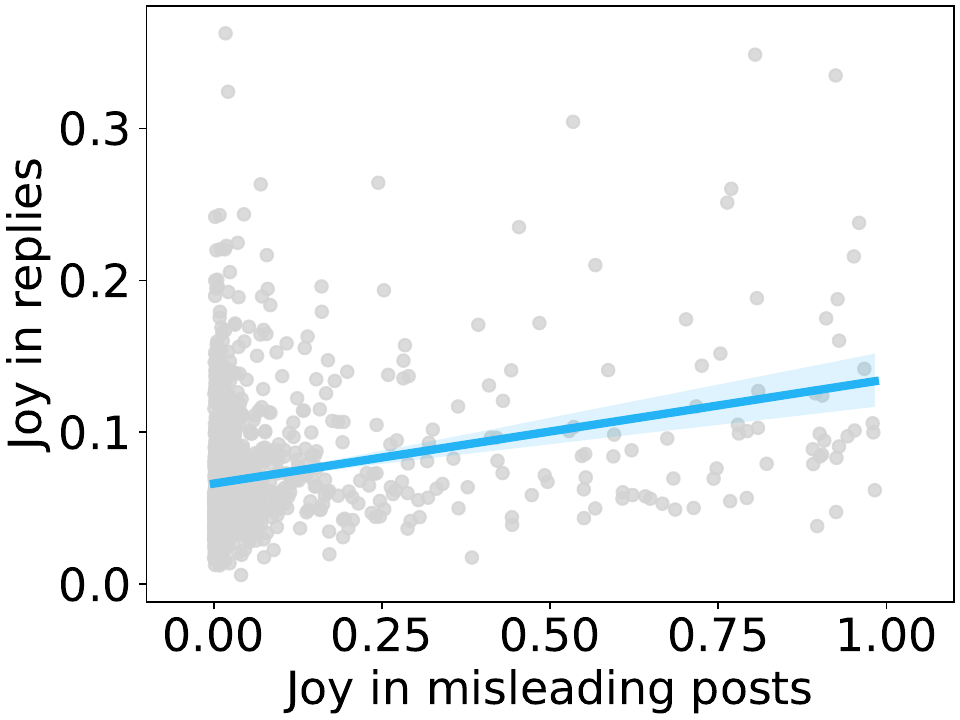}
\caption{}
\label{fig:joy_roberta}
\end{subfigure}
\hfill
\begin{subfigure}{0.32\textwidth}
\includegraphics[width=\textwidth]{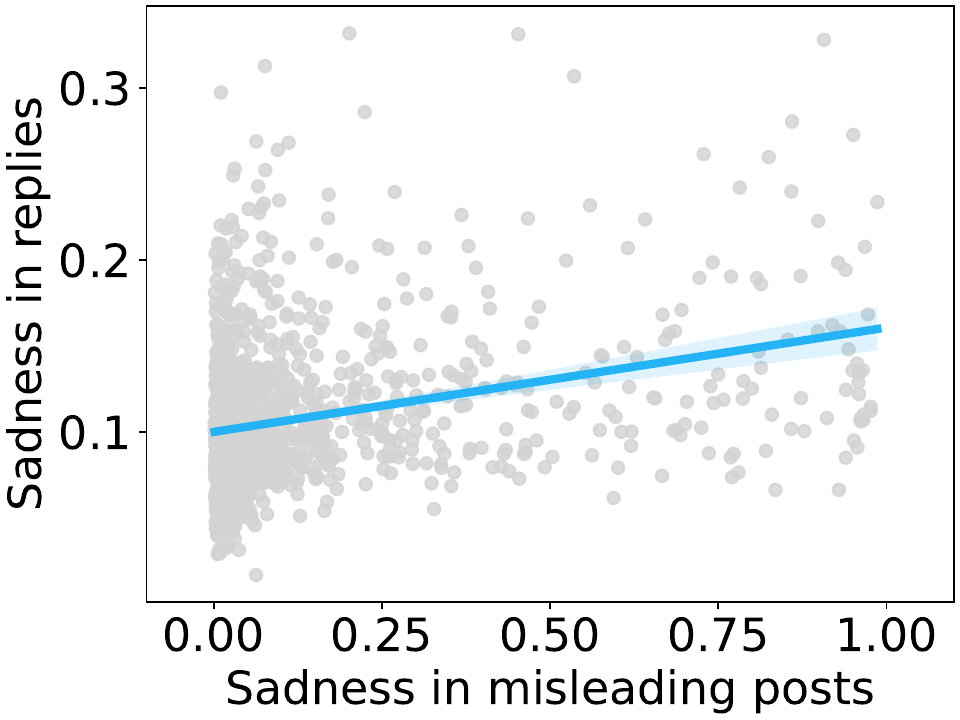}
\caption{}
\label{fig:sadness_roberta}
\end{subfigure}
\hfill
\begin{subfigure}{0.32\textwidth}
\includegraphics[width=\textwidth]{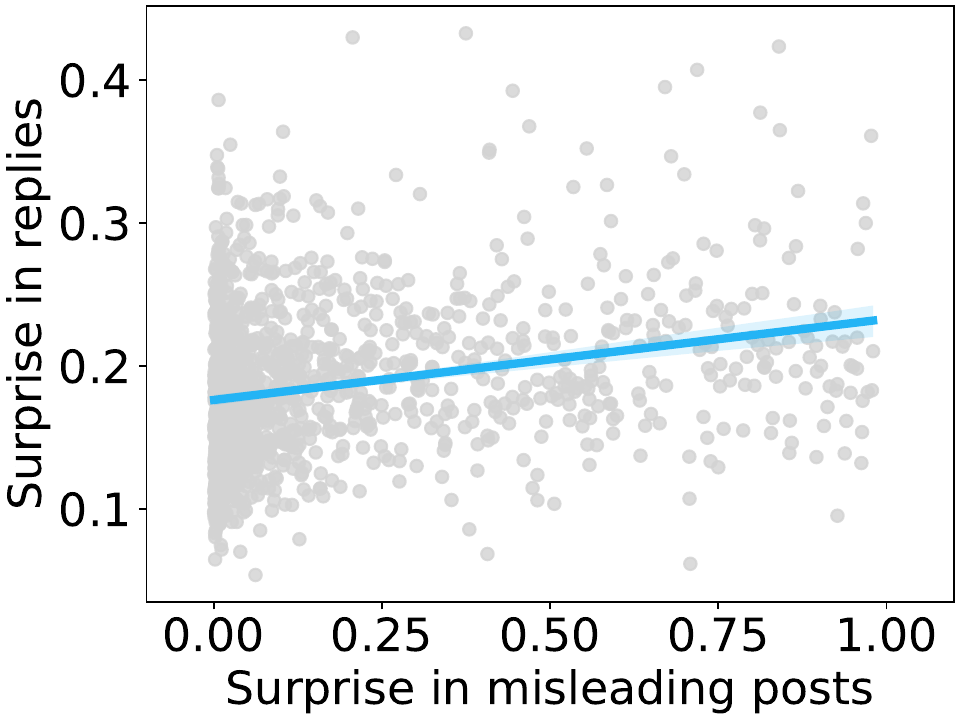}
\caption{}
\label{fig:surprise_roberta}
\end{subfigure}
\caption{The Pearson correlations between sentiments (emotions) in replies and source sentiments (emotions) in misleading posts. \subref{fig:positive_roberta} Positive sentiment ($r=0.365$, $p<0.001$).  \subref{fig:negative_roberta} Negative sentiment ($r=0.426$, $p<0.001$). \subref{fig:anger_roberta} Anger ($r=0.316$, $p<0.001$). \subref{fig:disgust_roberta} Disgust ($r=0.235$, $p<0.001$). \subref{fig:fear_roberta} Fear ($r=0.354$, $p<0.001$). \subref{fig:joy_roberta} Joy ($r=0.287$, $p<0.001$). \subref{fig:sadness_roberta} Sadness ($r=0.302$, $p<0.001$). \subref{fig:surprise_roberta} Surprise ($r=0.254$, $p<0.001$).}
\label{fig:corr_source_reply}
\end{figure}

Given the potential emotion transfers from source posts to their replies, we analyze the correlations between source sentiments and emotions in the misleading posts and the subsequent sentiments and emotions in the replies. Of note, the replies after the display of community notes are excluded in the analysis of emotion transfers to prevent the potential contamination of note display. We find that all the sentiments and emotions in replies are positively correlated to the corresponding source sentiments and emotions in the misleading posts (see Fig. \ref{fig:corr_source_reply}). 

\newpage
\subsection{Sentiments and Emotions in Replies Over Time}
\label{sec:rdd_observation}

\subsubsection{Pre-trends before note display.}
We analyze whether the sentiments and emotions in replies change following the spread of source posts (\ie, post age) before the display of community notes. Specifically, we take sentiments and emotions in replies before note display as dependent variables and examine the effect of the independent variable of $\var{PostAge}$. The estimation results are reported in Table \ref{tab:sentiments_pretrend} and Table \ref{tab:emotions_pretrend}. All the coefficient estimates of $\var{PostAge}$ are consistently not statistically significant across sentiments and emotions in replies before note display (each $p>0.05$). This suggests that the sentiments and emotions in replies remain stable over time following the spread of posts.

\subsubsection{Discontinuity around the display of community notes.}
We use an interrupted time series design to detect discontinuity around the cut-off point indicated by the display of community notes. The estimation results for sentiments and emotions in replies within the entire lifespan are reported in Table \ref{tab:sentiments_its} and Table \ref{tab:emotions_its}. The key variables are $\var{Displayed}$ and its interaction with $\var{PostAge}$. We find that the coefficient estimates of $\var{Displayed}$ are significantly significant for positive, negative, anger, and surprise, indicating significant changes after the display of community notes. Additionally, the coefficient estimates of $\var{Displayed} \times \var{PostAge}$ are not statistically significant across sentiments and emotions in replies (each $p>0.05$). This suggests that the sentiments and emotions in replies remain stable over time after the changes at the cut-off point. Moreover, the coefficient estimates of $\var{PostAge}$ are consistent with that in the analysis of pre-trends for sentiments and emotions in replies (each $p>0.05$). Taken together, these findings suggest the existence of discontinuities for certain sentiments and emotions at the cut-off point.

\begin{figure}[H]
\centering
\begin{subfigure}{0.32\textwidth}
\caption{}
\includegraphics[width=\textwidth]{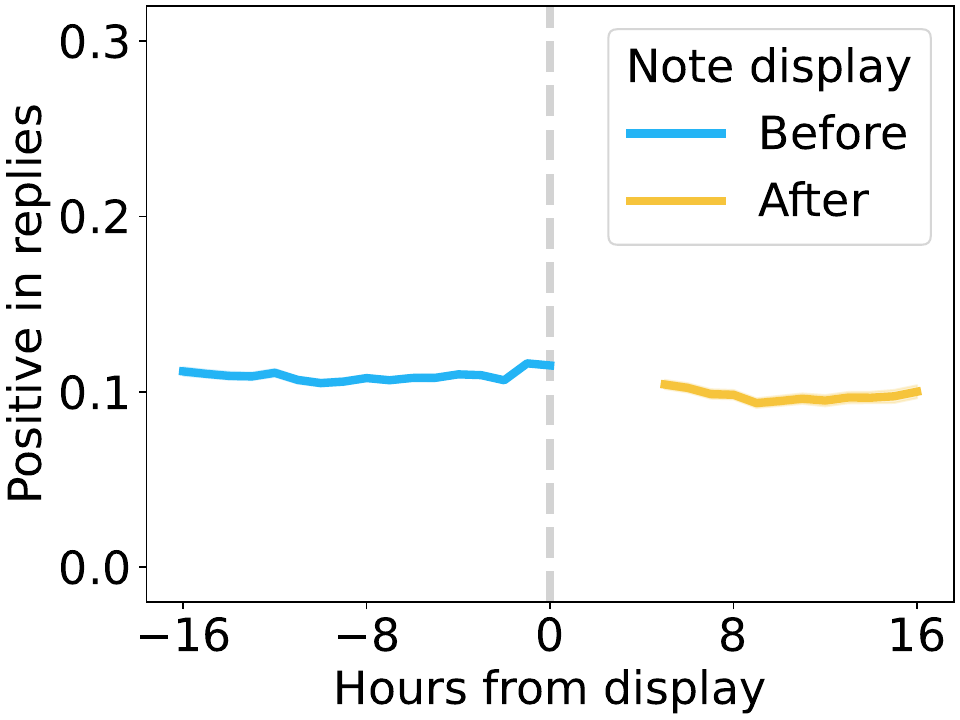}
\label{fig:positive_over_time}
\end{subfigure}
\hspace{1cm}
\begin{subfigure}{0.32\textwidth}
\caption{}
\includegraphics[width=\textwidth]{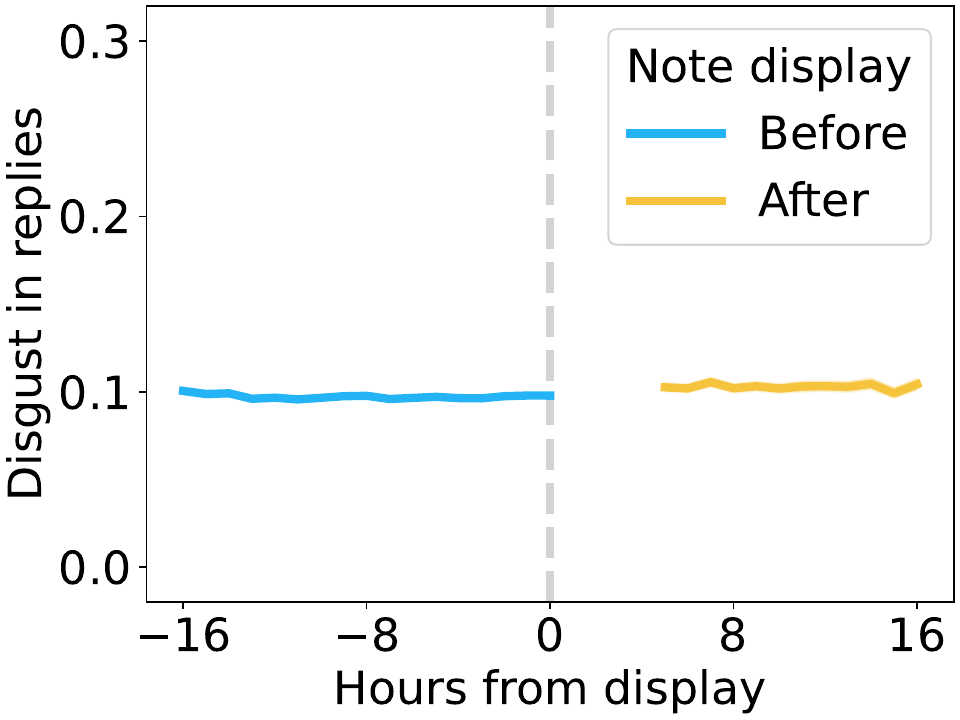}
\label{fig:disgust_over_time}
\end{subfigure}

\begin{subfigure}{0.32\textwidth}
\caption{}
\includegraphics[width=\textwidth]{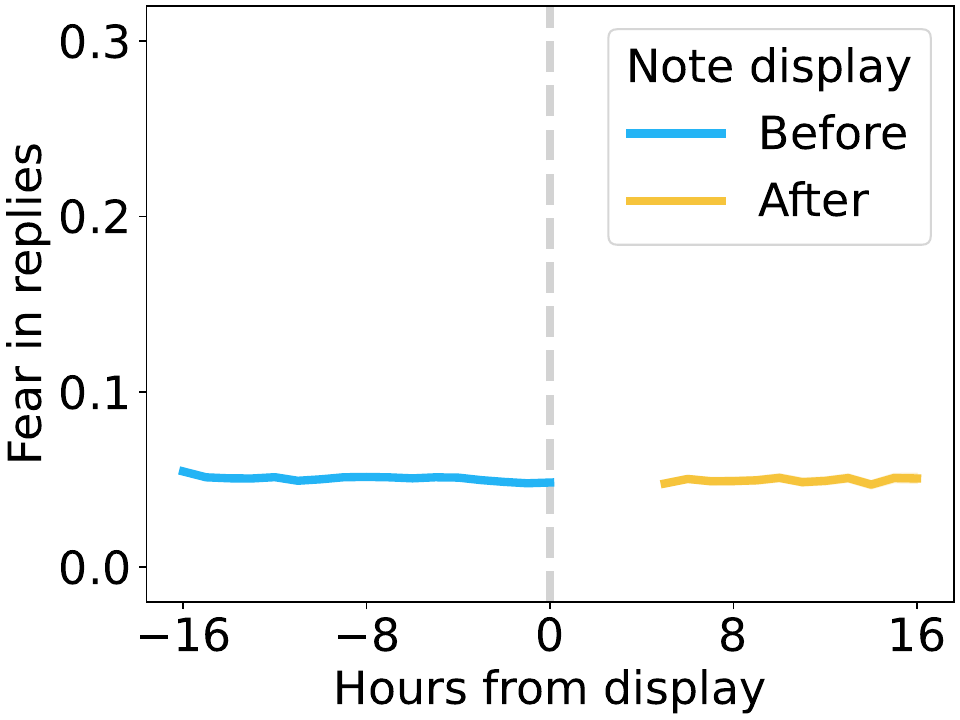}
\label{fig:fear_over_time}
\end{subfigure}
\hfill
\begin{subfigure}{0.32\textwidth}
\caption{}
\includegraphics[width=\textwidth]{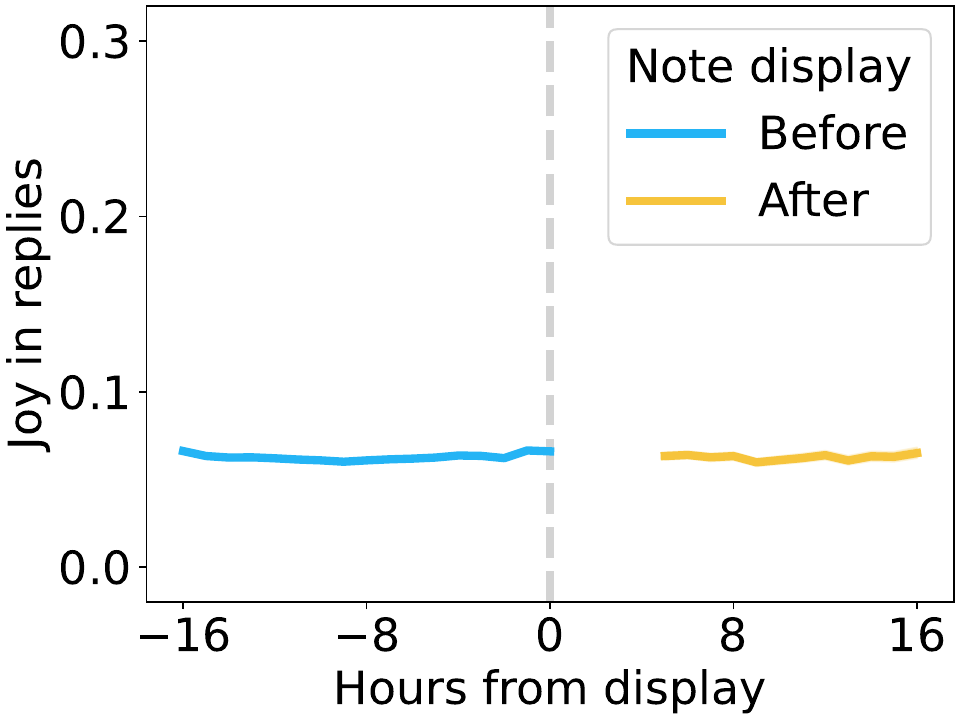}
\label{fig:joy_over_time}
\end{subfigure}
\hfill
\begin{subfigure}{0.32\textwidth}
\caption{}
\includegraphics[width=\textwidth]{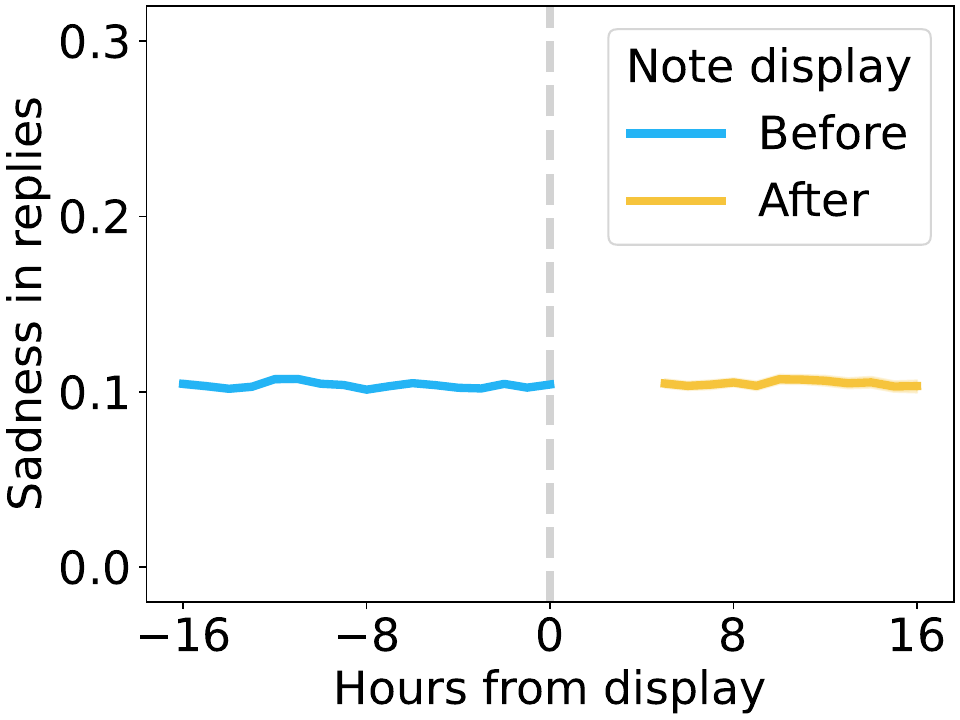}
\label{fig:sadness_over_time}
\end{subfigure}
\caption{The hourly averages in sentiments and emotions across hours from note display. \subref{fig:positive_over_time} The hourly averages of positive sentiment in replies across hours from note display. \subref{fig:disgust_over_time} The hourly averages of disgust in replies across hours from note display. \subref{fig:fear_over_time} The hourly averages of fear in replies across hours from note display. \subref{fig:joy_over_time} The hourly averages of joy in replies across hours from note display. \subref{fig:sadness_over_time} The hourly averages of sadness in replies across hours from note display. The error bands represent 95\% CIs.}
\label{fig:emotions_over_time}
\end{figure}

To better clarify the stable trends in sentiments and emotions over time and discontinuities at the cut-off point, we illustrate the hourly averages of sentiments and emotions in replies within the 16-hour bandwidth (see Fig. \ref{fig:emotions_over_time}). 

\newpage
\begin{table}[H]
\centering
% \footnotesize
\caption{Estimation results for positive [Column (1)] and negative [Column (2)] sentiments in replies before note display. Reported are coefficient estimates with post-clustered standard errors in parentheses. \sym{*} \(p<0.05\), \sym{**} \(p<0.01\), \sym{***} \(p<0.001\).}
\begin{tabularx}{\columnwidth}{@{\hspace{\tabcolsep}\extracolsep{\fill}}l*{2}{S}}
\toprule
&\multicolumn{1}{c}{(1)}&\multicolumn{1}{c}{(2)}\\
&\multicolumn{1}{c}{Positive}&\multicolumn{1}{c}{Negative}\\
\midrule
SourcePositive&       0.039\sym{*}  &       0.015         \\
            &     (0.017)         &     (0.020)         \\
SourceNegative&      -0.054\sym{***}&       0.122\sym{***}\\
            &     (0.016)         &     (0.022)         \\
PostAge     &      -0.001         &      -0.004         \\
            &     (0.006)         &     (0.007)         \\
Intercept      &       0.000         &      -0.000         \\
            &     (0.013)         &     (0.018)         \\
\midrule
\(R^{2}\)   &       0.007         &       0.013         \\
\midrule
\#Replies (\(N\))       &        \num{1688102}         &       \num{1688102}\\
\#Source posts      &        \num{1840}         &       \num{1840}\\
\bottomrule
\end{tabularx}
\label{tab:sentiments_pretrend}
\end{table}

\begin{table}[H]
\centering
% \footnotesize
\caption{Estimation results for anger [Column (1)], disgust [Column (2)], fear [Column (3)], joy [Column (4)], sadness [Column (5)], and surprise [Column (6)] in replies before note display. Reported are coefficient estimates with post-clustered standard errors in parentheses. \sym{*} \(p<0.05\), \sym{**} \(p<0.01\), \sym{***} \(p<0.001\).}
\begin{tabularx}{\columnwidth}{@{\hspace{\tabcolsep}\extracolsep{\fill}}l*{6}{S}}
\toprule
&\multicolumn{1}{c}{(1)}&\multicolumn{1}{c}{(2)}&\multicolumn{1}{c}{(3)}&\multicolumn{1}{c}{(4)}&\multicolumn{1}{c}{(5)}&\multicolumn{1}{c}{(6)}\\
&\multicolumn{1}{c}{Anger}&\multicolumn{1}{c}{Disgust}&\multicolumn{1}{c}{Fear}&\multicolumn{1}{c}{Joy}&\multicolumn{1}{c}{Sadness}&\multicolumn{1}{c}{Surprise}\\
\midrule
SourceAnger &       0.083\sym{***}&       0.010         &       0.017         &      -0.005         &       0.019         &      -0.039\sym{***}\\
            &     (0.014)         &     (0.011)         &     (0.009)         &     (0.009)         &     (0.015)         &     (0.011)         \\
SourceDisgust&      -0.000         &       0.073\sym{***}&      -0.004         &      -0.009         &      -0.010         &      -0.004         \\
            &     (0.008)         &     (0.015)         &     (0.004)         &     (0.005)         &     (0.006)         &     (0.007)         \\
SourceFear  &       0.013         &       0.011         &       0.067\sym{***}&      -0.023\sym{**} &       0.003         &       0.002         \\
            &     (0.015)         &     (0.012)         &     (0.011)         &     (0.009)         &     (0.010)         &     (0.014)         \\
SourceJoy   &       0.006         &       0.000         &       0.005         &       0.038\sym{**} &       0.024\sym{**} &       0.003         \\
            &     (0.014)         &     (0.009)         &     (0.006)         &     (0.014)         &     (0.008)         &     (0.011)         \\
SourceSadness&       0.009         &       0.001         &       0.005         &       0.010         &       0.058\sym{***}&      -0.012         \\
            &     (0.010)         &     (0.010)         &     (0.005)         &     (0.009)         &     (0.009)         &     (0.009)         \\
SourceSurprise&      -0.054\sym{***}&      -0.005         &      -0.001         &       0.000         &       0.012         &       0.065\sym{***}\\
            &     (0.011)         &     (0.013)         &     (0.007)         &     (0.010)         &     (0.013)         &     (0.013)         \\
PostAge     &      -0.004         &       0.011         &       0.004         &       0.001         &      -0.000         &       0.006         \\
            &     (0.004)         &     (0.009)         &     (0.004)         &     (0.005)         &     (0.003)         &     (0.004)         \\
Intercept      &      -0.000         &       0.000         &       0.000         &      -0.000         &      -0.000         &      -0.000         \\
            &     (0.012)         &     (0.010)         &     (0.008)         &     (0.009)         &     (0.012)         &     (0.010)         \\
\midrule
\(R^{2}\)   &       0.013         &       0.006         &       0.005         &       0.003         &       0.003         &       0.008         \\
\midrule
\#Replies (\(N\))       &        \num{1688102}         &       \num{1688102}&        \num{1688102}         &       \num{1688102}&        \num{1688102}         &       \num{1688102}\\
\#Source posts      &        \num{1840}         &       \num{1840}&        \num{1840}         &       \num{1840}&        \num{1840}         &       \num{1840}\\
\bottomrule
\end{tabularx}
\label{tab:emotions_pretrend}
\end{table}

\newpage
\begin{table}[H]
\centering
% \footnotesize
\caption{Estimation results for positive [Column (1)] and negative [Column (2)] sentiments in replies. Reported are coefficient estimates with post-clustered standard errors in parentheses. \sym{*} \(p<0.05\), \sym{**} \(p<0.01\), \sym{***} \(p<0.001\).}
\begin{tabularx}{\columnwidth}{@{\hspace{\tabcolsep}\extracolsep{\fill}}l*{2}{S}}
\toprule
&\multicolumn{1}{c}{(1)}&\multicolumn{1}{c}{(2)}\\
&\multicolumn{1}{c}{Positive}&\multicolumn{1}{c}{Negative}\\
\midrule
Displayed   &      -0.057\sym{**} &       0.117\sym{***}\\
            &     (0.018)         &     (0.024)         \\
PostAge     &      -0.004         &      -0.009         \\
            &     (0.013)         &     (0.015)         \\
Displayed $\times$ PostAge&       0.008         &       0.004         \\
            &     (0.013)         &     (0.014)         \\
SourcePositive&       0.039\sym{*}  &       0.015         \\
            &     (0.018)         &     (0.020)         \\
SourceNegative&      -0.047\sym{**} &       0.113\sym{***}\\
            &     (0.016)         &     (0.023)         \\
Intercept      &       0.013         &      -0.029         \\
            &     (0.013)         &     (0.018)         \\
\midrule
\(R^{2}\)   &       0.006         &       0.014         \\
\midrule
\#Replies (\(N\))       &        \num{2225260}         &       \num{2225260}\\
\#Source posts      &        \num{1841}         &       \num{1841}\\
\bottomrule
\end{tabularx}
\label{tab:sentiments_its}
\end{table}

\begin{table}[H]
\centering
% \footnotesize
\caption{Estimation results for anger [Column (1)], disgust [Column (2)], fear [Column (3)], joy [Column (4)], sadness [Column (5)], and surprise [Column (6)] in replies. Reported are coefficient estimates with post-clustered standard errors in parentheses. \sym{*} \(p<0.05\), \sym{**} \(p<0.01\), \sym{***} \(p<0.001\).}
\begin{tabularx}{\columnwidth}{@{\hspace{\tabcolsep}\extracolsep{\fill}}l*{6}{S}}
\toprule
&\multicolumn{1}{c}{(1)}&\multicolumn{1}{c}{(2)}&\multicolumn{1}{c}{(3)}&\multicolumn{1}{c}{(4)}&\multicolumn{1}{c}{(5)}&\multicolumn{1}{c}{(6)}\\
&\multicolumn{1}{c}{Anger}&\multicolumn{1}{c}{Disgust}&\multicolumn{1}{c}{Fear}&\multicolumn{1}{c}{Joy}&\multicolumn{1}{c}{Sadness}&\multicolumn{1}{c}{Surprise}\\
\midrule
Displayed   &       0.080\sym{***}&       0.021         &      -0.008         &      -0.017         &      -0.011         &      -0.049\sym{***}\\
            &     (0.012)         &     (0.011)         &     (0.009)         &     (0.011)         &     (0.011)         &     (0.010)         \\
PostAge     &      -0.009         &       0.025         &       0.009         &       0.001         &      -0.001         &       0.014         \\
            &     (0.009)         &     (0.019)         &     (0.010)         &     (0.010)         &     (0.008)         &     (0.010)         \\
Displayed $\times$ PostAge&       0.006         &      -0.028         &      -0.006         &       0.004         &       0.005         &      -0.008         \\
            &     (0.008)         &     (0.019)         &     (0.009)         &     (0.010)         &     (0.008)         &     (0.010)         \\
SourceAnger &       0.078\sym{***}&       0.008         &       0.016\sym{*}  &      -0.006         &       0.020         &      -0.037\sym{***}\\
            &     (0.012)         &     (0.011)         &     (0.008)         &     (0.008)         &     (0.014)         &     (0.010)         \\
SourceDisgust&       0.001         &       0.069\sym{***}&      -0.004         &      -0.011\sym{*}  &      -0.012\sym{*}  &      -0.004         \\
            &     (0.008)         &     (0.015)         &     (0.004)         &     (0.005)         &     (0.006)         &     (0.007)         \\
SourceFear  &       0.013         &       0.008         &       0.066\sym{***}&      -0.021\sym{**} &       0.006         &       0.004         \\
            &     (0.014)         &     (0.011)         &     (0.010)         &     (0.007)         &     (0.009)         &     (0.012)         \\
SourceJoy   &       0.009         &      -0.001         &       0.005         &       0.032         &       0.026\sym{***}&       0.000         \\
            &     (0.016)         &     (0.009)         &     (0.006)         &     (0.018)         &     (0.007)         &     (0.012)         \\
SourceSadness&       0.008         &      -0.002         &       0.006         &       0.009         &       0.056\sym{***}&      -0.008         \\
            &     (0.010)         &     (0.009)         &     (0.005)         &     (0.009)         &     (0.008)         &     (0.008)         \\
SourceSurprise&      -0.055\sym{***}&      -0.012         &      -0.002         &       0.000         &       0.016         &       0.068\sym{***}\\
            &     (0.011)         &     (0.012)         &     (0.007)         &     (0.009)         &     (0.012)         &     (0.013)         \\
Intercept      &      -0.020         &      -0.002         &       0.002         &       0.004         &       0.002         &       0.012         \\
            &     (0.011)         &     (0.010)         &     (0.008)         &     (0.009)         &     (0.012)         &     (0.010)         \\
\midrule
\(R^{2}\)   &       0.014         &       0.006         &       0.005         &       0.002         &       0.003         &       0.009         \\
\midrule
\#Replies (\(N\))       &        \num{2225260}         &       \num{2225260}&        \num{2225260}         &       \num{2225260}&        \num{2225260}         &       \num{2225260}\\
\#Source posts      &        \num{1841}         &       \num{1841}&        \num{1841}         &       \num{1841}&        \num{1841}         &       \num{1841}\\
\bottomrule
\end{tabularx}
\label{tab:emotions_its}
\end{table}

\newpage

\newpage
\subsection{Correlations Between Sentiments and Emotions}
The Pearson correlations between sentiments and emotions are shown in Table \ref{tab:corr_sentiment_emotion}. Anger, disgust, fear, and sadness are positively associated with negative sentiment, while joy and surprise are positively associated with positive sentiment.

\begin{table}[H]
\centering
\caption{The Pearson correlations between sentiments and emotions with two-sided p-values. \sym{*} \(p<0.05\), \sym{**} \(p<0.01\), \sym{***} \(p<0.001\).}
\begin{tabularx}{\columnwidth}{@{\hspace{\tabcolsep}\extracolsep{\fill}}l*{7}{S}}
\toprule
&\multicolumn{1}{c}{Anger}&\multicolumn{1}{c}{Disgust}&\multicolumn{1}{c}{Fear}&\multicolumn{1}{c}{Joy}&\multicolumn{1}{c}{Sadness}&\multicolumn{1}{c}{Surprise}\\
\midrule
Positive&-0.204\sym{***}&-0.175\sym{***}&-0.074\sym{***}&0.572\sym{***}&-0.119\sym{***}&0.044\sym{***}\\
Negative&0.406\sym{***}&0.323\sym{***}&0.089\sym{***}&-0.324\sym{***}&0.192\sym{***}&-0.141\sym{***}\\
\bottomrule
\end{tabularx}
\label{tab:corr_sentiment_emotion}
\end{table}

\subsection{Correlations Among Emotions}
The Pearson correlations among the six basic emotions in replies are reported in Table \ref{tab:corr_emotions}. Only anger and disgust are positively correlated, and all the other pairs of emotions are negatively correlated.

\begin{table}[H]
\centering
\caption{The Pearson correlations among emotions with two-sided p-values. \sym{*} \(p<0.05\), \sym{**} \(p<0.01\), \sym{***} \(p<0.001\).}
\begin{tabularx}{\columnwidth}{@{\hspace{\tabcolsep}\extracolsep{\fill}}l*{7}{S}}
\toprule
&\multicolumn{1}{c}{Anger}&\multicolumn{1}{c}{Disgust}&\multicolumn{1}{c}{Fear}&\multicolumn{1}{c}{Joy}&\multicolumn{1}{c}{Sadness}&\multicolumn{1}{c}{Surprise}\\
\midrule
Anger\\
Disgust&0.111\sym{***}\\
Fear&-0.040\sym{***}&-0.076\sym{***}\\
Joy&-0.195\sym{***}&-0.168\sym{***}&-0.085\sym{***}\\
Sadness&-0.147\sym{***}&-0.129\sym{***}&-0.062\sym{***}&-0.095\sym{***}\\
Surprise&-0.313\sym{***}&-0.241\sym{***}&-0.136\sym{***}&-0.066\sym{***}&-0.131\sym{***}\\
\bottomrule
\end{tabularx}
\label{tab:corr_emotions}
\end{table}

\newpage

\section{Descriptive Analysis}
\label{sec:descriptive_analysis}

\begin{figure}[H]
\centering
\begin{subfigure}{0.32\textwidth}
\caption{}
\includegraphics[width=\textwidth]{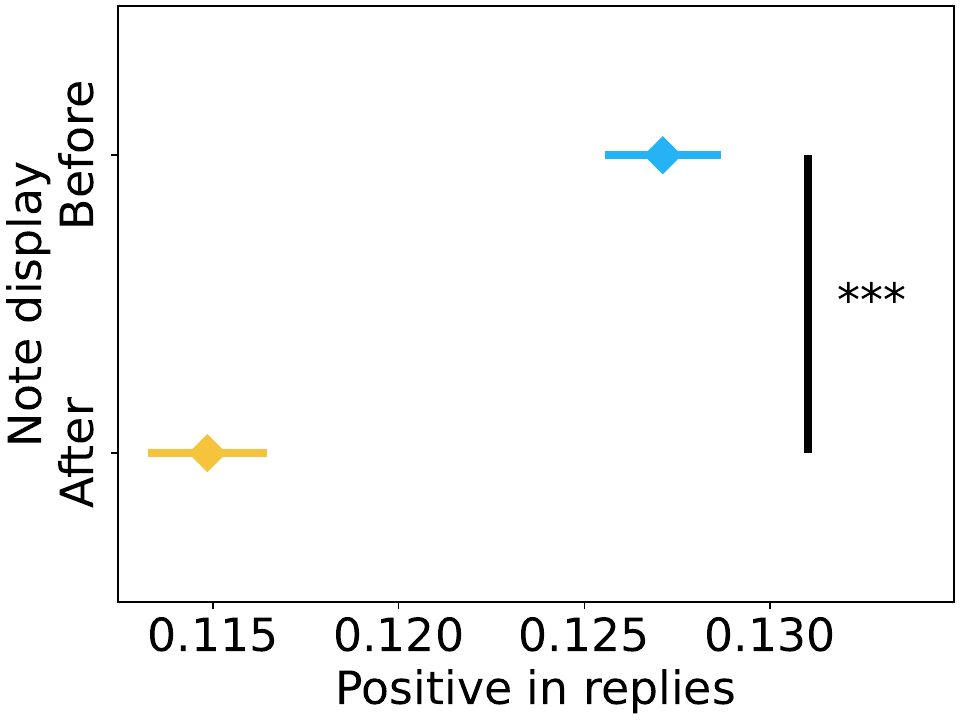}
\label{fig:positive_roberta_in_replies}
\end{subfigure}
\hspace{1cm}
\begin{subfigure}{0.32\textwidth}
\caption{}
\includegraphics[width=\textwidth]{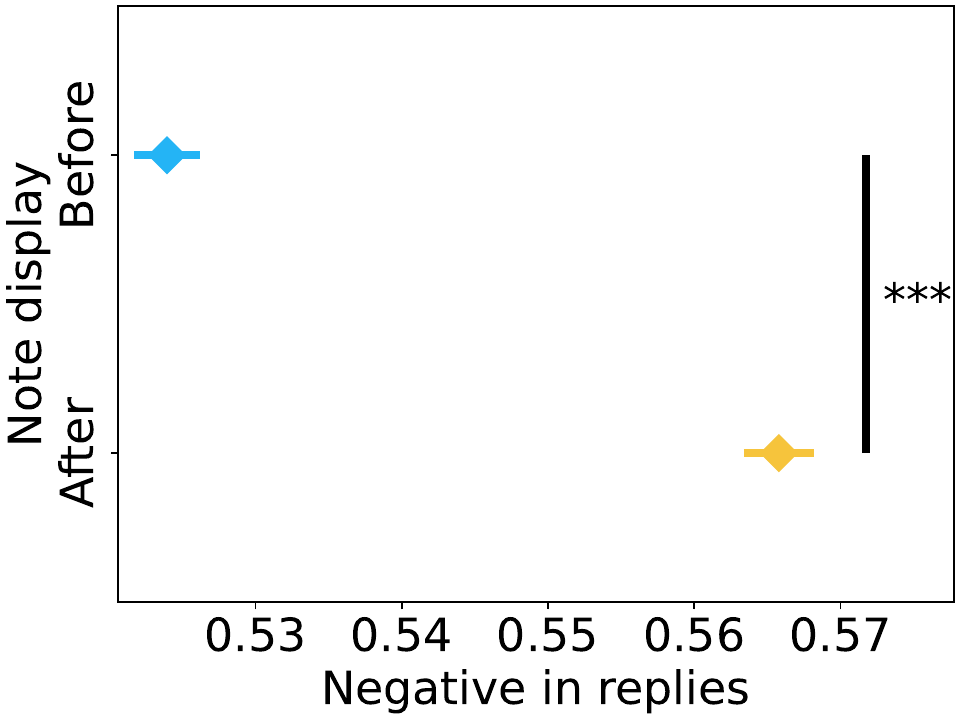}
\label{fig:negative_roberta_in_replies}
\end{subfigure}

\begin{subfigure}{0.32\textwidth}
\caption{}
\includegraphics[width=\textwidth]{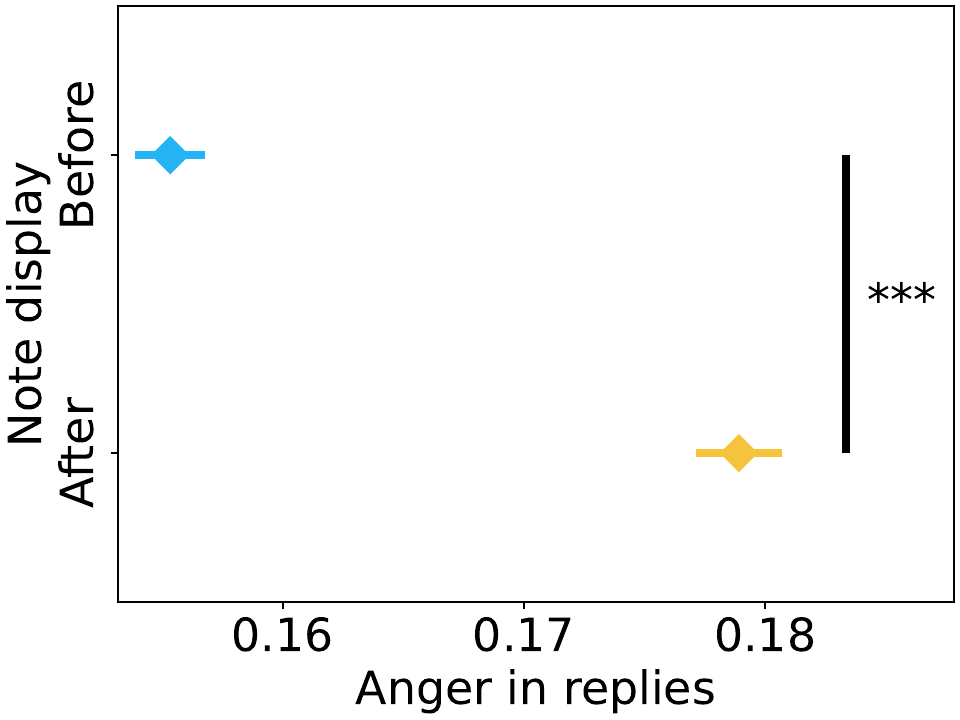}
\label{fig:anger_roberta_in_replies}
\end{subfigure}
\hfill
\begin{subfigure}{0.32\textwidth}
\caption{}
\includegraphics[width=\textwidth]{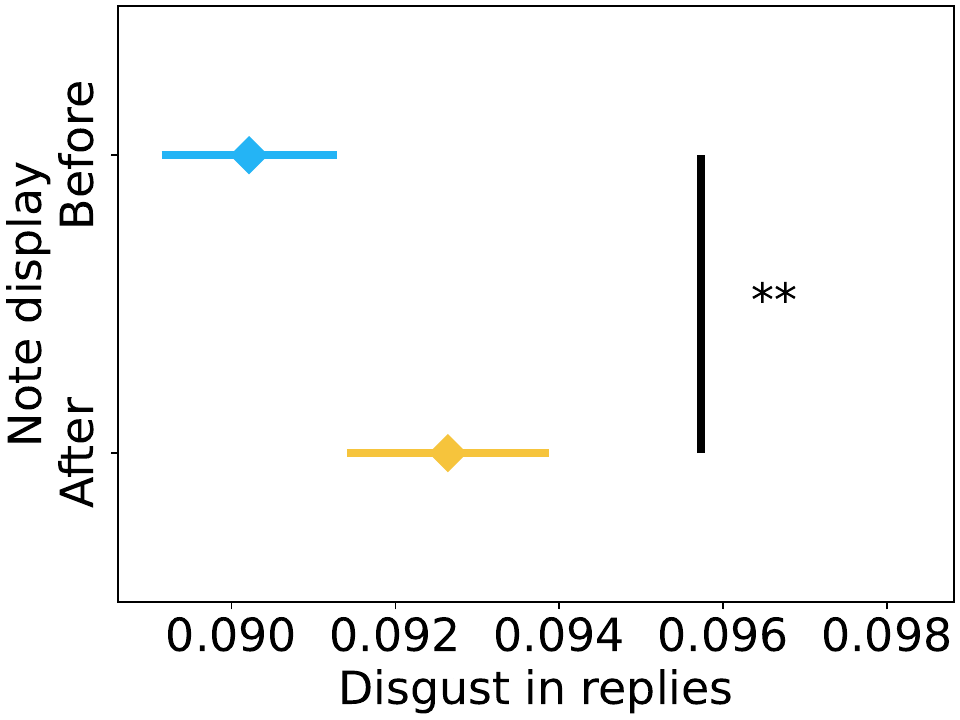}
\label{fig:disgust_roberta_in_replies}
\end{subfigure}
\hfill
\begin{subfigure}{0.32\textwidth}
\caption{}
\includegraphics[width=\textwidth]{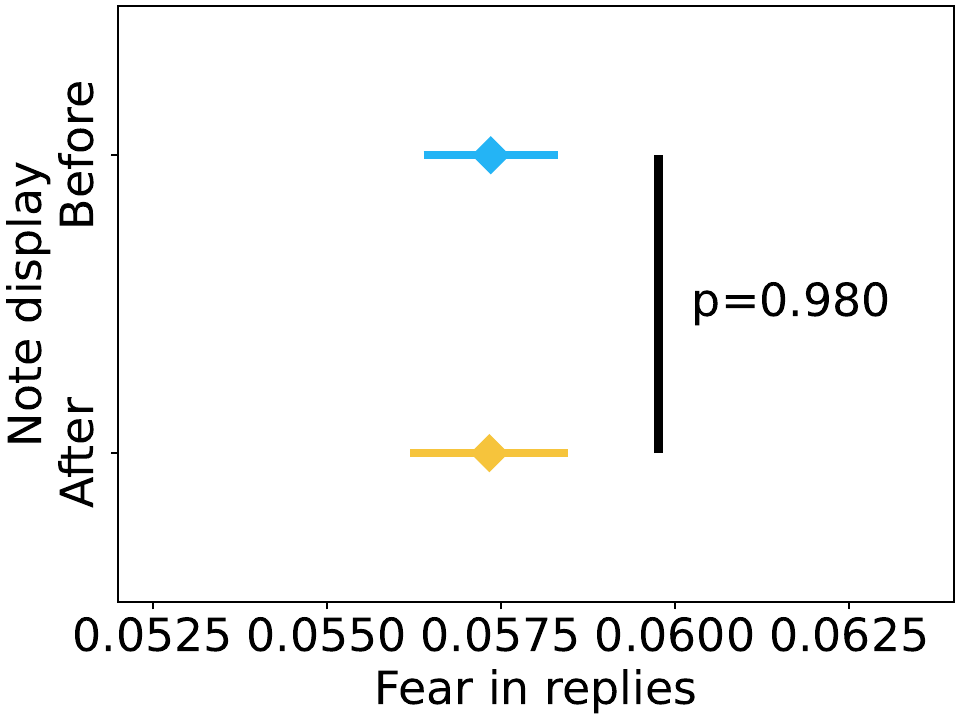}
\label{fig:fear_roberta_in_replies}
\end{subfigure}

\begin{subfigure}{0.32\textwidth}
\caption{}
\includegraphics[width=\textwidth]{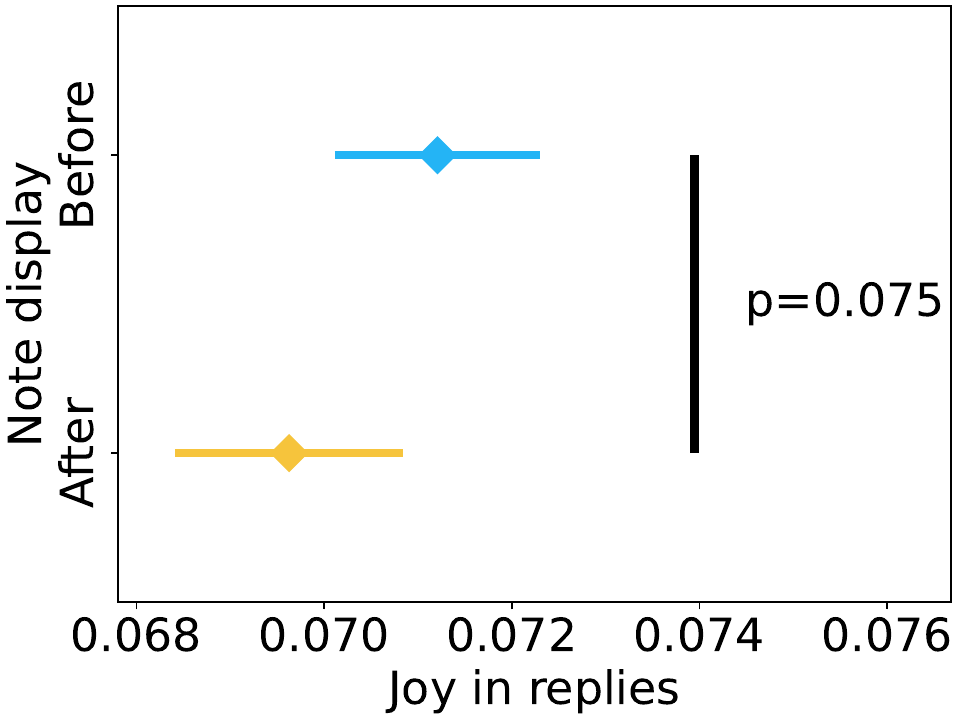}
\label{fig:joy_roberta_in_replies}
\end{subfigure}
\hfill
\begin{subfigure}{0.32\textwidth}
\caption{}
\includegraphics[width=\textwidth]{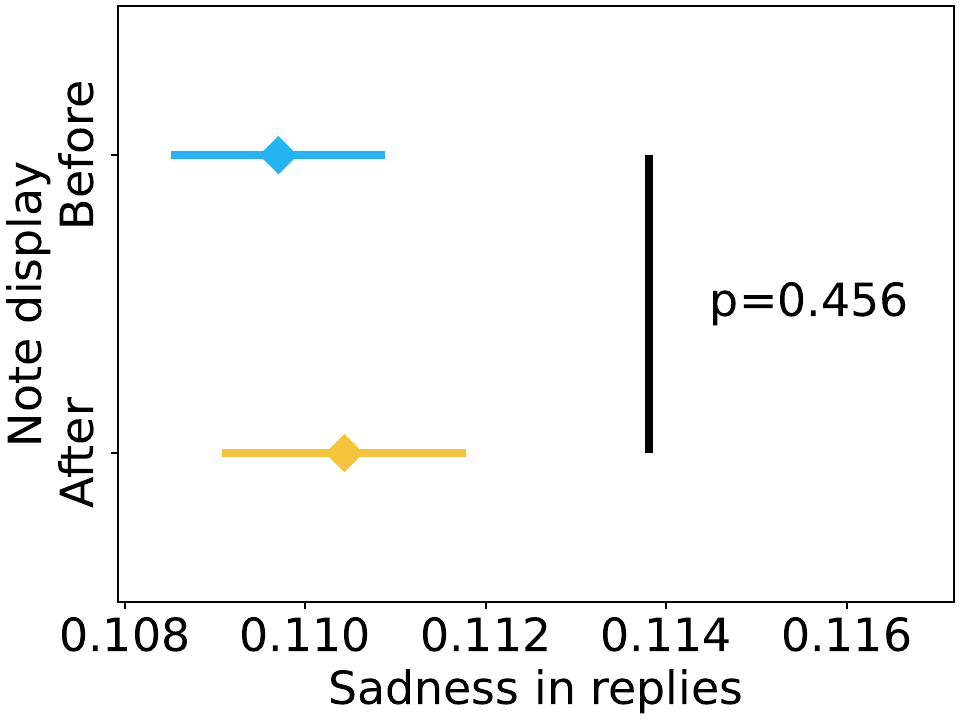}
\label{fig:sadness_roberta_in_replies}
\end{subfigure}
\hfill
\begin{subfigure}{0.32\textwidth}
\caption{}
\includegraphics[width=\textwidth]{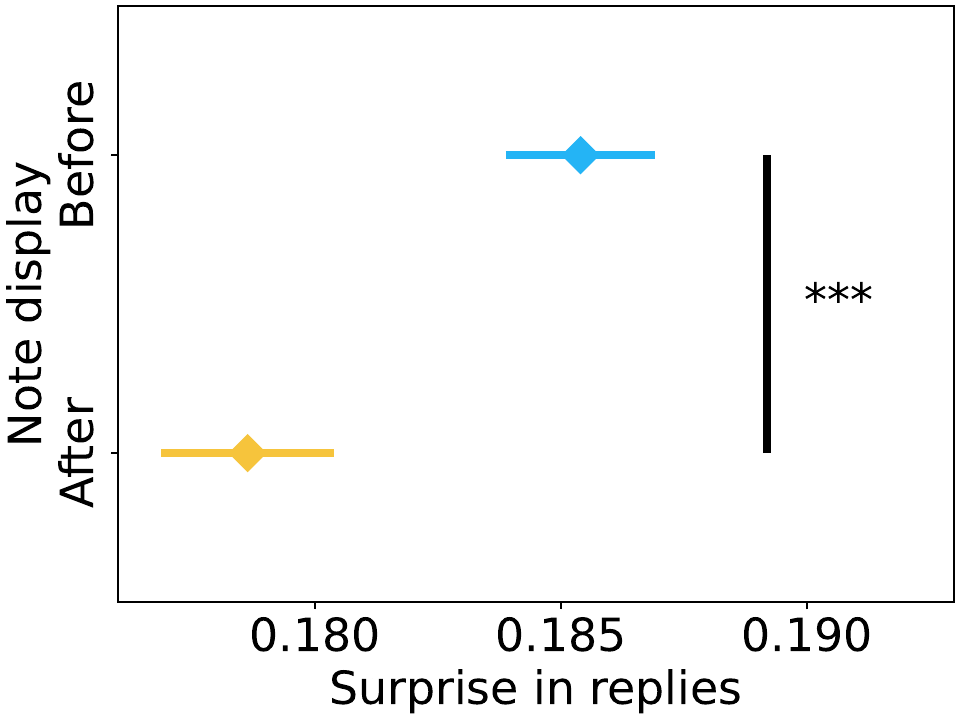}
\label{fig:surprise_roberta_in_replies}
\end{subfigure}
\caption{The sentiments and emotions in replies before and after the display of community notes. Shown are mean values with error bars representing 95\% CIs. \sym{*} \(p<0.05\), \sym{**} \(p<0.01\), \sym{***} \(p<0.001\). \subref{fig:positive_roberta_in_replies} Positive sentiment in replies. \subref{fig:negative_roberta_in_replies} Negative sentiment in replies. \subref{fig:anger_roberta_in_replies} Anger in replies. \subref{fig:disgust_roberta_in_replies} Disgust in replies. \subref{fig:fear_roberta_in_replies} Fear in replies. \subref{fig:joy_roberta_in_replies} Joy in replies. \subref{fig:sadness_roberta_in_replies} Sadness in replies. \subref{fig:surprise_roberta_in_replies} Surprise in replies.}
\label{fig:emotions_in_replies}
\end{figure}

Fig. \ref{fig:emotions_in_replies} shows the comparisons between means before note display and means after note display across sentiments and emotions. For sentiments, positive sentiment in replies after the display of community notes (mean of 0.115) is significantly lower than before the note display (mean of 0.127; $t=10.249$, $p<0.001$; Fig. \ref{fig:positive_roberta_in_replies}). In contrast, negative sentiment in replies after the display of community notes (mean of 0.566) is significantly higher than before the note display (mean of 0.524; $t=-23.748$, $p<0.001$; Fig. \ref{fig:negative_roberta_in_replies}). In terms of basic emotions, anger (mean of 0.179) and disgust (mean of 0.093) in replies after the display of community notes are significantly higher than anger (mean of 0.155; $t=-18.531$, $p<0.001$; Fig. \ref{fig:anger_roberta_in_replies}) and disgust (mean of 0.090; $t=-2.717$, $p<0.01$; Fig. \ref{fig:disgust_roberta_in_replies}) in replies before the note display, respectively. Fear (before: 0.057, after: 0.057; $t=0.025$, $p=0.980$; Fig. \ref{fig:fear_roberta_in_replies}), joy (before: 0.071, after: 0.070; $t=1.781$, $p=0.075$; Fig. \ref{fig:joy_roberta_in_replies}), and sadness (before: 0.110, after: 0.110; $t=-0.745$, $p=0.456$; Fig. \ref{fig:sadness_roberta_in_replies}) have no statistically significant changes after the display of community notes. Surprise in replies after the display of community notes (mean of 0.179) is significantly lower than that in replies before the note display (mean of 0.185; $t=5.325$, $p<0.001$; Fig. \ref{fig:surprise_roberta_in_replies}). 
% Taken together, our descriptive analysis suggests that replies after note display tend to be more negative and embed higher anger and disgust, compared to replies before the display of community notes

\newpage
\section{Estimation Results for Main Analysis}
\label{sec:sm_main_analysis}
%lifespan without donut

The estimation results for the main analysis on the changes in specific sentiments and emotions in replies after the display of community notes are reported in Table \ref{tab:sentiments_coefs}, Table \ref{tab:emotions_coefs_1}, and Table \ref{tab:emotions_coefs_2}. Here, we use all direct replies within the entire lifespan for coefficient estimation. Subsequently, we estimate the effect sizes of community notes on sentiments and emotions within a 16-hour bandwidth (see Fig. \ref{fig:predicted_effects}). The equivalence of RDD models across short and long bandwidths is validated in Suppl. \ref{sec:rdd_coefs_bandwidths}.

\begin{table}[H]
\centering
\caption{Estimation results for positive sentiment [Column (1)] and negative sentiment [Column (2)] in replies within reply lifespan. Reported are coefficient estimates with post-clustered standard errors in parentheses. \sym{*} \(p<0.05\), \sym{**} \(p<0.01\), \sym{***} \(p<0.001\).}
\begin{tabularx}{\columnwidth}{@{\hspace{\tabcolsep}\extracolsep{\fill}}l*{2}{S}}
\toprule
&\multicolumn{1}{c}{(1)}&\multicolumn{1}{c}{(2)}\\
&\multicolumn{1}{c}{Positive}&\multicolumn{1}{c}{Negative}\\
\midrule
Displayed   &      -0.055\sym{**} &       0.114\sym{***}\\
            &     (0.017)         &     (0.022)         \\
HoursFromDisplay&      -0.003         &      -0.001         \\
            &     (0.014)         &     (0.021)         \\
SourcePositive&       0.039\sym{*}  &       0.015         \\
            &     (0.019)         &     (0.021)         \\
SourceNegative&      -0.048\sym{**} &       0.112\sym{***}\\
            &     (0.017)         &     (0.023)         \\
PostAge     &       0.005         &      -0.005         \\
            &     (0.009)         &     (0.013)         \\
Intercept      &       0.014         &      -0.028         \\
            &     (0.014)         &     (0.018)         \\
\midrule
\(R^{2}\)   &       0.007         &       0.013         \\
\midrule
\#Replies (\(N\))       &        \num{2155873}         &       \num{2155873}\\
\#Source posts      &        \num{1339}         &       \num{1339}\\
\bottomrule
\end{tabularx}
\label{tab:sentiments_coefs}
\end{table}

\newpage
\begin{table}[H]
\centering
\caption{Estimation results for anger [Column (1)], disgust [Column (2)], and fear [Column (3)] in replies within reply lifespan. Reported are coefficient estimates with post-clustered standard errors in parentheses. \sym{*} \(p<0.05\), \sym{**} \(p<0.01\), \sym{***} \(p<0.001\).}
\begin{tabularx}{\columnwidth}{@{\hspace{\tabcolsep}\extracolsep{\fill}}l*{3}{S}}
\toprule
&\multicolumn{1}{c}{(1)}&\multicolumn{1}{c}{(2)}&\multicolumn{1}{c}{(3)}\\
&\multicolumn{1}{c}{Anger}&\multicolumn{1}{c}{Disgust}&\multicolumn{1}{c}{Fear}\\
\midrule
Displayed   &       0.078\sym{***}&       0.023\sym{*}  &      -0.007         \\
            &     (0.010)         &     (0.010)         &     (0.008)         \\
HoursFromDisplay&      -0.004         &      -0.006         &      -0.001         \\
            &     (0.018)         &     (0.010)         &     (0.005)         \\
SourceAnger &       0.078\sym{***}&       0.007         &       0.017\sym{*}  \\
            &     (0.012)         &     (0.011)         &     (0.008)         \\
SourceDisgust&       0.001         &       0.070\sym{***}&      -0.004         \\
            &     (0.008)         &     (0.015)         &     (0.004)         \\
SourceFear  &       0.013         &       0.008         &       0.066\sym{***}\\
            &     (0.015)         &     (0.011)         &     (0.010)         \\
SourceJoy   &       0.010         &      -0.000         &       0.004         \\
            &     (0.016)         &     (0.009)         &     (0.006)         \\
SourceSadness&       0.010         &      -0.005         &       0.005         \\
            &     (0.010)         &     (0.009)         &     (0.005)         \\
SourceSurprise&      -0.056\sym{***}&      -0.013         &      -0.001         \\
            &     (0.011)         &     (0.012)         &     (0.007)         \\
PostAge     &      -0.002         &       0.004         &       0.004         \\
            &     (0.010)         &     (0.008)         &     (0.004)         \\
Intercept      &      -0.019         &      -0.006         &       0.002         \\
            &     (0.012)         &     (0.010)         &     (0.008)         \\
\midrule
\(R^{2}\)   &       0.014         &       0.006         &       0.005         \\
\midrule
\#Replies (\(N\))       &        \num{2155873}         &       \num{2155873}    &       \num{2155873}\\
\#Source posts      &        \num{1339}         &       \num{1339}         &       \num{1339}\\
\bottomrule
\end{tabularx}
\label{tab:emotions_coefs_1}
\end{table}

\newpage
\begin{table}[H]
\centering
\caption{Estimation results for joy [Column (1)], sadness [Column (2)], and surprise [Column (3)] in replies within reply lifespan. Reported are coefficient estimates with post-clustered standard errors in parentheses. \sym{*} \(p<0.05\), \sym{**} \(p<0.01\), \sym{***} \(p<0.001\).}
\begin{tabularx}{\columnwidth}{@{\hspace{\tabcolsep}\extracolsep{\fill}}l*{3}{S}}
\toprule
&\multicolumn{1}{c}{(1)}&\multicolumn{1}{c}{(2)}&\multicolumn{1}{c}{(3)}\\
&\multicolumn{1}{c}{Joy}&\multicolumn{1}{c}{Sadness}&\multicolumn{1}{c}{Surprise}\\
\midrule
Displayed   &      -0.015         &      -0.008         &      -0.043\sym{***}\\
            &     (0.010)         &     (0.012)         &     (0.009)         \\
HoursFromDisplay&      -0.004         &      -0.004         &      -0.005         \\
            &     (0.010)         &     (0.007)         &     (0.010)         \\
SourceAnger &      -0.006         &       0.021         &      -0.036\sym{***}\\
            &     (0.008)         &     (0.014)         &     (0.010)         \\
SourceDisgust&      -0.011\sym{*}  &      -0.012         &      -0.003         \\
            &     (0.005)         &     (0.006)         &     (0.007)         \\
SourceFear  &      -0.021\sym{**} &       0.006         &       0.004         \\
            &     (0.008)         &     (0.009)         &     (0.012)         \\
SourceJoy   &       0.033         &       0.025\sym{***}&      -0.001         \\
            &     (0.019)         &     (0.007)         &     (0.012)         \\
SourceSadness&       0.009         &       0.052\sym{***}&      -0.007         \\
            &     (0.009)         &     (0.008)         &     (0.009)         \\
SourceSurprise&       0.000         &       0.016         &       0.069\sym{***}\\
            &     (0.010)         &     (0.013)         &     (0.013)         \\
PostAge     &       0.007         &       0.006         &       0.010         \\
            &     (0.006)         &     (0.004)         &     (0.006)         \\
Intercept      &       0.004         &       0.002         &       0.011         \\
            &     (0.009)         &     (0.012)         &     (0.010)         \\
\midrule
\(R^{2}\)   &       0.002         &       0.003         &       0.008         \\
\midrule
\#Replies (\(N\))&\num{2155873}&\num{2155873}&\num{2155873}\\
\#Source posts&\num{1339}&\num{1339}&\num{1339}\\
\bottomrule
\end{tabularx}
\label{tab:emotions_coefs_2}
\end{table}

\newpage
\section{RDD Estimates Across Different Bandwidths}
\label{sec:rdd_coefs_bandwidths}

We examine the robustness of RDD estimates (\ie, the coefficient estimates of $\var{Displayed}$) across three progressively extended bandwidths: 16 hours, one week, and the entire reply lifespan. Tables \ref{tab:sentiments_16hours}--\ref{tab:emotions_16hours} report the estimation results within 16-hour bandwidth, Tables \ref{tab:sentiments_1week}--\ref{tab:emotions_1week} report the estimation results within one-week bandwidth, and Tables \ref{tab:sentiments_lifespan_donut}--\ref{tab:emotions_lifespan_donut} report the estimation results within the entire reply lifespan. As shown in Fig. \ref{fig:predicted_effects}, the initial four hours after the display of community notes are omitted in the three cases to avoid potential cold-start contamination in the smaller bandwidth, such as the 16-hour window. We find that the RDD estimates across the three bandwidths are consistent, with no statistically significant differences observed.

\newpage
\begin{table}[H]
\centering
% \footnotesize
\caption{Estimation results for positive sentiment [Column (1)] and negative sentiment [Column (2)] in replies within 16-hour bandwidth. Reported are coefficient estimates with post-clustered standard errors in parentheses. \sym{*} \(p<0.05\), \sym{**} \(p<0.01\), \sym{***} \(p<0.001\).}
\begin{tabularx}{\columnwidth}{@{\hspace{\tabcolsep}\extracolsep{\fill}}l*{2}{S}}
\toprule
&\multicolumn{1}{c}{(1)}&\multicolumn{1}{c}{(2)}\\
&\multicolumn{1}{c}{Positive}&\multicolumn{1}{c}{Negative}\\
\midrule
Displayed   &      -0.049\sym{*}  &       0.096\sym{***}\\
            &     (0.024)         &     (0.028)         \\
HoursFromDisplay&      -0.037         &       0.145         \\
            &     (0.127)         &     (0.163)         \\
SourcePositive&       0.053\sym{*}  &       0.000         \\
            &     (0.025)         &     (0.023)         \\
SourceNegative&      -0.030         &       0.093\sym{***}\\
            &     (0.018)         &     (0.023)         \\
PostAge     &       0.008         &      -0.007         \\
            &     (0.013)         &     (0.018)         \\
Intercept      &      -0.002         &      -0.031         \\
            &     (0.016)         &     (0.022)         \\
\midrule
\(R^{2}\)   &       0.006         &       0.012         \\
\midrule
\#Replies (\(N\))       &        \num{1166861}         &       \num{1166861}\\
\#Source posts      &        \num{1335}         &       \num{1335}\\
\bottomrule
\end{tabularx}
\label{tab:sentiments_16hours}
\end{table}

\begin{table}[H]
\centering
% \footnotesize
\caption{Estimation results for anger [Column (1)], disgust [Column (2)], fear [Column (3)], joy [Column (4)], sadness [Column (5)], and surprise [Column (6)] in replies within 16-hour bandwidth. Reported are coefficient estimates with post-clustered standard errors in parentheses. \sym{*} \(p<0.05\), \sym{**} \(p<0.01\), \sym{***} \(p<0.001\).}
\begin{tabularx}{\columnwidth}{@{\hspace{\tabcolsep}\extracolsep{\fill}}l*{6}{S}}
\toprule
&\multicolumn{1}{c}{(1)}&\multicolumn{1}{c}{(2)}&\multicolumn{1}{c}{(3)}&\multicolumn{1}{c}{(4)}&\multicolumn{1}{c}{(5)}&\multicolumn{1}{c}{(6)}\\
&\multicolumn{1}{c}{Anger}&\multicolumn{1}{c}{Disgust}&\multicolumn{1}{c}{Fear}&\multicolumn{1}{c}{Joy}&\multicolumn{1}{c}{Sadness}&\multicolumn{1}{c}{Surprise}\\
\midrule
Displayed   &       0.059\sym{**} &       0.039\sym{*}  &       0.014         &      -0.011         &       0.016         &      -0.059\sym{***}\\
            &     (0.018)         &     (0.017)         &     (0.012)         &     (0.020)         &     (0.014)         &     (0.014)         \\
HoursFromDisplay&       0.087         &      -0.084         &      -0.080         &       0.001         &      -0.056         &       0.025         \\
            &     (0.092)         &     (0.078)         &     (0.059)         &     (0.078)         &     (0.076)         &     (0.066)         \\
SourceAnger &       0.068\sym{***}&       0.008         &       0.012         &      -0.007         &       0.018         &      -0.029\sym{**} \\
            &     (0.013)         &     (0.011)         &     (0.007)         &     (0.007)         &     (0.012)         &     (0.009)         \\
SourceDisgust&      -0.004         &       0.074\sym{***}&      -0.007         &      -0.008         &      -0.009         &       0.000         \\
            &     (0.009)         &     (0.017)         &     (0.005)         &     (0.006)         &     (0.005)         &     (0.007)         \\
SourceFear  &       0.017         &       0.010         &       0.064\sym{***}&      -0.023\sym{**} &       0.008         &       0.000         \\
            &     (0.016)         &     (0.011)         &     (0.011)         &     (0.007)         &     (0.009)         &     (0.009)         \\
SourceJoy   &       0.007         &      -0.001         &       0.004         &       0.028         &       0.023\sym{***}&      -0.002         \\
            &     (0.020)         &     (0.008)         &     (0.007)         &     (0.023)         &     (0.007)         &     (0.012)         \\
SourceSadness&       0.002         &      -0.004         &       0.006         &       0.007         &       0.046\sym{***}&      -0.001         \\
            &     (0.011)         &     (0.009)         &     (0.007)         &     (0.010)         &     (0.008)         &     (0.009)         \\
SourceSurprise&      -0.062\sym{***}&      -0.021         &      -0.003         &      -0.003         &       0.028         &       0.074\sym{***}\\
            &     (0.014)         &     (0.012)         &     (0.009)         &     (0.010)         &     (0.016)         &     (0.016)         \\
PostAge     &      -0.011         &       0.020         &       0.008         &       0.009         &       0.007         &       0.010         \\
            &     (0.011)         &     (0.015)         &     (0.009)         &     (0.011)         &     (0.009)         &     (0.009)         \\
Intercept      &      -0.014         &       0.008         &       0.008         &      -0.012         &      -0.010         &      -0.000         \\
            &     (0.015)         &     (0.013)         &     (0.009)         &     (0.011)         &     (0.013)         &     (0.011)         \\
\midrule
\(R^{2}\)   &       0.014         &       0.008         &       0.004         &       0.002         &       0.002         &       0.009         \\
\midrule
\#Replies (\(N\))       &        \num{1166861}         &       \num{1166861}&        \num{1166861}         &       \num{1166861}&        \num{1166861}         &       \num{1166861}         \\
\#Source posts      &        \num{1335}         &       \num{1335}&        \num{1335}         &       \num{1335}&        \num{1335}         &       \num{1335}         \\
\bottomrule
\end{tabularx}
\label{tab:emotions_16hours}
\end{table}

% one week
\newpage
\begin{table}[H]
\centering
% \footnotesize
\caption{Estimation results for positive sentiment [Column (1)] and negative sentiment [Column (2)] in replies within one-week bandwidth. Reported are coefficient estimates with post-clustered standard errors in parentheses. \sym{*} \(p<0.05\), \sym{**} \(p<0.01\), \sym{***} \(p<0.001\).}
\begin{tabularx}{\columnwidth}{@{\hspace{\tabcolsep}\extracolsep{\fill}}l*{2}{S}}
\toprule
&\multicolumn{1}{c}{(1)}&\multicolumn{1}{c}{(2)}\\
&\multicolumn{1}{c}{Positive}&\multicolumn{1}{c}{Negative}\\
\midrule
Displayed   &      -0.040         &       0.113\sym{***}\\
            &     (0.023)         &     (0.028)         \\
HoursFromDisplay&      -0.060         &       0.065         \\
            &     (0.035)         &     (0.043)         \\
SourcePositive&       0.036\sym{*}  &       0.016         \\
            &     (0.018)         &     (0.021)         \\
SourceNegative&      -0.047\sym{**} &       0.110\sym{***}\\
            &     (0.017)         &     (0.024)         \\
PostAge     &       0.005         &      -0.016         \\
            &     (0.011)         &     (0.018)         \\
Intercept      &       0.011         &      -0.027         \\
            &     (0.013)         &     (0.018)         \\
\midrule
\(R^{2}\)   &       0.007         &       0.014         \\
\midrule
\#Replies (\(N\))       &        \num{1952474}         &       \num{1952474}\\
\#Source posts      &        \num{1339}         &       \num{1339}\\
\bottomrule
\end{tabularx}
\label{tab:sentiments_1week}
\end{table}

\begin{table}[H]
\centering
% \footnotesize
\caption{Estimation results for anger [Column (1)], disgust [Column (2)], fear [Column (3)], joy [Column (4)], sadness [Column (5)], and surprise [Column (6)] in replies within one-week bandwidth. Reported are coefficient estimates with post-clustered standard errors in parentheses. \sym{*} \(p<0.05\), \sym{**} \(p<0.01\), \sym{***} \(p<0.001\).}
\begin{tabularx}{\columnwidth}{@{\hspace{\tabcolsep}\extracolsep{\fill}}l*{6}{S}}
\toprule
&\multicolumn{1}{c}{(1)}&\multicolumn{1}{c}{(2)}&\multicolumn{1}{c}{(3)}&\multicolumn{1}{c}{(4)}&\multicolumn{1}{c}{(5)}&\multicolumn{1}{c}{(6)}\\
&\multicolumn{1}{c}{Anger}&\multicolumn{1}{c}{Disgust}&\multicolumn{1}{c}{Fear}&\multicolumn{1}{c}{Joy}&\multicolumn{1}{c}{Sadness}&\multicolumn{1}{c}{Surprise}\\
\midrule
Displayed   &       0.067\sym{***}&       0.035\sym{*}  &      -0.008         &      -0.001         &       0.010         &      -0.047\sym{***}\\
            &     (0.012)         &     (0.017)         &     (0.008)         &     (0.015)         &     (0.013)         &     (0.014)         \\
HoursFromDisplay&       0.066\sym{***}&      -0.033         &       0.000         &      -0.044         &      -0.033         &      -0.020         \\
            &     (0.020)         &     (0.036)         &     (0.017)         &     (0.024)         &     (0.018)         &     (0.028)         \\
SourceAnger &       0.069\sym{***}&       0.006         &       0.016         &       0.000         &       0.025         &      -0.033\sym{**} \\
            &     (0.011)         &     (0.011)         &     (0.009)         &     (0.008)         &     (0.016)         &     (0.010)         \\
SourceDisgust&       0.002         &       0.072\sym{***}&      -0.004         &      -0.011\sym{*}  &      -0.011         &      -0.004         \\
            &     (0.008)         &     (0.015)         &     (0.004)         &     (0.005)         &     (0.006)         &     (0.007)         \\
SourceFear  &       0.018         &       0.012         &       0.069\sym{***}&      -0.024\sym{**} &       0.003         &      -0.002         \\
            &     (0.015)         &     (0.011)         &     (0.011)         &     (0.007)         &     (0.009)         &     (0.011)         \\
SourceJoy   &       0.008         &       0.001         &       0.005         &       0.033         &       0.025\sym{***}&       0.000         \\
            &     (0.016)         &     (0.008)         &     (0.007)         &     (0.018)         &     (0.007)         &     (0.012)         \\
SourceSadness&       0.008         &      -0.004         &       0.005         &       0.010         &       0.053\sym{***}&      -0.007         \\
            &     (0.010)         &     (0.009)         &     (0.005)         &     (0.009)         &     (0.009)         &     (0.009)         \\
SourceSurprise&      -0.055\sym{***}&      -0.010         &      -0.000         &       0.001         &       0.017         &       0.068\sym{***}\\
            &     (0.011)         &     (0.012)         &     (0.007)         &     (0.010)         &     (0.013)         &     (0.013)         \\
PostAge     &      -0.018\sym{*}  &       0.019         &       0.009         &       0.007         &       0.009         &       0.016         \\
            &     (0.009)         &     (0.013)         &     (0.009)         &     (0.008)         &     (0.006)         &     (0.009)         \\
Intercept      &      -0.019         &      -0.005         &       0.003         &       0.001         &       0.000         &       0.010         \\
            &     (0.011)         &     (0.010)         &     (0.008)         &     (0.009)         &     (0.013)         &     (0.010)         \\
\midrule
\(R^{2}\)   &       0.014         &       0.006         &       0.005         &       0.002         &       0.003         &       0.008         \\
\midrule
\#Replies (\(N\))       &        \num{1952474}         &       \num{1952474}&        \num{1952474}         &       \num{1952474}&        \num{1952474}         &       \num{1952474}         \\
\#Source posts      &        \num{1339}         &       \num{1339}&        \num{1339}         &       \num{1339}&        \num{1339}         &       \num{1339}         \\
\bottomrule
\end{tabularx}
\label{tab:emotions_1week}
\end{table}

% lifespan
\newpage
\begin{table}[H]
\centering
% \footnotesize
\caption{Estimation results for positive sentiment [Column (1)] and negative sentiment [Column (2)] in replies within reply lifespan. Reported are coefficient estimates with post-clustered standard errors in parentheses. \sym{*} \(p<0.05\), \sym{**} \(p<0.01\), \sym{***} \(p<0.001\).}
\begin{tabularx}{\columnwidth}{@{\hspace{\tabcolsep}\extracolsep{\fill}}l*{2}{S}}
\toprule
&\multicolumn{1}{c}{(1)}&\multicolumn{1}{c}{(2)}\\
&\multicolumn{1}{c}{Positive}&\multicolumn{1}{c}{Negative}\\
\midrule
Displayed   &      -0.071\sym{***}&       0.144\sym{***}\\
            &     (0.021)         &     (0.026)         \\
HoursFromDisplay&      -0.002         &      -0.002         \\
            &     (0.014)         &     (0.021)         \\
SourcePositive&       0.038\sym{*}  &       0.016         \\
            &     (0.018)         &     (0.021)         \\
SourceNegative&      -0.049\sym{**} &       0.113\sym{***}\\
            &     (0.017)         &     (0.024)         \\
PostAge     &       0.007         &      -0.008         \\
            &     (0.009)         &     (0.013)         \\
Intercept      &       0.014         &      -0.029         \\
            &     (0.014)         &     (0.018)         \\
\midrule
\(R^{2}\)   &       0.007         &       0.014         \\
\midrule
\#Replies (\(N\))       &        \num{2000658}         &       \num{2000658}\\
\#Source posts      &        \num{1339}         &       \num{1339}\\
\bottomrule
\end{tabularx}
\label{tab:sentiments_lifespan_donut}
\end{table}

\begin{table}[H]
\centering
% \footnotesize
\caption{Estimation results for anger [Column (1)], disgust [Column (2)], fear [Column (3)], joy [Column (4)], sadness [Column (5)], and surprise [Column (6)] in replies within reply lifespan. Reported are coefficient estimates with post-clustered standard errors in parentheses. \sym{*} \(p<0.05\), \sym{**} \(p<0.01\), \sym{***} \(p<0.001\).}
\begin{tabularx}{\columnwidth}{@{\hspace{\tabcolsep}\extracolsep{\fill}}l*{6}{S}}
\toprule
&\multicolumn{1}{c}{(1)}&\multicolumn{1}{c}{(2)}&\multicolumn{1}{c}{(3)}&\multicolumn{1}{c}{(4)}&\multicolumn{1}{c}{(5)}&\multicolumn{1}{c}{(6)}\\
&\multicolumn{1}{c}{Anger}&\multicolumn{1}{c}{Disgust}&\multicolumn{1}{c}{Fear}&\multicolumn{1}{c}{Joy}&\multicolumn{1}{c}{Sadness}&\multicolumn{1}{c}{Surprise}\\
\midrule
Displayed   &       0.096\sym{***}&       0.027\sym{*}  &      -0.004         &      -0.022         &      -0.004         &      -0.052\sym{***}\\
            &     (0.011)         &     (0.011)         &     (0.008)         &     (0.012)         &     (0.013)         &     (0.010)         \\
HoursFromDisplay&      -0.005         &      -0.006         &      -0.001         &      -0.004         &      -0.004         &      -0.005         \\
            &     (0.018)         &     (0.010)         &     (0.005)         &     (0.010)         &     (0.007)         &     (0.010)         \\
SourceAnger &       0.079\sym{***}&       0.006         &       0.017\sym{*}  &      -0.006         &       0.020         &      -0.037\sym{***}\\
            &     (0.012)         &     (0.011)         &     (0.008)         &     (0.008)         &     (0.015)         &     (0.010)         \\
SourceDisgust&       0.001         &       0.071\sym{***}&      -0.005         &      -0.010\sym{*}  &      -0.012         &      -0.003         \\
            &     (0.008)         &     (0.015)         &     (0.004)         &     (0.005)         &     (0.006)         &     (0.007)         \\
SourceFear  &       0.013         &       0.009         &       0.066\sym{***}&      -0.021\sym{**} &       0.005         &       0.004         \\
            &     (0.015)         &     (0.011)         &     (0.011)         &     (0.008)         &     (0.009)         &     (0.013)         \\
SourceJoy   &       0.010         &      -0.000         &       0.004         &       0.032         &       0.024\sym{**} &      -0.000         \\
            &     (0.016)         &     (0.009)         &     (0.006)         &     (0.018)         &     (0.007)         &     (0.012)         \\
SourceSadness&       0.010         &      -0.005         &       0.005         &       0.009         &       0.052\sym{***}&      -0.007         \\
            &     (0.010)         &     (0.009)         &     (0.005)         &     (0.009)         &     (0.008)         &     (0.009)         \\
SourceSurprise&      -0.056\sym{***}&      -0.011         &      -0.001         &       0.001         &       0.016         &       0.068\sym{***}\\
            &     (0.011)         &     (0.012)         &     (0.007)         &     (0.010)         &     (0.013)         &     (0.013)         \\
PostAge     &      -0.004         &       0.004         &       0.004         &       0.008         &       0.006         &       0.011         \\
            &     (0.010)         &     (0.008)         &     (0.004)         &     (0.006)         &     (0.005)         &     (0.006)         \\
Intercept      &      -0.020         &      -0.006         &       0.002         &       0.004         &       0.002         &       0.011         \\
            &     (0.011)         &     (0.010)         &     (0.008)         &     (0.009)         &     (0.012)         &     (0.010)         \\
\midrule
\(R^{2}\)   &       0.015         &       0.006         &       0.005         &       0.002         &       0.003         &       0.008         \\
\midrule
\#Replies (\(N\))       &        \num{2000658}         &       \num{2000658}&        \num{2000658}         &       \num{2000658}&        \num{2000658}         &       \num{2000658}         \\
\#Source posts      &        \num{1339}         &       \num{1339}&        \num{1339}         &       \num{1339}&        \num{1339}         &       \num{1339}         \\
\bottomrule
\end{tabularx}
\label{tab:emotions_lifespan_donut}
\end{table}

\newpage
\section{Sensitivity Analysis}
\label{sec:sm_sensitivity_analysis}
The estimation results for the sensitivity analysis across political and non-political misleading posts are reported in Table \ref{tab:sentiments_politics_coefs}, Table \ref{tab:emotions_politics_coefs_1}, and Table \ref{tab:emotions_politics_coefs_2}.

\begin{table}[H]
\centering
\caption{Estimation results for positive and negative sentiments in replies to political and non-political misleading source posts. Column (1) reports the estimation result for positive sentiment in replies to political misleading source posts. Column (2) reports the estimation result for positive sentiment in replies to non-political misleading source posts. Column (3) reports the estimation result for negative sentiment in replies to political misleading source posts. Column (4) reports the estimation result for negative sentiment in replies to non-political misleading source posts. Reported are coefficient estimates with post-clustered standard errors in parentheses. \sym{*} \(p<0.05\), \sym{**} \(p<0.01\), \sym{***} \(p<0.001\).}
\begin{tabularx}{\columnwidth}{@{\hspace{\tabcolsep}\extracolsep{\fill}}l*{4}{S}}
\toprule
&\multicolumn{1}{c}{(1)}&\multicolumn{1}{c}{(2)}&\multicolumn{1}{c}{(3)}&\multicolumn{1}{c}{(4)}\\
&\multicolumn{1}{c}{Positive: politics}&\multicolumn{1}{c}{Positive: non-politics}&\multicolumn{1}{c}{Negative: politics}&\multicolumn{1}{c}{Negative: non-politics}\\
\midrule
Displayed   &      -0.068\sym{**} &      -0.020         &       0.116\sym{***}&       0.075\sym{***}\\
            &     (0.024)         &     (0.016)         &     (0.025)         &     (0.017)         \\
HoursFromDisplay&       0.002         &      -0.006         &      -0.010         &       0.003         \\
            &     (0.020)         &     (0.010)         &     (0.027)         &     (0.013)         \\
SourcePositive&      -0.015         &       0.099\sym{***}&       0.020         &      -0.010         \\
            &     (0.016)         &     (0.019)         &     (0.022)         &     (0.024)         \\
SourceNegative&      -0.069\sym{**} &      -0.036         &       0.090\sym{***}&       0.137\sym{***}\\
            &     (0.023)         &     (0.019)         &     (0.027)         &     (0.028)         \\
PostAge     &       0.003         &       0.000         &      -0.001         &       0.005         \\
            &     (0.012)         &     (0.008)         &     (0.015)         &     (0.011)         \\
Intercept      &      -0.044\sym{*}  &       0.070\sym{***}&       0.102\sym{***}&      -0.146\sym{***}\\
            &     (0.019)         &     (0.016)         &     (0.022)         &     (0.021)         \\
\midrule
\(R^{2}\)   &       0.005         &       0.013         &       0.010         &       0.021         \\
\midrule
\#Replies (\(N\))&\num{1037340}&\num{1118533}&\num{1037340}&\num{1118533}\\
\#Source posts&\num{379}&\num{960}&\num{379}&\num{960}\\
\bottomrule
\end{tabularx}
\label{tab:sentiments_politics_coefs}
\end{table}

\newpage
\begin{table}[H]
\centering
\caption{Estimation results for anger, disgust, and fear in replies to political and non-political misleading source posts. Column (1) reports the estimation result for anger in replies to political misleading source posts. Column (2) reports the estimation result for anger in replies to non-political misleading source posts. Column (3) reports the estimation result for disgust in replies to political misleading source posts. Column (4) reports the estimation result for disgust in replies to non-political misleading source posts. Column (5) reports the estimation result for fear in replies to political misleading source posts. Column (6) reports the estimation result for fear in replies to non-political misleading source posts. Reported are coefficient estimates with post-clustered standard errors in parentheses. \sym{*} \(p<0.05\), \sym{**} \(p<0.01\), \sym{***} \(p<0.001\).}
\begin{tabularx}{\columnwidth}{@{\hspace{\tabcolsep}\extracolsep{\fill}}l*{6}{S}}
\toprule
&\multicolumn{2}{c}{Anger}&\multicolumn{2}{c}{Disgust}&\multicolumn{2}{c}{Fear}\\
\cmidrule(lr){2-3}\cmidrule(lr){4-5}\cmidrule(lr){6-7}
&\multicolumn{1}{c}{(1)}&\multicolumn{1}{c}{(2)}&\multicolumn{1}{c}{(3)}&\multicolumn{1}{c}{(4)}&\multicolumn{1}{c}{(5)}&\multicolumn{1}{c}{(6)}\\
&\multicolumn{1}{c}{Politics}&\multicolumn{1}{c}{Non-politics}&\multicolumn{1}{c}{Politics}&\multicolumn{1}{c}{Non-politics}&\multicolumn{1}{c}{Politics}&\multicolumn{1}{c}{Non-politics}\\
\midrule
Displayed   &       0.063\sym{***}&       0.070\sym{***}&       0.027\sym{**} &       0.012         &       0.005         &      -0.015         \\
            &     (0.014)         &     (0.015)         &     (0.010)         &     (0.016)         &     (0.009)         &     (0.011)         \\
HoursFromDisplay&      -0.012         &       0.000         &       0.000         &      -0.011         &       0.001         &      -0.001         \\
            &     (0.029)         &     (0.005)         &     (0.011)         &     (0.018)         &     (0.006)         &     (0.005)         \\
SourceAnger &       0.075\sym{***}&       0.062\sym{***}&       0.018         &      -0.006         &       0.018\sym{*}  &       0.018         \\
            &     (0.015)         &     (0.016)         &     (0.012)         &     (0.017)         &     (0.008)         &     (0.015)         \\
SourceDisgust&       0.011         &       0.009         &       0.085\sym{***}&       0.064\sym{**} &      -0.006         &      -0.005         \\
            &     (0.012)         &     (0.009)         &     (0.015)         &     (0.021)         &     (0.009)         &     (0.005)         \\
SourceFear  &       0.019         &       0.013         &       0.007         &       0.010         &       0.042\sym{***}&       0.084\sym{***}\\
            &     (0.026)         &     (0.014)         &     (0.013)         &     (0.016)         &     (0.011)         &     (0.014)         \\
SourceJoy   &       0.021         &      -0.014         &       0.016\sym{*}  &      -0.017         &       0.007         &       0.003         \\
            &     (0.015)         &     (0.012)         &     (0.007)         &     (0.012)         &     (0.008)         &     (0.009)         \\
SourceSadness&       0.022         &       0.003         &       0.017         &      -0.019         &       0.008         &       0.005         \\
            &     (0.016)         &     (0.012)         &     (0.012)         &     (0.012)         &     (0.010)         &     (0.006)         \\
SourceSurprise&      -0.051\sym{**} &      -0.057\sym{***}&       0.002         &      -0.023         &      -0.016         &       0.008         \\
            &     (0.017)         &     (0.011)         &     (0.014)         &     (0.018)         &     (0.009)         &     (0.009)         \\
PostAge     &       0.003         &       0.001         &      -0.004         &       0.010         &      -0.001         &       0.005         \\
            &     (0.017)         &     (0.004)         &     (0.006)         &     (0.014)         &     (0.004)         &     (0.005)         \\
Intercept      &       0.065\sym{***}&      -0.096\sym{***}&       0.005         &      -0.017         &      -0.028\sym{**} &       0.025         \\
            &     (0.017)         &     (0.011)         &     (0.012)         &     (0.014)         &     (0.009)         &     (0.013)         \\
\midrule
\(R^{2}\)   &       0.012         &       0.013         &       0.005         &       0.008         &       0.003         &       0.007         \\
\midrule
\#Replies (\(N\))&\num{1037340}&\num{1118533}&\num{1037340}&\num{1118533}&\num{1037340}&\num{1118533}\\
\#Source posts&\num{379}&\num{960}&\num{379}&\num{960}&\num{379}&\num{960}\\
\bottomrule
\end{tabularx}
\label{tab:emotions_politics_coefs_1}
\end{table}

\newpage
\begin{table}[H]
\centering
\caption{Estimation results for joy, sadness, and surprise in replies to political and non-political misleading source posts. Column (1) reports the estimation result for joy in replies to political misleading source posts. Column (2) reports the estimation result for joy in replies to non-political misleading source posts. Column (3) reports the estimation result for sadness in replies to political misleading source posts. Column (4) reports the estimation result for sadness in replies to non-political misleading source posts. Column (5) reports the estimation result for surprise in replies to political misleading source posts. Column (6) reports the estimation result for surprise in replies to non-political misleading source posts. Reported are coefficient estimates with post-clustered standard errors in parentheses. \sym{*} \(p<0.05\), \sym{**} \(p<0.01\), \sym{***} \(p<0.001\).}
\begin{tabularx}{\columnwidth}{@{\hspace{\tabcolsep}\extracolsep{\fill}}l*{6}{S}}
\toprule
&\multicolumn{2}{c}{Joy}&\multicolumn{2}{c}{Sadness}&\multicolumn{2}{c}{Surprise}\\
\cmidrule(lr){2-3}\cmidrule(lr){4-5}\cmidrule(lr){6-7}
&\multicolumn{1}{c}{(1)}&\multicolumn{1}{c}{(2)}&\multicolumn{1}{c}{(3)}&\multicolumn{1}{c}{(4)}&\multicolumn{1}{c}{(5)}&\multicolumn{1}{c}{(6)}\\
&\multicolumn{1}{c}{Politics}&\multicolumn{1}{c}{Non-politics}&\multicolumn{1}{c}{Politics}&\multicolumn{1}{c}{Non-politics}&\multicolumn{1}{c}{Politics}&\multicolumn{1}{c}{Non-politics}\\
\midrule
Displayed   &      -0.019         &       0.003         &      -0.014         &      -0.007         &      -0.025\sym{*}  &      -0.048\sym{***}\\
            &     (0.014)         &     (0.011)         &     (0.021)         &     (0.011)         &     (0.012)         &     (0.013)         \\
HoursFromDisplay&      -0.004         &      -0.005         &      -0.006         &      -0.002         &      -0.016         &       0.004         \\
            &     (0.017)         &     (0.007)         &     (0.016)         &     (0.004)         &     (0.020)         &     (0.007)         \\
SourceAnger &      -0.016         &       0.011         &       0.021         &       0.019         &      -0.050\sym{***}&      -0.017         \\
            &     (0.011)         &     (0.011)         &     (0.021)         &     (0.012)         &     (0.013)         &     (0.015)         \\
SourceDisgust&      -0.018\sym{*}  &      -0.011         &       0.003         &      -0.018\sym{**} &      -0.014         &      -0.002         \\
            &     (0.008)         &     (0.007)         &     (0.011)         &     (0.007)         &     (0.010)         &     (0.008)         \\
SourceFear  &      -0.017         &      -0.026\sym{**} &       0.008         &       0.004         &      -0.002         &       0.007         \\
            &     (0.012)         &     (0.009)         &     (0.014)         &     (0.011)         &     (0.022)         &     (0.012)         \\
SourceJoy   &       0.008         &       0.062\sym{***}&       0.033\sym{*}  &       0.017\sym{*}  &      -0.020         &       0.021         \\
            &     (0.016)         &     (0.016)         &     (0.014)         &     (0.007)         &     (0.012)         &     (0.013)         \\
SourceSadness&      -0.004         &       0.016         &       0.064\sym{***}&       0.044\sym{***}&      -0.023         &       0.002         \\
            &     (0.013)         &     (0.011)         &     (0.014)         &     (0.011)         &     (0.014)         &     (0.011)         \\
SourceSurprise&      -0.005         &       0.004         &       0.020         &       0.013         &       0.055\sym{***}&       0.078\sym{***}\\
            &     (0.014)         &     (0.012)         &     (0.013)         &     (0.020)         &     (0.015)         &     (0.018)         \\
PostAge     &       0.007         &       0.005         &       0.007         &       0.006         &       0.017         &       0.003         \\
            &     (0.010)         &     (0.005)         &     (0.009)         &     (0.003)         &     (0.012)         &     (0.005)         \\
Intercept      &      -0.020         &       0.029\sym{*}  &       0.013         &      -0.006         &      -0.022         &       0.042\sym{**} \\
            &     (0.013)         &     (0.012)         &     (0.019)         &     (0.014)         &     (0.014)         &     (0.013)         \\
\midrule
\(R^{2}\)   &       0.001         &       0.005         &       0.003         &       0.003         &       0.009         &       0.008         \\
\midrule
\#Replies (\(N\))&\num{1037340}&\num{1118533}&\num{1037340}&\num{1118533}&\num{1037340}&\num{1118533}\\
\#Source posts&\num{379}&\num{960}&\num{379}&\num{960}&\num{379}&\num{960}\\
\bottomrule
\end{tabularx}
\label{tab:emotions_politics_coefs_2}
\end{table}

\newpage
\section{Moral Outrage Analysis}
\label{sec:sm_moral_outrage}
The estimation results for the analysis of moral outrage in replies are reported in Table \ref{tab:moral_outrage}. Additionally, the estimation results for moral outrage predicted by Digital Outrage Classifier (DOC) \cite{brady2021social} and other-condemning emotions predicted by moral emotion classifier \cite{kim2024moral} are reported in Table \ref{tab:sm_outrage_othercondemning}. We find that the coefficient estimates of $\var{Displayed}$ for moral outrage represented by the product of anger and disgust and for moral outrage predicted by DOC are quantitatively identical based on $\var{SUEST}$ test ($\var{\chi^2}=0.00, p=0.957$).

\begin{table}[H]
\centering
\caption{Estimation results for moral outrage in replies across all misleading posts [Column (1)], political misleading posts [Column (2)], and non-political misleading posts [Column (3)]. Reported are coefficient estimates with post-clustered standard errors in parentheses. \sym{*} \(p<0.05\), \sym{**} \(p<0.01\), \sym{***} \(p<0.001\).}
\begin{tabularx}{\columnwidth}{@{\hspace{\tabcolsep}\extracolsep{\fill}}l*{3}{S}}
\toprule
&\multicolumn{1}{c}{(1)}&\multicolumn{1}{c}{(2)}&\multicolumn{1}{c}{(3)}\\
&\multicolumn{1}{c}{All}&\multicolumn{1}{c}{Politics}&\multicolumn{1}{c}{Non-politics}\\
\midrule
Displayed   &       0.067\sym{***}&       0.062\sym{***}&       0.060\sym{***}\\
            &     (0.009)         &     (0.013)         &     (0.012)         \\
HoursFromDisplay&      -0.000         &      -0.001         &      -0.000         \\
            &     (0.013)         &     (0.022)         &     (0.008)         \\
SourceAnger &       0.033\sym{**} &       0.030\sym{*}  &       0.025\sym{*}  \\
            &     (0.011)         &     (0.015)         &     (0.011)         \\
SourceDisgust&       0.027\sym{***}&       0.044\sym{*}  &       0.028\sym{***}\\
            &     (0.007)         &     (0.017)         &     (0.007)         \\
SourceFear  &       0.001         &      -0.007         &       0.011         \\
            &     (0.013)         &     (0.022)         &     (0.013)         \\
SourceJoy   &      -0.004         &       0.001         &      -0.016         \\
            &     (0.009)         &     (0.008)         &     (0.010)         \\
SourceSadness&      -0.002         &       0.007         &      -0.008         \\
            &     (0.008)         &     (0.012)         &     (0.009)         \\
SourceSurprise&      -0.037\sym{***}&      -0.030\sym{*}  &      -0.040\sym{***}\\
            &     (0.010)         &     (0.013)         &     (0.012)         \\
PostAge     &      -0.002         &      -0.003         &       0.003         \\
            &     (0.008)         &     (0.012)         &     (0.008)         \\
Intercept      &      -0.017         &       0.035\sym{*}  &      -0.062\sym{***}\\
            &     (0.009)         &     (0.014)         &     (0.010)         \\
\midrule
\(R^{2}\)   &       0.006         &       0.005         &       0.007         \\
\midrule
\#Replies (\(N\))&\num{2155873}&\num{1037340}&\num{1118533}\\
\#Source posts&\num{1339}&\num{379}&\num{960}\\
\bottomrule
\end{tabularx}
\label{tab:moral_outrage}
\end{table}

\newpage
\begin{table}[H]
\centering
\caption{Estimation results for moral outrage based on DOC [Column (1)] and other-condemning moral emotions [Column (2)] in replies. Reported are coefficient estimates with post-clustered standard errors in parentheses. \sym{*} \(p<0.05\), \sym{**} \(p<0.01\), \sym{***} \(p<0.001\).}
\begin{tabularx}{\columnwidth}{@{\hspace{\tabcolsep}\extracolsep{\fill}}l*{2}{S}}
\toprule
&\multicolumn{1}{c}{(1)}&\multicolumn{1}{c}{(2)}\\
&\multicolumn{1}{c}{Moral outrage (DOC)}&\multicolumn{1}{c}{Other-condemning moral emotions}\\
\midrule
Displayed   &       0.067\sym{***}&       0.132\sym{***}\\
            &     (0.013)         &     (0.015)         \\
HoursFromDisplay&      -0.004         &      -0.001         \\
            &     (0.010)         &     (0.017)         \\
SourceAnger &       0.055\sym{***}&       0.085\sym{***}\\
            &     (0.012)         &     (0.018)         \\
SourceDisgust&       0.017\sym{*}  &       0.004         \\
            &     (0.008)         &     (0.012)         \\
SourceFear  &       0.015         &       0.011         \\
            &     (0.013)         &     (0.018)         \\
SourceJoy   &       0.019         &       0.012         \\
            &     (0.025)         &     (0.028)         \\
SourceSadness&       0.008         &       0.015         \\
            &     (0.011)         &     (0.016)         \\
SourceSurprise&      -0.024         &      -0.045\sym{*}  \\
            &     (0.014)         &     (0.020)         \\
PostAge     &       0.003         &      -0.006         \\
            &     (0.007)         &     (0.011)         \\
Intercept      &      -0.017         &      -0.033         \\
            &     (0.012)         &     (0.017)         \\
\midrule
\(R^{2}\)   &       0.006         &       0.016         \\
\midrule
\#Replies (\(N\))&\num{2155873}&\num{2155873}\\
\#Source posts&\num{1339}&\num{1339}\\
\bottomrule
\end{tabularx}
\label{tab:sm_outrage_othercondemning}
\end{table}

\newpage
\section{Robustness Checks}
\label{sec:sm_robustness_checks}

\subsection{Variance Inflation Factors}
\label{sec:sm_vifs}
The Variance Inflation Factors (VIFs) for the independent variables are reported in Table \ref{tab:vifs}.

\begin{table}[H]
\centering
% \footnotesize
\caption{VIFs for independent variables. Column (1) reports the VIFs for independent variables in regressions for sentiments in replies. Column (2) reports the VIFs for independent variables in regressions for emotions in replies.}
\begin{tabularx}{\columnwidth}{@{\hspace{\tabcolsep}\extracolsep{\fill}}l*{2}{S}}
\toprule
&\multicolumn{1}{c}{(1)}&\multicolumn{1}{c}{(2)}\\
&\multicolumn{1}{c}{Sentiments}&\multicolumn{1}{c}{Emotions}\\
\midrule
Displayed   &1.11 &1.12\\
HoursFromDisplay&1.51&1.52\\
SourcePositive&1.48  & \\
SourceNegative&1.48 &\\
SourceAnger& & 1.38\\
SourceDisgust& &1.16\\
SourceFear& &1.34\\
SourceJoy& & 1.33\\
SourceSadness&&1.24\\
SourceSurprise&&1.57\\
PostAge     & 1.43&1.44\\
\midrule
Mean VIF&1.40&1.34\\
\bottomrule
\end{tabularx}
\label{tab:vifs}
\end{table}

\newpage
\subsection{Test for Potential Autocorrelation}
\label{sec:sm_autocorr}

\begin{figure}[H]
    \centering
    \includegraphics[width=\textwidth]{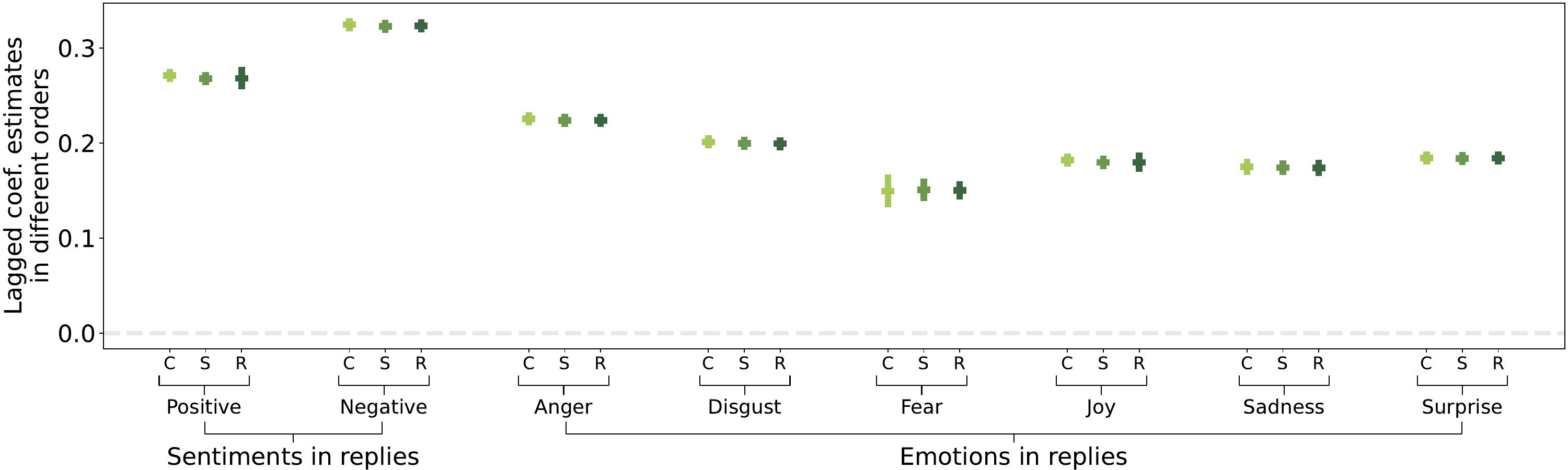}
    \caption{The lagged correlation estimates across sentiments and emotions in replies in Chronological (C), Shuffled (S), and Reversed (R) orders, respectively. Shown are mean values with error bars representing 95\% CIs. The full estimation results are reported in Table \ref{tab:lagged_coefs_orders}.}
    \label{fig:lagged_coefs_orders}
\end{figure}

We test the correlations of sentiments and emotions in replies in chronological, shuffled, and reversed orders, respectively. The estimation results are reported in Table \ref{tab:lagged_coefs_orders} and visualized in Fig. \ref{fig:lagged_coefs_orders}. Additionally, we incorporate lagged-dependent covariates, \ie, average sentiments/emotions in previous direct replies, into the regression models. Given the high positive correlations between sentiments/emotions in replies and their corresponding ones in source posts (Table \ref{tab:corr_emotions_lagged}), we use their product as combined variables to mitigate potential multicollinearity issues. The results remain consistent and support our findings (Tables \ref{tab:sentiments_lagged}--\ref{tab:emotions_lagged}).

\begin{table}[H]
\centering
% \footnotesize
\caption{The lagged coefficient estimates across sentiments and emotions in replies. $N=$ \num{1686262}. Reported are coefficient estimates with post-clustered standard errors in parentheses. \sym{*} \(p<0.05\), \sym{**} \(p<0.01\), \sym{***} \(p<0.001\).}
\begin{tabularx}{\columnwidth}{@{\hspace{\tabcolsep}\extracolsep{\fill}}l*{9}{S}}
\toprule
&\multicolumn{2}{c}{(1)}&\multicolumn{6}{c}{(2)}\\
&\multicolumn{2}{c}{Sentiments in replies}&\multicolumn{6}{c}{Emotions in replies}\\
\cmidrule(lr){2-3}\cmidrule(lr){4-9}
&{Positive}&{Negative}&{Anger}&{Disgust}&{Fear}&{Joy}&{Sadness}&{Surprise}\\
\midrule
Chronological&0.271\sym{***}&0.324\sym{***}&0.225\sym{***}&0.201\sym{***}&0.149\sym{***}&0.182\sym{***}&0.175\sym{***}&0.184\sym{***}\\
&(0.003)&(0.003)&(0.003)&(0.003)&(0.009)&(0.003)&(0.004)&(0.003)\\
\midrule
Shuffled&0.268\sym{***}&0.323\sym{***}&0.224\sym{***}&0.200\sym{***}&0.151\sym{***}&0.180\sym{***}&0.174\sym{***}&0.184\sym{***}\\
&(0.003)&(0.002)&(0.002)&(0.003)&(0.006)&(0.004)&(0.004)&(0.003)\\
\midrule
Reversed&0.268\sym{***}&0.323\sym{***}&0.224\sym{***}&0.199\sym{***}&0.150\sym{***}&0.180\sym{***}&0.174\sym{***}&0.184\sym{***}\\
&(0.006)&(0.002)&(0.002)&(0.003)&(0.005)&(0.005)&(0.004)&(0.003)\\
\bottomrule
\end{tabularx}
\label{tab:lagged_coefs_orders}
\end{table}

\begin{table}[H]
\centering
% \footnotesize
\caption{The Pearson correlations between source sentiments/emotions in original posts and lagged sentiments/emotions in replies with two-sided p-values. \sym{*} \(p<0.05\), \sym{**} \(p<0.01\), \sym{***} \(p<0.001\).}
\begin{tabularx}{\columnwidth}{@{\hspace{\tabcolsep}\extracolsep{\fill}}l*{9}{S}}
\toprule
&\multicolumn{1}{c}{$r$: Source -- Lagged}\\
\midrule
Positive&0.358\sym{***}\\
Negative&0.426\sym{***}\\
Anger&0.290\sym{***}\\
Disgust&0.228\sym{***}\\
Fear&0.342\sym{***}\\
Joy&0.246\sym{***}\\
Sadness&0.274\sym{***}\\
Surprise&0.230\sym{***}\\
\bottomrule
\end{tabularx}
\label{tab:corr_emotions_lagged}
\end{table}

\newpage
\begin{table}[H]
\centering
% \footnotesize
\caption{Estimation results for positive [Column (1)] and negative [Column (2)] sentiments in replies. The product terms representing the combination of source sentiments and lagged sentiments in replies are included. Reported are coefficient estimates with post-clustered standard errors in parentheses. \sym{*} \(p<0.05\), \sym{**} \(p<0.01\), \sym{***} \(p<0.001\).}
\begin{tabularx}{\columnwidth}{@{\hspace{\tabcolsep}\extracolsep{\fill}}l*{2}{S}}
\toprule
&\multicolumn{1}{c}{(1)}&\multicolumn{1}{c}{(2)}\\
&\multicolumn{1}{c}{Positive}&\multicolumn{1}{c}{Negative}\\
\midrule
Displayed   &      -0.053\sym{***}&       0.117\sym{***}\\
            &     (0.014)         &     (0.017)         \\
HoursFromDisplay&       0.004         &      -0.012         \\
            &     (0.018)         &     (0.027)         \\
PriorPositiveMultiply&       0.045\sym{***}&      -0.015         \\
            &     (0.010)         &     (0.011)         \\
PriorNegativeMultiply&       0.027         &      -0.078\sym{**} \\
            &     (0.019)         &     (0.030)         \\
PostAge     &       0.001         &       0.002         \\
            &     (0.010)         &     (0.015)         \\
Intercept      &      -0.006         &      -0.001         \\
            &     (0.015)         &     (0.017)         \\
\midrule
\(R^{2}\)   &       0.007         &       0.010         \\
\midrule
\#Replies (\(N\))       &        \num{2154534}         &       \num{2154534}\\
\#Source posts      &        \num{1339}         &       \num{1339}\\
\bottomrule
\end{tabularx}
\label{tab:sentiments_lagged}
\end{table}

\begin{table}[H]
\centering
% \footnotesize
\caption{Estimation results for anger [Column (1)], disgust [Column (2)], fear [Column (3)], joy [Column (4)], sadness [Column (5)], surprise [Column (6)], and moral outrage [Column (7)] in replies. The product terms representing the combination of source emotions and lagged emotions in replies are included. Reported are coefficient estimates with post-clustered standard errors in parentheses. \sym{*} \(p<0.05\), \sym{**} \(p<0.01\), \sym{***} \(p<0.001\).}
\begin{tabularx}{\columnwidth}{@{\hspace{\tabcolsep}\extracolsep{\fill}}l*{7}{S}}
\toprule
&\multicolumn{1}{c}{(1)}&\multicolumn{1}{c}{(2)}&\multicolumn{1}{c}{(3)}&\multicolumn{1}{c}{(4)}&\multicolumn{1}{c}{(5)}&\multicolumn{1}{c}{(6)}&\multicolumn{1}{c}{(7)}\\
&\multicolumn{1}{c}{Anger}&\multicolumn{1}{c}{Disgust}&\multicolumn{1}{c}{Fear}&\multicolumn{1}{c}{Joy}&\multicolumn{1}{c}{Sadness}&\multicolumn{1}{c}{Surprise}&\multicolumn{1}{c}{Moral outrage}\\
\midrule
Displayed   &       0.089\sym{***}&       0.031\sym{***}&      -0.003         &      -0.008         &      -0.003         &      -0.052\sym{***}&       0.074\sym{***}\\
            &     (0.014)         &     (0.009)         &     (0.007)         &     (0.009)         &     (0.012)         &     (0.012)         &     (0.009)         \\
HoursFromDisplay&      -0.001         &      -0.004         &      -0.003         &      -0.006         &      -0.007         &      -0.009         &       0.002         \\
            &     (0.018)         &     (0.009)         &     (0.004)         &     (0.007)         &     (0.006)         &     (0.012)         &     (0.012)         \\
PriorAngerMultiply&       0.038\sym{*}  &       0.008         &       0.003         &      -0.017\sym{**} &      -0.013         &      -0.032\sym{***}&       0.020         \\
            &     (0.017)         &     (0.007)         &     (0.005)         &     (0.006)         &     (0.009)         &     (0.009)         &     (0.012)         \\
PriorDisgustMultiply&      -0.002         &       0.031\sym{***}&      -0.005\sym{**} &      -0.004\sym{**} &      -0.005\sym{***}&      -0.001         &       0.009\sym{**} \\
            &     (0.004)         &     (0.005)         &     (0.002)         &     (0.002)         &     (0.001)         &     (0.003)         &     (0.003)         \\
PriorFearMultiply&      -0.006         &      -0.004         &       0.047\sym{***}&      -0.008\sym{**} &      -0.004         &      -0.005         &      -0.005         \\
            &     (0.005)         &     (0.005)         &     (0.008)         &     (0.003)         &     (0.003)         &     (0.004)         &     (0.004)         \\
PriorJoyMultiply&      -0.011\sym{**} &      -0.004         &      -0.000         &       0.025\sym{**} &       0.002         &       0.001         &      -0.006\sym{*}  \\
            &     (0.004)         &     (0.003)         &     (0.004)         &     (0.009)         &     (0.002)         &     (0.004)         &     (0.003)         \\
PriorSadnessMultiply&      -0.011         &      -0.009         &       0.001         &       0.005         &       0.029\sym{***}&       0.003         &      -0.009         \\
            &     (0.006)         &     (0.005)         &     (0.004)         &     (0.007)         &     (0.007)         &     (0.006)         &     (0.005)         \\
PriorSurpriseMultiply&      -0.022\sym{*}  &      -0.017         &      -0.007         &      -0.003         &       0.015         &       0.040\sym{**} &      -0.018         \\
            &     (0.011)         &     (0.009)         &     (0.005)         &     (0.005)         &     (0.017)         &     (0.014)         &     (0.009)         \\
PostAge     &      -0.005         &       0.002         &       0.004         &       0.007         &       0.006         &       0.013         &      -0.003         \\
            &     (0.010)         &     (0.007)         &     (0.004)         &     (0.005)         &     (0.004)         &     (0.008)         &     (0.008)         \\
Intercept      &      -0.021         &      -0.010         &      -0.013         &       0.009         &      -0.003         &       0.011         &      -0.018         \\
            &     (0.015)         &     (0.011)         &     (0.007)         &     (0.011)         &     (0.012)         &     (0.013)         &     (0.011)         \\
\midrule
\(R^{2}\)   &       0.004         &       0.005         &       0.006         &       0.002         &       0.002         &       0.004         &       0.002         \\
\midrule
\#Replies (\(N\))       &        \num{2154534}         &       \num{2154534}&        \num{2154534}         &       \num{2154534}&        \num{2154534}         &       \num{2154534}         &       \num{2154534}\\
\#Source posts      &        \num{1339}         &       \num{1339}&        \num{1339}         &       \num{1339}&        \num{1339}         &       \num{1339}         &       \num{1339}\\
\bottomrule
\end{tabularx}
\label{tab:emotions_lagged}
\end{table}

\newpage
\subsection{Shorter Bandwidth and Placebo Test}
\label{sec:short_band_placebo}

We repeat our analysis within a shorter bandwidth of 2 hours before and after the display of community notes. Specifically, the before-display period is 2 hours preceding the display of community notes, and the after-display period is from fifth to sixth hour (2 hours) after the note display. The initial four hours after the display of community notes are omitted to avoid potential cold-start contamination in the short bandwidth. The full estimation results are reported in Table \ref{tab:sentiments_2hours} and Table \ref{tab:emotions_2hours}. All the results, especially for anger and moral outrage, are robust and consistent with our main findings.

Additionally, we shift the cut-off point to 2 hours before the display of community notes and conduct placebo test based on replies within 2 hours before and after the shifted cut-off point. The full estimation results are reported in Table \ref{tab:sentiments_placebo} and Table \ref{tab:emotions_placebo}. All the RDD estimators, \ie, the coefficient estimates of $\var{Displayed}$ across sentiments and emotions in replies, are statistically not significant. This indicates that the changes in sentiments and emotions are specific to the display of community notes.

\newpage
\begin{table}[H]
\centering
% \footnotesize
\caption{Estimation results for positive sentiment [Column (1)] and negative sentiment [Column (2)] in replies within 2-hour bandwidth. Reported are coefficient estimates with post-clustered standard errors in parentheses. \sym{*} \(p<0.05\), \sym{**} \(p<0.01\), \sym{***} \(p<0.001\).}
\begin{tabularx}{\columnwidth}{@{\hspace{\tabcolsep}\extracolsep{\fill}}l*{2}{S}}
\toprule
&\multicolumn{1}{c}{(1)}&\multicolumn{1}{c}{(2)}\\
&\multicolumn{1}{c}{Positive}&\multicolumn{1}{c}{Negative}\\
\midrule
Displayed   &      -0.106         &       0.152\sym{*}  \\
            &     (0.056)         &     (0.063)         \\
HoursFromDisplay&       0.852         &      -0.792         \\
            &     (0.512)         &     (0.539)         \\
SourcePositive&       0.075         &      -0.026         \\
            &     (0.045)         &     (0.038)         \\
SourceNegative&      -0.020         &       0.071\sym{*}  \\
            &     (0.029)         &     (0.032)         \\
PostAge     &      -0.004         &      -0.016         \\
            &     (0.025)         &     (0.032)         \\
Intercept      &      -0.153         &       0.130         \\
            &     (0.094)         &     (0.103)         \\
\midrule
\(R^{2}\)   &       0.010         &       0.010         \\
\midrule
\#Replies (\(N\))       &        \num{278754}         &       \num{278754}\\
\#Source posts      &        \num{1280}         &       \num{1280}\\
\bottomrule
\end{tabularx}
\label{tab:sentiments_2hours}
\end{table}

\begin{table}[H]
\centering
% \footnotesize
\caption{Estimation results for anger [Column (1)], disgust [Column (2)], fear [Column (3)], joy [Column (4)], sadness [Column (5)], surprise [Column (6)], and moral outrage [Column (7)] in replies within 2-hour bandwidth. Reported are coefficient estimates with post-clustered standard errors in parentheses. \sym{*} \(p<0.05\), \sym{**} \(p<0.01\), \sym{***} \(p<0.001\).}
\begin{tabularx}{\columnwidth}{@{\hspace{\tabcolsep}\extracolsep{\fill}}l*{7}{S}}
\toprule
&\multicolumn{1}{c}{(1)}&\multicolumn{1}{c}{(2)}&\multicolumn{1}{c}{(3)}&\multicolumn{1}{c}{(4)}&\multicolumn{1}{c}{(5)}&\multicolumn{1}{c}{(6)}&\multicolumn{1}{c}{(7)}\\
&\multicolumn{1}{c}{Anger}&\multicolumn{1}{c}{Disgust}&\multicolumn{1}{c}{Fear}&\multicolumn{1}{c}{Joy}&\multicolumn{1}{c}{Sadness}&\multicolumn{1}{c}{Surprise}&\multicolumn{1}{c}{Moral outrage}\\
\midrule
Displayed   &       0.096\sym{*}  &       0.019         &      -0.002         &      -0.057         &       0.008         &      -0.093\sym{***}&       0.073\sym{**} \\
            &     (0.038)         &     (0.025)         &     (0.021)         &     (0.060)         &     (0.020)         &     (0.025)         &     (0.028)         \\
HoursFromDisplay&      -0.389         &      -0.055         &       0.038         &       0.607         &      -0.029         &       0.396         &      -0.329         \\
            &     (0.371)         &     (0.238)         &     (0.219)         &     (0.553)         &     (0.222)         &     (0.255)         &     (0.305)         \\
SourceAnger &       0.059\sym{***}&       0.005         &       0.019\sym{*}  &      -0.008         &       0.027\sym{*}  &      -0.021\sym{*}  &       0.016         \\
            &     (0.016)         &     (0.014)         &     (0.008)         &     (0.010)         &     (0.011)         &     (0.010)         &     (0.013)         \\
SourceDisgust&      -0.007         &       0.062\sym{***}&      -0.007         &      -0.002         &      -0.011\sym{*}  &       0.007         &       0.023\sym{*}  \\
            &     (0.012)         &     (0.012)         &     (0.005)         &     (0.009)         &     (0.005)         &     (0.009)         &     (0.011)         \\
SourceFear  &       0.013         &      -0.008         &       0.064\sym{***}&      -0.013         &       0.019\sym{*}  &       0.007         &      -0.007         \\
            &     (0.018)         &     (0.013)         &     (0.012)         &     (0.009)         &     (0.009)         &     (0.010)         &     (0.016)         \\
SourceJoy   &       0.006         &      -0.011         &       0.001         &       0.049         &       0.026\sym{***}&      -0.004         &      -0.012         \\
            &     (0.022)         &     (0.010)         &     (0.008)         &     (0.034)         &     (0.007)         &     (0.010)         &     (0.012)         \\
SourceSadness&       0.004         &      -0.014         &       0.007         &       0.013         &       0.048\sym{***}&      -0.004         &      -0.010         \\
            &     (0.015)         &     (0.010)         &     (0.008)         &     (0.014)         &     (0.010)         &     (0.011)         &     (0.013)         \\
SourceSurprise&      -0.056\sym{**} &      -0.027\sym{*}  &      -0.006         &       0.003         &       0.029\sym{*}  &       0.081\sym{***}&      -0.042\sym{**} \\
            &     (0.017)         &     (0.013)         &     (0.009)         &     (0.014)         &     (0.015)         &     (0.017)         &     (0.014)         \\
PostAge     &      -0.005         &       0.022         &       0.008         &       0.001         &      -0.001         &      -0.004         &       0.020         \\
            &     (0.018)         &     (0.017)         &     (0.013)         &     (0.015)         &     (0.013)         &     (0.013)         &     (0.016)         \\
Intercept      &       0.065         &       0.012         &      -0.018         &      -0.120         &      -0.015         &      -0.061         &       0.068         \\
            &     (0.070)         &     (0.043)         &     (0.040)         &     (0.098)         &     (0.043)         &     (0.047)         &     (0.057)         \\
\midrule
\(R^{2}\)   &       0.009         &       0.008         &       0.004         &       0.005         &       0.003         &       0.008         &       0.004         \\
\midrule
\#Replies (\(N\))       &        \num{278754}         &       \num{278754}&        \num{278754}         &       \num{278754}&        \num{278754}         &       \num{278754}         &       \num{278754}\\
\#Source posts      &        \num{1280}         &       \num{1280}&        \num{1280}         &       \num{1280}&        \num{1280}         &       \num{1280}         &       \num{1280}\\
\bottomrule
\end{tabularx}
\label{tab:emotions_2hours}
\end{table}

\newpage
\begin{table}[H]
\centering
% \footnotesize
\caption{Estimation results for positive sentiment [Column (1)] and negative sentiment [Column (2)] in replies within 2-hour bandwidth. The cut-off point is shifted to 2 hours before the display of community notes for placebo test. Reported are coefficient estimates with post-clustered standard errors in parentheses. \sym{*} \(p<0.05\), \sym{**} \(p<0.01\), \sym{***} \(p<0.001\).}
\begin{tabularx}{\columnwidth}{@{\hspace{\tabcolsep}\extracolsep{\fill}}l*{2}{S}}
\toprule
&\multicolumn{1}{c}{(1)}&\multicolumn{1}{c}{(2)}\\
&\multicolumn{1}{c}{Positive}&\multicolumn{1}{c}{Negative}\\
\midrule
Displayed   &       0.043         &      -0.039         \\
            &     (0.028)         &     (0.028)         \\
HoursFromDisplay&      -0.012         &       0.009         \\
            &     (0.011)         &     (0.010)         \\
SourcePositive&       0.064         &      -0.011         \\
            &     (0.044)         &     (0.037)         \\
SourceNegative&      -0.026         &       0.080\sym{*}  \\
            &     (0.028)         &     (0.032)         \\
PostAge     &      -0.001         &      -0.005         \\
            &     (0.010)         &     (0.013)         \\
Intercept      &      -0.017         &       0.015         \\
            &     (0.026)         &     (0.033)         \\
\midrule
\(R^{2}\)   &       0.007         &       0.008         \\
\midrule
\#Replies (\(N\))       &        \num{373248}         &       \num{373248}\\
\#Source posts      &        \num{1282}         &       \num{1282}\\
\bottomrule
\end{tabularx}
\label{tab:sentiments_placebo}
\end{table}

\begin{table}[H]
\centering
% \footnotesize
\caption{Estimation results for anger [Column (1)], disgust [Column (2)], fear [Column (3)], joy [Column (4)], sadness [Column (5)], surprise [Column (6)], and moral outrage [Column (7)] in replies within 2-hour bandwidth. The cut-off point is shifted to 2 hours before the display of community notes for placebo test. Reported are coefficient estimates with post-clustered standard errors in parentheses. \sym{*} \(p<0.05\), \sym{**} \(p<0.01\), \sym{***} \(p<0.001\).}
\begin{tabularx}{\columnwidth}{@{\hspace{\tabcolsep}\extracolsep{\fill}}l*{7}{S}}
\toprule
&\multicolumn{1}{c}{(1)}&\multicolumn{1}{c}{(2)}&\multicolumn{1}{c}{(3)}&\multicolumn{1}{c}{(4)}&\multicolumn{1}{c}{(5)}&\multicolumn{1}{c}{(6)}&\multicolumn{1}{c}{(7)}\\
&\multicolumn{1}{c}{Anger}&\multicolumn{1}{c}{Disgust}&\multicolumn{1}{c}{Fear}&\multicolumn{1}{c}{Joy}&\multicolumn{1}{c}{Sadness}&\multicolumn{1}{c}{Surprise}&\multicolumn{1}{c}{Moral outrage}\\
\midrule
Displayed   &      -0.002         &      -0.007         &       0.003         &       0.029         &      -0.012         &      -0.000         &      -0.013         \\
            &     (0.016)         &     (0.010)         &     (0.008)         &     (0.023)         &     (0.008)         &     (0.009)         &     (0.014)         \\
HoursFromDisplay&      -0.005         &       0.007         &      -0.007         &      -0.008         &       0.008         &       0.007         &       0.007         \\
            &     (0.007)         &     (0.005)         &     (0.005)         &     (0.008)         &     (0.005)         &     (0.006)         &     (0.007)         \\
SourceAnger &       0.061\sym{***}&       0.011         &       0.020\sym{*}  &      -0.008         &       0.019         &      -0.026\sym{*}  &       0.022         \\
            &     (0.015)         &     (0.013)         &     (0.008)         &     (0.010)         &     (0.011)         &     (0.010)         &     (0.013)         \\
SourceDisgust&      -0.008         &       0.079\sym{***}&      -0.008         &       0.001         &      -0.014\sym{*}  &       0.006         &       0.030\sym{**} \\
            &     (0.011)         &     (0.015)         &     (0.007)         &     (0.010)         &     (0.006)         &     (0.010)         &     (0.010)         \\
SourceFear  &       0.017         &       0.002         &       0.058\sym{***}&      -0.016         &       0.012         &       0.004         &       0.002         \\
            &     (0.017)         &     (0.012)         &     (0.012)         &     (0.009)         &     (0.009)         &     (0.010)         &     (0.015)         \\
SourceJoy   &       0.023         &      -0.004         &       0.001         &       0.047         &       0.027\sym{**} &      -0.012         &       0.001         \\
            &     (0.030)         &     (0.013)         &     (0.009)         &     (0.040)         &     (0.009)         &     (0.016)         &     (0.018)         \\
SourceSadness&       0.008         &      -0.014         &       0.010         &       0.020         &       0.047\sym{***}&      -0.004         &      -0.010         \\
            &     (0.016)         &     (0.010)         &     (0.009)         &     (0.016)         &     (0.011)         &     (0.013)         &     (0.012)         \\
SourceSurprise&      -0.052\sym{**} &      -0.022         &      -0.008         &       0.004         &       0.024         &       0.076\sym{***}&      -0.036\sym{**} \\
            &     (0.016)         &     (0.012)         &     (0.008)         &     (0.014)         &     (0.016)         &     (0.018)         &     (0.013)         \\
PostAge     &      -0.001         &       0.012         &      -0.001         &       0.002         &       0.004         &      -0.001         &       0.009         \\
            &     (0.007)         &     (0.006)         &     (0.005)         &     (0.006)         &     (0.005)         &     (0.005)         &     (0.006)         \\
Intercept      &       0.001         &       0.003         &      -0.001         &      -0.011         &       0.005         &       0.000         &       0.005         \\
            &     (0.018)         &     (0.012)         &     (0.009)         &     (0.018)         &     (0.011)         &     (0.012)         &     (0.014)         \\
\midrule
\(R^{2}\)   &       0.009         &       0.008         &       0.004         &       0.003         &       0.002         &       0.008         &       0.004         \\
\midrule
\#Replies (\(N\))       &        \num{373248}         &       \num{373248}&        \num{373248}         &       \num{373248}&        \num{373248}         &       \num{373248}         &       \num{373248}\\
\#Source posts      &        \num{1282}         &       \num{1282}&        \num{1282}         &       \num{1282}&        \num{1282}         &       \num{1282}         &       \num{1282}\\
\bottomrule
\end{tabularx}
\label{tab:emotions_placebo}
\end{table}

\newpage
\subsection{Analysis Without Limiting the Number of Replies}
\label{sec:sm_no5limit}

In the main analysis, we restrict our analysis to posts that receive at least 5 replies both before and after the display of community notes. This ensures that each post has sufficient replies both before and after note display for a meaningful comparison. To test the robustness of our findings, we repeat our analysis without this restriction. The estimation results for sentiments and emotions in replies are reported in Table \ref{tab:sentiments_no5limit} and Table \ref{tab:emotions_no5limit}. All the results remain consistent with our main findings.

\begin{table}[H]
\centering
% \footnotesize
\caption{Estimation results for positive [Column (1)] and negative [Column (2)] sentiments in replies without number limit. Reported are coefficient estimates with post-clustered standard errors in parentheses. \sym{*} \(p<0.05\), \sym{**} \(p<0.01\), \sym{***} \(p<0.001\).}
\begin{tabularx}{\columnwidth}{@{\hspace{\tabcolsep}\extracolsep{\fill}}l*{2}{S}}
\toprule
&\multicolumn{1}{c}{(1)}&\multicolumn{1}{c}{(2)}\\
&\multicolumn{1}{c}{Positive}&\multicolumn{1}{c}{Negative}\\
\midrule
Displayed   &      -0.059\sym{***}&       0.113\sym{***}\\
            &     (0.017)         &     (0.023)         \\
HoursFromDisplay&       0.003         &       0.007         \\
            &     (0.010)         &     (0.016)         \\
SourcePositive&       0.039\sym{*}  &       0.015         \\
            &     (0.018)         &     (0.020)         \\
SourceNegative&      -0.047\sym{**} &       0.113\sym{***}\\
            &     (0.016)         &     (0.023)         \\
PostAge     &       0.002         &      -0.008         \\
            &     (0.006)         &     (0.009)         \\
Intercept      &       0.014         &      -0.027         \\
            &     (0.013)         &     (0.018)         \\
\midrule
\(R^{2}\)   &       0.006         &       0.014         \\
\midrule
\#Replies (\(N\))       &        \num{2225260}         &       \num{2225260}\\
\#Source posts      &        \num{1841}         &       \num{1841}\\
\bottomrule
\end{tabularx}
\label{tab:sentiments_no5limit}
\end{table}

\newpage
\begin{table}[H]
\centering
% \footnotesize
\caption{Estimation results for anger [Column (1)], disgust [Column (2)], fear [Column (3)], joy [Column (4)], sadness [Column (5)], surprise [Column (6)], and moral outrage [Column (7)] in replies without number limit. Reported are coefficient estimates with post-clustered standard errors in parentheses. \sym{*} \(p<0.05\), \sym{**} \(p<0.01\), \sym{***} \(p<0.001\).}
\begin{tabularx}{\columnwidth}{@{\hspace{\tabcolsep}\extracolsep{\fill}}l*{7}{S}}
\toprule
&\multicolumn{1}{c}{(1)}&\multicolumn{1}{c}{(2)}&\multicolumn{1}{c}{(3)}&\multicolumn{1}{c}{(4)}&\multicolumn{1}{c}{(5)}&\multicolumn{1}{c}{(6)}&\multicolumn{1}{c}{(7)}\\
&\multicolumn{1}{c}{Anger}&\multicolumn{1}{c}{Disgust}&\multicolumn{1}{c}{Fear}&\multicolumn{1}{c}{Joy}&\multicolumn{1}{c}{Sadness}&\multicolumn{1}{c}{Surprise}&\multicolumn{1}{c}{Moral outrage}\\
\midrule
Displayed   &       0.079\sym{***}&       0.022\sym{*}  &      -0.009         &      -0.018         &      -0.010         &      -0.043\sym{***}&       0.069\sym{***}\\
            &     (0.010)         &     (0.010)         &     (0.008)         &     (0.010)         &     (0.011)         &     (0.009)         &     (0.008)         \\
HoursFromDisplay&       0.002         &       0.001         &       0.002         &       0.001         &      -0.004         &      -0.011         &       0.004         \\
            &     (0.013)         &     (0.006)         &     (0.004)         &     (0.007)         &     (0.008)         &     (0.008)         &     (0.009)         \\
SourceAnger &       0.078\sym{***}&       0.008         &       0.016\sym{*}  &      -0.006         &       0.020         &      -0.037\sym{***}&       0.033\sym{**} \\
            &     (0.013)         &     (0.011)         &     (0.008)         &     (0.008)         &     (0.014)         &     (0.010)         &     (0.011)         \\
SourceDisgust&       0.001         &       0.069\sym{***}&      -0.005         &      -0.011\sym{*}  &      -0.012\sym{*}  &      -0.003         &       0.027\sym{***}\\
            &     (0.008)         &     (0.015)         &     (0.004)         &     (0.005)         &     (0.006)         &     (0.007)         &     (0.007)         \\
SourceFear  &       0.013         &       0.008         &       0.066\sym{***}&      -0.021\sym{**} &       0.006         &       0.004         &       0.001         \\
            &     (0.014)         &     (0.011)         &     (0.010)         &     (0.007)         &     (0.009)         &     (0.012)         &     (0.012)         \\
SourceJoy   &       0.009         &      -0.002         &       0.005         &       0.032         &       0.026\sym{***}&       0.001         &      -0.006         \\
            &     (0.016)         &     (0.009)         &     (0.006)         &     (0.018)         &     (0.007)         &     (0.012)         &     (0.009)         \\
SourceSadness&       0.008         &      -0.002         &       0.006         &       0.009         &       0.057\sym{***}&      -0.008         &      -0.002         \\
            &     (0.010)         &     (0.009)         &     (0.005)         &     (0.009)         &     (0.008)         &     (0.008)         &     (0.008)         \\
SourceSurprise&      -0.055\sym{***}&      -0.013         &      -0.002         &       0.000         &       0.016         &       0.068\sym{***}&      -0.036\sym{***}\\
            &     (0.011)         &     (0.012)         &     (0.007)         &     (0.009)         &     (0.012)         &     (0.013)         &     (0.010)         \\
PostAge     &      -0.005         &       0.001         &       0.003         &       0.004         &       0.005         &       0.012\sym{*}  &      -0.003         \\
            &     (0.006)         &     (0.005)         &     (0.003)         &     (0.004)         &     (0.004)         &     (0.005)         &     (0.005)         \\
Intercept      &      -0.019         &      -0.005         &       0.002         &       0.004         &       0.002         &       0.011         &      -0.017         \\
            &     (0.011)         &     (0.010)         &     (0.008)         &     (0.009)         &     (0.012)         &     (0.010)         &     (0.009)         \\
\midrule
\(R^{2}\)   &       0.014         &       0.006         &       0.005         &       0.002         &       0.003         &       0.009         &       0.006         \\
\midrule
\#Replies (\(N\))       &        \num{2225260}         &       \num{2225260}&        \num{2225260}         &       \num{2225260}&        \num{2225260}         &       \num{2225260}         &       \num{2225260}\\
\#Source posts      &        \num{1841}         &       \num{1841}&        \num{1841}         &       \num{1841}&        \num{1841}         &       \num{1841}         &       \num{1841}\\
\bottomrule
\end{tabularx}
\label{tab:emotions_no5limit}
\end{table}

\newpage
\subsection{Month-Year Fixed Effects}
\label{sec:sm_month_year}
We consider the month-year fixed effects at two levels: post level and reply level. At the post level, we incorporate month-year fixed effects based on the creation time of misleading posts. The estimation results across sentiments and emotions are reported in Table \ref{tab:sentiments_month_year_tweet} and Table \ref{tab:emotions_month_year_tweet}. The estimation results for moral outrage are reported in Column (1) of Table \ref{tab:moral_outrage_robust}. At the reply level, we incorporate month-year fixed effects based on the creation time of replies. The estimation results across sentiments and emotions are reported in Table \ref{tab:sentiments_month_year_reply} and Table \ref{tab:emotions_month_year_reply}. The estimation results for moral outrage are reported in Column (2) of Table \ref{tab:moral_outrage_robust}. All the results are robust and consistent with our main findings.

\begin{table}[H]
\centering
% \footnotesize
\caption{Estimation results for positive sentiment [Column (1)] and negative sentiment [Column (2)] in replies. Post-level month-year fixed effects are included. Reported are coefficient estimates with post-clustered standard errors in parentheses. \sym{*} \(p<0.05\), \sym{**} \(p<0.01\), \sym{***} \(p<0.001\).}
\begin{tabularx}{\columnwidth}{@{\hspace{\tabcolsep}\extracolsep{\fill}}l*{2}{S}}
\toprule
&\multicolumn{1}{c}{(1)}&\multicolumn{1}{c}{(2)}\\
&\multicolumn{1}{c}{Positive}&\multicolumn{1}{c}{Negative}\\
\midrule
Displayed   &      -0.055\sym{***}&       0.114\sym{***}\\
            &     (0.016)         &     (0.020)         \\
HoursFromDisplay&      -0.003         &       0.000         \\
            &     (0.016)         &     (0.020)         \\
SourcePositive&       0.040\sym{*}  &       0.014         \\
            &     (0.018)         &     (0.021)         \\
SourceNegative&      -0.048\sym{**} &       0.113\sym{***}\\
            &     (0.016)         &     (0.023)         \\
PostAge     &       0.007         &      -0.006         \\
            &     (0.009)         &     (0.012)         \\
Post: MonthYear&\checkmark&\checkmark\\
Intercept      &      -0.017         &      -0.045         \\
            &     (0.018)         &     (0.037)         \\
\midrule
\(R^{2}\)   &       0.008         &       0.014         \\
\midrule
\#Replies (\(N\))&\num{2155873}&\num{2155873}\\
\#Source posts&\num{1339}&\num{1339}\\
\bottomrule
\end{tabularx}
\label{tab:sentiments_month_year_tweet}
\end{table}

\newpage
\begin{table}[H]
\centering
% \footnotesize
\caption{Estimation results for emotions in replies: anger [Column (1)], disgust [Column (2)], fear [Column (3)], joy [Column (4)], sadness [Column (5)], and surprise [Column (6)]. Post-level month-year fixed effects are included. Reported are coefficient estimates with post-clustered standard errors in parentheses. \sym{*} \(p<0.05\), \sym{**} \(p<0.01\), \sym{***} \(p<0.001\).}
\begin{tabularx}{\columnwidth}{@{\hspace{\tabcolsep}\extracolsep{\fill}}l*{6}{S}}
\toprule
&\multicolumn{1}{c}{(1)}&\multicolumn{1}{c}{(2)}&\multicolumn{1}{c}{(3)}&\multicolumn{1}{c}{(4)}&\multicolumn{1}{c}{(5)}&\multicolumn{1}{c}{(6)}\\
&\multicolumn{1}{c}{Anger}&\multicolumn{1}{c}{Disgust}&\multicolumn{1}{c}{Fear}&\multicolumn{1}{c}{Joy}&\multicolumn{1}{c}{Sadness}&\multicolumn{1}{c}{Surprise}\\
\midrule
Displayed   &       0.077\sym{***}&       0.023\sym{*}  &      -0.006         &      -0.015         &      -0.006         &      -0.043\sym{***}\\
            &     (0.010)         &     (0.009)         &     (0.007)         &     (0.010)         &     (0.011)         &     (0.009)         \\
HoursFromDisplay&      -0.004         &      -0.007         &      -0.001         &      -0.005         &      -0.002         &      -0.005         \\
            &     (0.017)         &     (0.010)         &     (0.005)         &     (0.011)         &     (0.008)         &     (0.010)         \\
SourceAnger &       0.076\sym{***}&       0.008         &       0.019\sym{*}  &      -0.008         &       0.020         &      -0.037\sym{***}\\
            &     (0.012)         &     (0.011)         &     (0.008)         &     (0.008)         &     (0.013)         &     (0.010)         \\
SourceDisgust&      -0.003         &       0.071\sym{***}&      -0.002         &      -0.010         &      -0.011         &      -0.003         \\
            &     (0.009)         &     (0.015)         &     (0.004)         &     (0.006)         &     (0.006)         &     (0.007)         \\
SourceFear  &       0.009         &       0.008         &       0.068\sym{***}&      -0.019\sym{**} &       0.007         &       0.004         \\
            &     (0.014)         &     (0.011)         &     (0.011)         &     (0.007)         &     (0.009)         &     (0.012)         \\
SourceJoy   &       0.004         &       0.000         &       0.007         &       0.032         &       0.026\sym{***}&       0.000         \\
            &     (0.015)         &     (0.008)         &     (0.006)         &     (0.017)         &     (0.007)         &     (0.011)         \\
SourceSadness&       0.004         &      -0.003         &       0.009         &       0.008         &       0.052\sym{***}&      -0.008         \\
            &     (0.010)         &     (0.009)         &     (0.006)         &     (0.009)         &     (0.008)         &     (0.009)         \\
SourceSurprise&      -0.059\sym{***}&      -0.010         &       0.002         &      -0.001         &       0.016         &       0.067\sym{***}\\
            &     (0.011)         &     (0.012)         &     (0.007)         &     (0.010)         &     (0.013)         &     (0.013)         \\
PostAge     &      -0.002         &       0.004         &       0.004         &       0.008         &       0.006         &       0.010         \\
            &     (0.010)         &     (0.008)         &     (0.004)         &     (0.007)         &     (0.005)         &     (0.006)         \\
Post: MonthYear&\checkmark&\checkmark&\checkmark&\checkmark&\checkmark&\checkmark\\
Intercept      &      -0.047\sym{*}  &       0.015         &       0.032         &      -0.022         &      -0.006         &       0.001         \\
            &     (0.020)         &     (0.021)         &     (0.022)         &     (0.012)         &     (0.020)         &     (0.017)         \\
\midrule
\(R^{2}\)   &       0.015         &       0.006         &       0.005         &       0.003         &       0.003         &       0.009         \\
\midrule
\#Replies (\(N\))&\num{2155873}&\num{2155873}&\num{2155873}&\num{2155873}&\num{2155873}&\num{2155873}\\
\#Source posts&\num{1339}&\num{1339}&\num{1339}&\num{1339}&\num{1339}&\num{1339}\\
\bottomrule
\end{tabularx}
\label{tab:emotions_month_year_tweet}
\end{table}

\newpage
\begin{table}[H]
\centering
% \footnotesize
\caption{Estimation results for positive sentiment [Column (1)] and negative sentiment [Column (2)] in replies. Reply-level month-year fixed effects are included. Reported are coefficient estimates with post-clustered standard errors in parentheses. \sym{*} \(p<0.05\), \sym{**} \(p<0.01\), \sym{***} \(p<0.001\).}
\begin{tabularx}{\columnwidth}{@{\hspace{\tabcolsep}\extracolsep{\fill}}l*{2}{S}}
\toprule
&\multicolumn{1}{c}{(1)}&\multicolumn{1}{c}{(2)}\\
&\multicolumn{1}{c}{Positive}&\multicolumn{1}{c}{Negative}\\
\midrule
Displayed   &      -0.057\sym{**} &       0.115\sym{***}\\
            &     (0.018)         &     (0.021)         \\
HoursFromDisplay&      -0.002         &      -0.001         \\
            &     (0.016)         &     (0.020)         \\
SourcePositive&       0.039\sym{*}  &       0.014         \\
            &     (0.018)         &     (0.022)         \\
SourceNegative&      -0.049\sym{**} &       0.113\sym{***}\\
            &     (0.017)         &     (0.023)         \\
PostAge     &       0.003         &      -0.002         \\
            &     (0.009)         &     (0.012)         \\
Reply: MonthYear&\checkmark&\checkmark\\
Intercept      &      -0.017         &      -0.050         \\
            &     (0.017)         &     (0.037)         \\
\midrule
\(R^{2}\)   &       0.008         &       0.014         \\
\midrule
\#Replies (\(N\))&\num{2155873}&\num{2155873}\\
\#Source posts&\num{1339}&\num{1339}\\
\bottomrule
\end{tabularx}
\label{tab:sentiments_month_year_reply}
\end{table}

\begin{table}[H]
\centering
% \footnotesize
\caption{Estimation results for emotions in replies: anger [Column (1)], disgust [Column (2)], fear [Column (3)], joy [Column (4)], sadness [Column (5)], and surprise [Column (6)]. Reply-level month-year fixed effects are included. Reported are coefficient estimates with post-clustered standard errors in parentheses. \sym{*} \(p<0.05\), \sym{**} \(p<0.01\), \sym{***} \(p<0.001\).}
\begin{tabularx}{\columnwidth}{@{\hspace{\tabcolsep}\extracolsep{\fill}}l*{6}{S}}
\toprule
&\multicolumn{1}{c}{(1)}&\multicolumn{1}{c}{(2)}&\multicolumn{1}{c}{(3)}&\multicolumn{1}{c}{(4)}&\multicolumn{1}{c}{(5)}&\multicolumn{1}{c}{(6)}\\
&\multicolumn{1}{c}{Anger}&\multicolumn{1}{c}{Disgust}&\multicolumn{1}{c}{Fear}&\multicolumn{1}{c}{Joy}&\multicolumn{1}{c}{Sadness}&\multicolumn{1}{c}{Surprise}\\
\midrule
Displayed   &       0.078\sym{***}&       0.023\sym{*}  &      -0.007         &      -0.016         &      -0.007         &      -0.043\sym{***}\\
            &     (0.011)         &     (0.009)         &     (0.007)         &     (0.010)         &     (0.011)         &     (0.009)         \\
HoursFromDisplay&      -0.005         &      -0.008         &      -0.001         &      -0.003         &      -0.002         &      -0.004         \\
            &     (0.017)         &     (0.010)         &     (0.005)         &     (0.012)         &     (0.008)         &     (0.010)         \\
SourceAnger &       0.076\sym{***}&       0.008         &       0.019\sym{*}  &      -0.008         &       0.020         &      -0.037\sym{***}\\
            &     (0.012)         &     (0.011)         &     (0.008)         &     (0.008)         &     (0.013)         &     (0.010)         \\
SourceDisgust&      -0.002         &       0.072\sym{***}&      -0.002         &      -0.012\sym{*}  &      -0.012\sym{*}  &      -0.004         \\
            &     (0.008)         &     (0.015)         &     (0.004)         &     (0.006)         &     (0.006)         &     (0.007)         \\
SourceFear  &       0.009         &       0.009         &       0.067\sym{***}&      -0.020\sym{**} &       0.007         &       0.003         \\
            &     (0.014)         &     (0.011)         &     (0.011)         &     (0.007)         &     (0.009)         &     (0.012)         \\
SourceJoy   &       0.004         &       0.001         &       0.007         &       0.030         &       0.024\sym{***}&      -0.000         \\
            &     (0.015)         &     (0.008)         &     (0.006)         &     (0.018)         &     (0.007)         &     (0.011)         \\
SourceSadness&       0.004         &      -0.002         &       0.008         &       0.006         &       0.052\sym{***}&      -0.008         \\
            &     (0.010)         &     (0.009)         &     (0.006)         &     (0.009)         &     (0.008)         &     (0.008)         \\
SourceSurprise&      -0.060\sym{***}&      -0.009         &       0.002         &      -0.002         &       0.016         &       0.067\sym{***}\\
            &     (0.011)         &     (0.012)         &     (0.007)         &     (0.010)         &     (0.013)         &     (0.012)         \\
PostAge     &      -0.001         &       0.003         &       0.005         &       0.005         &       0.006         &       0.011         \\
            &     (0.009)         &     (0.008)         &     (0.004)         &     (0.007)         &     (0.004)         &     (0.006)         \\
Reply: MonthYear&\checkmark&\checkmark&\checkmark&\checkmark&\checkmark&\checkmark\\
Intercept      &      -0.050\sym{*}  &       0.016         &       0.028         &      -0.024         &      -0.008         &       0.002         \\
            &     (0.020)         &     (0.021)         &     (0.022)         &     (0.012)         &     (0.019)         &     (0.017)         \\
\midrule
\(R^{2}\)   &       0.015         &       0.006         &       0.005         &       0.003         &       0.003         &       0.009         \\
\midrule
\#Replies (\(N\))&\num{2155873}&\num{2155873}&\num{2155873}&\num{2155873}&\num{2155873}&\num{2155873}\\
\#Source posts&\num{1339}&\num{1339}&\num{1339}&\num{1339}&\num{1339}&\num{1339}\\
\bottomrule
\end{tabularx}
\label{tab:emotions_month_year_reply}
\end{table}

\newpage
\begin{table}[H]
\centering
% \footnotesize
\caption{Estimation results for moral outrage across robustness checks: post-level month-year fixed effects [Column (1)], reply-level month-year fixed effects [Column (2)], alternative sentiment and emotion lexicons [Column (3)], and expansion with emotions in community notes [Column (4)]. Reported are coefficient estimates with post-clustered standard errors in parentheses. \sym{*} \(p<0.05\), \sym{**} \(p<0.01\), \sym{***} \(p<0.001\).}
\begin{tabularx}{\columnwidth}{@{\hspace{\tabcolsep}\extracolsep{\fill}}l*{4}{S}}
\toprule
&\multicolumn{1}{c}{(1)}&\multicolumn{1}{c}{(2)}&\multicolumn{1}{c}{(3)}&\multicolumn{1}{c}{(4)}\\
&\multicolumn{1}{c}{Post: month-year}&\multicolumn{1}{c}{Reply: month-year}&\multicolumn{1}{c}{Alternative lexicons}&\multicolumn{1}{c}{Emotions in community notes}\\
\midrule
Displayed   &       0.066\sym{***}&       0.066\sym{***}&       0.054\sym{***}&       0.068\sym{***}\\
            &     (0.008)         &     (0.009)         &     (0.009)         &     (0.008)         \\
HoursFromDisplay&      -0.001         &      -0.001         &      -0.021         &      -0.001         \\
            &     (0.013)         &     (0.013)         &     (0.016)         &     (0.013)         \\
NoteAnger   &&&&       0.001         \\
            &&&&     (0.011)         \\
NoteDisgust &&&&       0.002         \\
            &&&&     (0.008)         \\
NoteFear    &&&&       0.002         \\
            &&&&     (0.009)         \\
NoteJoy     &&&&       0.006         \\
            &&&&     (0.011)         \\
NoteSadness &&&&      -0.008         \\
            &&&&     (0.009)         \\
NoteSurprise&&&&      -0.023\sym{*}  \\
            &&&&     (0.010)         \\
Displayed $\times$ NoteAnger&&&&       0.021         \\
            &&&&     (0.012)         \\
Displayed $\times$ NoteDisgust&&&&      -0.010         \\
            &&&&     (0.008)         \\
Displayed $\times$ NoteFear &&&&       0.004         \\
            &&&&     (0.011)         \\
Displayed $\times$ NoteJoy  &&&&      -0.007         \\
            &&&&     (0.010)         \\
Displayed $\times$ NoteSadness&&&&       0.004         \\
            &&&&     (0.009)         \\
Displayed $\times$ NoteSurprise&&&&       0.008         \\
            &&&&     (0.011)         \\
SourceAnger &       0.032\sym{**} &       0.032\sym{**} &       0.032\sym{*}  &       0.030\sym{**} \\
            &     (0.011)         &     (0.011)         &     (0.013)         &     (0.011)         \\
SourceDisgust&       0.025\sym{***}&       0.026\sym{***}&       0.025\sym{**} &       0.024\sym{**} \\
            &     (0.007)         &     (0.008)         &     (0.008)         &     (0.007)         \\
SourceFear  &      -0.002         &      -0.001         &      -0.005         &       0.000         \\
            &     (0.012)         &     (0.012)         &     (0.007)         &     (0.012)         \\
SourceJoy   &      -0.008         &      -0.007         &       0.009         &      -0.007         \\
            &     (0.008)         &     (0.008)         &     (0.009)         &     (0.009)         \\
SourceSadness&      -0.005         &      -0.005         &       0.022         &      -0.001         \\
            &     (0.008)         &     (0.008)         &     (0.016)         &     (0.009)         \\
SourceSurprise&      -0.038\sym{***}&      -0.038\sym{***}&      -0.004         &      -0.035\sym{***}\\
            &     (0.010)         &     (0.010)         &     (0.005)         &     (0.010)         \\
PostAge     &      -0.001         &      -0.001         &       0.016         &      -0.001         \\
            &     (0.008)         &     (0.008)         &     (0.010)         &     (0.008)         \\
Intercept      &      -0.027         &      -0.026         &      -0.013         &      -0.017         \\
            &     (0.018)         &     (0.018)         &     (0.008)         &     (0.009)         \\
\midrule
\(R^{2}\)   &       0.006         &       0.006         &       0.003         &       0.006         \\
\midrule
\#Replies (\(N\))&\num{2155873}&\num{2155873}&\num{2155873}&\num{2155873}\\
\#Source posts&\num{1339}&\num{1339}&\num{1339}&\num{1339}\\
\bottomrule
\end{tabularx}
\label{tab:moral_outrage_robust}
\end{table}

\newpage
\subsection{Alternative Lexicon-Based Sentiments and Emotions}
\label{sec:sm_alternatives}
We use VADER sentiment lexicon \cite{hutto2014vader} and NRC emotion lexicon \cite{mohammad2013crowdsourcing} as alternative methods for measuring sentiments and emotions. The estimation results across sentiments and emotions based on the alternative approaches are reported in Table \ref{tab:emotions_lexicon_1} and Table \ref{tab:emotions_lexicon_2}. Additionally, the estimation results for moral outrage based on NRC emotion lexicon are reported in Column (3) of Table \ref{tab:moral_outrage_robust}. All the results are robust and consistent with our main findings.

\begin{table}[H]
\centering
% \footnotesize
\caption{Estimation results for sentiment polarity and emotions in replies: polarity [Column (1)], anger [Column (2)], disgust [Column (3)], and fear [Column (4)]. Reported are coefficient estimates with post-clustered standard errors in parentheses. \sym{*} \(p<0.05\), \sym{**} \(p<0.01\), \sym{***} \(p<0.001\).}
\begin{tabularx}{\columnwidth}{@{\hspace{\tabcolsep}\extracolsep{\fill}}l*{4}{S}}
\toprule
&\multicolumn{1}{c}{(1)}&\multicolumn{1}{c}{(2)}&\multicolumn{1}{c}{(3)}&\multicolumn{1}{c}{(4)}\\
&\multicolumn{1}{c}{Polarity}&\multicolumn{1}{c}{Anger}&\multicolumn{1}{c}{Disgust}&\multicolumn{1}{c}{Fear}\\
\midrule
Displayed   &      -0.044\sym{**} &       0.035\sym{**} &       0.062\sym{***}&      -0.006         \\
            &     (0.015)         &     (0.014)         &     (0.011)         &     (0.010)         \\
HoursFromDisplay&       0.001         &      -0.015         &      -0.029         &       0.002         \\
            &     (0.012)         &     (0.012)         &     (0.021)         &     (0.006)         \\
SourcePolarity&       0.072\sym{***}&                     &                     &                     \\
            &     (0.016)         &                     &                     &                     \\
SourceAnger &                     &       0.058\sym{***}&       0.015         &       0.026\sym{*}  \\
            &                     &     (0.012)         &     (0.018)         &     (0.013)         \\
SourceDisgust&                     &       0.004         &       0.069\sym{***}&      -0.004         \\
            &                     &     (0.007)         &     (0.018)         &     (0.007)         \\
SourceFear  &                     &       0.022\sym{**} &      -0.018         &       0.099\sym{***}\\
            &                     &     (0.008)         &     (0.010)         &     (0.010)         \\
SourceJoy   &                     &       0.004         &       0.004         &      -0.013         \\
            &                     &     (0.009)         &     (0.011)         &     (0.008)         \\
SourceSadness&                     &       0.010         &       0.006         &      -0.017\sym{**} \\
            &                     &     (0.012)         &     (0.016)         &     (0.006)         \\
SourceSurprise&                     &      -0.011         &      -0.006         &      -0.008         \\
            &                     &     (0.006)         &     (0.006)         &     (0.006)         \\
PostAge     &      -0.001         &       0.009         &       0.023         &      -0.000         \\
            &     (0.007)         &     (0.007)         &     (0.015)         &     (0.005)         \\
Intercept      &       0.011         &      -0.009         &      -0.016         &       0.001         \\
            &     (0.012)         &     (0.010)         &     (0.012)         &     (0.010)         \\
\midrule
\(R^{2}\)   &       0.006         &       0.005         &       0.007         &       0.013         \\
\midrule
\#Replies (\(N\))&\num{2155873}&\num{2155873}&\num{2155873}&\num{2155873}\\
\#Source posts&\num{1339}&\num{1339}&\num{1339}&\num{1339}\\
\bottomrule
\end{tabularx}
\label{tab:emotions_lexicon_1}
\end{table}

\newpage
\begin{table}[H]
\centering
% \footnotesize
\caption{Estimation results for emotions in replies: joy [Column (1)], sadness [Column (2)], and surprise [Column (3)]. Reported are coefficient estimates with post-clustered standard errors in parentheses. \sym{*} \(p<0.05\), \sym{**} \(p<0.01\), \sym{***} \(p<0.001\).}
\begin{tabularx}{\columnwidth}{@{\hspace{\tabcolsep}\extracolsep{\fill}}l*{3}{S}}
\toprule
&\multicolumn{1}{c}{(1)}&\multicolumn{1}{c}{(2)}&\multicolumn{1}{c}{(3)}\\
&\multicolumn{1}{c}{Joy}&\multicolumn{1}{c}{Sadness}&\multicolumn{1}{c}{Surprise}\\
\midrule
Displayed   &       0.007         &       0.010         &      -0.007         \\
            &     (0.012)         &     (0.016)         &     (0.013)         \\
HoursFromDisplay&       0.002         &       0.004         &      -0.000         \\
            &     (0.007)         &     (0.006)         &     (0.006)         \\
SourceAnger &      -0.005         &       0.003         &      -0.006         \\
            &     (0.006)         &     (0.011)         &     (0.007)         \\
SourceDisgust&       0.002         &       0.009         &      -0.005         \\
            &     (0.006)         &     (0.009)         &     (0.009)         \\
SourceFear  &      -0.011         &      -0.014         &      -0.013         \\
            &     (0.006)         &     (0.009)         &     (0.007)         \\
SourceJoy   &       0.074\sym{***}&       0.010         &      -0.009         \\
            &     (0.009)         &     (0.013)         &     (0.008)         \\
SourceSadness&       0.020\sym{**} &       0.056\sym{**} &       0.003         \\
            &     (0.007)         &     (0.017)         &     (0.010)         \\
SourceSurprise&       0.006         &      -0.016\sym{*}  &       0.047\sym{***}\\
            &     (0.006)         &     (0.006)         &     (0.010)         \\
PostAge     &       0.000         &      -0.004         &      -0.005         \\
            &     (0.005)         &     (0.005)         &     (0.005)         \\
Intercept      &      -0.002         &      -0.002         &       0.002         \\
            &     (0.008)         &     (0.014)         &     (0.010)         \\
\midrule
\(R^{2}\)   &       0.006         &       0.004         &       0.003         \\
\midrule
\#Replies (\(N\))&\num{2155873}&\num{2155873}&\num{2155873}\\
\#Source posts&\num{1339}&\num{1339}&\num{1339}\\
\bottomrule
\end{tabularx}
\label{tab:emotions_lexicon_2}
\end{table}

\newpage
\subsection{Expansion With Sentiments/Emotions in Community Notes}
\label{sec:sm_note_emotions}
We expand our models by incorporating sentiment (emotions) in community notes and their interactions with $\var{Displayed}$. The estimation results across sentiments and emotions are reported in Table \ref{tab:sentiments_notes} and Table \ref{tab:emotions_notes}. Additionally, the estimation results for moral outrage are reported in Column (4) of Table \ref{tab:moral_outrage_robust}. All the results are robust and consistent with our main findings. 

\begin{table}[H]
\centering
% \footnotesize
\caption{Estimation results for positive sentiment [Column (1)] and negative sentiment [Column (2)] in replies. Sentiments in community notes and their interactions with $\var{Displayed}$ are included. Reported are coefficient estimates with post-clustered standard errors in parentheses. \sym{*} \(p<0.05\), \sym{**} \(p<0.01\), \sym{***} \(p<0.001\).}
\begin{tabularx}{\columnwidth}{@{\hspace{\tabcolsep}\extracolsep{\fill}}l*{2}{S}}
\toprule
&\multicolumn{1}{c}{(1)}&\multicolumn{1}{c}{(2)}\\
&\multicolumn{1}{c}{Positive}&\multicolumn{1}{c}{Negative}\\
\midrule
Displayed   &      -0.054\sym{**} &       0.113\sym{***}\\
            &     (0.017)         &     (0.022)         \\
HoursFromDisplay&      -0.003         &      -0.002         \\
            &     (0.013)         &     (0.023)         \\
NotePositive&      -0.002         &       0.036\sym{*}  \\
            &     (0.011)         &     (0.016)         \\
NoteNegative&       0.023         &       0.031         \\
            &     (0.013)         &     (0.020)         \\
Displayed $\times$ NotePositive&       0.006         &      -0.003         \\
            &     (0.010)         &     (0.013)         \\
Displayed $\times$ NoteNegative&      -0.024         &       0.015         \\
            &     (0.013)         &     (0.019)         \\
SourcePositive&       0.040\sym{*}  &       0.017         \\
            &     (0.019)         &     (0.021)         \\
SourceNegative&      -0.049\sym{**} &       0.110\sym{***}\\
            &     (0.017)         &     (0.023)         \\
PostAge     &       0.005         &      -0.005         \\
            &     (0.008)         &     (0.014)         \\
Intercept      &       0.013         &      -0.028         \\
            &     (0.013)         &     (0.018)         \\
\midrule
\(R^{2}\)   &       0.007         &       0.015         \\
\midrule
\#Replies (\(N\))       &        \num{2155873}         &       \num{2155873}\\
\#Source posts      &        \num{1339}         &       \num{1339}\\
\bottomrule
\end{tabularx}
\label{tab:sentiments_notes}
\end{table}

\newpage
\begin{table}[H]
\centering
% \footnotesize
\caption{Estimation results for anger [Column (1)], disgust [Column (2)], fear [Column (3)], joy [Column (4)], sadness [Column (5)], and surprise [Column (6)] in replies. Emotions in community notes and their interactions with $\var{Displayed}$ are included. Reported are coefficient estimates with post-clustered standard errors in parentheses. \sym{*} \(p<0.05\), \sym{**} \(p<0.01\), \sym{***} \(p<0.001\).}
\begin{tabularx}{\columnwidth}{@{\hspace{\tabcolsep}\extracolsep{\fill}}l*{6}{S}}
\toprule
&\multicolumn{1}{c}{(1)}&\multicolumn{1}{c}{(2)}&\multicolumn{1}{c}{(3)}&\multicolumn{1}{c}{(4)}&\multicolumn{1}{c}{(5)}&\multicolumn{1}{c}{(6)}\\
&\multicolumn{1}{c}{Anger}&\multicolumn{1}{c}{Disgust}&\multicolumn{1}{c}{Fear}&\multicolumn{1}{c}{Joy}&\multicolumn{1}{c}{Sadness}&\multicolumn{1}{c}{Surprise}\\
\midrule
Displayed   &       0.078\sym{***}&       0.023\sym{*}  &      -0.005         &      -0.016         &      -0.010         &      -0.043\sym{***}\\
            &     (0.010)         &     (0.010)         &     (0.007)         &     (0.009)         &     (0.010)         &     (0.008)         \\
HoursFromDisplay&      -0.006         &      -0.007         &      -0.003         &      -0.003         &       0.001         &      -0.004         \\
            &     (0.018)         &     (0.009)         &     (0.004)         &     (0.010)         &     (0.007)         &     (0.009)         \\
NoteAnger   &       0.034\sym{**} &      -0.014         &      -0.017         &       0.026\sym{*}  &       0.010         &      -0.020\sym{*}  \\
            &     (0.011)         &     (0.013)         &     (0.009)         &     (0.011)         &     (0.012)         &     (0.009)         \\
NoteDisgust &      -0.005         &       0.002         &      -0.004         &       0.015         &      -0.009         &       0.009         \\
            &     (0.007)         &     (0.009)         &     (0.004)         &     (0.009)         &     (0.008)         &     (0.008)         \\
NoteFear    &       0.017         &      -0.012         &       0.042\sym{*}  &      -0.004         &      -0.007         &      -0.021\sym{*}  \\
            &     (0.011)         &     (0.012)         &     (0.018)         &     (0.011)         &     (0.010)         &     (0.010)         \\
NoteJoy     &       0.019         &       0.009         &      -0.012\sym{*}  &       0.006         &       0.002         &      -0.005         \\
            &     (0.012)         &     (0.014)         &     (0.006)         &     (0.008)         &     (0.009)         &     (0.011)         \\
NoteSadness &      -0.002         &      -0.008         &      -0.012\sym{*}  &       0.016         &       0.059\sym{***}&      -0.004         \\
            &     (0.010)         &     (0.013)         &     (0.006)         &     (0.009)         &     (0.012)         &     (0.011)         \\
NoteSurprise&      -0.033\sym{**} &      -0.020         &      -0.001         &       0.004         &      -0.008         &       0.026\sym{**} \\
            &     (0.012)         &     (0.011)         &     (0.006)         &     (0.007)         &     (0.010)         &     (0.009)         \\
Displayed $\times$ NoteAnger&       0.006         &       0.031\sym{*}  &       0.004         &      -0.026\sym{**} &      -0.009         &       0.001         \\
            &     (0.009)         &     (0.015)         &     (0.007)         &     (0.010)         &     (0.011)         &     (0.008)         \\
Displayed $\times$ NoteDisgust&       0.001         &      -0.011         &       0.005         &      -0.001         &       0.003         &      -0.003         \\
            &     (0.007)         &     (0.009)         &     (0.003)         &     (0.007)         &     (0.007)         &     (0.007)         \\
Displayed $\times$ NoteFear &       0.007         &       0.006         &      -0.003         &      -0.020         &      -0.002         &       0.012         \\
            &     (0.012)         &     (0.012)         &     (0.015)         &     (0.011)         &     (0.009)         &     (0.011)         \\
Displayed $\times$ NoteJoy  &      -0.000         &       0.001         &       0.007         &      -0.006         &       0.004         &      -0.007         \\
            &     (0.011)         &     (0.011)         &     (0.006)         &     (0.012)         &     (0.008)         &     (0.010)         \\
Displayed $\times$ NoteSadness&       0.007         &       0.003         &       0.005         &      -0.026\sym{**} &       0.003         &       0.010         \\
            &     (0.009)         &     (0.012)         &     (0.005)         &     (0.008)         &     (0.012)         &     (0.010)         \\
Displayed $\times$ NoteSurprise&       0.010         &       0.010         &       0.009         &       0.005         &      -0.003         &      -0.007         \\
            &     (0.012)         &     (0.012)         &     (0.007)         &     (0.010)         &     (0.011)         &     (0.010)         \\
SourceAnger &       0.066\sym{***}&       0.008         &       0.015\sym{*}  &      -0.009         &       0.013         &      -0.027\sym{**} \\
            &     (0.013)         &     (0.011)         &     (0.007)         &     (0.008)         &     (0.011)         &     (0.010)         \\
SourceDisgust&      -0.004         &       0.067\sym{***}&      -0.003         &      -0.013\sym{*}  &      -0.008         &      -0.000         \\
            &     (0.009)         &     (0.014)         &     (0.005)         &     (0.006)         &     (0.006)         &     (0.008)         \\
SourceFear  &       0.008         &       0.012         &       0.054\sym{***}&      -0.019\sym{*}  &       0.006         &       0.010         \\
            &     (0.012)         &     (0.011)         &     (0.008)         &     (0.008)         &     (0.008)         &     (0.011)         \\
SourceJoy   &       0.003         &      -0.002         &       0.004         &       0.032         &       0.026\sym{***}&       0.003         \\
            &     (0.013)         &     (0.008)         &     (0.007)         &     (0.018)         &     (0.007)         &     (0.010)         \\
SourceSadness&       0.010         &      -0.001         &       0.005         &       0.005         &       0.043\sym{***}&      -0.007         \\
            &     (0.010)         &     (0.010)         &     (0.006)         &     (0.008)         &     (0.009)         &     (0.009)         \\
SourceSurprise&      -0.052\sym{***}&      -0.009         &      -0.008         &       0.002         &       0.017         &       0.068\sym{***}\\
            &     (0.011)         &     (0.012)         &     (0.007)         &     (0.010)         &     (0.011)         &     (0.013)         \\
PostAge     &      -0.001         &       0.005         &       0.006         &       0.006         &       0.003         &       0.009         \\
            &     (0.010)         &     (0.008)         &     (0.004)         &     (0.006)         &     (0.004)         &     (0.006)         \\
Intercept      &      -0.019         &      -0.005         &       0.001         &       0.004         &       0.002         &       0.011         \\
            &     (0.011)         &     (0.010)         &     (0.007)         &     (0.009)         &     (0.011)         &     (0.009)         \\
\midrule
\(R^{2}\)   &       0.018         &       0.006         &       0.007         &       0.003         &       0.007         &       0.010         \\
\midrule
\#Replies (\(N\))       &        \num{2155873}         &       \num{2155873}&        \num{2155873}         &       \num{2155873}&        \num{2155873}         &       \num{2155873}\\
\#Source posts      &        \num{1339}         &       \num{1339}&        \num{1339}         &       \num{1339}&        \num{1339}         &       \num{1339}\\
\bottomrule
\end{tabularx}
\label{tab:emotions_notes}
\end{table}

\newpage
\subsection{Expansion With Helpfulness Scores of Community Notes}
\label{sec:sm_note_score}

Misleading posts that garner significant public concern are often subject to external factors such as heightened public criticism \cite{wojcik2022birdwatch,allen2022birds,chuai2024community.new}. These factors may contribute to the display of community notes with high helpfulness scores and increased negativity in replies, thereby introducing potential confounds. In addition to displaying helpful community notes on the corresponding misleading posts directly, Community Notes official X account (@HelpfulNotes) automatically quote community notes and associated misleading posts to promote the helpful context based on certain criteria, for example, having a high helpfulness score. This could influence how users interact with the original misleading posts and complicate the effect of community notes display. To mitigate these concerns, we repeat our analysis by including helpfulness scores of community notes. We first calculate community notes' helpfulness scores based on the same algorithm used in production (the open-source algorithm is available at \url{https://github.com/twitter/communitynotes/tree/main}). Subsequently, we incorporate the variable $\var{NoteScore}$ indicating helpfulness scores of community notes and its interaction with $\var{Displayed}$ in the RDD models. The estimation results for sentiments and emotions in replies are reported in Table \ref{tab:sentiments_notescore} and Table \ref{tab:emotions_notescore}. We find that the coefficient estimates of $\var{Displayed} \times \var{NoteScore}$ are not statistically significant across sentiments and emotions in replies (each $p>0.05$). This suggests that quoting community notes by Community Notes official X account has no significant effect on the sentiments and emotions in replies received by misleading posts.

\begin{table}[H]
\centering
% \footnotesize
\caption{Estimation results for positive [Column (1)] and negative [Column (2)] sentiments in replies. The control variable $\var{NoteScore}$ and its interaction with $\var{Displayed}$ is included. Reported are coefficient estimates with post-clustered standard errors in parentheses. \sym{*} \(p<0.05\), \sym{**} \(p<0.01\), \sym{***} \(p<0.001\).}
\begin{tabularx}{\columnwidth}{@{\hspace{\tabcolsep}\extracolsep{\fill}}l*{2}{S}}
\toprule
&\multicolumn{1}{c}{(1)}&\multicolumn{1}{c}{(2)}\\
&\multicolumn{1}{c}{Positive}&\multicolumn{1}{c}{Negative}\\
\midrule
Displayed   &      -0.048\sym{**} &       0.101\sym{***}\\
            &     (0.015)         &     (0.018)         \\
HoursFromDisplay&      -0.008         &       0.007         \\
            &     (0.011)         &     (0.016)         \\
NoteScore   &       0.057\sym{***}&      -0.098\sym{***}\\
            &     (0.013)         &     (0.019)         \\
Displayed $\times$ NoteScore&      -0.007         &       0.010         \\
            &     (0.015)         &     (0.021)         \\
SourcePositive&       0.050\sym{*}  &      -0.004         \\
            &     (0.021)         &     (0.023)         \\
SourceNegative&      -0.044\sym{**} &       0.104\sym{***}\\
            &     (0.016)         &     (0.021)         \\
PostAge     &       0.007         &      -0.008         \\
            &     (0.007)         &     (0.011)         \\
Intercept      &       0.012         &      -0.026         \\
            &     (0.013)         &     (0.017)         \\
\midrule
\(R^{2}\)   &       0.010         &       0.023         \\
\midrule
\#Replies (\(N\))       &        \num{2106396}         &       \num{2106396}\\
\#Source posts      &        \num{1281}         &       \num{1281}\\
\bottomrule
\end{tabularx}
\label{tab:sentiments_notescore}
\end{table}

\newpage
\begin{table}[H]
\centering
% \footnotesize
\caption{Estimation results for anger [Column (1)], disgust [Column (2)], fear [Column (3)], joy [Column (4)], sadness [Column (5)], surprise [Column (6)], and moral outrage [Column (7)] in replies. The control variable $\var{NoteScore}$ and its interaction with $\var{Displayed}$ is included. Reported are coefficient estimates with post-clustered standard errors in parentheses. \sym{*} \(p<0.05\), \sym{**} \(p<0.01\), \sym{***} \(p<0.001\).}
\begin{tabularx}{\columnwidth}{@{\hspace{\tabcolsep}\extracolsep{\fill}}l*{7}{S}}
\toprule
&\multicolumn{1}{c}{(1)}&\multicolumn{1}{c}{(2)}&\multicolumn{1}{c}{(3)}&\multicolumn{1}{c}{(4)}&\multicolumn{1}{c}{(5)}&\multicolumn{1}{c}{(6)}&\multicolumn{1}{c}{(7)}\\
&\multicolumn{1}{c}{Anger}&\multicolumn{1}{c}{Disgust}&\multicolumn{1}{c}{Fear}&\multicolumn{1}{c}{Joy}&\multicolumn{1}{c}{Sadness}&\multicolumn{1}{c}{Surprise}&\multicolumn{1}{c}{Moral outrage}\\
\midrule
Displayed   &       0.070\sym{***}&       0.026\sym{**} &      -0.006         &      -0.014         &      -0.010         &      -0.039\sym{***}&       0.066\sym{***}\\
            &     (0.009)         &     (0.010)         &     (0.007)         &     (0.010)         &     (0.012)         &     (0.009)         &     (0.009)         \\
HoursFromDisplay&       0.001         &      -0.006         &      -0.001         &      -0.007         &      -0.001         &      -0.006         &       0.003         \\
            &     (0.014)         &     (0.010)         &     (0.005)         &     (0.009)         &     (0.008)         &     (0.009)         &     (0.011)         \\
NoteScore   &      -0.064\sym{***}&       0.000         &       0.014         &       0.024\sym{**} &      -0.019         &       0.020         &      -0.035\sym{***}\\
            &     (0.011)         &     (0.010)         &     (0.009)         &     (0.009)         &     (0.012)         &     (0.011)         &     (0.010)         \\
Displayed $\times$ NoteScore&       0.006         &      -0.000         &      -0.009         &      -0.004         &       0.012         &       0.003         &       0.013         \\
            &     (0.011)         &     (0.011)         &     (0.008)         &     (0.011)         &     (0.012)         &     (0.010)         &     (0.009)         \\
SourceAnger &       0.075\sym{***}&       0.006         &       0.018\sym{*}  &      -0.004         &       0.020         &      -0.036\sym{***}&       0.030\sym{**} \\
            &     (0.011)         &     (0.011)         &     (0.008)         &     (0.008)         &     (0.014)         &     (0.010)         &     (0.011)         \\
SourceDisgust&      -0.004         &       0.069\sym{***}&      -0.003         &      -0.009         &      -0.013\sym{*}  &      -0.001         &       0.024\sym{**} \\
            &     (0.009)         &     (0.015)         &     (0.004)         &     (0.005)         &     (0.007)         &     (0.007)         &     (0.008)         \\
SourceFear  &       0.016         &       0.008         &       0.066\sym{***}&      -0.021\sym{**} &       0.007         &       0.003         &       0.003         \\
            &     (0.014)         &     (0.011)         &     (0.010)         &     (0.008)         &     (0.009)         &     (0.012)         &     (0.012)         \\
SourceJoy   &      -0.004         &      -0.005         &       0.006         &       0.038\sym{*}  &       0.023\sym{**} &       0.005         &      -0.014         \\
            &     (0.015)         &     (0.009)         &     (0.007)         &     (0.019)         &     (0.008)         &     (0.012)         &     (0.009)         \\
SourceSadness&       0.007         &      -0.005         &       0.006         &       0.010         &       0.051\sym{***}&      -0.006         &      -0.004         \\
            &     (0.010)         &     (0.009)         &     (0.005)         &     (0.009)         &     (0.008)         &     (0.009)         &     (0.008)         \\
SourceSurprise&      -0.055\sym{***}&      -0.015         &      -0.001         &       0.003         &       0.017         &       0.067\sym{***}&      -0.038\sym{***}\\
            &     (0.011)         &     (0.012)         &     (0.007)         &     (0.010)         &     (0.013)         &     (0.014)         &     (0.010)         \\
PostAge     &      -0.005         &       0.004         &       0.005         &       0.008         &       0.005         &       0.010         &      -0.003         \\
            &     (0.008)         &     (0.008)         &     (0.004)         &     (0.006)         &     (0.005)         &     (0.006)         &     (0.007)         \\
Intercept      &      -0.018         &      -0.007         &       0.000         &       0.004         &       0.003         &       0.009         &      -0.016         \\
            &     (0.011)         &     (0.010)         &     (0.008)         &     (0.010)         &     (0.013)         &     (0.010)         &     (0.009)         \\
\midrule
\(R^{2}\)   &       0.018         &       0.006         &       0.005         &       0.003         &       0.003         &       0.009         &       0.007         \\
\midrule
\#Replies (\(N\))       &        \num{2106396}         &       \num{2106396}&        \num{2106396}         &       \num{2106396}&        \num{2106396}         &       \num{2106396}         &       \num{2106396}\\
\#Source posts      &        \num{1281}         &       \num{1281}&        \num{1281}         &       \num{1281}&        \num{1281}         &       \num{1281}         &       \num{1281}\\
\bottomrule
\end{tabularx}
\label{tab:emotions_notescore}
\end{table}

\newpage
\subsection{Robustness to the Development of Community Notes Program}
\label{sec:sm_mfro}

We incorporate a continuous variable, $\var{MFRO}$ (\ie, Months From the Roll-Out for source posts or replies), to investigate how the effect of community notes changes over time as the program develops. Table \ref{tab:sentiments_mfro_tweet} and Table \ref{tab:emotions_mfro_tweet} report estimates results with $\var{MFRO}$ based on post creation. Table \ref{tab:sentiments_mfro_reply} and Table \ref{tab:emotions_mfro_reply} report estimates results with $\var{MFRO}$ based on reply creation. All coefficient estimates of $\var{Displayed} \times \var{MFRO}$ are not statistically significant (each $p>0.05$). This suggests that the effect of community notes is stable relative to the development of Community Notes program.

\begin{table}[H]
\centering
% \footnotesize
\caption{Estimation results for positive [Column (1)] and negative [Column (2)] sentiments in replies. The control variable $\var{MFRO_{post}}$ and its interaction with $\var{Displayed}$ is included. Reported are coefficient estimates with post-clustered standard errors in parentheses. \sym{*} \(p<0.05\), \sym{**} \(p<0.01\), \sym{***} \(p<0.001\).}
\begin{tabularx}{\columnwidth}{@{\hspace{\tabcolsep}\extracolsep{\fill}}l*{2}{S}}
\toprule
&\multicolumn{1}{c}{(1)}&\multicolumn{1}{c}{(2)}\\
&\multicolumn{1}{c}{Positive}&\multicolumn{1}{c}{Negative}\\
\midrule
Displayed   &      -0.053\sym{***}&       0.110\sym{***}\\
            &     (0.015)         &     (0.019)         \\
HoursFromDisplay&      -0.003         &       0.001         \\
            &     (0.014)         &     (0.020)         \\
$\var{\text{MFRO}_\text{post}}$   &       0.006         &      -0.022         \\
            &     (0.017)         &     (0.019)         \\
Displayed $\times$ $\var{\text{MFRO}_\text{post}}$&      -0.017         &       0.044         \\
            &     (0.019)         &     (0.023)         \\
SourcePositive&       0.039\sym{*}  &       0.016         \\
            &     (0.019)         &     (0.021)         \\
SourceNegative&      -0.048\sym{**} &       0.112\sym{***}\\
            &     (0.017)         &     (0.023)         \\
PostAge     &       0.006         &      -0.006         \\
            &     (0.008)         &     (0.012)         \\
Intercept      &       0.014         &      -0.029         \\
            &     (0.014)         &     (0.018)         \\
\midrule
\(R^{2}\)   &       0.007         &       0.014         \\
\midrule
\#Replies (\(N\))       &        \num{2155873}         &       \num{2155873}\\
\#Source posts      &        \num{1339}         &       \num{1339}\\
\bottomrule
\end{tabularx}
\label{tab:sentiments_mfro_tweet}
\end{table}

\newpage
\begin{table}[H]
\centering
% \footnotesize
\caption{Estimation results for anger [Column (1)], disgust [Column (2)], fear [Column (3)], joy [Column (4)], sadness [Column (5)], surprise [Column (6)], and moral outrage [Column (7)] in replies. The control variable $\var{MFRO_{post}}$ and its interaction with $\var{Displayed}$ is included. Reported are coefficient estimates with post-clustered standard errors in parentheses. \sym{*} \(p<0.05\), \sym{**} \(p<0.01\), \sym{***} \(p<0.001\).}
\begin{tabularx}{\columnwidth}{@{\hspace{\tabcolsep}\extracolsep{\fill}}l*{7}{S}}
\toprule
&\multicolumn{1}{c}{(1)}&\multicolumn{1}{c}{(2)}&\multicolumn{1}{c}{(3)}&\multicolumn{1}{c}{(4)}&\multicolumn{1}{c}{(5)}&\multicolumn{1}{c}{(6)}&\multicolumn{1}{c}{(7)}\\
&\multicolumn{1}{c}{Anger}&\multicolumn{1}{c}{Disgust}&\multicolumn{1}{c}{Fear}&\multicolumn{1}{c}{Joy}&\multicolumn{1}{c}{Sadness}&\multicolumn{1}{c}{Surprise}&\multicolumn{1}{c}{Moral outrage}\\
\midrule
Displayed   &       0.076\sym{***}&       0.020\sym{*}  &      -0.006         &      -0.014         &      -0.007         &      -0.040\sym{***}&       0.065\sym{***}\\
            &     (0.011)         &     (0.010)         &     (0.008)         &     (0.009)         &     (0.011)         &     (0.009)         &     (0.009)         \\
HoursFromDisplay&      -0.004         &      -0.007         &      -0.001         &      -0.004         &      -0.002         &      -0.005         &      -0.001         \\
            &     (0.018)         &     (0.010)         &     (0.004)         &     (0.010)         &     (0.007)         &     (0.010)         &     (0.013)         \\
$\var{\text{MFRO}_\text{post}}$   &      -0.005         &       0.007         &       0.002         &      -0.002         &      -0.015         &      -0.001         &       0.006         \\
            &     (0.012)         &     (0.009)         &     (0.007)         &     (0.011)         &     (0.012)         &     (0.010)         &     (0.009)         \\
Displayed $\times$ $\var{\text{MFRO}_\text{post}}$&       0.014         &       0.010         &      -0.002         &      -0.005         &       0.006         &      -0.012         &       0.007         \\
            &     (0.011)         &     (0.009)         &     (0.007)         &     (0.012)         &     (0.009)         &     (0.008)         &     (0.010)         \\
SourceAnger &       0.078\sym{***}&       0.006         &       0.017\sym{*}  &      -0.006         &       0.021         &      -0.036\sym{***}&       0.032\sym{**} \\
            &     (0.012)         &     (0.011)         &     (0.008)         &     (0.008)         &     (0.014)         &     (0.010)         &     (0.011)         \\
SourceDisgust&       0.002         &       0.069\sym{***}&      -0.005         &      -0.011\sym{*}  &      -0.011         &      -0.003         &       0.027\sym{***}\\
            &     (0.008)         &     (0.015)         &     (0.004)         &     (0.005)         &     (0.006)         &     (0.007)         &     (0.008)         \\
SourceFear  &       0.013         &       0.007         &       0.066\sym{***}&      -0.020\sym{**} &       0.007         &       0.004         &       0.000         \\
            &     (0.015)         &     (0.011)         &     (0.011)         &     (0.007)         &     (0.009)         &     (0.012)         &     (0.012)         \\
SourceJoy   &       0.009         &      -0.002         &       0.004         &       0.033         &       0.026\sym{***}&       0.000         &      -0.005         \\
            &     (0.016)         &     (0.008)         &     (0.006)         &     (0.018)         &     (0.007)         &     (0.011)         &     (0.009)         \\
SourceSadness&       0.010         &      -0.005         &       0.005         &       0.009         &       0.053\sym{***}&      -0.007         &      -0.003         \\
            &     (0.010)         &     (0.009)         &     (0.005)         &     (0.009)         &     (0.009)         &     (0.009)         &     (0.008)         \\
SourceSurprise&      -0.055\sym{***}&      -0.013         &      -0.001         &       0.001         &       0.018         &       0.069\sym{***}&      -0.037\sym{***}\\
            &     (0.011)         &     (0.012)         &     (0.007)         &     (0.010)         &     (0.013)         &     (0.013)         &     (0.010)         \\
PostAge     &      -0.003         &       0.005         &       0.005         &       0.007         &       0.005         &       0.010         &      -0.001         \\
            &     (0.010)         &     (0.008)         &     (0.004)         &     (0.006)         &     (0.005)         &     (0.006)         &     (0.008)         \\
Intercept      &      -0.020         &      -0.005         &       0.002         &       0.004         &       0.001         &       0.011         &      -0.016         \\
            &     (0.012)         &     (0.010)         &     (0.008)         &     (0.009)         &     (0.012)         &     (0.010)         &     (0.009)         \\
\midrule
\(R^{2}\)   &       0.014         &       0.006         &       0.005         &       0.002         &       0.003         &       0.008         &       0.006         \\
\midrule
\#Replies (\(N\))       &        \num{2155873}         &       \num{2155873}&        \num{2155873}         &       \num{2155873}&        \num{2155873}         &       \num{2155873}         &       \num{2155873}\\
\#Source posts      &        \num{1339}         &       \num{1339}&        \num{1339}         &       \num{1339}&        \num{1339}         &       \num{1339}         &       \num{1339}\\
\bottomrule
\end{tabularx}
\label{tab:emotions_mfro_tweet}
\end{table}

\begin{table}[H]
\centering
% \footnotesize
\caption{Estimation results for positive [Column (1)] and negative [Column (2)] sentiments in replies. The control variable $\var{MFRO_{reply}}$ and its interaction with $\var{Displayed}$ is included. Reported are coefficient estimates with post-clustered standard errors in parentheses. \sym{*} \(p<0.05\), \sym{**} \(p<0.01\), \sym{***} \(p<0.001\).}
\begin{tabularx}{\columnwidth}{@{\hspace{\tabcolsep}\extracolsep{\fill}}l*{2}{S}}
\toprule
&\multicolumn{1}{c}{(1)}&\multicolumn{1}{c}{(2)}\\
&\multicolumn{1}{c}{Positive}&\multicolumn{1}{c}{Negative}\\
\midrule
Displayed   &      -0.053\sym{***}&       0.110\sym{***}\\
            &     (0.015)         &     (0.019)         \\
HoursFromDisplay&      -0.003         &       0.001         \\
            &     (0.014)         &     (0.020)         \\
$\var{\text{MFRO}_\text{reply}}$   &       0.006         &      -0.022         \\
            &     (0.017)         &     (0.020)         \\
Displayed $\times$ $\var{\text{MFRO}_\text{reply}}$&      -0.017         &       0.044         \\
            &     (0.019)         &     (0.023)         \\
SourcePositive&       0.039\sym{*}  &       0.016         \\
            &     (0.019)         &     (0.021)         \\
SourceNegative&      -0.048\sym{**} &       0.112\sym{***}\\
            &     (0.017)         &     (0.023)         \\
PostAge     &       0.006         &      -0.007         \\
            &     (0.009)         &     (0.013)         \\
Intercept      &       0.014         &      -0.030         \\
            &     (0.014)         &     (0.018)         \\
\midrule
\(R^{2}\)   &       0.007         &       0.014         \\
\midrule
\#Replies (\(N\))       &        \num{2155873}         &       \num{2155873}\\
\#Source posts      &        \num{1339}         &       \num{1339}\\
\bottomrule
\end{tabularx}
\label{tab:sentiments_mfro_reply}
\end{table}

\newpage
\begin{table}[H]
\centering
% \footnotesize
\caption{Estimation results for anger [Column (1)], disgust [Column (2)], fear [Column (3)], joy [Column (4)], sadness [Column (5)], surprise [Column (6)], and moral outrage [Column (7)] in replies. The control variable $\var{MFRO_{reply}}$ and its interaction with $\var{Displayed}$ is included. Reported are coefficient estimates with post-clustered standard errors in parentheses. \sym{*} \(p<0.05\), \sym{**} \(p<0.01\), \sym{***} \(p<0.001\).}
\begin{tabularx}{\columnwidth}{@{\hspace{\tabcolsep}\extracolsep{\fill}}l*{7}{S}}
\toprule
&\multicolumn{1}{c}{(1)}&\multicolumn{1}{c}{(2)}&\multicolumn{1}{c}{(3)}&\multicolumn{1}{c}{(4)}&\multicolumn{1}{c}{(5)}&\multicolumn{1}{c}{(6)}&\multicolumn{1}{c}{(7)}\\
&\multicolumn{1}{c}{Anger}&\multicolumn{1}{c}{Disgust}&\multicolumn{1}{c}{Fear}&\multicolumn{1}{c}{Joy}&\multicolumn{1}{c}{Sadness}&\multicolumn{1}{c}{Surprise}&\multicolumn{1}{c}{Moral outrage}\\
\midrule
Displayed   &       0.076\sym{***}&       0.021\sym{*}  &      -0.006         &      -0.014         &      -0.007         &      -0.040\sym{***}&       0.065\sym{***}\\
            &     (0.011)         &     (0.010)         &     (0.008)         &     (0.009)         &     (0.011)         &     (0.009)         &     (0.009)         \\
HoursFromDisplay&      -0.004         &      -0.007         &      -0.001         &      -0.004         &      -0.002         &      -0.005         &      -0.001         \\
            &     (0.018)         &     (0.010)         &     (0.004)         &     (0.010)         &     (0.007)         &     (0.010)         &     (0.013)         \\
$\var{\text{MFRO}_\text{reply}}$   &      -0.005         &       0.007         &       0.002         &      -0.002         &      -0.015         &      -0.001         &       0.006         \\
            &     (0.012)         &     (0.009)         &     (0.007)         &     (0.011)         &     (0.012)         &     (0.010)         &     (0.009)         \\
Displayed $\times$ $\var{\text{MFRO}_\text{reply}}$&       0.015         &       0.009         &      -0.003         &      -0.005         &       0.006         &      -0.012         &       0.006         \\
            &     (0.012)         &     (0.009)         &     (0.007)         &     (0.012)         &     (0.009)         &     (0.008)         &     (0.010)         \\
SourceAnger &       0.078\sym{***}&       0.006         &       0.017\sym{*}  &      -0.006         &       0.021         &      -0.036\sym{***}&       0.032\sym{**} \\
            &     (0.012)         &     (0.011)         &     (0.008)         &     (0.008)         &     (0.014)         &     (0.010)         &     (0.011)         \\
SourceDisgust&       0.002         &       0.069\sym{***}&      -0.005         &      -0.011\sym{*}  &      -0.011         &      -0.003         &       0.027\sym{***}\\
            &     (0.008)         &     (0.015)         &     (0.004)         &     (0.005)         &     (0.006)         &     (0.007)         &     (0.008)         \\
SourceFear  &       0.013         &       0.007         &       0.066\sym{***}&      -0.020\sym{**} &       0.007         &       0.005         &       0.000         \\
            &     (0.015)         &     (0.011)         &     (0.011)         &     (0.007)         &     (0.009)         &     (0.012)         &     (0.012)         \\
SourceJoy   &       0.009         &      -0.002         &       0.004         &       0.033         &       0.026\sym{***}&       0.000         &      -0.005         \\
            &     (0.016)         &     (0.008)         &     (0.006)         &     (0.018)         &     (0.007)         &     (0.011)         &     (0.009)         \\
SourceSadness&       0.010         &      -0.005         &       0.005         &       0.009         &       0.053\sym{***}&      -0.007         &      -0.003         \\
            &     (0.010)         &     (0.009)         &     (0.005)         &     (0.009)         &     (0.009)         &     (0.009)         &     (0.008)         \\
SourceSurprise&      -0.055\sym{***}&      -0.013         &      -0.001         &       0.001         &       0.018         &       0.069\sym{***}&      -0.037\sym{***}\\
            &     (0.011)         &     (0.012)         &     (0.007)         &     (0.010)         &     (0.013)         &     (0.013)         &     (0.010)         \\
PostAge     &      -0.003         &       0.004         &       0.005         &       0.007         &       0.006         &       0.011         &      -0.002         \\
            &     (0.010)         &     (0.008)         &     (0.004)         &     (0.006)         &     (0.004)         &     (0.006)         &     (0.008)         \\
Intercept      &      -0.020         &      -0.006         &       0.002         &       0.004         &       0.001         &       0.011         &      -0.017         \\
            &     (0.012)         &     (0.010)         &     (0.008)         &     (0.009)         &     (0.012)         &     (0.010)         &     (0.009)         \\
\midrule
\(R^{2}\)   &       0.014         &       0.006         &       0.005         &       0.002         &       0.003         &       0.008         &       0.006         \\
\midrule
\#Replies (\(N\))       &        \num{2155873}         &       \num{2155873}&        \num{2155873}         &       \num{2155873}&        \num{2155873}         &       \num{2155873}         &       \num{2155873}\\
\#Source posts      &        \num{1339}         &       \num{1339}&        \num{1339}         &       \num{1339}&        \num{1339}         &       \num{1339}         &       \num{1339}\\
\bottomrule
\end{tabularx}
\label{tab:emotions_mfro_reply}
\end{table}

\end{document}